\RequirePackage{fix-cm}
\documentclass[pdflatex]{sn-jnl} 
\usepackage{graphicx}
\usepackage{natbib}
\usepackage{tabularx}
\usepackage{rotating}
\usepackage{textcomp}
\usepackage{amsmath}
\usepackage{amssymb}
\usepackage{pdfpages}
\usepackage{verbatim}
\usepackage{gensymb}
\begin{document}

\title{Simulation Models for Exploring Magnetic Reconnection}

\author*[1]{\fnm{Michael} \sur{Shay}}\email{shay@udel.edu}

\author[2]{\fnm{Subash} \sur{Adhikari}}

\author[3]{\fnm{Naoki} \sur{Beesho}}

\author[4]{\fnm{Joachim} \sur{Birn}}

\author[5]{\fnm{J\"org} \sur{B\"uchner}}

\author[2]{\fnm{Paul} \sur{Cassak}}

\author[6]{\fnm{Li-Jen} \sur{Chen}}

\author[7]{\fnm{Yuxi} \sur{Chen}}

\author[8]{\fnm{Giulia} \sur{Cozzani}}

\author[9,10]{\fnm{Jim} \sur{Drake}}

\author[11]{\fnm{Fan} \sur{Guo}}

\author[12]{\fnm{Michael} \sur{Hesse}}

\author[13]{\fnm{Neeraj} \sur{Jain}}

\author[8]{\fnm{Yann} \sur{Pfau-Kempf}}

\author[14]{\fnm{Yu} \sur{Lin}}

\author[15]{\fnm{Yi-Hsin} \sur{Liu}}

\author[16]{\fnm{Mitsuo} \sur{Oka}}

\author[4,17]{\fnm{Yuri A.} \sur{Omelchenko}}

\author[8]{\fnm{Minna} \sur{Palmroth}}

\author[18]{\fnm{Oreste} \sur{Pezzi}}

\author[19]{\fnm{Patricia H.} \sur{Reiff}}

\author[10]{\fnm{Marc} \sur{Swisdak}}

\author[19]{\fnm{Frank} \sur{Toffoletto}}

\author[20]{\fnm{Gabor} \sur{Toth}}

\author[19]{\fnm{Richard A.} \sur{Wolf}}



\affil*[1]{\orgdiv{Bartol Research Institute, Department of Physics and Astronomy}, \orgname{University of Delaware}, \orgaddress{\city{Newark}, \postcode{19716}, \state{DE}, \country{USA}}}

\affil[2]{\orgdiv{Department of Physics and Astronomy}, \orgname{West Virginia University}, \orgaddress{\city{Morgantown}, \postcode{26506}, \state{WV}, \country{USA}}}

\affil[3]{\orgdiv{Department of Astronomy}, \orgname{University of Maryland}, \orgaddress{\city{College Park}, \postcode{20742}, \state{MD}, \country{USA}}}

\affil[4]{\orgdiv{Center for Space Plasma Physics}, \orgname{Space Science Institute}, \orgaddress{\city{Boulder}, \postcode{80301}, \state{CO}, \country{USA}}}

\affil[5]{\orgname{Max Planck Institute for Solar System Research}, \orgaddress{\city{G\"ottingen}, \postcode{27077}, \country{Germany}}}

\affil[6]{\orgname{NASA Goddard Space Flight Center}, \orgaddress{\city{Greenbelt}, \postcode{20771}, \state{MD}, \country{USA}}}

\affil[7]{\orgdiv{Center for Space Physics and Department of Astronomy}, \orgname{Boston University}, \orgaddress{\city{Boston}, \postcode{02215}, \state{MA}, \country{USA}}}

\affil[8]{\orgdiv{Department of Physics}, \orgname{University of Helsinki}, \orgaddress{\city{P.O. Box 68}, \postcode{00014}, \state{Uusimaa}, \country{Finland}}}

\affil[9]{\orgdiv{Department of Physics}, \orgname{University of Maryland}, \orgaddress{\city{College Park}, \postcode{20740}, \state{MD}, \country{USA}}}

\affil[10]{\orgdiv{Institute for Research in Electronics and Applied Physics}, \orgname{University of Maryland}, \orgaddress{\city{College Park}, \postcode{20740}, \state{MD}, \country{USA}}}

\affil[11]{\orgname{Los Alamos National Laboratory}, \orgaddress{\city{Los Alamos}, \postcode{87545}, \state{NM}, \country{USA}}}

\affil[12]{\orgname{NASA Ames Research Center}, \orgaddress{\city{Moffett Field}, \postcode{94035}, \state{CA}, \country{USA}}}

\affil[13]{\orgdiv{Center for Astronomy and Astrophysics}, \orgname{Technical University Berlin}, \orgaddress{\city{Berlin}, \postcode{10623}, \country{Germany}}}

\affil[14]{\orgdiv{Physics Department}, \orgname{Auburn University}, \orgaddress{\city{Auburn}, \postcode{36832}, \state{AL}, \country{USA}}}

\affil[15]{\orgdiv{Department of Physics and Astronomy}, \orgname{Dartmouth College}, \orgaddress{\city{Hanover}, \postcode{03750}, \state{NH}, \country{USA}}}

\affil[16]{\orgdiv{Space Sciences Laboratory}, \orgname{University of California}, \orgaddress{\city{Berkeley}, \postcode{94720}, \state{CA}, \country{USA}}}

\affil[17]{\orgname{Trinum Research Inc.}, \orgaddress{\city{San Diego}, \postcode{92126}, \state{CA}, \country{USA}}}

\affil[18]{\orgdiv{Istituto per la Scienza e Tecnologia dei Plasmi (ISTP)}, \orgname{Consiglio Nazionale delle Ricerche}, \orgaddress{\city{Bari}, \postcode{I-70126}, \country{Italy}}}

\affil[19]{\orgdiv{Department of Physics and Astronomy}, \orgname{Rice University}, \orgaddress{\city{Houston}, \postcode{77005}, \state{TX}, \country{USA}}}

\affil[20]{\orgname{University of Michigan}, \orgaddress{\city{Ann Arbor}, \postcode{48109}, \state{MI}, \country{USA}}}




\date{Received: date / Accepted: date}

\maketitle

\begin{abstract} 

Simulations have played a critical role in the advancement of our knowledge of magnetic reconnection. However, due to the inherently multiscale nature of reconnection, it is impossible to simulate all physics at all scales. For this reason, a wide range of simulation methods have been crafted to study particular aspects and consequences of magnetic reconnection. This chapter reviews many of these methods, laying out critical assumptions, numerical techniques, and giving examples of scientific results.  Plasma models described include magnetohydrodynamics (MHD), Hall MHD, Hybrid, kinetic particle-in-cell (PIC), kinetic Vlasov, Fluid models with embedded PIC, Fluid models with direct feedback from energetic populations, and the Rice Convection Model (RCM).


\end{abstract}

\section{Introduction}
\label{sec:introduction}

Numerical computation has always played an important role in science. The term ``computer'' was used during the Renaissance to describe a person who performed mathematical calculations, and such computers were used extensively to calculate the positions of the planets. However, with the advent of digital computers last century, the role of such computation has exploded and revolutionized science in general. The study of magnetic reconnection has seen such a revolution in the last several decades as both numerical power has increased and numerical techniques have become more sophisticated. 

Magnetic reconnection is considered a multiscale process because it allows physics that emerges at very small length and time scales to have global consequences in the system. A straightforward example of this large separation of scales is  magnetic reconnection on the sun side of Earth's magnetosphere. In this region, magnetic field lines are finally broken on a length scale of the order of $5\,\mathrm{km}$ which is the electron inertial length $d_e \equiv c/\omega_{pe}$ . However, the dynamical effects of this breaking of field lines include driving global convection of the magnetosphere, a system spanning 100s of Earth radii ($R_e$) which is hundreds of thousands of $d_e.$  A grid scale of $1\,c/w_{pe}$ over 100 Earth Radii requires about 100,000 spatial grid points in only 1 dimension. Clearly, accurately resolving the physics breaking the frozen-in constraint while simulating global scales is impossible. 

The impossibility of globally simulating the whole 3D system and resolving all scales has led to the generation of a wide range of simulation models, each of which has its own strengths of weaknesses. Through many decades of research, scientists have carefully crafted these models for the particular application or applications they are studying.  Typically, the more realistic physics that is included in the simulation, the more computationally expensive it is. Studies of the basic physics of magnetic reconnection~\citep{biskamp1996magnetic} have very often used kinetic PIC simulations which include all relevant physics, but require a simplified geometry and boundary conditions. Global magnetospheric simulations include the complex boundaries associated with the solar wind and the ionosphere, but until recently were required to be fluid models due to the cost of including kinetic effects. 

In this paper we will provide an overview of the primary simulation models that are currently being used to study magnetic reconnection. Please note, however, that the topic of plasma simulation is extremely complex and detailed and cannot be fully covered in a single book, much less a single chapter. If the reader wishes to dive deeper into a particular model, there are many references available, many of which are cited in the individual sections of this paper. There are also excellent books devoted to the subject (e.g.,~\cite{buchner2003space}). 

In the field of magnetic reconnection research, more than one system of units is used. As of this writing, one can generally say that scientists specializing in theory/simulation primarily use cgs units and scientist specializing in observational analysis use SI units. We have chosen as much as possible to use cgs units in this paper, although the chapter on the Rice Convection Model has been left in SI units. For an excellent description on how to convert units between cgs and SI, please see the \textit{NRL Plasma Formulary}~\citep{huba1998nrl}. 

For the organization of the paper, we choose to move generally from fluid models to kinetic models; we start with magnetohydrodynamics (MHD) and gradually increase in physical complexity until reaching fully kinetic simulations. We end with the Rice Convection Model (RCM), a widely used model for the inner magnetosphere, which acts as an inner boundary for magnetic field lines which are reconnecting in the magnetosphere. Section 2 describes MHD. Section 3 describes Hall MHD. Section 4 describes Hybrid Simulations. Section 5 describes kinetic particle-in-cell simulations. Section 6 describes embedding PIC codes into fluid models like MHD. Section 7 describes Kglobal, an MHD model which self consistently evolves energetic particles. Section 8 describes kinetic Vlasov models. And Section 9 describes the Rice Convection Model. 

\section{MHD} \label{sec:MHD}

\subsection{Equations of MHD}\label{sec:MHD_equations}

Magnetohydrodynamics (MHD) is the simplest fluid model used to study large scale plasma dynamics. MHD is based on the assumption that the characteristic length and time scales of the system under study are much larger than the length and time scales of the plasma species, usually Debye length ($\lambda_D$) or gyroradius and gyroperiod. Therefore, MHD represents the slow evolution of plasmas, often electrons and ions as a single fluid. The macroscopic behavior of the fluid in presence of a magnetic field is described by MHD using hydrodynamics and Maxwell’s equations.

Let us consider a fluid (in this case a plasma with ions and electrons), moving with a flow velocity $\mathbf{u}$, characterized by a mass density $\rho = m_i n_i$ where $m_i$ is the mass of protons ($m_i\gg m_e$) and $n_i$ is the number density of protons (with quasi-neutrality $n_i=n_e=n$), thermal pressure $p$, and a magnetic field $\mathbf{B}$. The evolution of these fields in space and time are governed by the MHD equations given by

\begin{equation}\label{eqn:continuity}
    \frac{\partial \rho}{\partial t} = -\nabla \cdot (\rho \mathbf{u}),
\end{equation}
\begin{equation}\label{eqn:momentum}
    \rho \frac{\partial \mathbf{u}}{\partial t} + \rho (\mathbf{u}\cdot \nabla) \mathbf{u} = -\nabla p + \frac{1}{4\pi}(\nabla \times \mathbf{B}) \times \mathbf{B} + \nu \nabla^2 \mathbf{u},
\end{equation}
\begin{equation}\label{eqn:adiabatic}
    \frac{d}{dt}\left(\frac{p}{\rho^\gamma}\right) = 0,
\end{equation}
\begin{equation}\label{eqn:faraday}
    \frac{\partial \mathbf{B}}{\partial t} = \nabla \times (\mathbf{u}\times \mathbf{B}) + \frac{\eta c^2}{4\pi} \nabla^2 \mathbf{B},
\end{equation}
where $\gamma$ is the adiabatic index (usually $5/3$), $\nu$ is the dynamic viscosity, $\eta$ is the resistivity and $\eta c^2/4\pi$ collectively is known as the magnetic diffusivity. Here Eq.~\ref{eqn:continuity} is the continuity equation representing conservation of mass density, Eq.~ (\ref{eqn:momentum}) is the momentum conservation equation, Eq.~ (\ref{eqn:adiabatic})
is the simple adiabatic gas equation representing the conservation of energy and Eq.~\ref{eqn:faraday} is the induction equation, where the first term on the right is the advection term and the second one represents diffusion. Note, as a result of MHD approximations, the displacement current term is ommitted in the induction equation. In an ideal situation, there are no dissipative processes and therefore $\nu$ and $\eta=0$ gives ideal MHD equations. Studies have shown that ideal MHD description is a very good approximation to study dynamical properties of strongly magnetized plasmas.

\subsection{Regional MHD Simulations}
Several large-scale MHD approaches do not model the entire magnetosphere but only sections of it, such as certain magnetopause regions (dealt with elsewhere in this volume) or the magnetotail. Here we focus particularly on the magnetotail. The basic numerical approach used in regional MHD simulations is essentially similar to that used in (some) global simulations. It is typically based on explicit finite difference methods to solve the MHD equations. Minor differences might exist in adding resistive terms, which are usually necessary in local MHD to initiate reconnection.

The main difference, however, consists of the setup or initialization. Whereas global simulations typically involve of period of interaction with the solar wind to create a realistic magnetotail, local MHD simulations generally start from some equilibrium or near-equilibrium that models the stretched magnetotail (e.g.,~\citet{schindler1972self}). This approach provides more flexibility in treating different scenarios, for instance, varying the tail flaring between $y$ and $z$~\citep{birn2000large} or including a local $B_z$ hump~\citep{merkin2016stability, birn2018mhd}. This flexibility has also proven useful in PIC simulations that go beyond the commonly used initial 1D Harris sheet, most notably in addressing the holy grail of reconnection onset (e.g.,~\citet{liu2014onset}; see also~\cite{liu2024ohm}).

On the other side, interactions with the ionosphere or the solar wind are incorporated only in some ad hoc fashion, if at all. Regional magnetotail MHD simulations therefore have been most successful in treating dynamic tail phenomena on relatively short time scales that are typically substorm related. The successes include
\begin{enumerate}
    \item The demonstration that x-line formation and plasmoid ejection can be part of a 2D or 3D tearing-type instability of the tail (e.g.,~\citet{birn1981three}).
    \item The demonstration that the build-up of the substorm current wedge, involving dipolarization and Region-1-type field-aligned currents~\citep{mcpherron1973satellite}, can be due to the braking and azimuthal diversion of earthward flow from a near-tail reconnection site~\citep{birn1991substorm, scholer1991magnetotail}. This basic picture has been modified more recently, most notably by the addition of a Region-2 current system connecting to the ionosphere at lower latitude~\citep{birn2014substorm, kepko2015substorm}, in agreement with observations~\citep{sergeev2014testing}. While the buildup of the SCW in the simulations is based on the shear and vorticity of the earthward flow, the persistence of the currents relies on the changes of the magnetic flux and pressure patterns brought about by the severance of a plasmoid and the resulting los of entropy and redistribution of the pressure.
    It is noteworthy that these features can be, and have been, found also in global simulations. In the regional simulations, however, they arise as consequences of an instability without involvement of external driving or feedback from the ionosphere. This would be harder to extract from the global simulations.
    \item Regional MHD simulations have also been used to address the evolution prior to the onset of reconnection in the tail, demonstrating, specifically, the formation of a thin concentrated current sheet embedded in the near-tail plasma sheet. These approaches have included interaction with the solar wind in two complementary, ad-hoc ways. In one approach magnetic flux is added to the tail lobes~\citep{birn2002thin}, the other is based on low-latitude magnetic flux reduction from convection around the Earth toward the dayside~\citep{hsieh2014influence,hsieh2015thin}. Both mechanisms are expected from solar wind interaction. For more details, see the review by~\citet{sitnov2019explosive}.
    \item Recently, regional MHD simulations have also demonstrated that the magnetotail may become unstable even under ideal 2D MHD constraints, when it includes a region of inverse (i.e. tailward) gradient of the normal magnetic field $B_z$ (denoted ‘$B_z$ hump’ instability; ~\citep{merkin2016stability, birn2018mhd}).
\end{enumerate}

\subsection{Global MHD}

Global MHD models representing the (outer) magnetosphere of Earth typically extend around $100-200\,R_E$ in the flank and tail directions, and around $30\,R_E$ (beyond the bow shock) towards the Sun, where the solar wind is coming from. When, occasionally, the solar wind becomes sub-Alfvenic, the bow shock disappears and Alfv\'en wings form. 
In this case the upstream boundary has to be moved much further to minimize the boundary effects. It is computationally very demanding to obtain an accurate solution in such a large domain while resolving various structures such as the current sheets at the dayside magnetopause and in the tail. 
There are various approaches to overcome this difficulty, including the block-adaptive grid of the BATSRUS code \citep{Powell:1999, Toth:2012swmf}, the stretched Cartesian grid of OpenGGCM \citep{Raeder:1997} or the stretched spherical grid of LFM \citep{lyon2004lyon}. 

Another interesting aspect of the upstream condition is the handling of the divergence of the magnetic field. Typically, we have observations at a single point near L1, and assume that the solar wind and IMF have no variation in the transverse direction. If $B_x$ varies, and it certainly does, these assumptions lead to a finite $\nabla \cdot \mathbf B = \partial B_x/\partial x$ propagating into the domain. There are various approaches to handle this situation. One is to ignore the problem and propagate the finite $\nabla\cdot\mathbf B$ with the flow using some variation of the 8-wave scheme \citep{Powell:1994}. Another common approach is to set $B_x$ to a constant value, for example 0. Finally, one can relax the condition that the transverse gradients in the $y$ and $z$ directions are zero, and guess those gradients from the temporal evolution of $B_x$. Unfortunately this is an underspecified problem, so there is no unique solution. A typical approach is to smooth $B_x$ in time, and apply some minimum variance constraint. While theoretically nice, in practice the minimum variance approach does not work great. The likely reason is that the magnetic field is turbulent, so local changes in $B_x$ are not representative of the large scale tilt of the propagation plain.

The inner boundary conditions are usually applied 
at a sphere of radius 1.5\,$R_E$ to 3\,$R_E$ surrounding the Earth. One can use a semi-empirical electrodynamic solver~\citep{Ridley:2004}, or a fully empirical model \citep{Weimer:1996, Weimer:2001} to calculate the $\mathbf E\times\mathbf B$ drift velocity at the inner boundary. Another important use of an electrodynamic solver is that it can also provide the  $\mathbf E\times\mathbf B$ drift to an inner magnetosphere model \citep{Wolf:1982, Toffoletto:2003, Buzulukova:2010, Liemohn:2001b, Jordanova:1994, Zaharia:2006}, which can calculate realistic ring current and associated pressure (and density) in the closed field line region during geomagnetic storms \citep{Liemohn:2018}. The global MHD model can then relax its pressure (and density) towards the values supplied by the inner magnetosphere model. In return, the global MHD model can supply the plasma boundary conditions for the inner magnetosphere model at the edge of the closed field region as well as the magnetic field configuration \citep{DeZeeuw:2004, Meng:2013}.

Global models can properly represent the overall dynamics of the interaction of the solar wind with the magnetosphere, including the formation of the bow shock and the magnetopause, as well as the main current sheet in the magnetotail. Magnetic reconnection will happen on the dayside magnetopause and in the magnetotail in agreement with the theory of the Dungey cycle \citep{Dungey:1961}. The magnetic reconnection in the MHD simulation is not represented by the actual kinetic physics but it is approximated by numerical diffusion or artificial resistivity. Despite these caveats, global MHD models generate reconnection sites where the magnetic field changes sign and the reconnection rate is approximately correct (see Appendix A of \cite{Wang:2022a}).

When the grid is relatively coarse, the numerical diffusion will easily adjust to reconnect the incoming magnetic flux carried by the solar wind. For a constant solar wind and IMF driving the simulation will settle to a steady state solution.
Using fine grids in combination with low dissipation numerical methods can lead to a more dynamic reconnection process in the model. On the dayside, simulations can produce Flux Transfer Events \citep{Raeder:2006}
and in the tail flux ropes can be produced even by ideal MHD simulations. If the flux ropes in the tail are triggered by sign changes of the IMF $B_z$, the MHD simulations can match observations very well. A more challenging problem is reproducing substorms and sawtooth events. 
MHD models cannot do this well \citep{Haiducek:2020}, and one needs to add either ionospheric outflow to regulate the reconnection rate
\citep{Brambles:2013, Zhang:2020}, or kinetic reconnection physics \citep{Wang:2022b}
to produce the typical spatial and temporal scales of sawtooth oscillations.

\subsection{Test Particles in MHD Simulations}
Test particle approaches consist of tracing charged particle orbits in electromagnetic fields that are either prescribed in some plausible fashion or obtained from a simulation that typically does not contain individual particle information, most commonly based on MHD. The approach bridges the gap between large-scale MHD and small-scale particle simulations. In contrast to the latter, it can treat together realistic 3D space and large evolution time scales and realistic electron mass. However, it is not self-consistent and relies on whether the MHD model or the postulated $\mathbf{E}, \mathbf{B}$ fields capture the main physics. But that may also be considered an advantage as it permits studying the effects of large-scale fields in isolation.

In the magnetospheric context ions are usually treated by integration of the full orbit
\begin{equation}\label{eqn:lorentz_eqn}
    \frac{D\mathbf{u}}{D t} = \frac{e}{m}\Bigg( \mathbf{E} + \frac{1}{\gamma c} \textbf{u} \times \textbf{B}\Bigg).
\end{equation}
Here $\gamma$ is the relativistic factor, which may be more relevant for electrons, $\mathbf{u} = \gamma \mathbf{w}$, where $\mathbf{w}$ is the particle velocity, $c$ is the speed of light and $D/Dt = \partial/\partial t + \mathbf{w}\cdot \nabla$ denotes the derivative along the full orbit. Full integration of Equation~\ref{eqn:lorentz_eqn} over extended orbits is not practical for electrons, as it is more time consuming and might accumulate too large errors.  Also, the adiabatic drift approximation, based on conservation of the magnetic moment $\mu$, is valid over larger areas in the magnetosphere and can be adequate for identifying typical acceleration mechanisms (e.g.,~\cite{delcourt1994plasma, li1998simulation, zaharia2000particle, gabrielse2012effects}).

However, when the full history of electron orbits is considered; this may involve encounters of the reconnection site and low magnetic field, or high curvature regions, where the conservation of $\mu$ breaks down, and full orbit integration is required. Consequently, several codes have been developed that involve a transition between full orbits, integrated by eqn.~\ref{eqn:lorentz_eqn}, and drift orbits (e.g.,~\citet{birn2004electron, schriver2005modeling, ashour2011observations, sorathia2017energetic}). The drift is described by the guiding center drift velocity (e.g.,~\citet{birn2004electron})
\begin{equation}\label{eqn:guiding_center_velocity}
    \mathbf{v}_d = \mathbf{v}_E -\frac{\mu c}{\gamma e} \frac{\mathbf{B}\times \nabla B}{B^2}-\frac{\gamma m_e cv_\parallel}{e} \frac{\mathbf{B}}{B^2}\times \frac{d\mathbf{b}}{dt} -\frac{m_e c}{e}\frac{\mathbf{B}}{B^2}\times \frac{d(\gamma \mathbf{v}_E)}{dt}
\end{equation}
where $\mu$ is the (relativistic) magnetic moment, $\mathbf{v}_E=\mathbf{E}\times \mathbf{B}/B^2$ and $\mathbf{b}=\mathbf{B}/B$. In addition, the field-aligned velocity is advanced by
\begin{equation}\label{eqn:field_aligned_velocity}
    \frac{du_\parallel}{dt} = -\frac{e}{m_e}E_\parallel - \frac{\mu}{\gamma m_e}\frac{\partial B}{\partial s} - \big(\mathbf{u}_E+\mathbf{u}_{\nabla B}\big)\cdot \frac{d\mathbf{b}}{\partial t}
\end{equation}
where now $\mathbf{u} = \gamma \mathbf{v}$ and $\mathbf{v} = \mathbf{v}_E+\mathbf{v}_d+\mathbf{v}_\parallel$ describes the guiding center velocity, $\mathbf{v}_{\nabla B}$ is the grad B drift, given by the second term on the right side of eqn.~\ref{eqn:guiding_center_velocity}, and $d/dt$ is the derivative along the guiding center path. The transition between full orbit and drift orbit is typically determined from an adiabaticity criterion that is based on the ratio between the local field line curvature radius and the gyro radius based on the local magnetic field strength (e.g.,~\citet{buchner:1989}). On the switch from drift to full orbit a phase has to be generated, which is typically chosen randomly. Although this can alter individual orbits and make them not reversible, it was found to have no significant effect on general conclusions about sources and properties of distributions (e.g.,~\citet{birn2004electron}).

Two different techniques are used, tracing particle motion either forward or backward in time. Forward tracing requires larger numbers of particles, sometimes comparable to those in PIC simulations, to obtain sufficient numbers at the points of interest (e.g.,~\citet{scholer1987particle, sachsenweger1989test, peroomian2008storm, ukhorskiy2017ion, ukhorskiy2018ion}). However, since the particles are not interacting this approach is even more suitable for parallel processing than full particle simulations. In principle, this approach can also include wave scattering and collisions (albeit in an ad-hoc non self-consistent manner) to add to the simple collisionless advance.

Backward tracing is generally based on Liouville’s theorem of the conservation of phase space density $f$ to map $f$ from source locations to the final location of interest (e.g.,~\citet{curran1989particle, birn1994particle, birn2004electron}). It requires fewer orbits to identify properties at selected final locations, but relies on the validity of Liouville’s theorem, i.e. the absence of collisions. Backward tracing permits an easier identification of different sources contributing to the final population. Thus, sometimes a combination of both techniques is employed (e.g.,~\citet{ashour2011observations}).

Further complications are related to the use of MHD simulation results. Since the fields are given only on a finite grid, they have to be interpolated in space and time.
\begin{enumerate}
    \item The advance of the drift equations~\ref{eqn:guiding_center_velocity} and~\ref{eqn:field_aligned_velocity} requires a third order spatial interpolation in B for continuous transition between grid cells, which could lead to spurious maxima or minima. This can be avoided, however, by employing a monotonicity algorithm~\citep{hyman1983accurate,birn2004electron}.
    \item Simple interpolation of the electric field could also yield spurious parallel components. This can be avoided, however, by various techniques, for instance, by interpolating $E_\parallel$ and $E_\perp$ separately (e.g.,~\citet{birn2022electron}).
\end{enumerate}

\section{Hall MHD} \label{sec:Hall}

\subsection{Introduction}
\label{sec:HallIntro}

The ideal-MHD model, as discussed in Sec.~\ref{sec:MHD_equations}, is well-suited for magnetized plasmas when the dynamics is slow compared to the gyration time of charged particles around the magnetic fields and the length scales over which quantities vary is much larger than the gyroradius of the charged particles. However, going back many years in the study of neutral fluids, fluid models can lead to incorrect and paradoxical results at boundaries layers [{\it e.g.,} d'Alembert's paradox (Sec.~4.7 of \citet{Choudhuri98})].  In a magnetized plasma, these problems can occur where plasmas of two different origins abut against each other (such as at Earth's magnetopause), at shocks and discontinuities such as Earth's bow shock, and at localized regions where the magnetic field goes to zero, such as in the solar corona near sunspots.

Magnetic reconnection, in particular, occurs at a boundary layer at a region where at least two components of the magnetic field go to zero, so it is a key example of a physical process that cannot be faithfully modeled by ideal-MHD. Often in numerical simulations, ideal-MHD is used anyway, with numerical dissipation allowing reconnection to occur with the hope that it mimics the actual process.  Another approach employs resistivity to model the effect of collisions; this is a useful approach in systems for which collisions are dynamically relevant, but many settings where reconnection occurs -- especially in space and the solar corona -- are weakly collisional or effectively collisionless \citep{Priest00,Cassak12}.  There are examples where either approach can be good enough for the questions being asked. For other questions that rely on a faithful representation of the physics in the regions where ideal- and resistive-MHD break down, a new model is necessary.  In this section, we discuss a number of approaches within the fluid description that are used to go beyond ideal- and resistive-MHD simulations. In later sections, simulation techniques using the kinetic theory of gases are discussed.  There are previous review papers discussing Hall-MHD and numerical approaches \citep{Vasyliunas75,Huba95,Huba03c,Gomez06}.

\subsection{The Hall-MHD Model}
\label{sec:HallMHDModel}

The equations of Hall-MHD are similar to those of MHD with one key difference. In resistive-MHD, Ohm's law is given (in cgs units) by 
\begin{equation}
{\bf E} + \frac{{\bf u} \times {\bf B}}{c} = \eta {\bf J}, \label{eq:resistiveOhms}
\end{equation}
where ${\bf E}$ and ${\bf B}$ are the electric and magnetic fields, ${\bf u}$ is the single fluid bulk flow velocity, $\eta$ is the resistivity, and ${\bf J} = (c/4\pi) \nabla \times {\bf B}$ is the current density; in ideal-MHD, $\eta$ is set to zero. To go beyond this model, we revisit where Ohm's law comes from. 

The equation of motion of an electron fluid ({\it i.e.,} Newton's 2nd law) in a fully ionized plasma (in cgs units) is \citep{braginskii65a}
\begin{equation}
    m_{e} \frac{d{\bf u}_e}{dt} = -e \left({\bf E} + \frac{{\bf u}_e \times {\bf B}}{c}\right) - \frac{1}{n_e} \nabla \cdot {\bf p}_e + {\bf R}_e, \label{eq:newtone}
\end{equation}
where $m_e$ is the electron mass, ${\bf u}_e$ is the electron bulk flow velocity, $-e$ is the electron charge, $n_e$ is the electron density, ${\bf p}_e$ is the electron pressure which we write more generally as a tensor for now, and ${\bf R}_e$ represents the rate of change of momentum resulting from collisions between electrons and other electrons or other charged or neutral particles in the plasma.  There are rigorous ways to determine the role of collisions ({\it e.g.,}~\citet{braginskii65a}) that we do not employ here.  Instead, we use the often used simpler approach that assumes ${\bf R}_e = m_e \nu_{ei} ({\bf u}_i - {\bf u}_e)$, where $\nu_{ei}$ is the electron-ion collision frequency, and ${\bf u}_i$ is the ion bulk flow velocity, and for simplicity we assume the plasma has only electrons and ions (it is fully ionized).  

To recover the resistive-MHD Ohm's law from this equation, first the ``electron inertia term'' $m_e d{\bf u}_e / dt$ and the ``electron pressure gradient term'' $-(\nabla \cdot {\bf p}_e) / n_e$ are ignored for reasons we return to in Sec.~\ref{sec:extensions}. Second, the single fluid bulk flow velocity used in MHD is ${\bf u} = (m_i n_i {\bf u}_i + m_e n_e {\bf u}_e)/(m_i n_i + m_e n_e)$ and the current density ${\bf J} = n_i q_i {\bf u}_i - n_e e {\bf u}_e \simeq n_e e ({\bf u}_i - {\bf u}_e)$, where the latter form uses the assumption of quasi-neutrality $n_i q_i - n_e e \simeq 0$. Using these expressions to write ${\bf u}_e$ in terms of ${\bf u}$ and ${\bf J}$ gives ${\bf u}_e = {\bf u} - ({\bf J}/ne) [m_i n_i / (m_i n_i + m_e n_e)] \simeq {\bf u} - ({\bf J}/ne)$, where in the latter form, we use the approximation that $m_i \gg m_e$, since it is at least 1836 in an electron-ion plasma. Using these approximations in Eq.~(\ref{eq:newtone}) and dividing by $e$ gives
\begin{equation}
    {\bf E} + \frac{{\bf u} \times {\bf B}}{c} = \frac{{\bf J} \times {\bf B}}{n_e e c} + \eta {\bf J},
\end{equation}
where the resistivity $\eta$ is defined as $m_e \nu_{ei} / n_e e^2$. If one additionally ignores the ${\bf J} \times {\bf B} / n_e e c$ term, what remains is the resistive-MHD Ohm's law in Eq.~(\ref{eq:resistiveOhms}). If instead, one ignores the resistive term $\eta {\bf J}$, the result is
\begin{equation}
    {\bf E} + \frac{{\bf u} \times {\bf B}}{c} = \frac{{\bf J} \times {\bf B}}{n_e e c}. \label{eq:hallohmslaw}
\end{equation}
The term on the right is called the ``Hall electric field'' ${\bf E}_H$ (or simply the ``Hall term''), and Eq.~(\ref{eq:hallohmslaw}) is called the ``Hall-MHD Ohm's law''.  Simply coupling this equation to the rest of the ideal-MHD equations gives the Hall-MHD model:
\begin{eqnarray}
\frac{\partial \rho}{\partial t} + \nabla \cdot (\rho {\bf u}) & = & 0, \label{eq:continuity} \\
\rho \left[\frac{\partial {\bf u}}{\partial t} + ({\bf u} \cdot \nabla) {\bf u} \right] & = & -\nabla p + \frac{{\bf J} \times {\bf B}}{c}, \label{eq:momentum} \\
\frac{\partial p}{\partial t} + ({\bf u} \cdot \nabla) p & = & -\gamma p (\nabla \cdot {\bf u}) \\
\frac{\partial {\bf B}}{\partial t} & = & -c \nabla \times {\bf E}, \label{eq:faraday} \\
{\bf E} + \frac{{\bf u} \times {\bf B}}{c} & = & \frac{{\bf J} \times {\bf B}}{n_e e c} \label{eq:hallohmslaw2} \\
\nabla \times {\bf B} & = & \frac{4\pi {\bf J}}{c} \label{eq:ampere2}
\end{eqnarray}
with the auxiliary equation $\nabla \cdot {\bf B} = 0$, and where $\gamma$ is the single fluid ratio of specific heats, typically taken to be 5/3.  Note $n_e$ in Eq.~(\ref{eq:hallohmslaw2}) is related to the MHD mass density via $\rho = m_i n_i + m_e n_e$, so $n_e = \rho / (m_i e / q_i + m_e) \simeq Z \rho / m_i$, where we again assume $m_i \gg m_e$, we use quasi-neutrality to write $q_i n_i \simeq e n_e$, and define $Z = q_i / e$ as the degree of ionization. With these assumptions, Eqs.~(\ref{eq:continuity}) - (\ref{eq:ampere2}) form a closed set of equations, and therefore can be used to model physical systems.  Technically, these equations actually give the ``ideal Hall-MHD model'' since resistivity is not retained here.

It is important to note a confusing aspect of these equations.  The Hall term is proportional to ${\bf J} \times {\bf B}$, and Eq.~(\ref{eq:momentum}) also contains a term including ${\bf J} \times {\bf B}$. It is tempting to draw relations between the two terms because of this outward similarity, but this should not be done. The two terms have different dimensions: ${\bf J} \times {\bf B} / c$ is a force density and ${\bf J} \times {\bf B} / n_e e c$ is an electric field. They have completely different manifestations and impacts on the physics, and therefore actually are not related despite their similar forms.

\subsection{Hall-MHD Physics}
\label{sec:HallMHDPhysics}

The only difference between the ideal-MHD and Hall-MHD models is the Hall electric field, and here we investigate the physics introduced by this term. Let $\tilde{L}$ be a characteristic length scale over which the plasma properties vary, and let $\tilde{r}_{Li}$ be the characteristic (Larmor) radius of the ions as they gyrate around a magnetic field of characteristic strength $\tilde{B}$. The Hall electric field is small and can be neglected if $\tilde{L} \gg \tilde{r}_{Li}$, and doing so brings us back to ideal-MHD.  It is important to retain the Hall electric field, and therefore use Hall-MHD, when $\tilde{L} \lesssim \tilde{r}_{Li}$.   

\begin{figure}[t]
\centering
\includegraphics[width=3.4in]{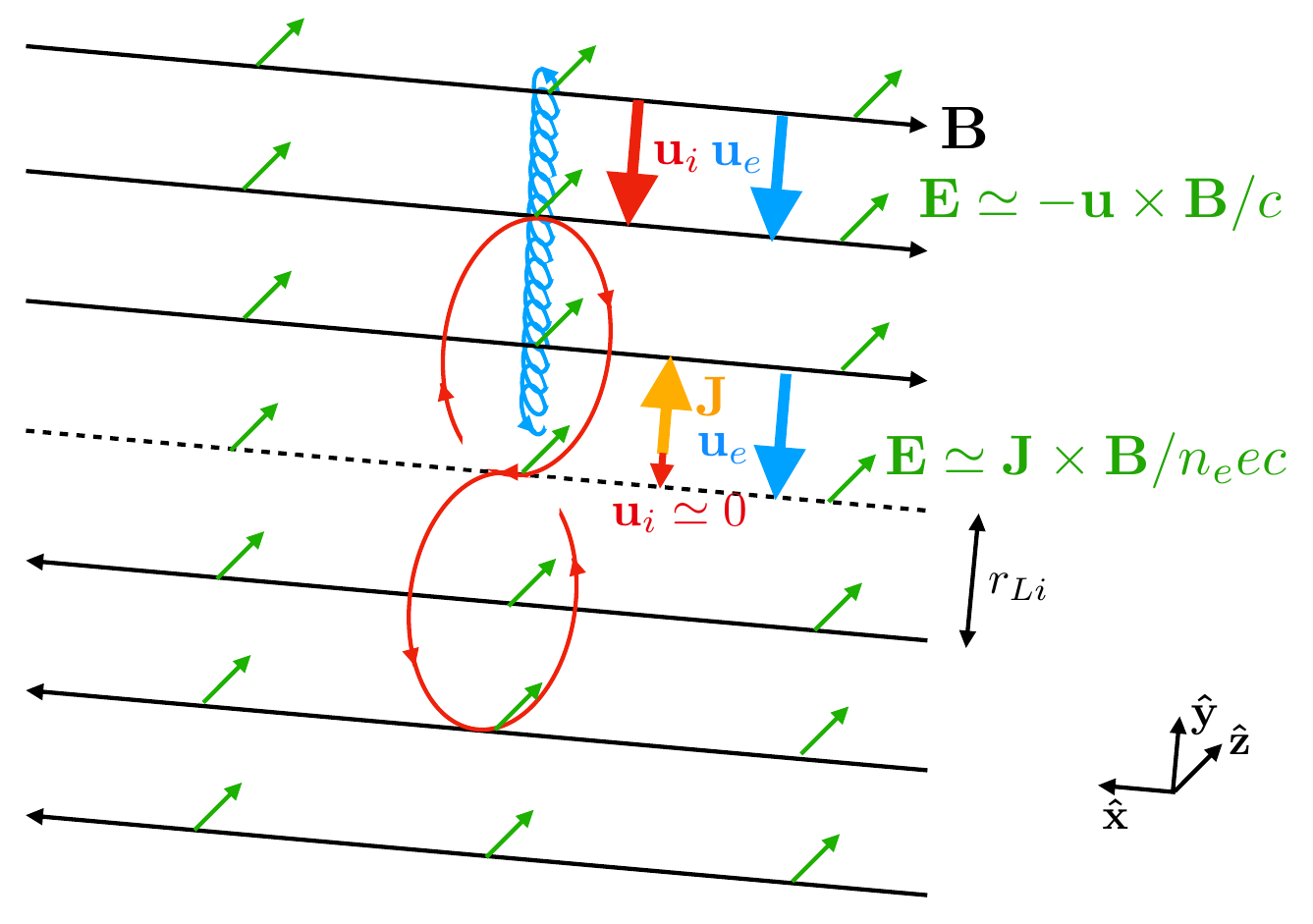} 
\caption { {\bf Physics of the Hall effect in reversing magnetic fields.} A reversing anti-parallel magnetic field in the $\pm {\bf \hat{x}}$ direction is drawn using black arrows, with the neutral line as the dashed line.  A uniform electric field in the ${\bf \hat{z}}$ direction is drawn in green.  The smaller blue trajectory is that of an electron $E \times B$ drifting towards the neutral line with bulk velocity ${\bf u}_e$.  An ion $E \times B$ drifting far from the neutral line has bulk ion velocity ${\bf u}_i$ identical to ${\bf u}_e$, so the electric field in this region is given by ${\bf E} = -{\bf u} \times {\bf B} / c$, where ${\bf u}$ is the single fluid (MHD) bulk flow velocity.  Within an ion gyroradius $r_{Li}$ of the neutral line, the electron continues with bulk flow ${\bf u}_e$, while the ion demagnetizes (in the red trajectory) and ${\bf u}_i$ becomes small. In this region, there is a non-zero current density ${\bf J}$ (orange), and the electric field in this region is predominantly given by the Hall electric field ${\bf E} = {\bf J} \times {\bf B} / n_e e c$.} 
\label{fig:FiguresHallMHD1}
\end{figure}

To see why structure below the ion gyroscale gives rise to the Hall effect, consider a magnetic field ${\bf B}$ that reverses direction over a scale $\tilde{r}_{Li}$ or less, but on larger scales than the characteristic electron gyroradius $\tilde{r}_{Le}$. For specificity, consider a magnetic field pointing in the $\pm {\bf \hat{x}}$ direction, as sketched as the black arrows in Fig.~\ref{fig:FiguresHallMHD1}.  The neutral line, where the magnetic field strength vanishes, is the dashed black line.  Suppose also there is a uniform electric field ${\bf E}$ pointing everywhere in the ${\bf \hat{z}}$ direction, as sketched as the green arrows.  (We use a reversed magnetic field and uniform electric field for illustrative purposes due to its relation to the reconnection process, but the Hall effect is important for any magnetic field configuration that varies on $\tilde{r}_{Li}$ scales.)  

At distances from the magnetic field reversal that greatly exceed $\tilde{r}_{Li}$, ions and electrons undergo the $E \times B$ drift that gives rise to a bulk flow towards the neutral line; the bulk flow velocity of ions and electrons, ${\bf u}_i$ and ${\bf u}_e$, respectively, are identical.  The $E \times B$ drift of the electrons above the neutral line is sketched as the blue curve.  In the region farther from the neutral line than $\tilde{r}_{Li}$, the electric field is given by ${\bf E} = -{\bf u} \times {\bf B} / c$, where ${\bf u} = {\bf u}_i = {\bf u}_e$ is the MHD bulk flow velocity.

As the ions reach a distance from the neutral line that is equal to its gyroradius, the ions cross the neutral line and are immersed in a magnetic field pointing in the opposite direction.  Their gyromotion changes direction, and they make figure 8 orbits, sketched as the red curve.  (They also accelerate in the ${\bf \hat{z}}$ direction due to the electric field, but this is omitted from the sketch.)  Consequently, their bulk velocity becomes small, ${\bf u}_i \simeq 0$.

Since we assumed $\tilde{L} > \tilde{r}_{Le}$, the electrons have a smaller gyroradius and do not see the magnetic field reversal, so they continue to undergo the $E \times B$ drift towards the neutral line.  The key is that the ions and electrons are undergoing different dynamics between distances from the neutral line of $\tilde{r}_{Li}$ and $\tilde{r}_{Le}$!

In this region, the difference in the bulk motion between ions and electrons implies there is a net current density ${\bf J}$, sketched as the orange arrow, and called the Hall current.  The current density is perpendicular to the magnetic field ${\bf B}$.  Then, between $\tilde{r}_{Li}$ and $\tilde{r}_{Le}$ from the neutral line, the electric field is given by the Hall electric field ${\bf E}_H = {\bf J} \times {\bf B} / n_e e c$.  This exemplifies why the Hall electric field is important between ion and electron gyroscales.

This situation in a plasma is analogous to the Hall effect in condensed matter physics, where it was originally discovered by Edwin Hall (a graduate student) in 1879 \citep{hall1879new}.  It has extensive applications in that field of study.  The derivation generalizing shear Alfv\'en waves to include the Hall effect happened as early as 1954 by Jim Dungey \citep{Dungey54}, the same person who first understood magnetic reconnection and gave the process its name.

\subsection{Linear Waves in ideal Hall-MHD}
\label{sec:HallMHDWaves}

It would take us too far afield to elucidate how the Hall electric field modifies all the physics of ideal-MHD.  Rather, we highlight one important example -- linear waves. In ideal-MHD, there are three propagating linear waves available to a uniform plasma: the shear Alfv\'en wave and the fast and slow magnetosonic waves. The Hall electric field introduces two wave modes that become important between ion and electron gyroscales -- the whistler wave and the kinetic Alfv\'en wave. 

\begin{figure*}[t]
\centering
\includegraphics[width=4.5in]{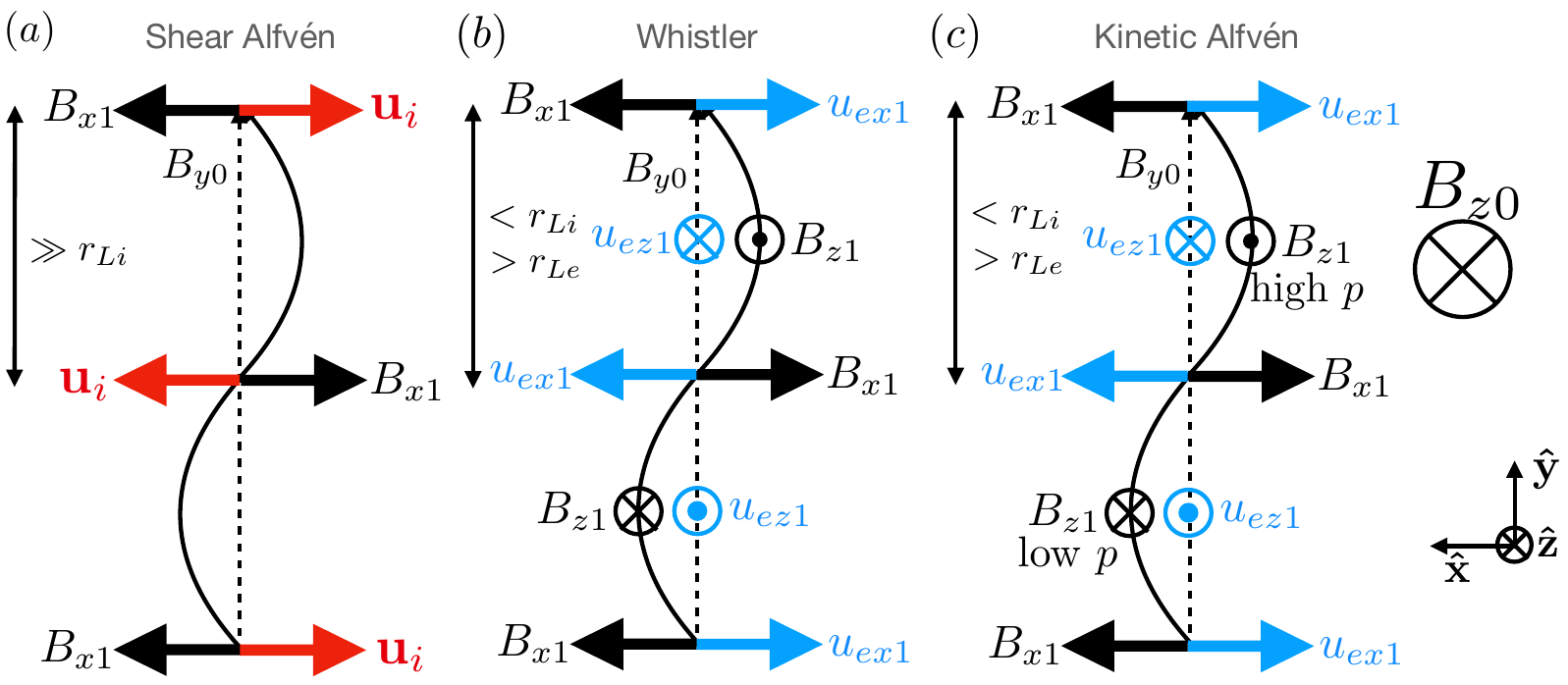} 
\caption { {\bf Sketch of how the Hall effect impacts linear waves in MHD.} The wave structure for (a) shear Alfv\'en waves, (b) parallel propagating whistler waves, and (c) kinetic Alfv\'en waves.  The equilibrium magnetic field $B_{y0}$ is the dashed black arrow.  The perturbed magnetic field ${\bf B}_1$ are the black arrows.  The perturbed ion bulk flow ${\bf u}_i$ are the red arrows, which occur in (a) because the wavelength is much larger than the ion gyroradius. In (b) and (c), the wavelength is at or below ion gyroscales, so the bulk flow is due to electron motion in the blue arrows.  In (c), there is a large out-of-plane equilibrium magnetic field $B_{z0}$, so the magnetic perturbation $B_{z1}$ changes the magnetic pressure which to first order and requires a change to the gas pressure $p$.} 
\label{fig:FiguresHallMHD2}
\end{figure*}

\subsubsection{The Whistler Wave}
\label{sec:whistler}

First, we consider transverse waves propagating along a uniform background magnetic field of strength $B_0$, which without loss of generality we take to be in the $y$ direction, in a plasma of equilibrium mass density $\rho_0 \simeq m_i n_i$ and with no equilibrium bulk flow or current ${\bf u}_0 = 0$ and ${\bf J}_0 = 0$.  The wave vector ${\bf k}$ is also in the $y$ direction. Linearizing the Hall-MHD equations (\ref{eq:continuity})-(\ref{eq:ampere}) about the given equilibrium with ${\bf k}$ parallel to ${\bf B}_0$ and solving for the dispersion relation gives 
\begin{equation}
    \omega^2 = k^2 c_{A0}^2 \left( 1 + \frac{k^2 d_{i0}^2}{2} + \sqrt{k^2 d_{i0}^2 + \frac{k^4 d_{i0}^4}{4}}\right), \label{eq:alfwhistler}
\end{equation}
where $\omega$ is the wave frequency, $c_{A0} = B_0 / (4 \pi \rho_0)^{1/2}$ is the Alfv\'en speed, and $d_{i0} = (m_i c^2 / 4 \pi n_0 q_i^2)^{1/2}$ is the ion inertial length.

In the limit of $k d_{i0} \rightarrow 0$, this dispersion relation reduces to $\omega^2 \rightarrow k^2 c_{A0}^2,$ which is simply the shear Alfv\'en wave from ideal-MHD.  In the other extreme, consider the limit $k d_{i0} \rightarrow \infty$.  The 1 in Eq.~(\ref{eq:alfwhistler}) becomes negligible compared to $k^2 d_{i0}^2/2$ outside the square root, and the $k^2 d_{i0}^2$ is negligible compared to $k^4 d_{i0}^4/4$ inside the square root, so to low order the dispersion relation becomes
\begin{equation}
    \omega^2 \rightarrow k^4 c_{A0}^2 d_{i0}^2 \hspace{1cm} ({\rm as} \ k d_{i0} \rightarrow \infty). \label{eq:disprelwhistler}
\end{equation}
This is the dispersion relation for the so-called ``whistler wave.''

Physically, the whistler wave is the sub-ion gyroscale counterpart of the shear Alfv\'en wave. The plasma properties in a shear Alfv\'en wave are sketched in Fig.~\ref{fig:FiguresHallMHD2}(a). The equilibrium magnetic field $B_{y0}$ is the dashed black arrow. A magnetic perturbation transverse to the equilibrium $B_{x1}$ is the large black arrow.  Since the size of the wave is far larger than the ion gyroradius, the frozen-in ions and electrons feel a restoring force analogous to plucking a guitar string, generating bulk flows in the $x$ direction (shown for ions as the red arrows). The shear Alfv\'en wave is a linearly polarized wave.  

The whistler wave, sketched in panel (b), occurs when the magnetic field varies between ion and electron gyroscales. Then, the ions are not frozen-in and do not respond to the plucked magnetic field line, but the electrons are frozen in and feel a restoring force.  The $x$ component of the perturbed electron flow $u_{ex1}$ is sketched as the blue arrows.  Since the electrons have a bulk velocity but the ions do not, this means there is a net current density, which induces a oscillating magnetic field out of the plane, labeled as $B_{z1}$.  Similarly, the varying $B_{x1}$ requires a $u_{ez1}$ to sustain the current.  This turns the wave into a circularly polarized wave.  The polarization is right handed, {\it i.e.,} a receiver sees the magnetic field rotate in a counterclockwise direction.   

There is a crucial difference between the shear Alfv\'en wave and the whistler wave.  The sheer Alfv\'en wave has a phase speed $\omega / k = c_{A0}$, which is a constant independent of $\omega$ and $k$, so a shear Alfv\'en wave travels at the same speed regardless of its wavelength.  It is an example of a non-dispersive wave that retains its waveform as it propagates (like a light wave in vacuum or a sound wave in a neutral fluid). In contrast, the phase speed for the whistler wave, from Eq.~(\ref{eq:disprelwhistler}), is $\omega / k = k c_{A0} d_{i0}$. Thus, the phase speed is faster for shorter wavelength waves.  This is an example of a dispersive wave, since a wave packet does not retain its shape.  The dispersive nature is what gives the wave its name; it makes a characteristic ``whistle'' from high frequencies descending to low frequencies when it is detected at a location away from where it was generated. For example, when a lightning strike in one hemisphere excites a whistler wave on Earth's magnetic field, the high frequencies arrive before the low frequencies in the other hemisphere. 

This analysis reveals the critical scale at which the Hall effect becomes important for transverse bending of the magnetic field.  The three terms in Eq.~(\ref{eq:disprelwhistler}) proportional to $d_{i0}$ are absent if the Hall term is left out of the governing equations ({\it i.e.,} in ideal-MHD). By comparing the first term (1) to the other terms, we see they become important when $k d_{i0}$ is on the order of 1.  This means that the Hall term becomes important at length scales below $d_{i0}$.  

The ion inertial scale is related to a form of the ion gyroradius.  To see this, note that an ion moving at the Alfv\'en speed $c_{A0}$ that gyrates around a magnetic field of strength $B_0$ has a gyroradius $r_{Li}$ given by
\begin{equation}
    r_{Li} = \frac{c_{A0}}{\Omega_{ci0}} = d_{i0},
\end{equation}
where $\Omega_{ci0} = q_i B_0 / m_i c$ is the ion gyrofrequency. Thus, the scale at which the Hall effect becomes important for transverse perturbations is when gradient scales are comparable to the ion inertial scale. Using characteristic scales for Earth's dayside magnetosheath, $d_{i0} \simeq$ 70 km; for the solar corona, $d_{i0} \simeq$ 2 m.  Thus, we have a feel for length scales over which we need to use Hall-MHD instead of ideal-MHD in two important space applications.

\subsubsection{The Kinetic Alfv\'en Wave}
\label{sec:kaw}

Now, we consider the kinetic Alfv\'en wave. These waves are almost completely longitudinal, but not perfectly longitudinal. To allow for a strong analogy to the whistler wave, consider a uniform magnetic field that has a very large $z$ component $B_{z0} \gg 0$ and very small $y$ component $B_{y0}$, as sketched in Fig.~\ref{fig:FiguresHallMHD2}(c). As in panels (a) and (b), the wave propagates in the $y$ direction. If $B_{z0}$ were zero, it would be the whistler wave. Physically, if the wavelength is between ion and electron gyroscales, this perturbation reacts similar to a whistler in that it sets up an electron flow that bends the magnetic field out of the plane. The perturbed magnetic field therefore has a component in the same or opposite direction as $B_{z0}$.  Where $B_{z0}$ and the perturbed magnetic field are parallel, the magnetic pressure $(B_{z0} + B_{z1})^2/8\pi$ is greater than the initial pressure $B_{z0}^2/8\pi$ to first order in the perturbed field. (Note, for the whistler wave, the pressure difference $B_{z1}^2/8\pi$ is second order in the perturbed magnetic field, so it is negligible.)  Similarly, where $B_{z1}$ opposes $B_{z0}$, the magnetic pressure decreases to first order.  Now, if the plasma is overall low $\beta$ (since we took $B_{z0}$ to be large), this magnetic pressure imbalance cannot be maintained, so the gas pressure has to change in order to balance pressure. Electrons move along the magnetic field and ions move across the magnetic field to move from the high magnetic pressure region to the low magnetic pressure region, setting up a high and low gas pressure region as denoted in panel (c).  This describes the physics of the kinetic Alfv\'en wave.

To find the dispersion relation of the kinetic Alfv\'en wave, we back up to the full dispersion relation of any wave in Hall-MHD.  Linearizing Eqs.~(\ref{eq:continuity}) - (\ref{eq:ampere}) around a uniform magnetic field, density, and pressure with an arbitrary linear perturbation and solving gives the following dispersion relation \citep{Rogers01}:
\begin{eqnarray}
    & \omega^6 & - (c_{ms}^2 + c_{Ak}^2 + k^2 c_{Ak}^2 d_{i0}^2) k^2 \omega^4 + \label{eq:disprelhall} \\ & & [c_{ms}^2 + c_s^2 (1 + k^2 d_{i0}^2)] k^4 c_{Ak}^2 \omega^2 - k^6 c_{Ak}^4 c_s^2 = 0, \nonumber
\end{eqnarray}
where $c_{A}^2 = B_0^2/(4\pi\rho_0)$ is the total Alfv\'en speed, $c_s^2 = \gamma p_0 / \rho_0$ is the sound speed, $c_{ms}^2 = c_s^2 + c_{A}^2$ is the total fast magnetosonic speed, and $c_{Ak}^2 = B_{y0}^2 / 4 \pi \rho_0$ is the Alfv\'en speed based only on $B_{y0}$.  In the $k^2 d_{i0} \rightarrow 0$ long wavelength limit, this equation reduces to the dispersion relation for ideal-MHD waves.  In the limit of large $k d_{i0}$, one solution is a high frequency solution which is approximately given by balancing the first two terms in Eq.~(\ref{eq:disprelhall}), which gives $\omega^2 \simeq k^4 c_{Ak}^4 d_{i0}^2$, the whistler wave dispersion relation in Eq.~(\ref{eq:disprelwhistler}). There is also a medium frequency solution which arises from balancing the middle two terms in Eq.~(\ref{eq:disprelhall}).  This ratio in general is
\begin{equation}
\omega^2 \simeq \frac{c_{ms}^2 + c_s^2 (1 + k^2 d_{i0}^2)}{c_{ms}^2 + c_{Ak}^2 + k^2 c_{Ak}^2 d_{i0}^2} k^2 c_{Ak}^2.
\end{equation}
In the limit in which $c_{ms}^2 \ll c_s^2 k^2 d_{i0}^2$ and $c_{ms}^2 \gg c_{Ak}^2 + k^2 c_{Ak}^2 d_{i0}^2$, the resulting dispersion relation is $\omega^2 \simeq (c_s^2/c_{ms}^2) k^4 d_{i0}^2 c_{Ak}^2$, which in the $c_{A}^2 \gg c_s^2$ (low $\beta$) limit gives $\omega^2 \simeq (c_s^2/c_A^2) k^4 d_{i0}^2 c_{Ak}^2$.  Since $k^2 c_{Ak}^2 = k_\|^2 c_{A}^2$, this becomes
\begin{equation}
\omega^2 \simeq k_\|^2 k^2 c_A^2 \rho_{s}^2, \end{equation}
where $\rho_s^2 = c_s^2 / \Omega_{ci}^2$ is the ion Larmor radius based on the sound speed. This is the dispersion relation for the kinetic Alfv\'en wave.  As with the whistler wave, the kinetic Alfv\'en wave is dispersive with $\omega / k \propto k_\|$, so it gets faster for smaller wavelengths.  The length scale at which the Hall term becomes important for the kinetic Alfv\'en wave is $\rho_s$ (as opposed to $d_{i0}$ for the whistler wave).

\subsection{A Numerical Algorithm for Hall-MHD}

Including the Hall electric field has a significant impact on numerical simulations relative to MHD simulations.  One way to think of why this is the case is that the waves in ideal-MHD are non-dispersive, so waves at any scale from the large scale down to the computational grid scale travel at the same speed. In Hall-MHD, as discussed in the previous section, both whistler and kinetic Alfv\'en waves are dispersive, so waves at the grid scale (which needs to be sub-ion gyroscale to capture the Hall electric field) are considerably faster than waves at the large scale. This makes the Hall-MHD equations ``stiff'' -- one must use a much smaller time step in Hall-MHD than in ideal-MHD, leading to a significant increase in the run time and expense of the simulation.

There are numerous algorithms that can be used to numerically evolve the equations in time.  We provide one in detail, and mention references to other algorithms that have been used.  We highlight the F3D code \citep{shay04a}, which has been used for many years to study magnetic reconnection.  First, the evolution equations are written in conservative form as
\begin{eqnarray}
\frac{\partial n}{\partial t} + \nabla \cdot {\bf J}_i & = & 0, \label{eq:continuitycons} \\
\frac{\partial {\bf J}_i}{\partial t} + \nabla \cdot \left(\frac{{\bf J}_i {\bf J}_i}{n} + \frac{p{\bf I}}{m_i} + \frac{{\bf B} {\bf B}}{4 \pi m_i} - \frac{B^2 {\bf I}}{8\pi m_i} \right) & = & 0, \label{eq:momentumcons} \\
\frac{\partial p}{\partial t} + \nabla \cdot ({\bf u} p) + (\gamma - 1) p (\nabla \cdot {\bf u}) & = & 0, \label{eq:pressurecons} \\
\frac{\partial {\bf B}}{\partial t} + c \nabla \times {\bf E} & = & 0, \label{eq:faradaycons}
\end{eqnarray}
with auxiliary equations ${\bf E} = {\bf J} \times {\bf B} / n e c - {\bf J}_i \times {\bf B} / n c$ and ${\bf J} = (c/4\pi)\nabla \times {\bf B}$. Here, $n \simeq \rho / m_i$ is the number density of ions (approximately equal to the number density of electrons due to quasi-neutrality) and ${\bf J}_i = n {\bf u}$ is the ion flux density. 

In F3D, these equations are stepped forward using the trapezoidal leapfrog technique, a predictor-corrector method that is well-equipped to handle conservative equations \citep{Zalesak79,Zalesak81}. Each of the above equations can be written as a conservative equation of the form
\begin{equation}
    \frac{\partial \psi}{\partial t} + \nabla \cdot {\bf {\cal F}} - D \nabla^2 \psi + F(\xi) = 0,  \label{eq:trapleap}
\end{equation}
where $\psi$ is the plasma variable in question, ${\cal F}$ is a suitably defined flux, $D$ is a second order diffusion coefficient which can be added to the equations to represent resistivity, viscosity, or a numerical dissipation to improve code stability, and $F(\xi)$ is a suitably defined source term in terms of any other plasma variables $\xi$.  (Equation~\ref{eq:faradaycons} is not exactly in this form, but an analogous expression holds.)  

To write the numerical algorithm, we use the standard notation where a superscript $n$ on a plasma variable refers to the time step in question, so the initial values are set at $n = 0$, the first time step is $n = 1$, and so on.  To evolve $\psi^n$ to $\psi^{n+1}$ in a time step $\Delta t$, the trapezoidal leapfrog algorithm is
\begin{eqnarray}
\psi^{n+1/2} & = & \frac{\psi^{n-1} + \psi^n}{2} + \Delta t \left[ - \nabla \cdot {\cal F}^n \right. \nonumber \\   & & + \left. D \nabla^2 \psi^{n-1} - F(\xi^n) \right] \\
\psi^{n+1} & = & \psi^n \Delta t \left[ - \nabla \cdot {\cal F}^{n+1/2} \right. \nonumber \\ & & + \left. D \nabla^2 \psi^{n} - F(\xi^{n+1/2}) \right].
\end{eqnarray}
The first equation uses the data at the $n$'th time step and the data at the previous $n-1$'st time step to ``predict'' $\psi$ half a time step in the future. Then, the flux and source terms are evaluated at this intermediate time step to evolve $\psi$ the next half time step to the desired step $n + 1$.  To go from $n = 0$ to $n = 1$, data is needed at $n = -1$, which is simply taken to be the same as the data at $n = 0$.  This algorithm is second order in the time step $\Delta t$, meaning that the error from the algorithm is approximately a coefficient times the square of the time step.  This is an example of an ``explicit'' time stepping algorithm because the data at the future time step is found completely using known data at the current or previous time steps.

The above shows how the equation is stepped forward temporally, but one also needs to calculate spatial derivatives.  The approach F3D uses to calculate spatial derivatives is with a finite difference technique, which means the spatial derivatives are simply approximated by the derivative over the size of a grid scale instead of over an infinitesimal distance.  For example, one approximation for the partial derivative in the $x$ direction on a grid with grid scale $\Delta x$ is $\partial \psi_j / {\partial x} \simeq (\psi_{j+1} -\psi_{j-1})/2\Delta x$, where the $j$ subscript refers to the index of the spatial cell for which the derivative is desired.  This is a second-order scheme because the error relative to the exact derivative scales like $\Delta x^2$.  The F3D code uses the approximation
\begin{equation}
\frac{\partial \psi_j}{\partial x} \simeq \frac{2}{3\Delta x} (\psi_{j+1} -\psi_{j-1}) - \frac{1}{12 \Delta x} (\psi_{j+2}-\psi_{j-2}),
\end{equation}
which is fourth order (the error scales like $\Delta x^4$), and the cost for this higher order derivative is that it requires data from two adjacent cells on each side rather than one.  Analogous expressions hold for derivatives in the $y$ and $z$ directions.

The F3D code is written in Fortran 90 and is parallelized using Message Passing Interface (MPI) for use on supercomputers.  The computational domain is rectangular with a fixed, regular grid.  It can be run in two or three dimensions.  When in two dimensions, the vectors can have an out of plane component even though all quantities are invariant in the out-of-plane direction; this is often referred to in the literature as ``2.5 dimensional''.  The F3D code does not explicitly enforce that $\nabla \cdot {\bf B} = 0$, but it has been demonstrated that the value is small when the initial magnetic field is divergence free.  Numerous aspects of the F3D code are on user-controlled switches that can turn terms or effects on or off, and the initial plasma variable profiles and values are controlled by the user.  Other features of F3D that are used to go beyond the Hall-MHD model will be treated in Sec.~\ref{sec:extensions}.

To run the F3D code, the user chooses the simulation domain size and desired grid scale to resolve the relevant physics, which is typically at least 5 times smaller than the relevant ion gyroradius.  As F3D is an explicit finite difference code, the time step $\Delta t$ can be no larger than allowed by the so-called Courant-Friedrichs-Lewy (CFL) condition \citep{press92a}, which requires $\Delta t \leq \Delta x / v_{{\rm fastest}}$, where $v_{{\rm fastest}}$ is the fastest speed that can occur in the system.  For Hall-MHD, the fastest speed is typically the whistler or kinetic Alfv\`en wave speed at or near the grid scale, but can be the fast magnetosonic speed for some ambient plasma conditions.  F3D simulations are typically performed with time step about 80\% of the CFL condition.  

Any employed diffusion coefficients then need to be chosen. Often, the resistivity is not used for Hall-MHD, but a fourth-order diffusion numerical dissipation of the form $+D_4 \nabla^2 \nabla^2$ on the right hand side of Eq.~(\ref{eq:trapleap}) is included to damp structures at the grid scale while minimally affecting larger scale structures.  It is handled numerically using the same finite difference approach as the other terms.  The appropriate diffusion coefficient scales with $D_4 \sim v_{{\rm fastest}} [\pi / (\Delta x)]^3$.  An appropriate value for this coefficient is when it is large enough to control numerical issues at the grid scale while not impacting the large scale physics.  A good approach to optimize this value is to run multiple simulations with only varying $D_4$, and finding a range of values for which the numerics are good and the large scale features are only weakly dependent on $D_4$; it is typically within an order of magnitude of the scaling prediction.

We have focused on F3D as an example of a Hall-MHD code because of its algorithmic simplicity and because it has long been used to study reconnection. There are drawbacks to the code.  As a finite difference code, it does not capture shocks, and therefore if the number density gets fairly small in any given simulation the code typically crashes.  The code performs well up to about 1,000-2,000 processors on high powered supercomputers; MHD codes without the Hall effect can be made to scale much better to 10s of thousands of processors.  It is also restricted to a rectangular geometry with a regular grid.

There are a number of other Hall-MHD codes that have been used to study magnetic reconnection, some of which we gather here.  Some codes used to study Hall-MHD reconnection in a rectangular domain have included VOODOO \citep{Huba03c}, HMHD \citep{Lottermoser97}, another code called HMHD \citep{Huang11}, and the UI Hall-MHD code \citep{Ma01}.  Codes that have been used to study Hall-MHD in the context of planetary magnetospheres (including Earth's) are a multi-fluid code \citep{Winglee04}, BATS-R-US \citep{Toth08}, and Gkyell \citep{Wang18}.  Codes used in the tokamak geometry include NIMROD \citep{Glasser99} and M3D-C1 \citep{Jardin08}.

\subsection{Boundary Conditions in Hall-MHD}

Boundary conditions for Hall-MHD codes can be challenging.  The boundary conditions are set using ``guard cells'' (sometimes called ``ghost cells'').  For example, if the desired spatial domain for a simulation has 100 cells in a particular direction, the array is padded by some number of cells to execute the boundary conditions.  For example, the fourth-order finite difference technique used for spatial derivatives in F3D requires two cells on either side of the grid point in question. Thus, for a cell on the boundary, there needs to be two cells worth of information to evolve those cells.  This is accomplished by using two guard cells on each side, so a 100 cell domain requires an array of 104 cells.  For a second-order finite difference, only a single guard cell is needed on each side.

The simplest set of boundary conditions is to use periodic boundaries, which is the technique employed for most F3D studies.  Suppose we have $N$ physical grid cells.  Let the cell index $j$ go over all spatial cells including the guard cells.  Then, the two guard cells have indices $j = 1, 2, N+3$, and $N+4$, and the physical cells have $j = 3, 4, \ldots, N+2$.  For a variable $\psi$, periodic boundary conditions are implemented using
\begin{eqnarray}
    \psi_1^{n+1} & = & \psi_{N+1}^{n+1}, \\
    \psi_2^{n+1} & = & \psi_{N+2}^{n+1}, \\
    \psi_{N+3}^{n+1} & = & \psi_{1}^{n+1}, \\
    \psi_{N+4}^{n+1} & = & \psi_{2}^{n+1},
\end{eqnarray}
where the superscript $n+1$ again refers to the time slice.  The advantage of periodic boundary conditions that there is no additional error introduced to the plasma variables at the boundaries as a result of the boundary conditions.  The primary disadvantage is that plasma from one side comes back through the other side which may not be realistic for the intended application.

Another boundary condition is zero derivative, {\it i.e.}, Neumann, boundary conditions.  To implement this, the guard cells are defined as 
\begin{eqnarray}
    \psi_1^{n+1} & = & \psi_{5}^{n+1}, \\
    \psi_2^{n+1} & = & \psi_{4}^{n+1}, \\
    \psi_{N+3}^{n+1} & = & \psi_{N+1}^{n+1}, \\
    \psi_{N+4}^{n+1} & = & \psi_{N}^{n+1}.
\end{eqnarray}
This boundary condition can be successful if the dynamics take place away from the boundaries.

Other choices for boundary conditions can be employed, but we finish our treatment with a discussion of ``conducting wall'' boundary conditions.  These boundary conditions are common for ideal MHD simulations.  Consider a boundary with its normal in the $y$ direction, representing the upstream/inflow direction.  In MHD, a conducting wall has a vanishing electric field in the plane of the boundary, $E_x = E_z = 0$ (Dirichlet boundary conditions), but the normal electric field $E_y$ can be nonzero (Neumann boundary conditions).  The magnetic field can be tangential ($B_x$ and $B_z$ are Neumann), but Faraday's law implies the normal magnetic field cannot change in time ($B_y$ is Dirichlet) because doing so would generate a tangential electric field.  With this magnetic field, there is no curl of $B_y$, so $J_x$ and $J_z$ must vanish at the boundary (Dirichlet), but there can be a non-zero $J_y$ (Neumann).  Finally, the bulk flow velocity can be tangential ($v_x$ and $v_z$ are Neumann) while the normal bulk flow must vanish ($v_y$ is Dirichlet) for Ohm's law to be satisfied at the boundary. This set of boundary conditions is consistent with the governing equations.

In Hall-MHD, these conducting wall boundary conditions do not work. The reason is that instead of Ohm's law being ${\bf E} + {\bf v} \times {\bf B} / c = 0$, it is ${\bf E} + {\bf v} \times {\bf B} / c = {\bf J} \times {\bf B} / n_e e c$. The boundary conditions in the previous paragraph for ${\bf v}$ and ${\bf J}$ are opposite.  Thus, there is no way to have the Hall term be consistent with the requirement that the electric field can only be normal to the boundary.

\subsection{Further Extensions of Hall-MHD}
\label{sec:extensions}

Here, we briefly discuss fluid model extensions beyond ideal-MHD that contain the Hall electric field and other terms.  We start with terms that were dropped from Eq.~(\ref{eq:newtone}).  

\subsubsection{Hall-MHD with Electron Inertia}

One extension of Hall-MHD is to retain the electron inertia term $m_e d{\bf u}_e/dt$ from Eq.~(\ref{eq:newtone}). We call this model ``Hall-MHD with electron inertia;'' it is often called the ``two-fluid model'' in the literature, but we refrain from this nomenclature since a completely different set of equations is also typically given the same name. 

Using the electron inertia term as is would lead to a new variable ${\bf u}_e$ with a time derivative in the model. Instead, the standard approach is to recognize that the prefactor $m_e$ is small for electron-ion plasmas, so the only way this term important is if the electrons are moving very fast. The ions would be too slow to keep up in such a case, so on time scales where this term is important, we can treat the ions as approximately stationary. Then, the electron bulk flow velocity ${\bf u}_e$ is related to the current density via ${\bf u}_e = - {\bf J} / n_e e$.  Since ${\bf J}$ and $n_e$ are already included in the Hall-MHD description, the set of equations remains closed. Such a simplification is convenient but not absolutely necessary. For example, Sec.~\ref{subsec:hybridfinitemass} describes a hybrid code in which the ion flows and density effects in the electron inertia term are included.

Analytically, the inertial electric field from dividing Eq.~(\ref{eq:newtone}) by $-e$ is given by $-(m_e/e)d{\bf u}_e/dt$, and replacing ${\bf u}_e$ by $-{\bf J}/n_e e$ gives $(m_e/e^2)d({\bf J}/n_e)/dt$.  Then, Ohm's law in Hall-MHD with electron inertia becomes
\begin{equation}
    {\bf E} + \frac{{\bf u} \times {\bf B}}{c} = \frac{{\bf J} \times {\bf B}}{n_e e c} + \frac{m_e}{e^2}\frac{d}{dt}\left(\frac{{\bf J}}{n_e}\right). \label{eq:hallohmslawwithme}
\end{equation}
It is important to note that the same approximation ${\bf u}_e \simeq -{\bf J} / n_e e$ is used in the convective derivative term, so that $d/dt \simeq \partial / \partial t - (1/n_e e){\bf J} \cdot \nabla$.  

One needs to treat the $\partial / \partial t$ term on the right hand side of Eq.~(\ref{eq:hallohmslawwithme}).  To do so, we eliminate ${\bf E}$ in Faraday's law using Eq.~(\ref{eq:hallohmslawwithme}); some algebra reveals that the equation becomes
\begin{equation}
\frac{\partial {\bf B}^\prime}{\partial t} = - c \nabla \times {\bf E}^\prime, \label{eq:faradayeinertia}
\end{equation}
where ${\bf B}^\prime = (1-d_e^2 \nabla^2){\bf B}$ is an auxiliary magnetic field and ${\bf E}^\prime = {\bf J} \times {\bf B}^\prime / nec -{\bf J}_i \times {\bf B} / nec$ is an auxiliary electric field, and the factor of $n$ in the inertia term is treated as a constant on the time scales of interest.  Coupling this equation with Eqs.~(\ref{eq:continuitycons})-(\ref{eq:pressurecons}) gives a closed set of equations.   

To solve these equations, we note that Eq.~(\ref{eq:faradayeinertia}) looks just like the usual Faraday's law except for the primes, so the same numerical technique can be used to solve for ${\bf B}^\prime$.  Once ${\bf B}^\prime$ is found, one uses that variable as the known source term in the equation $(1 - d_e^2 \nabla^2) {\bf B} = {\bf B}^\prime$ to solve for ${\bf B}$. This is an elliptic differential equation with many algorithms that can be used to solve it.  The simplest may be a relaxation technique \citep{press92a}, but it is relatively slow, especially when used for codes that have been parallelized for use on a supercomputer.  A faster version of relaxation is called multigrid \citep{Trottenberg00}.  F3D employs the Fast Fourier Transform approach.

We now briefly discuss the physics introduced by the electron inertia term and the advantages for including it in simulations. By comparing the electron inertia term to the Hall term, we determine the condition under which it is important to retain the electron inertia term.  We know the Hall electric field is important at scales below the ion inertial scale $d_i$, but vanishes at the X-line where ${\bf B} = 0$.  The electron inertia term can be important at small scales, so we seek the scale at which it becomes comparable to the Hall electric field. Setting them equal gives $J_y B_x / nec \sim (m_e/e^2) [({\bf J}/ne) \cdot \nabla] (J_z/n)$.  In the scaling sense, we use $B_x \sim B_{up}, J_z \sim cB_{up}/4 \pi \delta$ where $\delta$ is the scale at which the two terms are equal, so $J_y B_{up} / nec \sim (m_e/e^2) (J_y/ne \delta) (cB_{up}/4 \pi \delta n)$, which simplifies to $\delta^2 \sim d_e^2$, where $d_e^2 = m_e c^2/ 4 \pi n e^2$ is the electron inertial scale.  Thus, at length scales below $d_e$, the electron inertia term can be important.  

Consequently, one reason researchers include the electron inertia term into their Hall-MHD model is to aim to capture electron scale physics more accurately than without it.  We point out, however, that the MHD model itself was derived with the assumption that $m_e / m_i$ is small, and therefore including the electron inertia term as we have done does not actually provide a self-consistent treatment of sub-$d_e$ scale physics.  The two fluid model or the kinetic approach is needed to more accurately capture electron scale physics.  

Then why include electron inertia?  The answer is that it helps with the numerics.  To see this, we consider waves in the Hall-MHD with electron inertia system, generalizing the treatment in Secs.~\ref{sec:whistler} and \ref{sec:kaw}.  The dispersion relation for perfectly parallel propagating waves in Hall-MHD with electron inertia, generalizing Eq.~(\ref{eq:alfwhistler}), becomes
\begin{equation}
    \omega^2 = \frac{k^2 c_{A0}^2}{D_e} \left( 1 + \frac{k^2 d_{i0}^2}{2 D_e} + \sqrt{\frac{k^2 d_{i0}^2}{D_e} + \frac{k^4 d_{i0}^4}{4 D_e^2}}\right), \label{eq:alfwhistlerwithme}
\end{equation}
where $D_e = 1 + k^2 d_e^2$.  When $k d_e$ is negligible, this dispersion relation reduces to the Hall-MHD result in Eq.~(\ref{eq:alfwhistler}).  When $k d_e \gg 1$, the waves become electron cyclotron waves with $\omega^2 = \Omega_{ce}^2$, where $\Omega_{ce} = e B_0 / m_e c$.  The reason this is useful numerically is that the whistler wave is dispersive, and in Hall-MHD the phase speed goes to infinity as the wavelength goes to zero.  In Hall-MHD with electron inertia, the whistler rolls over to the electron cyclotron wave which does not propagate, so there is a maximum speed of the waves.  This means that the time step required to run the simulation does not go to zero, and the simulations with very high resolution are less stiff and therefore cheaper to carry out.

\subsubsection{Electron Magnetohydrodynamics (EMHD)}
\label{subsubsec:fluid-emhd}

The electron-MHD (EMHD) model \citep{Kingsep90} is used when the large (MHD) scales are not of interest, and only the scales betwen electron and ion scales are of interest.  In such a limit, the MHD terms concerning ion velocity are dropped.  Because of this, the density and pressure can no longer change (on the time scales of interest), so the only remaining governing equation is Faraday's law, with the electric field given solely by the Hall electric field:
\begin{eqnarray}
\frac{\partial {\bf B}}{\partial t} & = & - c \nabla \times {\bf E}, \label{eq:faradayconsemhd} \\
{\bf E} & = & \frac{{\bf J} \times {\bf B}}{n e c}.
\end{eqnarray}
Using Amp\`ere's law ${\bf J} = (c/4\pi) \nabla \times {\bf B}$, it is common to combine these equations into a single equation for ${\bf B}$,  
\begin{equation}
\frac{\partial {\bf B}}{\partial t} = - \frac{c}{4\pi ne} \nabla \times \left[ (\nabla \times {\bf B}) \times {\bf B} \right].
\end{equation}
This closed vector equation comprises the EMHD model.  One can include electron inertia in a manner analogous to the full Hall-MHD model: $\partial {\bf B}^\prime / \partial t = - (1/n_e e) \nabla \times ({\bf J} \times {\bf B}^\prime)$, where ${\bf J} = (c/4\pi) \nabla \times {\bf B}$ and ${\bf B}^\prime = (1 - d_e^2 \nabla^2) {\bf B}$.

\subsubsection{The Electron Pressure Term in Hall-MHD}

We now consider the electron pressure term in Eq.~(\ref{eq:newtone}). There, electron pressure is written as a tensor ${\bf p}_e$, which is the most general form following directly from the Vlasov/Boltzmann equation in kinetic theory.  Retaining it in Ohm's law would give a term on the right hand side of Eq.~(\ref{eq:hallohmslawwithme}) of the form $-(1/n_e e) \nabla \cdot {\bf p}_e$. We now consider a few commonly used approximations to simplify this. 

\paragraph{Isotropic Electron Pressure - }  In MHD, the total pressure is assumed to be isotropic, so that ${\bf p} = p {\bf I}$. In this case, the term entering Ohm's law is $-(1/n_e e) \nabla p_e$. When substituted into Faraday's law [Eq.~(\ref{eq:faraday})], this term gives a contribution of $(c/e) \nabla \times (\nabla p_e/n) = (c/e) \nabla (1/n_e) \times \nabla p_e = -(c/en_e^2) \nabla n_e \times \nabla p_e$.  This contribution to the electric field is called the ``Biermann battery'' \citep{Biermann50} and it is often of great importance in reconnection in high energy density plasmas \citep{Fox12}. In most space applications that employ a fluid model, the electrons are assumed to be adiabatic with $p_e/n_e^\gamma$ equal to a constant or isothermal with $p_e/n_e$ equal to a constant. In either limit, the Biermann battery term vanishes identically. 

One may presume from this result that the electron pressure gradient term does not have any contribution to Hall-MHD, but this is not true. Including the scalar electron pressure gradient in the Hall-MHD Ohm's law gives
\begin{equation}
{\bf E} + \frac{{\bf u}_e \times {\bf B}}{c} = -\frac{1}{n_e e} \nabla p_e, \label{eq:hallohmslawpress}
\end{equation}
where we write ${\bf u}_e \simeq {\bf u} - {\bf J} / n_e e$ for the bulk electron velocity.  Taking the cross product of this equation with ${\bf B}$ and solving for the bulk flow velocity ${\bf u}_{e,\perp}$ perpendicular to the magnetic field ${\bf B}$ gives
\begin{equation}
{\bf u}_{e \perp} = c \frac{{\bf E} \times {\bf B}}{B^2} + \frac{c}{n_e e B^2} \nabla p_e \times {\bf B}. \label{eq:diamagnetic}
\end{equation}
This equation implies the electron perpendicular bulk flow velocity is a sum of the $E \times B$ drift $c {\bf E} \times {\bf B} / B^2$ and the electron diamagnetic drift speed ${\bf u}_{*e} = c \nabla p_e \times {\bf B} / n_e e B^2$.  This important result shows that the electron diamagnetic drift is captured in Hall-MHD provided the electron pressure is non-zero.  An important implication of this is that magnetic flux convects at the sum of the $E \times B$ and electron diamagnetic drift speed when the electron pressure is retained in Ohm's law. 

\paragraph{Gyrotropic Electron Pressure - } The next level of approximation for the electron pressure tensor is motivated by the fact that electrons in a magnetic field often have a different temperature in the directions parallel and perpendicular to the magnetic field, {\it i.e.,} the electrons are gyrotropic.  The electron pressure parallel to the magnetic field is $p_{e,\|}$ and perpendicular to the magnetic field is $p_{e,\perp}$.  The general way to write a gyrotropic electron pressure tensor ${\bf p}_{e,g}$ for a magnetic field in an arbitrary direction ${\bf \hat{b}} = {\bf B} / B$ is \citep{PhysRev.107.924}
\begin{equation}
    {\bf p}_{e,g} = p_{e,\perp} {\bf I} + (p_{e,\|}-p_{e,\perp}) {\bf \hat{b}} {\bf \hat{b}}.
\end{equation}
Using this form in the generalized Ohm's law for the electron pressure, and an analogous term in the pressure gradient force in the momentum equation allows for the modeling of a plasma with a gyrotropic pressure.

Having a closed set of equations requires closures on the parallel and perpendicular pressures. The most widely known closure is the Chew-Goldberger-Low (CGL) closure, which assumes that there is no heat flux and the plasma is magnetized \citep{CGL}. A direct calculation using the Vlasov equation and these assumptions gives the CGL ``double adiabatic laws''
\begin{equation}
    \frac{d}{dt} \left( \frac{p_\perp}{n B} \right) = 0, \hspace*{0.5cm} \frac{d}{dt} \left( \frac{p_\| B^2}{n^3} \right) = 0, 
\end{equation}

A second model of great importance to reconnection is the Egedal closure \citep{Egedal13}.  This arises in the upstream and downstream regions of magnetic reconnection regions. The key physics is that the magnetic fields in this region are mirror fields that can trap electrons. The presence of an electric field heats them parallel to the electric field, leading to elongated gyrotropic distributions in the parallel direction.  For large magnetic field strength, the equations reduce to the isothermal equation of state.  For small magnetic field strength, the equations reduce to the CGL equations.  The two limits were interpolated to find a closure that could be implemented into a fluid code; see \cite{Egedal13} for details.

\begin{figure}[t]
\centering
\includegraphics[width=3.4in]{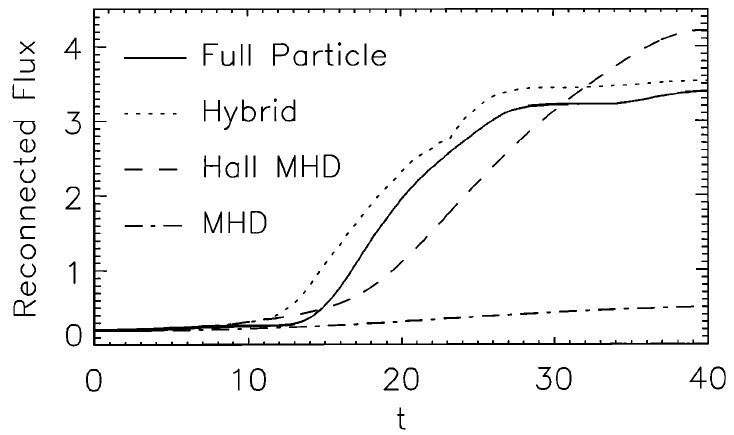} 
\caption { {\bf ``GEM Challenge'' result showing Hall-MHD simulations faithfully obtain the rate of change of reconnected flux obtained in kinetic models.}  The legend describes the simulation approach for each curve; resistive-MHD is far slower than the other models.  Adapted from \cite{birn:2001}.} 
\label{fig:birn01}
\end{figure}

\subsection{Examples of Reconnection Simulation Results with the Hall Electric Field}
\label{sec:examplesims}

\begin{figure}[t]
\centering
\includegraphics[width=3.4in]{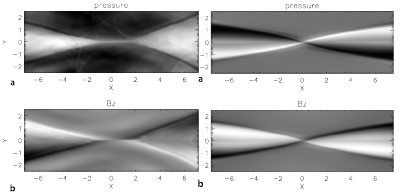} 
\caption { {\bf Pressure and out-of-plane magnetic field in Hall MHD reconnection simulations.}  The legend describes the simulation approach for each curve; resistive-MHD is far slower than the other models.  Adapted from \cite{rogers03a}.} 
\label{fig:rogers03}
\end{figure}

The Hall-MHD model holds a significant place of importance in the history of reconnection simulations.  From the earliest days of reconnection research, it was known that the Sweet-Parker model \citep{Sweet58,Parker57} was too slow to explain solar flares \citep{Parker63}, and it is also too slow to explain magnetotail reconnection \citep{Parker73} and the sawtooth crash in tokamaks \citep{Edwards86,Yamada94}.  It was discovered that using a localized resistivity leads to reconnection fast enough to explain the observed reconnection rates \citep{Ugai77,Sato79}, but the resistivity model was not derived from first principles and it remains inconclusive whether it can be.  Then, reconnection with the Hall electric field was found to also produce rates comparable to observed values without relying on ad hoc terms \citep{Aydemir92,Wang93,Kleva95,Ma96}. The ``GEM Challenge'' study (\citep{Birn01} and references therein), one of the most cited papers ever about reconnection, compared comparable simulations with different simulation models, and all models containing the Hall electric field led to fast reconnection with a reconnection rate approximately 0.1, as shown in Fig.~\ref{fig:birn01}.  It was only recently that an explanation of why the Hall electric field contributes to make the reconnection rate 0.1 was presented \citep{Liu22}.  Thus, Hall-MHD represents the minimal first-principles physics model that reproduces the reconnection rate achieved in kinetic models.

Another characteristic feature of collisionless reconnection is the quadrupolar structure of the out-of-plane magnetic field.  For anti-parallel reconnection, it was posited that the Hall electric field would cause there to be a quadrupolar out-of-plane magnetic field due to the in-plane currents within the ion diffusion region (where the ions decouple from the magnetic field but the electrons remain frozen-in) \citep{sonnerup79a}.  Although this is a completely nonlinear effect, it is analogous to the out-of-plane magnetic field generation in the whistler wave shown in Fig.~\ref{fig:FiguresHallMHD2}(b) \citep{Drake07}.  It can also be thought of as the out-of-plane current dragging the reconnecting field out of the reconnection plane \citep{Mandt94}.  Hall-MHD simulations of reconnection produce a quadrupolar out-of-plane magnetic field \citep{Huba02}, as shown in the lower left plot in Fig.~\ref{fig:rogers03} from simulations in \cite{rogers03a}.  When there is a strong out-of-plane (guide) magnetic field, the quadrupolar structure persists, but a quadrupolar structure in the gas pressure also arises with opposite polarity \citep{Kleva95}.  The physical reason is analogous to the gas pressure perturbation formation in kinetic Alfv\'en waves as shown in Fig.~\ref{fig:FiguresHallMHD2}(c). These quadrupolar structures arise in Hall-MHD simulations of reconnection, as shown in the right two panels of Fig.~\ref{fig:rogers03} from a simulation with guide field three times as strong as the reconnecting magnetic field. Despite the ability to produce quadrupolar structure in these quantities, it is now known that the detailed structure of the quadrupolar structures is not precisely the same as observed in kinetic simulations \citep{shay07a,Karimabadi07} or magnetospheric observations \citep{Phan07}.  It was shown that the inclusion of an electron pressure anisotropy leads to better agreement with kinetic modeling \citep{Ohia12}.  An electron pressure anisotropy with the CGL relations and without the Hall term can also produce reconnection rates comparable to Hall reconnection \citep{Cassak15}.

Another aspect of Hall-MHD reconnection that is not captured in collisional (Sweet-Parker) reconnection is how reconnection that is localized in the out-of-plane direction spreads in that direction.  Spreading has been studied in a number of Hall-MHD studies \citep{Huba02,Shay03,Huba03,Karimabadi04,Nakamura12,Shepherd12,Arencibia21,Arencibia22} and EMHD studies \citep{Jain13} and they agree with kinetic simulations of reconnection spreading \citep{Lapenta06}.  In particular, it was shown that anti-parallel reconnection does not spread in resistive-MHD, but it does spread in Hall-MHD and kinetic models \citep{Nakamura12,Arencibia21}.

An important issue related to Hall-MHD reconnection is its relation to the production of secondary islands.  Secondary islands are rapidly produced during Sweet-Parker reconnection for large enough Lundquist number \citep{Biskamp86,Loureiro07}.  It was argued on the basis of Hall-MHD simulations that the Hall effect prevents collisional reconnection effects from taking place \citep{cassak05a}, and suppresses secondary islands via collisional effects once it starts \citep{Shepherd10}, which was also reported in kinetic simulations \citep{Daughton09}. However, in Hall-MHD studies using island coalescence instead of a double Harris sheet, it was argued that collisionless secondary islands occur even after Hall-MHD effects take place \citep{Huang11}.  Collisonless secondary islands also arise in particle-in-cell kinetic simulations \citep{daughton:2006}. 

\subsection{The Future of Hall-MHD}

Interestingly, many computational plasma physicists in the reconnection community are moving away from the Hall-MHD model and its fluid extensions to study the small-scale properties of reconnection.  The Hall-MHD model was crucial for understanding the minimal physics that gives rise to a 0.1 reconnection rate.  As questions have moved to other aspects of reconnection including heating and particle acceleration, many researchers opt for kinetic models to more realistically capture the small scale physics than can be done with Hall-MHD.  Treatments of particle acceleration and heating at large scales \citep{arnold2021electron} do not require the Hall electric field.  

As example of an avenue of modern reconnection research and modeling where the Hall-MHD model remains highly beneficial is in the MHD-EPIC approach to global magnetospheric modeling \citep{daldorff}, as is discussed more fully in Sec.~\ref{sec:embedPIC}.  In this approach, the fluid model is used in regions of the magnetosphere where no small important scale physics takes place, which allows for faster run times.  In regions where small scale physics does take place, the code couples to a particle-in-cell code that captures this physics.  The numerical results between the two models are passed back and forth to each other across their boundaries.  The Hall-MHD model is well suited to be used in the transition region between the PIC and MHD models to facilitate a more accurate transition between the two models.  The Hall-MHD model has been used to study global magnetospheric systems for Earth and other planets and moons to great effect \citep{Paty04,Dorelli15,Dong19,li2023global}, and it is anticipated that further advances will continue to be made with the Hall-MHD approach for systems too large to employ global kinetic codes.

\section{Hybrid Simulations} \label{sec:hybrid}

The Earth's magnetosphere is a complex plasma system characterized by a multitude
of multiscale processes governing the interaction of the solar wind with
the Earth's dipole magnetic field. Modeling small-scale turbulent processes
in the foreshock and magnetosheath requires inclusion of ion kinetic effects
and Hall physics \citep{homa14, Omelchenko:2021a}. Furthermore, kinetic treatment
of hot and cold ion populations is greatly needed for improved modeling of
ionospheric outflows and their impact on the magnetopause and magnetotail.
To study magnetic reconnection, one also needs to incorporate finite electron-mass
effects \cite[e.g.][]{Biskamp2000,Birn2007,Gonzalez2016}, or mimic these
effects with {\it ad hoc} (resistivity) models.

The necessity to account for smaller and faster scales in kinetic simulations
in a manner that would guarantee their numerical accuracy and computational
efficiency creates challenges in global modeling of the Earth's magnetosphere.
Since describing plasma kinetics with pure ``first-principles'' models is
still not feasible, various approximations have been developed as candidates
for future ``beyond MHD'' operational modeling, with multiple levels of physical
fidelity included. However, to what degree kinetic processes may influence
the ``fluid-like'' behavior of the magnetosphere on global scales still remains
an open question. As we argue below, many of these challenges can be addressed
by self-consistent hybrid modeling, where Maxwell's equations are solved
in the quasi-neutral Darwin limit, ion species are treated kinetically, and
the plasma electrons are approximated as an inertialess fluid. These hybrid
models can be broken into two categories, according to the computational
techniques used to represent kinetic ions: Particle-in-Cell (PIC) models
\citep{Winske2003,Lipatov:2002} and Vlasov models \citep{Vlasiator2014}. In
what follows we discuss only the hybrid-PIC approach because it has already
been applied successfully to perform three-dimensional (3D) simulations of
global plasma systems that range from the Earth's magnetosphere \cite[e.g.][]{Lin2005,Omelchenko:2021a},
planets \citep{Hercik2013,Jarvinen:2020}, and small space bodies \citep{Fatemi:2017,
Kallio:2019} to compact laboratory plasmas \citep{Omelchenko:1997,Lin2008,Thoma:2013,Omelchenko:2015,Omelchenko:2022spr}.
The hybrid-Vlasov approach \citep{Vlasiator2014} is relatively new and considerably
more computationally expensive, with production runs being still restricted
to quasi-3D setups \citep{Pfau:2020}. 

An important issue to grasp reconnection correctly is the consideration of the finite
electron inertia, as it has been shown by EMHD simulations. This could reveal not
only the properties of electron-only reconnection, proposed by \citep{Jain2009} and 
recently discovered by MMS in the magnetsheath \citep{Phan2018}, 
but also the transition from electron- to
ion reconnection. These effects can be treated by hybrid-approach with kinetic ions 
and an inertial electron fluid~\ref{subsec:hybridfinitemass}.

\subsection{Model Equations - massless electrons}\label{subsec:hybridmassless}

The standard hybrid model \citep{Winske2003} assumes plasma quasi-neutrality,
neglects the displacement current in Maxwell's equations, and treats ions
as full-orbit macro-particles (in the PIC approach) moving in self-consistent
electric and magnetic fields. The plasma electrons are approximated as an
inertialess fluid with scalar pressure described by either an adiabatic law
or evolution equation. Together with a self-consistent PIC method for the
ion components, this leads to a set of hybrid equations that include Ampere's
law in the magnetostatic limit, Faraday's law, and an algebraic expression
for electric field (generalized Ohm's law) with the Hall, electron pressure
gradient, and resistive terms \cite[e.g.][]{Omelchenko2012a}: 

\begin{equation} \label{eq_amp}
\nabla\times {\bf B}=\frac{4\pi}{c}{\bf J}, ~{\bf J}={\bf J}_e+{\bf J}_i,
\end{equation} 
\begin{equation} \label{eq_far}
\frac{\partial {\bf B}}{\partial t}=-c\nabla\times{\bf E},
\end{equation}
\begin{equation} \label{eq_ohm}
{\bf E} = \frac{{\bf J}_e\times {\bf B}_t}{en_ec}
-\frac{\nabla p_e}{en_e}+{\eta {\bf J}}, 
~{\bf B}_{t}={\bf B}+{\bf B}_{ext},
\end{equation} 
\begin{equation} \label{eq_quasi}
en_e=\rho_i,
\end{equation}
\begin{equation} \label{eq_press}
p_e=n_eT_e\sim n_e^\gamma.
\end{equation}

In Eqs.~(\ref{eq_ohm}-\ref{eq_press}) $n_e,{\bf J}_e$ are the electron number
and current density, respectively; $p_e$ is the electron pressure, here assumed
to governed by Eq.~(\ref{eq_press})) with an adiabatic constant of $\gamma$;
$T_e$ is the electron temperature; $\rho_i,{\bf J}_i$ are the total ion charge
and current density (found by the PIC method); ${\bf E}$ is the electric
field; ${\bf B},{\bf B}_{ext}$ are the ``self-generated'' (${\bf B}\rvert_{t=0}=0$)
and ``external'' (steady state) magnetic fields, respectively. 

The applied plasma resistivity, $\eta$ is either constant or chosen to be
a function of plasma parameters \cite[e.g.][]{Lin2007, Omelchenko:2021a}.
The resistive term in the generalized Ohm's law (Eq.~(\ref{eq_ohm})) may
(i) describe finite conductivity of plasma or space bodies (e.g., the Moon
\citep{Fatemi:2017, Omelchenko:2021b}), (ii) imitate finite electron inertia
effects in magnetic reconnection events, and (iii) enable fast magnetic field
diffusion at low-density (``vacuum'') cells, $n_e \leq n_{min}$, where $n_{min}$
is a small density cutoff value \citep{Omelchenko:2021b}. Failure to properly
treat low-density regions in hybrid simulations may lead to non-physical
 results \citep{Omelchenko:2021b,Poppe2019}.

\subsection{Key Physics Beyond MHD in Hybrid Models}

The ``mesoscale'' hybrid model occupies the middle ground between the ``large-scale''
fluid and ``micro-scale'' first-principles modeling paradigms. For global
magnetospheric simulations, the hybrid model enables a number of ``beyond
MHD'' capabilities, as explained below.

{\it \textbf{Modeling turbulent processes in the foreshock and magnetosheath.}}
Unless large resistive damping (or smoothing) is applied, the hybrid model
accurately captures the Hall physics for mesh cell sizes, $\Delta\leq ~d_i$,
where $d_i=c/\omega_{pi}$ and $\omega_{pi}$ are the local ion inertial length
and plasma frequency, respectively. The Hall effects phase out on coarser
meshes, $\Delta\gg d_i$, where the Alfv{\'e}n term becomes greater than the
Hall term in Eq.~(\ref{eq_ohm}) and the whistler mode frequency, $\propto
\Delta^{-2}$ becomes lower than the Alfv{\'e}n mode frequency, $\propto \Delta^{-1}$.
In fact, in this case the Hall term can completely be removed from the electric
field in Faraday's law (Eq.~(\ref{eq_far})) and kept only in the equations
of ion motion \citep{Karimabadi2004}. The ability of a hybrid code with the
Hall term to run stably on coarser meshes ($\Delta\gtrsim d_i$) may depend
on the numerical implementation of Faraday's law \citep{Omelchenko2012a}.

Global hybrid codes have been used to address the ultra-low frequency (ULF)
physics of the curved bow shock on the ion inertial/Larmor radius scales
as the physics of the bow shock is predominantly determined by kinetic physics
associated with charged particles from the solar wind. Of particular interest
are the foreshock waves and diffuse ion distributions \citep{Wang2009} and
transient perturbations originating from the wave-particle processes in the
quasi-parallel shock or due to the shock interaction with incoming solar
wind discontinuities, including hot flow anomalies \citep{Lin2002, Lin2022},
foreshock bubbles \citep{Omidi2010, Wang2020_foreshock}, foreshock cavities \citep{Lin2005,
Blanco2011}, and high-speed jets \citep{Omelchenko2021, palmroth2018}. The
3D hybrid simulations 
with ANGIE3D 
link the foreshock perturbations to the surface perturbations and kinetic-scale
shear Alfv\'en waves (KAWs) at the magnetopause through mode conversion from
the incoming compressional waves \citep{Lin2005, Shi2013}, as well as the
subsequent excitation of toroidal-mode field line resonances in the magnetosphere
\citep{Shi2021}. It has also been shown that 3D models are essential for addressing
the nonlinear physics of mode coupling and ion diffusion at the magnetopause
\citep{Lin2012}.

The whistler mode plays a significant role in regulating turbulence in the
magnetosheath and mediating magnetic reconnection in the Earth's magnetosphere
\cite[e.g.][]{Dorelli:2003, Drake:2008}. Hybrid simulations generally have
to resolve the quadratic dispersion of this mode, $\omega\propto k^2$. This
requirement may create computational bottlenecks in simulations of strongly
inhomogeneous magnetospheric and laboratory plasmas, where whistler timescales
typically span several orders of magnitude \citep{Omelchenko2012a,Omelchenko:2022spr}).
If not accurately integrated in time (or resistively damped), the spurious
short-wavelength oscillations may grow explosively unstable from noise and
terminate simulation \citep{Lin2008}. It should also be noted that although
particle noise in hybrid-PIC simulations typically degrades their physical
resolution, the Lagrangian (particle) approach enables transport of ion species
with less numerical diffusion compared to the Eulerian approach to solving
the Vlasov equation on velocity meshes \citep{Vlasiator2014}.  

{\it \textbf{Collisionless reconnection at the magnetopause and in the tail
plasma sheet.}}
The physics of magnetic reconnection in the magnetosphere can be investigated
by carrying out global hybrid simulations with an ad-hoc current-dependent
resistivity. For the dayside magnetopause, the modeling topics include the
structures of ion diffusion region and outflow regions \citep{Tan2011}, global
evolution of flux transfer events (FTEs) and magnetic flux ropes \citep{Omidi2007,
Guo2020, GuoJ2021}, propagation of kinetic Alfv\'en waves and Poynting flux
from reconnection \citep{Wang2019}, ion cusp precipitation and energy spectrum
\citep{Omidi2007, Tan2012}, the triggering of reconnection  by solar wind
discontinuities \citep{Omidi2009, Pang2010, Guo2021_JGR}, and magnetosheath turbulence
\citep{Chen2021, Ng:2021}. Likewise, high-latitude reconnection tailward of
the cusp under northward IMF has also been simulated \citep{Lin2006, Guo2021three}.
The 3D global physics of storm-time magnetotail reconnection, fast flow and
entropy bubbles, the Hall-effects control of dawn-dusk asymmetry \citep{Lin2014,
Lu2016, Lin2017}, and the associated global Alfv\'enic coupling between the
magnetotail and the ionosphere under southward IMF have been simulated using
the ANGIE3D code \citep{Cheng2020}. Attempt has also been made to investigate
the subsequent connection of fast flows to the ring current and radiation
belt by combining ANGIE3D with the CIMI inner magnetosphere model \citep{Lin2021}.

{\it \textbf{Inclusion of multi-species plasma ion populations of  solar
wind origin and improved representation of ionospheric outflow.}} In general,
global hybrid models may include multiple ion species for representing solar
wind and ionospheric outflow plasmas. Ionospheric outflow ions should be
treated kinetically and self-consistently in order to properly account for
their impact on the Earth's magnetosphere. Multi-fluid MHD models do not
account for ion resonance acceleration and cyclotron effects, especially
for heavy ions \citep{Toledo:2021}. Self-consistent 3D hybrid simulations
of the impact of oxygen outflow on the magnetotail configuration and stability
have recently been performed with the HYPERS code \citep{Mouikis:2021, Omelchenko:2023}.

{\it \textbf{Modeling local reconnection and electron scale physics.}} In
MHD simulations, magnetic reconnection is often a result of mesh-dependent
diffusion that is difficult to control numerically. The hybrid-PIC model
is inherently more robust in this regard because ions are modeled as Lagrangian
particles. As a result, reconnection onset and dynamics are controlled by
the Hall physics and parameter-dependent resistivity. The hybrid-PIC model
is also known to accurately reproduce reconnection rate when the ion inertial
and cyclotron scales are properly resolved \citep{Stanier2015}. Further modifications
of the hybrid model, which for instance may incorporate finite electron mass
effects \cite[e.g.][]{Omelchenko:2021c}, could increase physical fidelity
of reconnection modeling in the future.  
 
{\it \textbf{Modeling non-MHD waves in a global context.}} The standard hybrid
model supports ion cyclotron, whistler, and kinetic-Alfv{\'e}n wave modes,
which play an important role in regulating plasma turbulence in the Earth's
magnetosphere and impact its global behavior. Modern observations report
streams of non-Maxwellian ions that excite plasma turbulence through numerous
kinetic instabilities that cannot be modeled within MHD. Global hybrid modeling
naturally incorporates the ion kinetic effects into global models of the
Earth's magnetosphere, which helps advance our understanding of the effects
of turbulent plasma dynamics on global physical processes.

{\it \textbf{Other applications.}} 3D hybrid codes in space physics
have been used to simulate shock-driven ion acceleration \citep{Caprioli:2014,Guo_Gia:2013},
solar wind turbulence \citep{franci2018, Royter:2015}, the Moon's wake \citep{Fatemi:2017,Kallio:2019,Omelchenko:2021b},
planetary magnetospheres \citep{Jarvinen:2020}, and small space bodies \citep{Alho:2019,
Delamere:2009}. Comparing results from full-scale 3D hybrid simulations of
small space bodies with satellite observations provides yet another important
route for validation and further extension of the hybrid approach to plasma
modeling. Importantly, the recent advances in ionosphere-magnetosphere coupling
\citep{Lin_Wang:2022}, code optimization \citep{Dong_atal:2021} and multiscale
computing \citep{Omelchenko:2021a} have greatly improved the prospects for
hybrid simulations to become a key factor to consider in the overall theory
of global solar wind-magnetosphere-ionosphere interactions.

\subsection{Numerics}
Eqs.~(\ref{eq_ohm}-\ref{eq_press}), together with the self-consistent equations
of motion for the ion macro-particles, may present computational challenges
when used for modeling complex 3D plasma systems, such as the Earth's magnetosphere.
Below we discuss some recent computational advances aimed at overcoming these
problems.    

\subsubsection{Spatial Scales}
Hybrid-PIC simulations typically intend to resolve the spatial scales of
the order of the ion inertial length, $d_i$ and ion cyclotron radius, $r_{ci}$.
The physical validity regime of the hybrid model ranges from large MHD scales
down to $k r_{ci} \sim 1$ and $\omega t \sim 1$. The actual physical resolution
of a hybrid simulation is largely determined by (i) how well these characteristic
lengths are resolved on a mesh, (ii) how many particles are used. For the
typical solar wind proton inertial length, $d_i \sim 100~km$, the Earth's
radius, $R_E\sim 60d_i$. To encompass the whole magnetosphere, the computational
mesh in a global simulation should cover the magnetopause with a typical
standoff distance, $R_{MP}\sim 10R_E \sim 600d_i$ and the magnetotail stretching
from the Earth to far distances, $R\sim 100R_E \sim 6000d_i$. The need to
accurately account for the ``far-field'' inflow and outflow boundary conditions
in the presence of a magnetic dipole may additionally require multiplying
these dimensions by a factor of 2-3. Approximating such large 3D domains
with uniform meshes with cell sizes of the order of $\sim 1d_i$ is prohibitively
expensive for the hybrid model because of the need to advance ions and fields
at all cells on kinetic scales. 

To overcome these restrictions, several options are available. First, one
may increase the cell size beyond $1d_i$ at the expense of lower accuracy
in resolving the Hall physics \citep{Omelchenko2012a}. Second, one may artificially
increase the ratio between the solar wind ion inertial length $d_i$ ($d_i\sim
r_{ci}$ for the outer magnetosphere regions with ion $\beta \sim 1$) and
the magnetopause distance $R_{MP}$, in order to better accommodate the available
computation resources while still choosing a sufficiently large value of
$R_{MP}/d_i$ for assuring the separation between the global and local-kinetic
scales \citep{Omidi2004}. Both approaches efficiently ``downscale'' the Earth's
magnetosphere. For instance, ANGIE3D \citep{Lin2014,Lin2005} does it by artificially
reducing the solar wind plasma density, i.e. by inflating the characteristic
inertial ion length and proportionally increasing the Alfv{\'e}n speed. Alternatively,
H3D \citep{homa14}, hybrid-VPIC \citep{Dong_atal:2021}, and HYPERS \citep{Omelchenko2012a}
employ the physical ion inertial length but scale the realistic magnetopause
standoff distance down by a factor of 4-6 by using a weaker magnetic dipole.

Regardless of a chosen magnetosphere scaling method, present-day 3D hybrid
codes typically use $R_{MP}/d_i\geq 100$. One of the largest 3D hybrid simulations
to date was performed with HYPERS for $R_{MP}/d_i \simeq 160$ on a uniform
mesh with approximately $1000 \times 2000 \times 2000$ cells \citep{Omelchenko:2021b}.
To speed up global 3D simulations, hybrid codes may employ nonuniform meshes,
among which `stretched'' (logically mapped) Cartesian meshes are the simplest.
Nonuniform meshes typically maintain high resolution in a central domain
of interest, while expanding cells towards domain boundaries \citep{Omelchenko:2021a,Lin2005}
to guarantee that the dipole field vanishes at the inflow/outflow (GSM X)
and lateral (GSM Y and Z) boundaries so that robust local boundary conditions
can be implemented. Sometimes, for simplicity, the lateral domain boundaries
may be assumed to be periodic (\cite[e.g.][]{Turc:2015,Muller:2011}). This
simplification, however, makes a global simulation valid for shorter simulation
periods, until reflected particles or electromagnetic perturbations reach
the periodic boundaries. To improve the counting statistics for macro-particles,
splitting techniques may be used to enhance energetic particle distributions
in the dayside magnetosphere \citep{Omelchenko2021} and magnetotail \citep{Lin2007,
Wang2009}.

To further reduce the number of computational cells in global simulations,
one may employ curvilinear (e.g., spherical) meshes (\cite[e.g.][]{Dyadechkin2013,Guo_Lin:2021}).
For example, to capture the short-wavelength physics along the shock normal
and the magnetopause, early hybrid simulations, focusing on the dayside regions,
employed cylindrical (2D) \citep{Swift:1996, Lin1996, Lin2002} and spherical
(3D) \citep{Lin2005} coordinate systems. The spherical coordinate lines, however,
have a singularity on the polar axis, which was handled by rotating the polar
coordinates to the equator and omitting a conic region around them, while
keeping the physical polar regions inside the domain \citep{Lin2005}. A 2D
hybrid simulation of the magnetotail also used curvilinear coordinates to
accommodate the tail geometry \citep{swift2001substorm, Lin2002}. Similar
to the need of assuring proper numerical resolution for resolving the kinetic
scales along the curved or oblique boundary surfaces on the Cartesian meshes,
special care is also necessary for the curvilinear meshes, especially when
their coordinate lines are not orthogonal \citep{swift2001substorm}. Compared
to Cartesian meshes, curvilinear meshes may introduce additional discretization
errors due to their (i) typically lower orders of numerical approximation,
(ii) anisotropic particle-mesh weighting. These errors lead to various numerical
artefacts and non-conservation of particle momentum (''self-forces''). As
a result, it is necessary to benchmark results from simulations obtained
with curvilinear meshes with similar simulations performed with Cartesian
meshes \citep{Dyadechkin2013}. 

Some global hybrid codes employ adaptive mesh refinement (AMR) \citep{Leclercq:2016,Muller:2011}.
Hybrid AMR simulations, however, may suffer from spurious particle ``self-forces''
and  wave reflections that occur at the mesh refinement interfaces. To mitigate
these artefacts, AMR algorithms are typically complemented with smoothing
procedures, which, however, should be performed with caution in order to
avoid affecting underlying physics in the regions of interest.

\subsubsection{Temporal Scales}

In addition to the ''slow'' Alfv{\'e}nic (MHD) timescales, hybrid codes need
to follow the ``fast'' ion kinetic, ion cyclotron and whistler timescales.
This requirement typically makes global hybrid simulations 
numerically ``stiff'' in the near-Earth space, where timesteps, required
for numerical accuracy and stability, may become prohibitively small \citep{Omelchenko2012a}.
To partially mitigate these effects, the kinetic ions may be replaced in
this region by a dense fluid \citep{Swift:1996, Lin_Wang:2021}. 
Hybrid-PIC simulations inherently generate spurious oscillations with large
wave numbers, $k\sim 1/\Delta$ and high frequencies, $\omega\sim k^2\sim
1/\Delta^2$. If not properly integrated or resistively damped, these noisy
oscillations may explosively grow and abort simulation \citep{Lin2008}. Deleterious
instabilities may be avoided by applying ``noise filtering'' (smoothing)
or/and various ``flux-limiting'' techniques for electric and magnetic fields.
These modifications, however, need to be implemented with caution, as the
may produce artificial solutions not supported by the hybrid model. 

For accuracy, typical full-orbit particle solvers (``pushers'') require that
timesteps, $\Delta t_p$ should be small enough that $\Omega\Delta t_p \ll
1$, where $\Omega$ is the local ion gyro-frequency. Using the same timestep
for all particles may create another numerical bottleneck in global hybrid
simulations of the Earth's magnetosphere. For instance, particle timesteps
of the order of $\Omega_0\Delta t_p \sim 0.05$ (where $\Omega_0$ is the ion
gyro-frequency computed with respect to the IMF strength, $B_{IMF}$)  fairly
well describe ion gyro-motion in the solar wind (\cite[e.g.][]{Turc:2015,Guo_Lin:2021}).
At the same time, gyro-orbits and drifts of ions with $\Omega \gtrsim 10~\Omega_0$
(e.g. found in the cusp or some parts of the magnetosheath), will not be
reproduced with accuracy. As a remedy, in some simulations, sub gyro-orbit
time steps may be employed in these (large magnetic field) regions of the
magnetosphere \citep{Lin2014}.

To summarize, predicting optimum global timesteps for the particles and fields
in global hybrid simulations is difficult in practice. This challenge has
been addressed by replacing time stepping with an asynchronous approach to
time integration, which combines discrete-event simulation (DES) with elements
of artificial intelligence: Event-driven Multi-Agent Planning System (EMAPS)
\citep{Omelchenko:2006, Omelchenko:2022spr}. EMAPS enables time advance of
individual particles and local fields on meshes of arbitrary topology by
integrating them on their self-adaptive timescales in a ``game of life''
 fashion. Thus, EMAPS effectively performs the role of an intelligent ``simulation
time operating system''. This approach was first applied to model 1D collisionless
plasma shocks \citep{Omelchenko:2006a} and fluids \citep{Omelchenko:2006, Omelchenko:2007}.
Implemented in HYPERS \citep{Omelchenko2012a}, EMAPS has enabled efficient
and accurate global 3D hybrid simulations of the Earth's magnetosphere \citep{Omelchenko:2021a,
Omelchenko:2021b, Omelchenko:2023}. 

Global 3D hybrid simulations of the Earth's magnetosphere are typically performed
for simulation periods, $\Omega_0t\sim 100-500$, where $\Omega_0$ is the
IMF based proton cyclotron frequency. Assuming $B_{IMF}=5~nT$, these simulations
formally span relatively short (compared to MHD) magnetospheric times, $t
< 20$~min. For convenience, in order to present physical results in ``magnetospheric
hours'', some modelers multiply this simulation time by a model scaling factor
\citep{Lin_Wang:2022}. Although this scaling is useful for comparing ``macro-scale''
simulation phenomena with observations, it is not appropriate for describing
ion kinetic effects, e.g. those that drive the ``magnetokinetic'' formation
of high-speed jets \citep{Omelchenko:2021a}. 

\subsubsection{Plasmasphere and Ionosphere}

Currently, hybrid codes cannot model global magnetospheric convection lasting
many hours or days, e.g. a steady-state process of magnetotail loading and
unloading. Therefore, global hybrid models typically assume a simple perfectly
conducting or resistive ionosphere, where dipole magnetic field lines are
``tied up'' to the inner boundary (zero electric field) or allowed to diffuse
due to its finite resistivity, respectively. 
model \cite[e.g.][]{Lin_Wang:2021},

To avoid computing fast kinetic timescales, a cold, incompressible, dense
ion fluid may be assumed to co-exist together with low-density particle ions
in the inner magnetosphere within the distance of plasmasphere, where the
plasma density is high \citep{Swift:1996, Lin2005, Lin2014}. In ANGIE3D, this
region is bounded by the near-Earth (inner) boundary, which is located at
a radial distance at $r \simeq 3.5 \ R_E$ in the inner magnetosphere. The
field-aligned currents, calculated near this inner boundary and mapped along
the geomagnetic dipole field lines down to the ionospheric altitude (1,000
km), are used as input to the ionospheric potential equation solved on a
sphere {\citep{Lin2014, Lin2021}}:
\begin{equation} \label{ }
\nabla \cdot (-{\bf \Sigma} \cdot \nabla \Phi ) = J_\parallel sin I,
\end{equation} 
where $\Sigma$ is the conductance tensor, $\Phi$ is the electric potential,
$J_\parallel$ is the mapped field-aligned current density, and $I$ is the
inclination of the dipole field at the ionosphere. The static analytical
model of Hall and Pederson conductance that accounts for EUV and diffuse
auroral contributions can be used for the conductance tensor. Similarly to
the global MHD models, the ionospheric electric field is mapped along the
dipole lines back to the inner magnetospheric boundary, to serve as a boundary
condition for the cold ion fluid \citep{Lin_Wang:2021}.

\subsection{Hybrid simulations with inertial electrons and EMHD simulations} 
\label{subsec:hybridfinitemass}

The MMS mission investigates physical processes like magnetic reconnection, shock waves and turbulence, 
which span from ion to  electron scales.  
Simulation studies of these processes should ideally cover full kinetic physics from ion to electron scales for which the necessary present and near-future computational resources are prohibitively expensive. Therefore simulation models, which cover different scale ranges and physical phenomena, are used.

Hybrid-kinetic plasma simulation model, introduced in the previous section \ref{subsec:hybridmassless} , 
treats ions as kinetic species and electrons as a massless fluid.
This restricts their applicability to physical processes in which not only electron kinetic effects are not important but also 
to the scales exceeding by far the electron scales. 
Hybrid-kinetic codes with inertia-less electrons, discussed in section \ref{subsec:hybridmassless}, can, therefore, be used  to simulate global phenomena and in some cases for specifically limited physics studies of magnetic reconnection, plasma turbulence and shock waves.
 
The validity of hybrid-kinetic model can, however, be extended down to electron length scales, viz., to electron inertial length by considering electrons as an inertial fluid \citep{Jain2023}. Since the electron kinetic physics is still ignored such plasma model might computationally be more feasible than the fully kinetic model and describe larger scale phenomena and plasma process like magnetic reconnection, plasma turbulence and shock formation in collisionless plasmas, in which electron scale structures develop.

\subsubsection{Model equations - inertial electrons}
\label{subsubsec:equations}
Hybrid-kinetic model treats ions as kinetic species and electrons as an inertial fluid. In hybrid-kinetic simulation codes, ion dynamics can be described by solving either the ion's Vlasov equation using Eulerian methods or the equations of motion for ion macro-particles using semi-Lagrangian Particle-in-Cell (PIC) method. Solving Vlasov equation is computationally more expensive. Here we discuss the hybrid-PIC codes which treat ions as Lagrangian macro-particles modelled via the PIC method. Following are the governing equations of hybrid-PIC  model.

\begin{align}
  \frac{d\mathbf{x}_i}{dt}    & =\mathbf{v}_i, \label{eq:xp}  \\
  m_i\frac{d\mathbf{v}_i}{dt} & =e(\mathbf{E} + \frac{\mathbf{v}_i\times \mathbf{B}}{c}), \label{eq:vp}\\
  \nabla \times \mathbf{E} & = -\frac{1}{c}\frac{\partial \mathbf{B}}{\partial t}, \label{eq:faraday_inertial} \\
  \nabla \times \mathbf{B} & = \frac{4\pi}{c} \mathbf{J}, \label{eq:ampere}\\
  \mathbf{J} & = e (n_i\mathbf{u}_i-n_e\mathbf{u}_e), \label{eq:J}\\
  n_i&=n_e \label{eq:quasi_neutrality}\\
  \mathbf{E} & = -\frac{\mathbf{u}_e\times \mathbf{B}}{c} -\frac{1}{en}\nabla p_e - \frac{m_e}{e}\left(\frac{\partial \mathbf{u}_e}{\partial t}+(\mathbf{u}_e\cdot\mathbf{\nabla})\mathbf{u}_e\right)  + \eta \mathbf{\jmath}, \label{eq:e_ohm}\\
  p_e&=C n_e^{\gamma} \label{eq:e_eos}
\end{align}

The electric and magnetic fields ($\mathbf{E}$ and $\mathbf{B}$ respectively) in Maxwell's equations, Eqs.~\eqref{eq:faraday_inertial} and \eqref{eq:ampere}, are coupled to the plasma dynamics via the total current density $\mathbf{J}=e (n_i\mathbf{u}_i-n_e\mathbf{u}_e)$ resulting from the bulk motion of ions (number density $n_i$, bulk velocity $\mathbf{u}_i$) and electrons (number density $n_e$, bulk velocity $\mathbf{u}_e$). Ion's number density $n_i$ and the bulk velocity $\mathbf{u}_i$ is obtained from their positions $\mathbf{x}_i$ and velocities $\mathbf{v}_i$ governed by Eqs. \eqref{eq:xp} and \eqref{eq:vp}. Electron dynamics is governed by quasi-neutrality condition, Eq. \eqref{eq:quasi_neutrality}, momentum equation of the inertial electron fluid, Eq. \eqref{eq:e_ohm}, and equation of state relating electron scalar pressure $p_e$ with electron number density $n_e$, Eq. \eqref{eq:e_eos}. Here, $e$ is the fundamental charge, $m_i$ ion mass, $m_e$ electron mass, $\mu_0$ magnetic permeability of vacuum, $\eta$ collisional resistivity, $\gamma$ the adiabatic constant and $C$ is a constant (to be determined from initial conditions).  
Eqs. \eqref{eq:xp}-\eqref{eq:e_eos} are the fundamental equations of the hybrid-kinetic model with inertial electron fluid \citep{Jain2023}. These equations differ from the equations of hybrid-kinetic model with inertia-less electron fluid only by the electron inertial terms proportional to $m_e$ on the RHS of Eq. \eqref{eq:e_ohm}. Addition of electron inertial terms in  Eq. \eqref{eq:e_ohm} makes the numerical solution of these equations much more involved in comparison to the case of inertia-less electron fluid. The algebraic calculation of electric field from Eq. \eqref{eq:e_ohm}  is not as straightforward as in the case of the inertia-less electron fluid. One needs to now calculate time derivative of $\mathbf{u}_e$ or find some other way to obtain electric field. The calculation of magnetic field also now requires numerical solution of additional elliptic partial differential equations arising because of the finite electron inertia. 

In majority of the hybrid-kinetic codes with inertial electrons, evolution of magnetic field is followed by solving  an evolution equation for the generalized vorticity $\mathbf{W}=\nabla\times\mathbf{u}_e-e\mathbf{B}/m_e c$  obtained by taking curl of Eq. \eqref{eq:e_ohm} and using Eq. \eqref{eq:faraday_inertial}. This equation is,
\begin{align} \label{eq:curl_emom}
	\frac{\partial \mathbf{W}}{\partial t}
	  & = \mathbf{\nabla}\times\left [\mathbf{u}_e\times \mathbf{W}\right]-\mathbf{\nabla}\times\left(\frac{\mathbf{\nabla} p_e}{m_en}\right)- \mathbf{\nabla}\times\left(\frac{e \eta}{m_e}\mathbf{\jmath}\right).
\end{align}
 The magnetic field is then calculated by solving an elliptic partial differential equation (PDE) which is obtained by substituting for $\mathbf{u}_e$ from Eq. \eqref{eq:ampere} and \eqref{eq:J}, $\mathbf{u}_e=\mathbf{u}_i-c\nabla\times\mathbf{B}/(4\pi e n)$, in the expression for $\mathbf{W}=\nabla\times\mathbf{u}_e-e\mathbf{B}/m_ec$.

\begin{align}
	\frac{c}{4 \pi e}\mathbf{\nabla}\times\left(\frac{\mathbf{\nabla}\times\mathbf{B}}{n}\right)+\frac{e\mathbf{B}}{m_ec} & = \mathbf{\nabla}\times\mathbf{u}_i-\mathbf{W}. \label{eq:elliptic}
\end{align}

Some of the hybrid-kinetic codes make approximations of electron inertial terms in Eq. \eqref{eq:e_ohm} and \eqref{eq:elliptic} to simplify their numerical solutions \citep{Lipatov2002,Shay1998,Kuznetsova1998}. Spatial density variations are neglected in Eq. \eqref{eq:elliptic} \citep{Shay1998,Kuznetsova1998}.  
The electric field was then calculated from the generalized
Ohm's law by neglecting the electron inertial term with
time derivatives of the electron fluid velocity~\citep{Kuznetsova1998}. Some codes neglected even the convective electron acceleration term~\citep{Shay1998}. These approximations are valid when length scale of variations is much larger than the electron inertial length. For a detailed discussion of these approximations, see Munoz et al. (2018) \citep{Munoz2018}. These hybrid-kinetic codes which partially included electron inertial effects have mainly been used to study collisionless magnetic reconnection~\citep{Shay1999,Kuznetsova2000,Kuznetsova2001}. In particular Shay et al. (1998) used an evolution equation for a scalar electron pressure~\citep{Shay1998} while Kuznetsova et al. (1998) included the  full electron pressure tensor to take into account the non-gyrotropic  effects~\citep{Kuznetsova1998} .

More recently a hybrid-kinetic code CHIEF (Code Hybrid With Inertial Electron Fluid) was developed \citep{Munoz2018}. This code solves Eqs. \eqref{eq:e_ohm} and \eqref{eq:elliptic} without making any of the electron inertia elated approximations used by other codes. The details of the numerical algorithm to solve Eqs. \eqref{eq:xp}-\eqref{eq:e_eos} are discussed by Munoz et al. (2018) \citep{Munoz2018}. CHIEF was used to simulate kinetic plasma turbulence and it was found that the electron inertia related approximations are not valid in electron scale current sheets formed in the turbulence \citep{Jain2022,Munoz2023}. 

Some hybrid-kinetic codes with electron inertia calculate electric field from  an elliptic PDE instead of Eq. \eqref{eq:e_ohm}  \citep{Amano2014,valentini2007}.  The  elliptic PDE for the electric field is obtained by taking curl of Faraday's law, Eq. \eqref{eq:faraday_inertial}, and utilizing Eqs. \eqref{eq:ampere} and \eqref{eq:e_ohm}. Electron inertia effects were considered in the elliptic equation for the electric field while, still, the electron inertia term was ignored that contains the divergence of the electric field. Two dimensional simulations of kinetic plasma turbulence have shown that this approximation is not valid from ion to electron scales \citep{Jain2022}.

\subsubsection{EMHD model: A special case of Hybrid-kinetic model with electron inertia in the limit of static ions \label{sec:emhd}}

Electron-magnetohydrodynamic (EMHD) model can be obtained as a special case of hybrid-kinetic model by taking the limit of stationary ions. However, it can also be obtained in the same manner from the Hall MHD equations (see Sec.~\ref{subsubsec:fluid-emhd}). In this model, ions provide a stationary charge neutralizing background for the electron dynamics. It can be applied to non-kinetic plasma processes at electron space and time scales during which ions do not respond or even at scales larger than electron scales as long as ions are stationary by some mechanism, e.g., fixed in lattice \citep{Gordeev1994}. EMHD model has been applied in the past to electron scale magnetic reconnection \citep{Bulanov1992,Mandt1994,Drake1994,Drake1997,Attico2000,Chacon2007,zocco2009,Jain2009,Jain2012,Jain2015,Jain2015a,Jain2015b} and is applicable to the electron-only reconnection recently discovered   by MMS observations \citep{Phan2018}.

Equations of EMHD can be derived from the equations of hybrid kinetic model with electron inertia, Eqs. \eqref{eq:xp}-\eqref{eq:e_eos}, by setting $\mathbf{u}_i=0$. In this limit, Eq. \eqref{eq:ampere} and \eqref{eq:elliptic} become,
\begin{align}
\mathbf{u}_e   &=-\frac{c}{4\pi n e} \nabla\times\mathbf{B} \label{eq:ampere_emhd}, \\
  \mathbf{W} &=	-\frac{c}{4\pi e}\mathbf{\nabla}\times\left(\frac{\mathbf{\nabla}\times\mathbf{B}}{n}\right)-\frac{e\mathbf{B}}{m_e c}. \label{eq:elliptic_emhd} 
\end{align}  

Here $n=n_i=n_e$ is constant as ions are stationary. Equations \eqref{eq:curl_emom}, \eqref{eq:ampere_emhd} and \eqref{eq:elliptic_emhd} form a closed set of equations for the evolution of magnetic field in EMHD model.

\subsubsection{Electron only reconnection: importance of electron inertia \label{subsubsec:importance_electron_inertia}}
In many of electron only reconnection events observed by MMS, guide magnetic field is significantly larger than the asymptotic value of the reconnecting component of magnetic field \citep{Phan2018,Zhou2021,Man2020,Stawarz2022}. In the statistical survey of electron only reconnection events in Earth's magnetosheath, guide field in majority of the events is  1 to 10 times larger than the reconnecting component of the magnetic field \citep{Stawarz2022}. MMS observations of electron scale reconnection in Earth's magnetosheath show that the electron non-gyrotropy, which balances the reconnection electric field in weak or zero guide field case, reduces with the increasing strength of the guide magnetic field \citep{Wilder2018}.  No evidence of agyrotropy was found in another MMS observations of large guide field reconnection \citep{Eriksson2016}. This is consistent with the fact that a strong guide magnetic field will magnetize electrons  in the electron diffusion region not allowing non-gyrotropy to develop if the scale length of the diffusion region is larger than the electron Larmor radius in the guide field.

In the case of the large guide magnetic field and thus weak or zero non-gyrotropy, the  reconnection electric field is expected to be balanced by  electron inertial terms in generalized Ohm's law. Indeed, PIC simulations of magnetic reconnection have shown that  electron inertial terms are significant to balance the reconnection electric field when guide field is large and/or current sheet has electron scale thickness \citep{Hesse1998,Hesse2016,Liu2013,Pritchett2005}. For the electron only events observed with large guide magnetic field and/or absence of non-gyrotropy,  electrons can be modeled as an inertial electron fluid as is done in hybrid kinetic simulations with electron inertia which are computationally less expensive in comparison to fully kinetic simulations.

\subsubsection{Magnetic reconnection through electron scale current sheets: EMHD simulations}
\label{EMHD}


Hybrid kinetic simulations of turbulence with electron inertia  show that the physics of electron only reconnection can be described by equations of an Electron-magnetohydrodynamic (EMHD) model which is a special case of hybrid-kinetic model with electron inertia in the limit of stationary and un-magnetized ions (see section \ref{sec:emhd}). This is expected as the ions do not couple in electron only reconnection events. Collisionless magnetic reconnection in electron scale current sheets has been investigated using EMHD model for several decades before the observational discovery of electron-only reconnection by MMS \citep{Bulanov1992,Mandt1994,Drake1994,Drake1997,Attico2000,Chacon2007,zocco2009,Jain2009,Jain2012,Jain2015,Jain2015a,Jain2015b}. It was, however, not termed as electron only reconnection as it was not conceived at that time that reconnection without ion participation is possible. It was rather termed as ``early phase of reconnection'' \citep{Jain2009,Jain2012,Jain2015a,Jain2015b}. Later, electron only reconnection in Earth's magnetotail was also interpreted as an early phase of the standard reconnection \citep{Hubbert2022,Farrugia2021} and there is ongoing discussion if the observed reconnection events without ion coupling are electron-only reconnection events or just an early phase of standard reconnection \citep{Lu2020,Lu2022,Wang2020,Yi2022}.
 
\cite{Jain2009} were the first to propose that the early phase of reconnection after its onset in electron scale current sheets will be dominated by electron dynamics without coupling to ion dynamics and carried out EMHD simulations to study the physics of the early phase. This study was followed by further theoretical and simulation studies, both in 2-D and 3-D, using EMHD model and comparison with space observations by Cluster spacecraft \citep{Jain2012,Jain2015a,Jain2015b,Jain2014,Jain2014a,Jain2015,Jain2017,jain2017a}. These and other EMHD studies are also relevant for electron only reconnection.

\begin{figure*}
  \includegraphics[width=\textwidth]{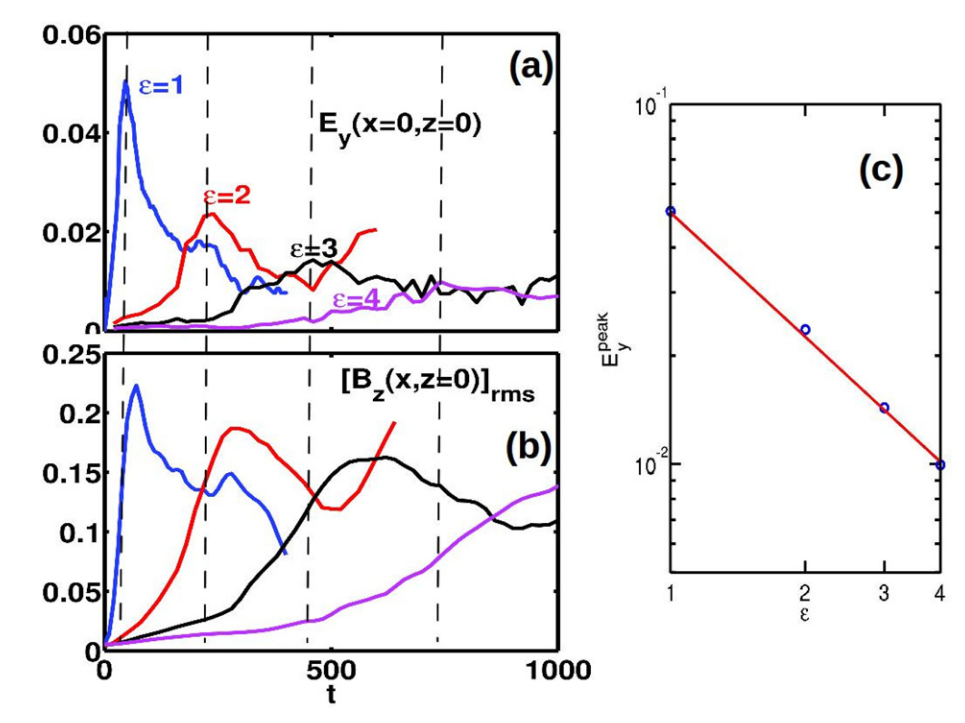}
  \caption{Evolution of (a) out-of-plane electric field $E_y$  at the X-point ($x=0,z=0$) and (b) root-mean-square value of the normal component of magnetic field $B_z$ evaluated over the length of the reconnecting current sheet (between two outflow regions)
    on the line $z=0$.
    Vertical dashed lines mark the times at which $E_y$ attains its peak value $E_{y}^{peak}$ for different values of the current sheet half-thickness $\epsilon$. (c) Scaling of $E_y^{peak}$ with $\epsilon$ (in log -scale) from simulations (blue circles) and a fit $E_y^{peak}=0.05/\epsilon^{1.15}$ (red line). Adapted from~\cite{Jain2015b}.  \label{fig:emhd_reconnection_rate}}
\end{figure*}

Fig. \ref{fig:emhd_reconnection_rate} shows results from the 2-D EMHD (x-z plane) simulations of magnetic reconnection in electron scale current sheets of different half-thicknesses ($\epsilon$) \citep{Jain2015b}. In these simulations, the equilibrium current density is $\mathbf{J}=-n_0eu_{ey0}\hat{y}=(B_{\infty}\,c/(4\pi\,\epsilon) \,\mathrm{sech}^2(z/\epsilon)\hat{y}$ corresponding to the equilibrium anti-parallel magnetic field $\mathbf{B}=B_{\infty}\tanh(z/\epsilon) \hat{x}$. The density $n_0$ is uniform and thus the current is due to the electron flow $u_{ey0}$. The results are shown in the normalized units: length by electron inertial length $d_e=c/\omega_{pe}=c/(4\pi n_0 e^2/m_e)^{1/2}$, magnetic field by $B_{\infty}$ and time by $\omega_{ce}^{-1}=(eB_{\infty}/m_ec)^{-1}$. Reconnection rate, measured by out-of-plane electric field $E_y$ at the X-point and shown in Fig. \ref{fig:emhd_reconnection_rate}a, reaches its peak value when the growth of the rms values of $B_z$ begins to slow down, consistent with the Faraday law which gives $\partial E_y/\partial x=-1/c\partial B_z/\partial t$. The peak reconnection rate $E_y^{peak}$ scales with $\epsilon$ as $E_y^{peak}=0.05/\epsilon^{1.15}$ and drops from $E_y^{peak}=0.05\, v_{Ae}B_{\infty}/c$ to $E_y^{peak}=0.01\, v_{Ae}B_{\infty}/c$ as the $\epsilon$ increases from $\epsilon=d_e$ to $\epsilon=4\,d_e$. This range of reconnection rate in units based on ion Alfv\'en speed is $E_y^{peak}=0.43-2.15\, v_{Ai}B_{\infty}/c$ (using ion $m_i/m_e=1836$) which is much larger than the value of the reconnection rate ($0.1\,v_{Ai}B_{\infty}$/c) for the standard ion-coupled reconnection. Reconnection rates for the electron only reconnection have been reported in the similar range by MMS observations \citep{burch2020} and PIC simulations \citep{Pyakurel2019}.

Note that the EMHD simulation results in Fig. \ref{fig:emhd_reconnection_rate} are independent of the strength of the guide magnetic field because a uniform guide field does not appear in 2-D EMHD equations \citep{Jain2015}. However, in 3-D, guide field can introduce current aligned instabilities in addition to the tearing instability as has been predicted by  3-D EMHD eigen value analysis \citep{Jain2015} and simulations \citep{jain2017a}. In 3-D, current aligned electron Kelvin-Helmholtz instabilities can grow in electron scale current sheets even in the absence of guide magnetic field \citep{Jain2014,Jain2014a,Greess2021}. These EMHD studies are relevant for the MMS observations of the electron shear flow generated electron Kelvin-Helmholtz vortices within the diffusion region of collisionless magnetic reconnection \citep{Zhong2022,Zhong2018,Hwang2019}.

\subsubsection{Outlook \label{sec:outlook}}
Hybrid-kinetic simulations with electron inertia provide a computationally less expensive (in comparison to fully kinetic simulations) tool to study magnetic reconnection with guide field in which bulk electron inertia is  the dominant mechanism breaking the frozen-in condition of magnetic field. There have been some hybrid-kinetic simulations studies with electron inertia of guide field magnetic reconnection \citep{Kuznetsova1998,Kuznetsova2000,Shay1998,Califano2020,Munoz2023}. More studies are, however, required to address still many open questions about the guide field magnetic reconnection. The EMHD limit (stationary ions) of the hybrid-kinetic model with electron inertia is particularly useful to study the nature of reconnection in electron scale current sheets and is relevant for the  recently discovered electron-only reconnection \citep{Phan2018}. At the same time, simulations of kinetic plasma turbulence from ion to electron scales using hybrid-kinetic model with electron inertia will shed light on the conditions under which electron scale current sheets form and reconnect with or without ion coupling in the turbulence.

\section{Fully Kinetic Particle-in-Cell Simulations}
\label{sec:PIC}

\subsection{Introduction}
\begin{figure}[t] 
\begin{center}
\includegraphics[width=1.00\textwidth]{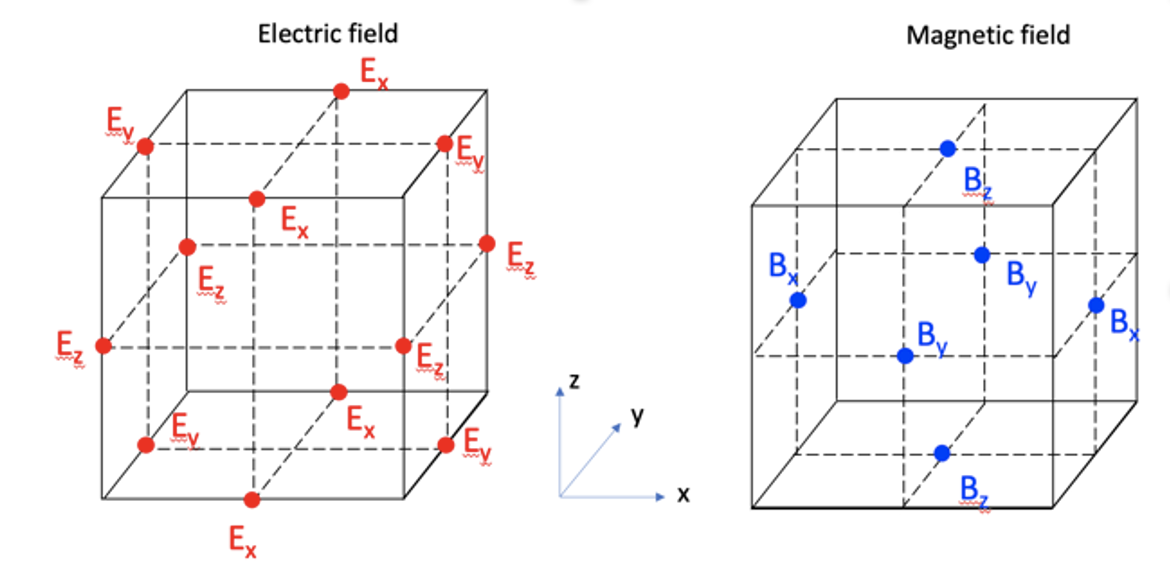}   
\caption{Yee lattice and electric and magnetic fields in a cell in a 3D case, where the length of each side is $\Delta x$. Electric fields $\mbox{\boldmath$E$}$ are defined at the midpoint of each side of the cube, while magnetic fields $\mbox{\boldmath$B$}$ are defined at the center of each face of the cube. In a 2D case, all the quantities are defined in the $x$-$y$ plane, projecting each position onto the cell in the $x$-$y$ plane.}
\label{figure-yee}
 \end{center} 
   \end{figure}

The particle-in-cell (PIC) method is a simulation method in which the plasma is treated as a collection of particles (electrons and ions), where each species is typically composed of up to $10^{12}$ particles in 3D cases. In PIC simulations, the motions of individual particles and the evolution of electric and magnetic fields are solved self-consistently. The electric and magnetic fields are defined on discrete grid points. In this subsection, we will explain an explicit PIC simulation, where all the quantities are updated based on the quantities obtained in the previous time step. Let us assume that the total particle number in a simulation is $N_p$ (in other words, $N_p/2$ for ions, and $N_p/2$ for electrons). The equation of motion for the $j$-th particle’s position $\mbox{\boldmath$x$}_j(t)$ and momentum $\mbox{\boldmath$p$}_j(t),$ where $j$ represents an integer between $1$ to $N_p$, is discretized in time, while Maxwell’s equations for electromagnetic fields $\mbox{\boldmath$E$}(\mbox{\boldmath$x$},t)$ and $\mbox{\boldmath$B$}(\mbox{\boldmath$x$},t)$ are discretized in both space and time, using a grid spacing $\Delta x$ (assuming that the grids are uniform in all the coordinates, i.e. $\Delta x=\Delta y= \Delta z$) and a time step $\Delta t,$ respectively. They are given as
\begin{equation}
    \frac{\mbox{\boldmath$x$}_j(t_n)-\mbox{\boldmath$x$}_j(t_{n-1})}{\Delta t}=\frac{\mbox{\boldmath$p$}_j(t_{n-1/2})}{\gamma_j(t_{n-1/2})},
    \label{eq-x-pic}
\end{equation}
\begin{equation}
    \frac{\mbox{\boldmath$p$}_j(t_{n+1/2})-\mbox{\boldmath$p$}_j(t_{n-1/2})}{\Delta t}=q_j\left[\mbox{\boldmath$E$}(\mbox{\boldmath$x$}_j(t_n),t_n)+\frac{\mbox{\boldmath$p$}_j(t_{n})}{c\gamma_j(t_{n})}\times\mbox{\boldmath$B$}(\mbox{\boldmath$x$}_j(t_n),t_n)\right],
    \label{eq-p-pic}
\end{equation}
\begin{equation}
    \frac{\mbox{\boldmath$E$}(\mbox{\boldmath$x$},t_{n+1})-\mbox{\boldmath$E$}(\mbox{\boldmath$x$},t_n)}{\Delta t}=-4\pi\mbox{\boldmath$J$}(\mbox{\boldmath$x$},t_{n+1/2})+c\nabla_f\times\mbox{\boldmath$B$}(\mbox{\boldmath$x$},t_{n+1/2}),
    \label{eq-e-pic}
\end{equation}
\begin{equation}
    \frac{\mbox{\boldmath$B$}(\mbox{\boldmath$x$},t_{n+3/2})-\mbox{\boldmath$B$}(\mbox{\boldmath$x$},t_{n+1/2})}{\Delta t}=-c\nabla_f\times\mbox{\boldmath$E$}(\mbox{\boldmath$x$},t_{n+1}),
    \label{eq-b-pic}
\end{equation}
where $\gamma_j$ is the Lorentz factor, $q_j$ is a
charge, $c$ is the speed of light, and the operator $\nabla_f \times$ represents the finite difference version of the curl operation. The time is discretized to be $t_a=a \Delta t$, where $a$ represents an integer $n$ or a half integer $n+1/2$. Note that $\mbox{\boldmath$B$}(\mbox{\boldmath$x$}_j(t_n), t_n)$ in the right-hand side of Eq. (\ref{eq-p-pic}) represents the mean of $\mbox{\boldmath$B$}(\mbox{\boldmath$x$}_j(t_n), t_{n+1/2})$ and $\mbox{\boldmath$B$}(\mbox{\boldmath$x$}_j(t_n), t_{n-1/2})$. As seen in Eqs. (\ref{eq-x-pic}) and (\ref{eq-p-pic}), the position $\mbox{\boldmath$x$}_j$ is computed at integer time, $t=t_n$, while the
momentum $\mbox{\boldmath$p$}_j$ is computed at half-integer time, $t=t_{n+1/2}$. This time staggering gives the second-order accuracy, i.e. the error is $O(\Delta t^2)$. In the same way, for the spatial discretization for $\mbox{\boldmath$E$}(\mbox{\boldmath$x$},t)$ and $\mbox{\boldmath$B$}(\mbox{\boldmath$x$},t)$, the Yee lattice \citep{yee1966} is used, in which each component of electric and magnetic fields is defined as in Fig. \ref{figure-yee}. Also, each component of the current density $\mbox{\boldmath$J$}(\mbox{\boldmath$x$},t)$ is defined at the same position as $\mbox{\boldmath$E$}(\mbox{\boldmath$x$},t)$. These fields, which are
spatially and temporarily staggered, are advanced using Eqs. (\ref{eq-e-pic}) and (\ref{eq-b-pic}), keeping the second-order accuracy in space and time. Eq. (\ref{eq-b-pic}) guarantees that the Gauss’s law for the magnetic field, $\nabla_f\cdot \mbox{\boldmath$B$}=0$, where $\nabla_f \cdot$ represents the finite difference version of the divergence, is satisfied when it is 
satisfied at $t=0$.

To solve Eq. (\ref{eq-p-pic}), the Boris method \citep{boris1970} is commonly used. This method has three steps: (1) The momentum is updated from $\mbox{\boldmath$p$}_j(t_{n-1/2})$ to $\mbox{\boldmath$p$}_j(t_n)^*$, using only the electric field
$\mbox{\boldmath$E$}(\mbox{\boldmath$x$}_j(t_n),t_n)$ for a half time step $\Delta t/2$. (2) The momentum vector $\mbox{\boldmath$p$}_j(t_n)^*$ is rotated to be $\mbox{\boldmath$p$}_j(t_n)^{**}$ using only the magnetic field $\mbox{\boldmath$B$}(\mbox{\boldmath$x$}_j(t_n),t_n)$ for a full time step $\Delta t$. (3) The rotated momentum is further updated from $\mbox{\boldmath$p$}_j(t_n)^{**}$ to $\mbox{\boldmath$p$}_j(t_{n+1/2})$, using the electric field $\mbox{\boldmath$E$}(\mbox{\boldmath$x$}_j(t_n), t_n)$ for another half time step $\Delta t/2$.

In the PIC method, each particle is not a point particle, but it has a finite size to reduce noise. The shape of a particle depends on simulation codes, but the most-commonly used shape function is a triangular function, $S_x(x-x_j)=(1-\mid x-x_j\mid /
\Delta x)$ when $|x-x_j|/\Delta x<1$ and zero otherwise, where only the $x$ component is considered. In 2D and 3D simulations, the $y$ and $z$ components of the shape functions are multiplied, as $S(\mbox{\boldmath$x$}-\mbox{\boldmath$x$}_j)=S_x(x-x_j)S_y(y-y_j)$ for 2D and $S(\mbox{\boldmath$x$}-\mbox{\boldmath$x$}_j)=S_x(x-x_j)S_y(y-y_j)S_z(z-z_j)$ for 3D. Using these shape functions, the charge density $\rho(\mbox{\boldmath$x$},t)$ is computed as $\rho (\mbox{\boldmath$x$},t)=\sum_jq_jS(\mbox{\boldmath$x$}-\mbox{\boldmath$x$}_j)$. This way of charge assignment is reversed to compute the electric field exerted from each grid point to a particle’s position. To avoid the self-force (the force due to the electric field generated by the particle itself), we must first average the electric fields defined on half-integer grids to obtain the mean electric field on each integer grid, before assigning the electric fields to the particle. The magnetic fields are assigned from grids to the particle position in the same way.

The current density can also be calculated using
$\mbox{\boldmath$J$}(\mbox{\boldmath$x$},t)=\sum_jq_j\mbox{\boldmath$v$}_jS(\mbox{\boldmath$x$}-\mbox{\boldmath$x$}_j)$, where $\mbox{\boldmath$v$}_j$ is the velocity, but the calculated $\mbox{\boldmath$J$}(\mbox{\boldmath$x$},t)$ does not satisfy the continuum equation, $[\rho(\mbox{\boldmath$x$},t_{n+1})-\rho(\mbox{\boldmath$x$},t_n)]/\Delta t+\nabla _f \cdot \mbox{\boldmath$J$}(\mbox{\boldmath$x$},t_{n+1/2})=0$;
therefore, the electric field calculated using Eq. (\ref{eq-e-pic}) with this $\mbox{\boldmath$J$}(\mbox{\boldmath$x$},t)$ does not satisfy the Gauss’s law, $\nabla_f \cdot \mbox{\boldmath$E$}(\mbox{\boldmath$x$},t_{n+1})=4\pi \rho(\mbox{\boldmath$x$},t_{n+1})$. This means that we must either correct the electric field $\mbox{\boldmath$E$}(\mbox{\boldmath$x$},t_{n+1})$ to satisfy the Gauss’s law, or use another
method to compute $\mbox{\boldmath$J$}(\mbox{\boldmath$x$},t)$. A technique for the former is explained in \cite{birdsall1991}. For the latter, for example, \cite{villasenor1992} developed a rigorous charge conservation method for
2D and 3D PIC simulations, which guarantees that both the Gauss’ law and the continuum equation are satisfied at the same time, when they are satisfied at $t=0$. 

The time step $\Delta t$ must satisfy the Courant–Friedrichs–Lewy condition, $\Delta t<\Delta x/(cN^{1/2}_d)$, where $N_d$ represents the dimensionality ($N_d=1, 2$, or 3). Also, the grid
spacing $\Delta x$ should be close to the Debye length $\lambda_D$, otherwise strong numerical heating occurs. Even when those conditions are satisfied, if particles are relativistic, a numerical Cherenkov
instability can occur and the noise field becomes extremely large. When this occurs, a noise reduction method such as by \cite{Godfrey1980} can be used.

Various boundary conditions can be implemented including periodic, conducting wall, and open boundaries \citep{daughton2006, ohtani2009}.

\subsection{Magnetotail reconnection}
In the Earth’s magnetotail, the strength of magnetic field across the current sheet is symmetric, and the guide field ($B_y$ field in the GSM coordinates) is small in many reconnection events. Many authors have been studying symmetric magnetic reconnection with zero guide-field, using the Earth’s magnetotail parameters \citep{hoshino1987,pritchett1991,horiuchi:1994,dreher1996,Zhu1996,hesse1996}. In these simulations, the initial plasma is set up based on a Harris equilibrium \citep{harris1962equilibrium}: the magnetic field $B_x=B_0\mbox{tanh}(z/w)$, and the density $n=n_0\mbox{sech}^2(z/w)+n_b$, where $B_0$ is the asymptotic magnetic field, $w$ is the sheet thickness, $n_0$ is the peak density of the current sheet, and $n_b$ is the background density. Also the conditions for the Harris equilibrium, $B_0^2/(8\pi)=n_0(T_i+T_e)$ and $|V_{di}-V_{de}|=[2c/(weB_0)](T_i+T_e)$, and $V_{di}/V_{de} =- T_i /T_e$, are satisfied, where $T_i$ and $T_e$ are the ion and electron temperatures, respectively, $e$ is the elementary charge, and $V_{di}$ and $V_{de}$ are the $y$-directional drift velocity in the current sheet component of ions and electrons, respectively. 

The following describes an example of a 2D PIC simulation of magnetotail reconnection, \cite{hesse2018a}. The system size is $L_x\times L_z=102.4d_i \times 51.2d_i$, where $d_i$ is the ion skin depth, $c/\omega_{pi}$ with $\omega_{pi}$ being the ion plasma frequency based on $n_0$ ($\omega_{pi}=(4\pi n_0e^2/m_i)^{1/2}$), and 3200 $\times$ 3200 grids are used. The mass ratio is $m_i/m_e=100$, the sheet thickness is $w=0.5d_i$, the temperature ratio is $T_i/T_e =5$, the density ratio $n_b/n_0=0.2$, and the ratio of the plasma frequency (based on $n_0$) to the electron cyclotron frequency (based on $B_0$) is $\omega_{pe}/\Omega_e=(4\pi n_0e^2/m_e)^{1/2}/[eB_0/(m_ec)]=2.0$, which gives the ratio of the light speed to the Alfv\'en speed (based on $B_0$ and $n_0$, $v_{A0}=B_0/(4\pi m_in_0)^{1/2}$) to be $c/v_{A0}=20.0$. The $x$ boundaries are periodic, and the $z$ boundaries are conducting walls. To initiate magnetic reconnection, a perturbation is added to the magnetic field as $\delta B_x=(a_0\pi/L_z)\cos(2\pi x/L_x)\sin(\pi z/L_z)$ and $\delta B_z=-(a_02\pi/L_z)\sin(2\pi x/L_x)\cos(\pi z/L_z)$, which gives a reconnection X-line at the origin $x=0$ and $z=0$. The total number of particles used in the simulation is $7 \times 10^{10}$.

\begin{figure}[!p] 
\begin{center}
\includegraphics[width=1.00\textwidth]{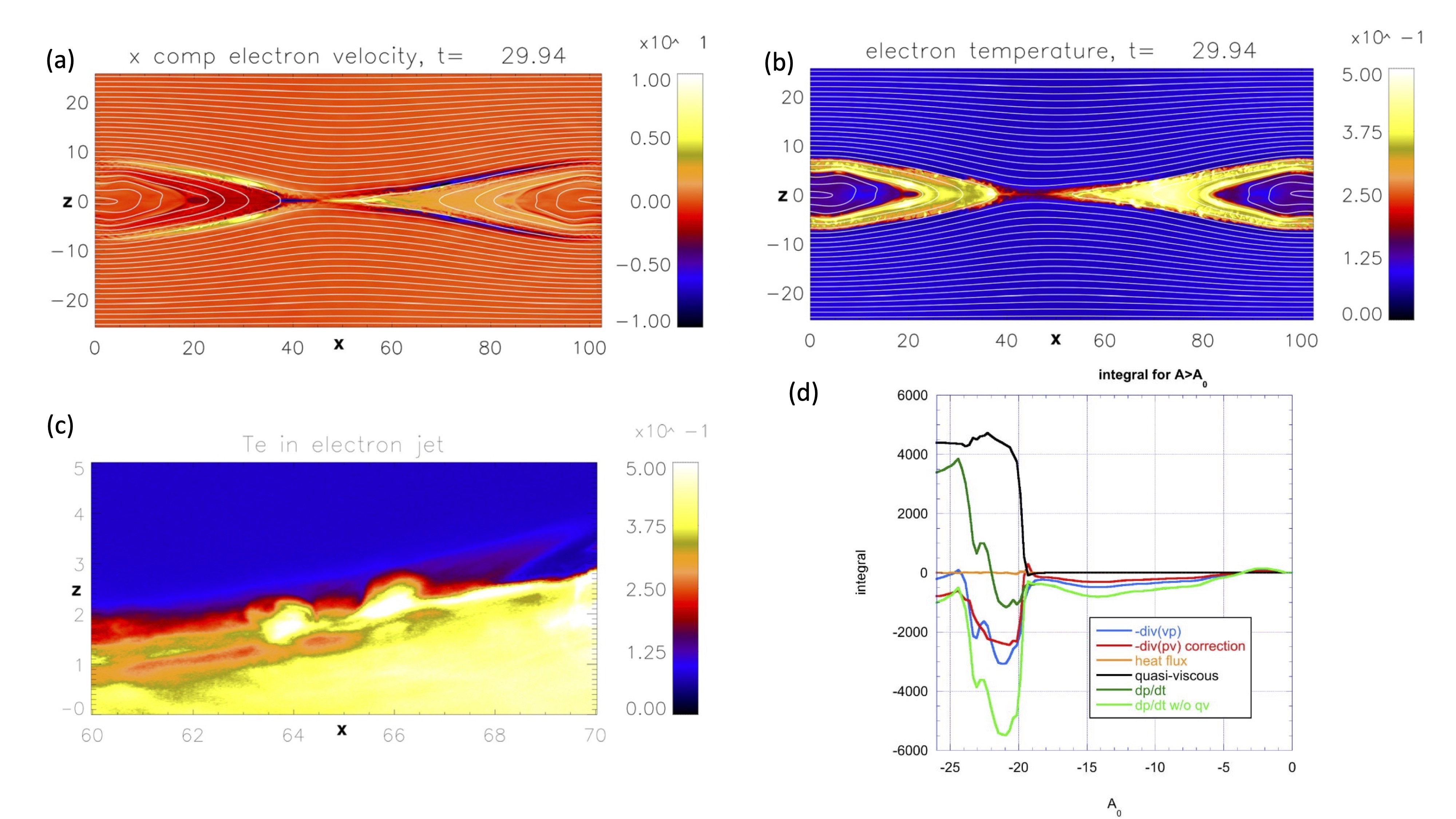}   
\caption{2D PIC simulation result for magnetotail reconnection. (a) Electron fluid velocity $V_{ex}$, (b) electron temperature $T_e$, (c) zoom-in view of $T_e$, and (d) the heating term in Eq.(\ref{eq-quasi-p}). Adapted from \cite{hesse2018a,hesse2019}.}
\label{figure-pic-mtail}
 \end{center} 
   \end{figure}

After the simulation starts, the reconnection electric field $E_y$ is generated in the diffusion region near the X line, where both the ion and electron motions are decoupled from the magnetic field line motion, which allows a pair of magnetic field lines across the current sheet, one is in the positive $z$ region ($B_x>0$) and the other is in the negative $z$ rection ($B_x<0$), are reconnected and energy conversion occurs from the magnetic energy to the kinetic and thermal energies of ions and electrons. The reconnection rate is measured as $E_y/(B_0v_{A0}/c)$, and in this simulation it is around 0.2 \citep{hesse2018b}. Both ion and electron outflows are produced from the X line toward the positive and negative $x$ directions. Fig. \ref{figure-pic-mtail}(a) shows the electron fluid velocity $V_{ex}$, where the bipolar positive and negative $V_{ex}$ peaks appear along the $z=0$ in $40<x/d_i<60$, and each peak value reaches near the electron Alfv\'en speed, $v_{Ae}=B_0/(4\pi m_en_0)^{1/2}=(m_i/m_e)^{1/2}v_{A0}$. Outside the region of $40<x/d_i<60$, there are strong inflows, which also reach near $v_{Ae}$, toward the X line along the separatrices. Because of these strong counter streaming electron flows (outflows and inflows), an electrostatic instability occurs that produces waves propagating along the separatrices toward the X line, and electrons are heated due to wave-particle interactions. Fig. \ref{figure-pic-mtail}(b) and (c) show the electron temperature in the entire box and a zoom-in view that includes a separatrix. Electrons are heated inside the separatrices (panel (b)), and the zoom-in view (panel (c)) shows that there are two solitary structures due to the nonlinear evolution of the wave (at $x=64d_i$ and $66d_i$ along the separatrix) where electron temperature significantly enhances. 

The locations of the instability, along the separatrices, correspond to the boundary of the high electron temperature, which suggests the importance of the electrostatic instability to heat electrons. To understand the effect of the instability on the heating, \cite{hesse2018a} analyzed the pressure equation:
\begin{equation}
\frac{\partial p}{\partial t}=-\nabla\cdot(\mbox{\boldmath$V$}p)-\frac{2}{3}\sum_lP_{ll}\frac{\partial}{\partial x_l}V_{l}-\frac{1}{3}\sum_{l,j}\frac{\partial}{\partial x_i}Q_{lii}-\frac{2}{3}\sum_{l,i (l\neq i)}P_{li}\frac{\partial}{\partial x_i}V_l,
\label{eq-quasi-p}
\end{equation}
where all the quantities are for electrons (the subscript $e$ is omitted): $p$ is the scalar pressure, \mbox{\boldmath$V$} is the fluid velocity, and $P_{ij}$ and $Q_{ijl}$ are the pressure tensor and the heat tensor, respectively. The first two terms represent the compression effect, the third term is due to the heat flux, and the fourth term represents the quasi-viscous effect due to the off-diagonal components of the pressure tensor. Fig. \ref{figure-pic-mtail}(d) shows the contribution of each term in Eq.(\ref{eq-quasi-p}), integrated over the region of $A>A_0$, where $A$ is the $y$ component of the mangnetic flux function (i.e., $B_x=\partial A/\partial z$ and $B_z=-\partial A/\partial x$). Note that $A=0$ at the outermost $z$ boundaries, and $A$ is decreasing as we approach $z=0$. Fig. \ref{figure-pic-mtail}(d) indicates that the quasi-viscous term (the fourth term) is the dominat term to provide the pressure increase, leading heating in the reconnection region.
\subsection{Magnetopause reconnection}

PIC simulations of magnetopause reconnection can include certain challenges due to asymmetries in the densities, temperatures and magnetic field strengths of the abutting plasmas \citep{sonnerup81a,cassak07a}.  Specifically, magnetospheric plasma is usually relatively sparse, hot, and threaded by a strong magnetic field, while magnetosheath plasma (which arises from shocked solar wind) is denser, cooler, and includes a somewhat weaker field.

From a simulation perspective, the density asymmetry -- which can exceed an order of magnitude -- can be particularly problematic. PIC simulations are inherently noisy.  The random fluctuations tend to follow Poissonian statistics with an amplitude scaling as $1/\sqrt{N_{pc}}$, with $N_{pc}$ the number of (macro) particles per computational cell.  If variations in the number of macroparticles directly translate to variations in density, a $16:1$ ratio between the magneosheath and magnetospheric plasma densities will produce noise levels an unaceptable four times larger in the latter than the former.  One obvious approach -- throwing more particles at the problem -- can quickly become computationally burdensome. An alternative is the use of particle weighting, in which each particle is assigned a weight $w$ (ranging, say, from $0$ to $255$) that determines its significance in the calculation of particle moments (e.g., charge and current density).  Doing so allows for an initially uniform distribution of particles with a roughly constant noise level. More sophisticated algorithms allow for the splitting and joining of particles as the simulation progresses to account for the development of density variations and to address computational load imbalances.

A 2D simulation of the magnetopause with {\tt p3d}, a PIC code employing weighted particles \citep{zeiler02a}, was presented in
\citet{swisdak18a}.  In its normalization, a reference magnetic field strength $B_0$ and density $n_0$ define the velocity unit
$v_{A0}=B_0/(4\pi m_in_0)^{1/2}$.  Times are normalized to the inverse ion cyclotron frequency $\Omega_{i0}^{-1}=m_ic/(eB_0)$, lengths to the ion inertial length $d_{i0} =c/\omega_{pi0}$ (where $\omega_{pi0} = (4\pi n_0 e^2/m_i)^{1/2}$ is the ion plasma frequency), electric fields to $v_{A0}B_0/c$, and temperatures to $m_iv_{A0}^2$. 

The initial conditions closely mimic those observed during the
diffusion region encounter described in \citet{burch16a}.  In the system considered here, $B_0$ and $n_0$ correspond to their asymptotic magnetosheath values: $B_0 = 23\text{ nT}$ and $n_0 = 11.3\text{ cm}^{-3}$. The simulation uses an $LMN$ coordinate system in which the reconnecting field parallels the $L$ axis (roughly north-south), the $M$ axis runs roughly east-west, with dawnward positive, and the $N$ axis points radially away from Earth and completes the right-handed triad.  The computational domain has dimensions $(L_L,L_N) = (40.96,20.48)$ with periodic boundary conditions used in all directions.  While particles can move in the $M$ direction, variations in physical quantities are not permitted: $\partial/\partial M = 0$.

The reconnecting component of the field $B_L$ and the ion and electron temperatures, $T_i$ and $T_e$, vary as functions of $N$ with hyperbolic tangent profiles of width 1. The asymptotic values of $n$, $B_L$, $T_i$, and $T_e$ in code units are 1.0, 1.0, 1.37, and 0.12 in the magnetosheath and 0.06, 1.70, 7.73, and 1.28 in the magnetosphere. Pressure balance determines the initial density profile. The guide field $B_M=0.099$ is much smaller than $B_L$ (i.e., the reconnection is nearly anti-parallel) and initially uniform.  While not an exact
kinetic equilibrium, the unperturbed configuration is in force balance and would not undergo significant evolution during the timescales of interest.  Instead, a small initial perturbation is introduced to trigger reconnection onset.

The ion-to-electron mass ratio is chosen to be $100$, which is
sufficient to separate the electron and ion scales (the electron
inertial length $d_{e0} = 0.1d_{i0}$).  The normalized speed of light is $c=15$ so that $\omega_{pe}/\Omega_{e}=1.5$ in the asymptotic magnetosheath and $\approx 0.2$ in the asymptotic magnetosphere; the observed ratios are larger, $\approx 46$ and $7$, and as a consequence the simulation's Debye length is larger than in the real system. However, since the development of reconnection does not appreciably depend on physical effects at the Debye scale the expected impact is minimal.  The spatial grid has resolution $\Delta = 0.01$ in normalized units while the Debye length in the simulation's magnetosheath, $\approx 0.03$, is the smallest physical scale. To ameliorate numerical noise, particularly in the low-density magnetosphere, each grid cell initially contains 3000 weighted macroparticles.

\begin{figure}
\includegraphics[width=\textwidth]{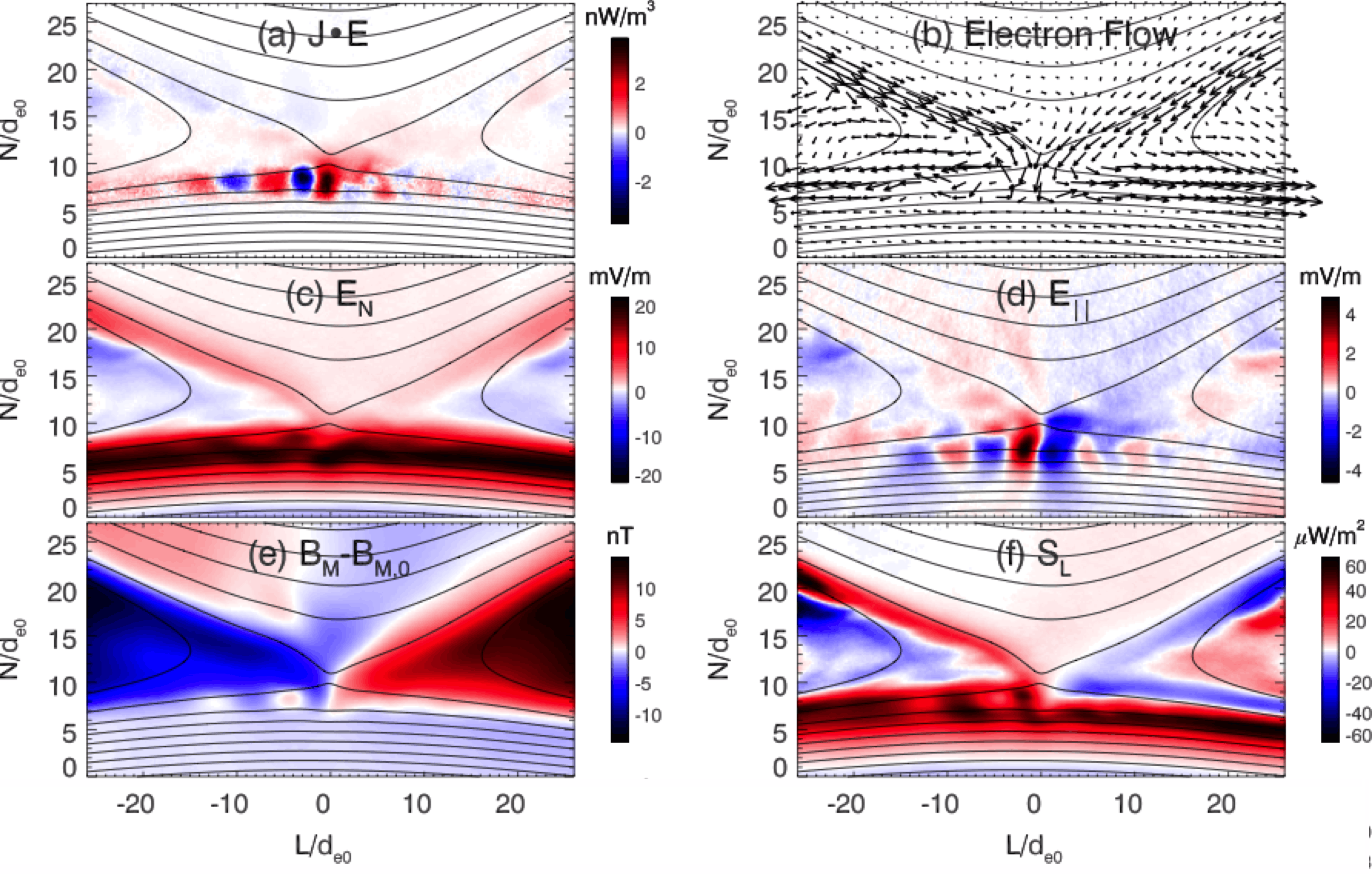}
\caption{\label{jdotefig} Simulation results overplotted with magnetic field lines. (a) The $\mathbf{J}\boldsymbol{\cdot}\mathbf{E}$ term from Poynting's theorem; (b) In-plane electron flow field; (c) $E_N$, the normal component of the electric field; (d) $E_{\parallel}$, the component of the electric field parallel to the magnetic field; (e) $B_M-B_{M,0}$, the change in the out-of-plane component of the magnetic field from its (spatially constant) initial value; (f) $S_L$, the horizontal component of the Poynting flux. Adapted from \cite{swisdak18a}.}
\end{figure}

Figure \ref{jdotefig} shows results.  The asymmetry in the field
strength is apparent in the distribution of the field lines, with the separatrices extending much farther (in the $N$ direction) into the plasma of the magnetosheath (top) than the magnetosphere (bottom). The Hall electric and magnetic fields (panels c and e) differ significantly from the case of symmetric reconnection, with the former concentrated almost exclusively on the magnetospheric side while the latter is almost completely dipolar rather than quadrupolar. Due to the use of weighted particles, the numerical noise is similar on both sides.

\subsection{The 3D nature of magnetic reconnection}

The third dimension out of the 2D reconnection plane introduces numerous additional plasma instabilities (e.g. \citet{daughton:2011}). In this subsection, we will focus more on the inherent 3D nature of reconnection X-line itself, including the effect of limited X-line extent, its tendency of spreading, and its orientation
preference.

\begin{figure}
\begin{center}
 \includegraphics[width=0.70\textwidth]{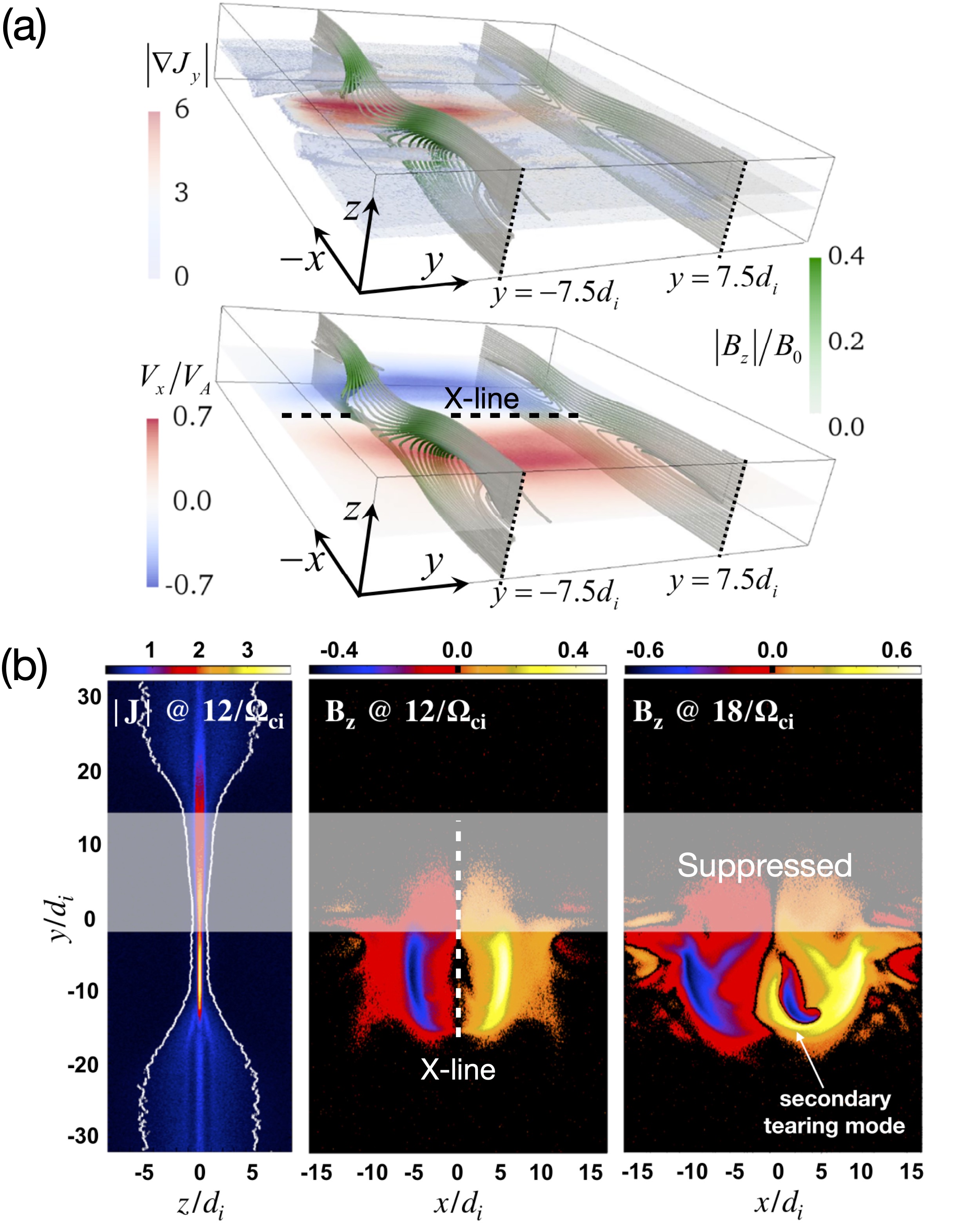}   
\caption{3D PIC simulation results. (a) 3D view of reconnection with a limited X-line extent, where the thin current sheet region extends $30d_i$ in $y$ \citep{huang2020}. The mass ratio in the simulation is 25. (b) The current density on the $x=0$ plane (left) and magnetic field $B_z$ on the $z=0$ plane (middle and right) \citep{liu2019}. The mass ratio is 75. The gray shaded area represents the ``suppressed reconnecting region". Adapted from \cite{huang2020} and \cite{liu2019}.}
\label{fig-pic-3d}
\end{center}
\end{figure}

To sustain a current sheet, electrons and ions drift in the opposite directions. This fact introduces the asymmetry along the X-line (current) direction. To reveal this effect, \cite{liu2019} and \cite{huang2020} studied magnetic reconnection with the X-line being spatially confined in the current direction. They included thick current layers to prevent reconnection from spreading out of the two ends of a thin current sheet that has a thickness on an ion inertial ($d_i$) scale. The $x$ component of the magnetic field is given as $B_x=B_0\mbox{tanh}[z/L(y)]$, where the half-thickness $L(y)=L_{min}+(L_{max}-L_{min})[1-f(y)]$ and $f(y)=[\mbox{tanh}((y+w_0)/S)-\mbox{tanh}((y-w_0)/S)]/[2\mbox{tanh}(w_0/S)]$, $L_{min}=0.5d_i$, $L_{max}=4d_i$, and $S=5d_i$. The parameter $w_0$, which controls the $y$-extent of the thin current region ($L_{y-thin}$), is varied from $w_0=2d_i$ to $20d_i$, corresponding to the $y$-extent of the thin current region from $L_{y-thin}\sim 4d_i$ ($w_0=2d_i$) to $L_{y-thin}\sim 30d_i$ ($w_0=20d_i$). The density is $n=n_0\mbox{sech}^2[z/L(y)]+n_b$, and $n_b=0.3n_0$. The system size is $L_x\times L_y\times Lz=32d_i\times 64d_i\times 16d_i$. The periodic boundary condition is used for $x$ and $y$, and the conducting walls are placed in the $z$ boundaries. Over $2.6\times 10^{10}$ particles for each species are used.

The resulting reconnection is shown in Fig. \ref{fig-pic-3d}, which is for  $L_{y-thin}\sim 30d_i$. \cite{liu2019} found that the reconnection rate and the outflow speed drop significantly when the extent of the thin current sheet, $L_{y-thin}$, is less than $\mathcal{O}(10d_i)$. When the thin current sheet extent is long enough, it consists of two distinct regions: a suppressed reconnecting region (on the ion-drifting side) exists adjacent to the active region where reconnection proceeds normally as in a 2D case with a typical fast rate value $\approx 0.1$. The extent of this suppression region
is $\mathcal{O}(10d_i)$, and it suppresses reconnection when $L_{y-thin}$ is comparable or shorter. 
The time scale of current sheet thinning toward fast reconnection can be translated into the spatial scale of this suppression region, because the electron drifts inside the ion diffusion region transport the reconnected magnetic flux (that is critical in driving outflows and furthers the current sheet thinning) away from this region. This is a consequence of the Hall effect in 3D.

\cite{huang2020} incorporated the length scale of this suppression region $\mathcal{O}(10d_i)$ to quantitatively model the reduction of the reconnection rate and the maximum outflow speed observed in the short X-line limit. The average reconnection rate drops because of the limited active region (where the current sheet thins down to the electron inertial scale) within the X-line. The outflow speed reduction correlates with the decrease of the $J\times B$ force, which can be modeled by the phase shift between the $J$ and $B$ profiles, also as a consequence of the flux transport out of the reconnection plane.

While the existence of this suppression region may explain the shortest possible azimuthal extent of dipolarizing flux bundles at Earth \citep{liu2015}, it may also explain the dawn-dusk asymmetry observed at the magnetotail of Mercury \citep{Sun2016,Sun2022}, which has a global dawn-dusk extent much shorter than that of Earth.

\subsection{Particle acceleration}

There have been quite remarkable advances in using PIC simulations to understand particle acceleration processes in magnetic reconnection, discussed in \citet{Oka2023}, Drake et al., and \citet{Guo2023} of this issue. The simulation provided energetic particle flux, spectra and even detailed distributions that can be compared with in situ observations. We introduce several key diagnostics recently used for gaining insight in particle energization.

\begin{figure}
\begin{center}
 \includegraphics[width=1.00\textwidth]{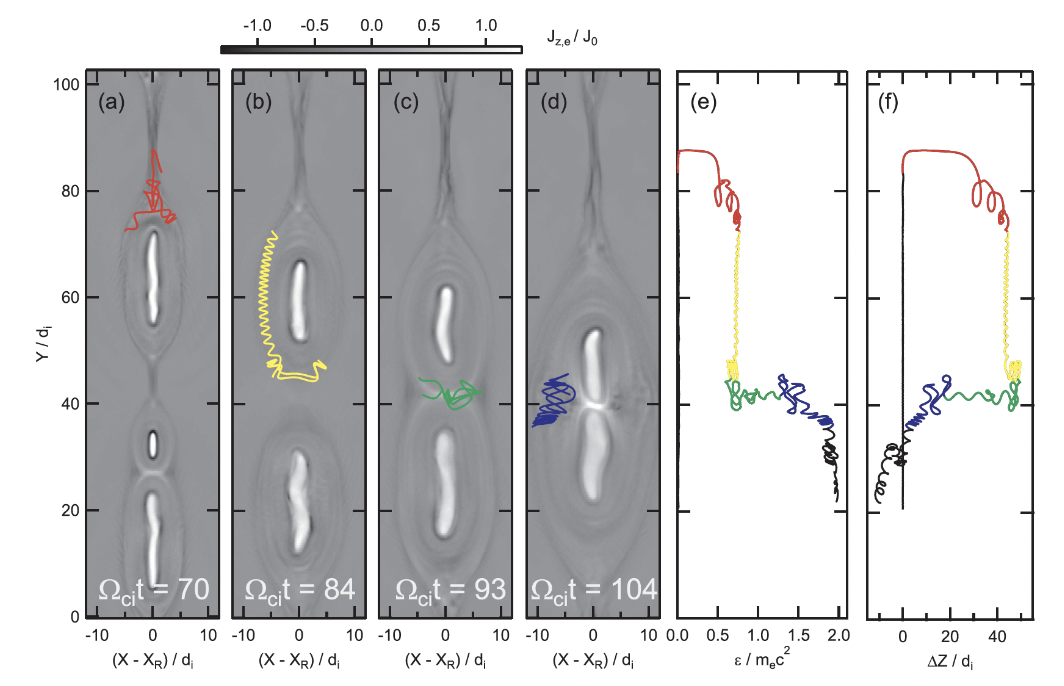}   
\end{center}
\caption{An example of particle trajectory analysis. Adapted from~\cite{Oka2010}. \label{figure-acceleration1}}
\end{figure}

First, it has been a common practice to output particle trajectories to study the acceleration process \citep[e.g.,][]{Hoshino2001,Drake2006,Fu2006,Oka2010,Guo2015}. These have led to the identification of different acceleration mechanisms, as discussed in \citet{Oka2023} and \citet{Guo2023} of this issue. Fig. \ref{figure-acceleration1} shows a representative particle trajectory adapted from \cite{Oka2010}. This particle is first accelerated by an X line (a), then further energized due to electric field during anti-reconnection between two merging island (c). The acceleration persists after the particle is ejected out of the X-line region. In addition, one can output the electric and magnetic fields and other quantities associated with particles, to complement the understanding of acceleration mechanisms.

The limitation of just showing several particle trajectories, even with the best effort, is that the “representative” examples are usually cherry-picking results and it is difficult to evaluate the relative importance of each mechanism. There has been recent effort to evaluate the acceleration processes  over a large number of trajectories, \citep{Guo2019,Guo2021,Kilian2020,French2022,Li2023}.

\begin{figure}
\begin{center}
 \includegraphics[width=1.00\textwidth]{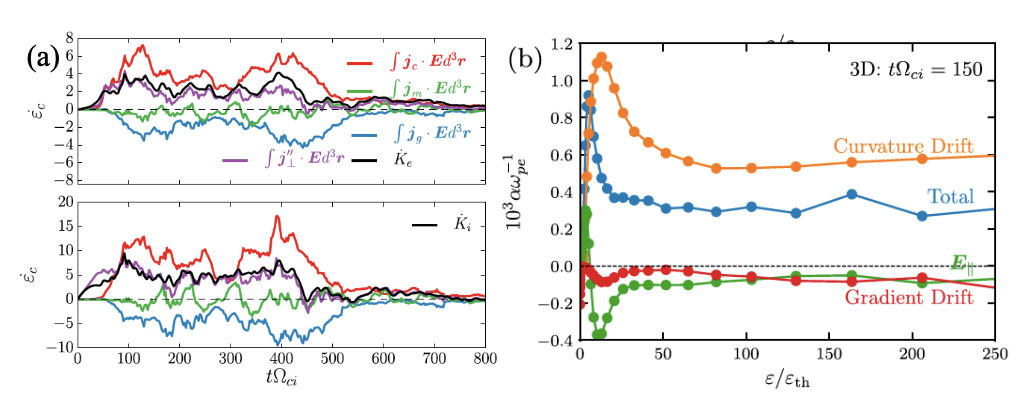}   
\end{center}
\caption{(a) An example of guiding-center drift analysis. Particle energization due to different drift currents for electrons (top) and ions (bottom). $j_c$ is due to particle curvature drift. $j_g$ is due to particle grad-$B$ drift. $j_m$ is due to magnetization. $j\prime\prime = j_c +j_g +j_m$. $\dot{K_e}$ and $\dot{K_i}$ are the energy change rates for electrons and ions, respectively. They are all normalized
by $m_ec^2\omega_{pe}$. (b) Similar to (a) but shows energy dependent values \citep{Li2019} at a given time. Adapted from \cite{Li2017}. \label{figure-acceleration2}}
\end{figure}

Another method, developed and widely used over the last decade, is to study the collective energy gain, such as guiding center and pressure-restrained terms (discussed in \citet{Oka2023}, this issue) using ensemble averaged moments \citep{Dahlin2014,Dahlin2015,Li2015,Li2017,Li2018,Li2019,Du2018}. For example, the acceleration due to the curvature (Fermi) and gradient (betatron) drifts can be evaluated under guiding center approximation. Fig. \ref{figure-acceleration2} shows an example under the guiding-center approximation, and shows the curvature drift term is the main acceleration term. Moreover, it is possible to collect the energy dependent acceleration rates by considering particles with different energy, so the energization can be studied in a energy-dependent fashion~\citep{dahlin2017_role,Guo2014,Li2018,Li2019}. Fig. \ref{figure-acceleration2}b shows such an example.

\subsection{Simulations of magnetic reconnection in shock waves}
Magnetic reconnection can occur in current sheets generated in plasma turbulence, and PIC simulations have also been applied to turbulent environments \citep{wu2013,Matthaeus2016,Hhggerty2017,Shay2018,vega2020,adhikari2021,rueda2021}, including turbulence in Kelvin Helmholtz vortices in the magnetopause flank region \citep{nakamura-t2014,nakamura-t2017,nakamura-t2022}, and the transition region in shock waves \citep{matsumoto2015,bohdan2017,bohdan2020,bessho2019,bessho2020,bessho2022,bessho2023,ng2022}. 

Here, let us review 2D and 3D PIC simulation studies of magnetic reconnection in the shock turbulence. \cite{bessho2019} used a 2D domain to study a quasi-parallel shock under the parameters in the Earth’s bow shock. The size of the simulation domain is $L_x\times L_y=375d_i\times 51.2d_i$, where the ion skin depth $d_i$ has 40 grids. The plasma is uniform at $t=0$, both ions and electrons are Maxwellian with their temperatures $T_i$ and $T_e$, respectively, and the magnetic field is given as $\mbox{\boldmath$B$}=[B_0\cos\theta, B_0\sin\theta, 0]$, where $\theta$ is the shock angle with respect to the $x$ axis.  Periodic boundaries are used in the $y$ direction, and conducting walls are placed in the $x$ direction. To all the plasma particles, a negative drift speed, $-v_d$, in the $x$ direction is given, and a uniform positive $z$ component of electric field, as $E_z=v_dB_0\sin\theta/c$, is set in the domain. At the right boundary, $x=L_x$, new particles for both ions and electrons are injected, using the same temperatures as the initially loaded particles, with the negative drift speed $-v_d$. At the left boundary, $x=0$, all the particles are specularly reflected, and the incident particles and the reflected particles generate counter-streaming beams, which cause a beam instability. As a result, a non-linear wave grows near the left boundary, and a wave steepening occurs. Eventually, a shock wave forms, propagating toward the positive $x$ direction. 

In the 2D simulation, the following parameters are used: the electron and ion beta $\beta_e=\beta_i=1$, the ratio of the plasma frequency to the electron cyclotron frequency $\omega_{pe}/ \Omega_e=4$, the shock angle $\theta=25^{\circ}$, and the mass ratio $m_i/m_e=200 $. 
With these parameters, the electron thermal speed becomes $v_{Te}=14.4v_A$. 
The drift speed is set to be $v_d=9v_A$. In the simulation (the downstream rest frame), the shock speed is $2.4v_A$, which corresponds to the Alfv\'en Mach number of the shock wave $M_A=11.4$. In other words, the shock speed in the laboratory frame is $11.4v_A$, which is less than $v_{Te}$, consistent with the Earth’s bow shock.

\begin{figure}
\begin{center}
 \includegraphics[width=1.00\textwidth]{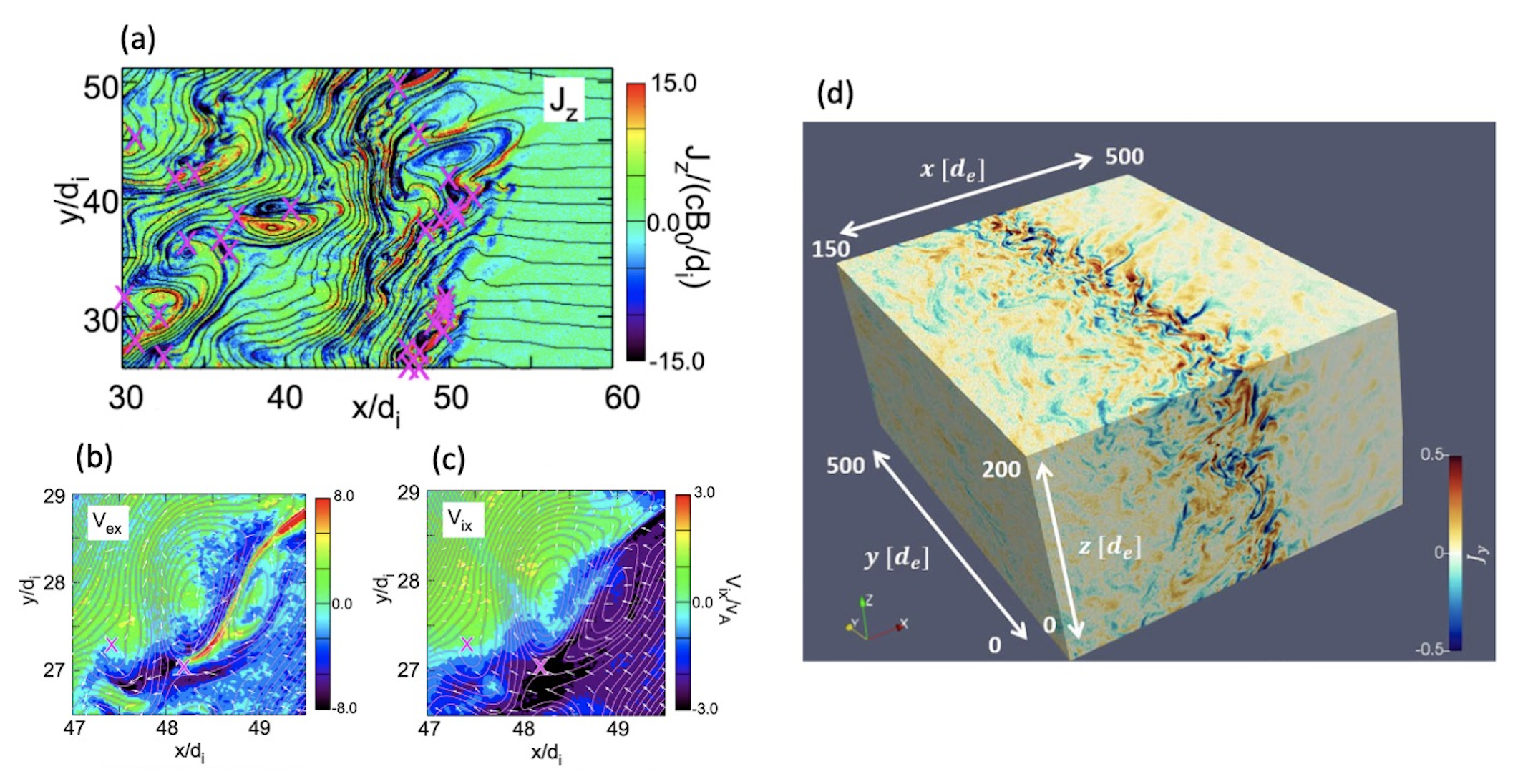}   
\caption{PIC simulations of reconnection in shocks. (a) 2D simulation domain and the current density $J_z$. (b) Electron fluid velocity $V_{ex}$. (c) Ion fluid velocity $V_{ix}$. (d) 3D simulation domain and the current density $J_z$. Adapted from ~\cite{bessho2019} and~\cite{ng2022}.  \label{fig-pic-shock}}
\end{center}
\end{figure}

In the simulation, the shock transition region shows a non-resonant ion-ion beam instability due to the interactions between the ions reflected by the shock and the incident ions, and many current sheets are generated, some of which show signatures of magnetic reconnection. In Fig. \ref{fig-pic-shock}(a), the color shows the current density $J_z$ in the 2D simulation domain, where the black curves are magnetic field lines projected onto the $x$-$y$ plane, and magenta X marks represent the positions of reconnection X lines. One of the reconnecting current sheet is zoomed up in Fig. \ref{fig-pic-shock}(b) and (c), where the electron fluid velocity $V_{ex}$ and the ion fluid velocity $V_{ix}$ are shown. There is one magnetic island above the current sheet, and there are bipolar electron jets generated from the X line. In contrast, the ion velocity plot does not show ion jet structures, and the ions are passing through the reconnection region with a negative $V_{ix}$. Therefore, this region is a site of electron-only reconnection, where only electrons are participating in reconnection, while ions cannot respond to the strong gradient of magnetic fields in the thin current sheet, whose thickness is less than the ion skin depth $d_i$. Electron-only reconnection has been observed in the Earth's magnetosheath \citep{Phan2018,gingell2021,Stawarz2022} and the transition region of the Earth's bow shock \citep{Wang2019,Gingell2019, Gingell2020}. Note that the shock transition region has a negative $B_z$ magnetic field, $B_z \sim-4B_0$ (not shown), and the reconnecting magnetic field is the same order. Therefore, in the 2D simulation, reconnection in the shock transition region is guide-field reconnection.

\cite{ng2022} performed a 3D PIC simulation to study reconnection in the shock transition region. The simulation parameters are: $\beta_e=\beta_i=1.41$, $\omega_{pe}/\Omega_e=4$, $m_i/m_e=100$, $\theta=30^{\circ}$, $v_d=10v_A$, and the system size $L_x\times L_y\times L_z=200d_i\times 50d_i \times 20d_i$. The $z$ direction is set to be a periodic boundary. Fig.~\ref{fig-pic-shock}(d) shows the current density $J_z$. In the 3D simulation, the current direction can be not only in the $z$ direction, but also in the $y$ direction; therefore, the reconnection plane does not have to be in the $x$-$y$ plane as in the 2D simulation, and some current sheets show reconnection with a weak guide field, even though the shock transition region has a large negative $B_z$.


\section{Embedded PIC: MHD-AEPIC}\label{sec:embedPIC}

\subsection{Overview}
Due to the large separation between the kinetic scales and the size of Earth's magnetosphere, it is highly computationally expensive to apply a purely kinetic code for simulating global magnetospheric dynamics. Various hybrid methods have been proposed to incorporate kinetic physics into global simulations while keeping the computational costs feasible. Traditional hybrid codes model the electron species as a fluid and simulate the ions with either macro-particles or a grid-based Vlasov solver. These hybrid models reduce the separation between the kinetic scales and the global scale by removing the electron kinetic scales from the model so that it becomes feasible to apply them to Earth's magnetosphere. 

 \begin{figure}[b]
     \begin{center}
     \vspace{0.25cm}  
     \includegraphics[width=1\textwidth, trim=0cm 8cm 0cm 3cm]{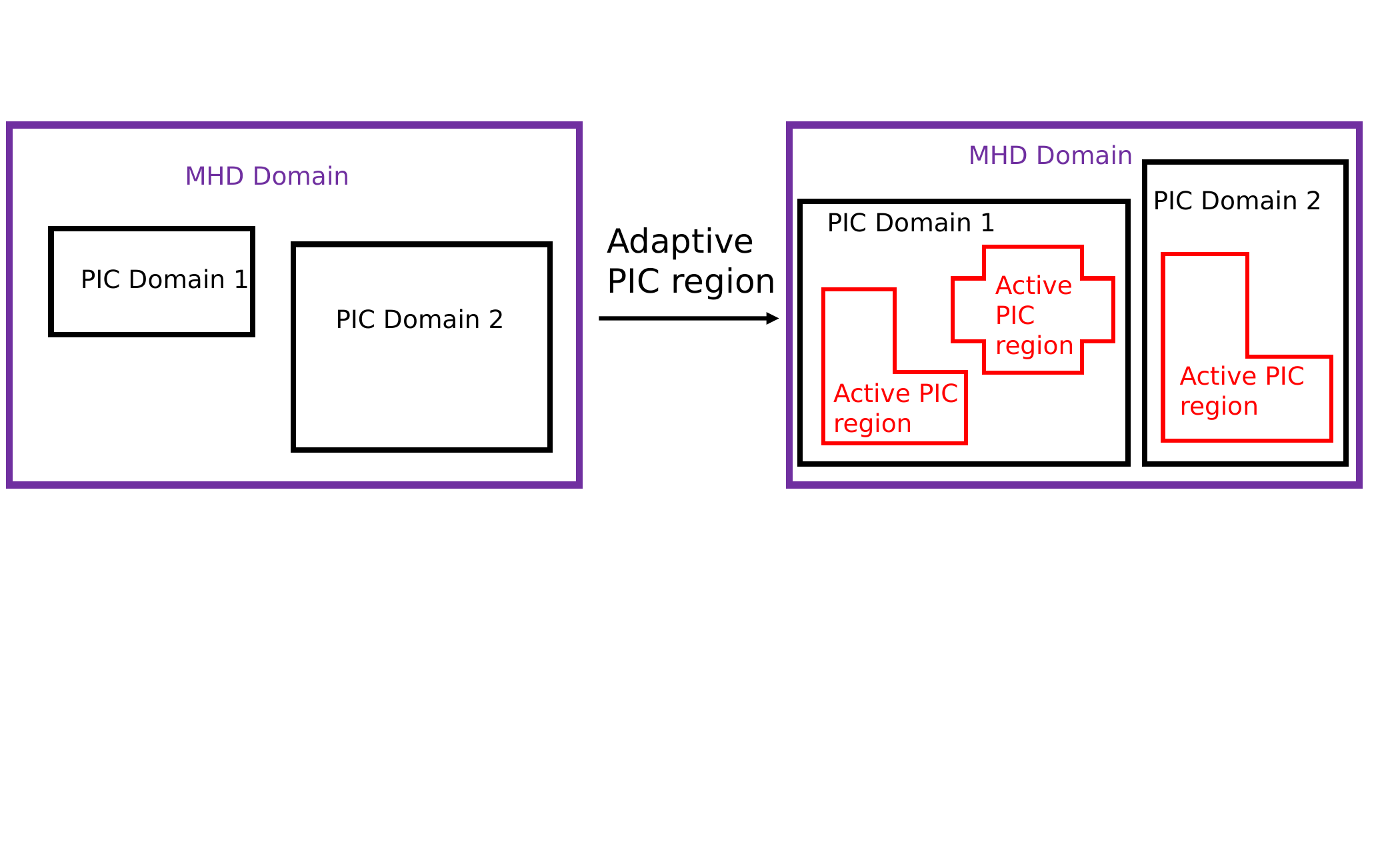}
         \caption{A schematic shows the improvement of the MHD-AEPIC (right) model from the MHD-EPIC (left) model. Aadapted from~\citet{Chen2023}.\label{fig:mhdepic_concept}}
     \end{center}
 \end{figure}

The Magnetohydrodynamic with Adaptively Embedded Particle-in-Cell (MHD-AEPIC) model represents another type of hybrid approach to incorporate kinetic effects into global models ~\citep{Daldorff:2014, Shou2021, Chen2023}. This type of hybrid model couples a kinetic code with a global fluid model, and only applies the kinetic code to simulate part of the simulation domain, where kinetic physics is crucial while using the fluid model to simulate the rest of the domain. Compared to a purely kinetic model, this type of hybrid model reduces the computational cost by reducing the domain size for the kinetic code, and it is best suited for applications where the important kinetic physics is localized. 
The MHD-AEPIC model, and its precursor, the Magnetohydrodynamic with Embedded Particle-in-Cell (MHD-EPIC) model, are the first two-way coupled models that work for global applications. Since then, similar coupled models have been developed by different independent teams. For example, ~\citet{Makwana:2017} also developed a model that couples a PIC code with an MHD code, and ~\citet{Rieke2015} tried to couple a Vlasov solver with a two-fluid code. 

\subsection{Methodology}
\subsubsection{Development History}
The original MHD-EPIC model was developed by~\citet{Daldorff:2014}, in which the semi-implicit particle-in-cell code iPIC3D~\citep{Markidis:2010} is coupled with the global fluid model BATS-R-US~\citep{Powell:1999} through the Space Weather Modeling Framework (SWMF)~\citep{Toth:2005swmf}. The model has been successfully applied to study magnetic reconnections in the magnetospheres of Ganymede~\citep{Toth:2016,Zhou2019,Zhou2020}, Earth~\citep{Chen:2017,Chen2020,Wang2021a,Wang2022}, Mercury \citep{Chen_mercury:2019} 
and Mars~\citep{Ma:2018}. A PIC region has to be a box in the MHD-EPIC model. To cover the kinetic regions of interest, the MHD-EPIC model supports applying multiple independent kinetic regions~\citep{Toth:2016} in the same simulation domain, and it also allows rotating a box so that the corresponding PIC region does not have to be aligned with the global grid~\citep{Chen2020}. These two features expand the capabilities of the MHD-EPIC model. However, not all the kinetic regions of interest can be covered by 
one or a few boxes. If the kinetic region moves at the global spatial scale during a simulation, the PIC box has to be very large to cover the whole region of interest, which is computationally expensive. To overcome these difficulties, the MHD-AEPIC model has been developed, which allows a dynamic PIC region of any shape (see Figure~\ref{fig:mhdepic_concept}).

To support dynamic PIC regions, two different PIC codes, the Adaptive Mesh Particle Simulator (AMPS)~\citep{Shou2021} and the FLexible Exascale Kinetic Simulator (FLEKS)~\citep{Chen2023}, have been developed as the PIC component of the MHD-AEPIC model. Since FLEKS is more widely used for MHD-AEPIC simulations, we focus on the FLEKS code in this section.

\subsubsection{Coupling Algorithm}
MHD-AEPIC supports coupling with single-fluid MHD, multi-species MHD, multi-ion MHD~\citep{Glocer:2009multifluid}, and five- and six-moment multi-fluid models~\citep{Huang:2019}. The most widely used fluid component is the single-fluid Hall MHD with a separate electron pressure equation, and we briefly describe the coupling algorithm for this case here. 

In an MHD-AEPIC simulation, the PIC code covers part of the whole simulation domain. The MHD model provides the initial conditions for the PIC code at the beginning of the coupled simulation. Once the initialization is done, both the PIC and MHD models update independently for one or a few time steps until the next coupling time point is reached. 
During the coupling, the MHD model provides the boundary conditions for the PIC code, and the PIC code provides the updated magnetic field and plasma quantities to overwrite the overlapped MHD region. 
Since the MHD and PIC codes solve different sets of equations, 
conversion between the MHD and PIC variables is needed. When calculating PIC variables from MHD variables, 
we need densities, velocities, and pressures for both electron and ion species, and they are calculated as follows:
\begin{itemize}
    \item Charge neutrality is assumed, so both the electron and ion densities can be easily obtained from total MHD density. 
    \item From the MHD magnetic field, the current density can be calculated. Since the sum of electron and ion momentum is the total MHD momentum, and the velocity difference between electrons and ions produces the current, the electron and ion velocities can be obtained. 
    \item Since we usually solve both ion and electron pressure equations on the MHD side, the electron and ion pressures can be obtained directly to initialize thermal PIC macro-particles.
\end{itemize}
Once the electron velocity is obtained, it is used to calculate the electric field $\mathbf E$ for PIC from the generalized Ohm's law: 
\begin{equation}
    \mathbf{E} = -\frac{\mathbf{U}_e \times \mathbf{B}}{c}.
    \label{eq:ohm}
\end{equation}
where $\mathbf B$ is the MHD magnetic field and $\mathbf U_e$ is the electron bulk velocity including the Hall term. We note that no matter Hall physics is included or not into the MHD model, the equation above is applied to calculate the initial and boundary electric field for PIC. 

Calculating MHD variables from PIC variables is more straightforward: we simply sum up the mass, momentum and energy of the electron and ion macro-particles to obtain the plasma variables required. We refer the readers to~\citet{Daldorff:2014} for more details. 
Currently, the PIC codes used for MHD-EPIC/MHD-AEPIC coupling have to use a Cartesian mesh, but the MHD model BATSU-R-US can use non-uniform Cartesian or non-Cartesian grids. The interpolation between the PIC and MHD grids is done by a second-order linear interpolation.

\subsubsection{Particle-In-Cell Algorithm}

From iPIC3D for MHD-EPIC to FLEKS for MHD-AEPIC, all the PIC codes are semi-implicit~\citep{Brackbill:2008, Lapenta:2017, Chen:2019}, meaning that the electric field is solved for by an implicit scheme. We choose the semi-implicit PIC algorithm because it has a relaxed stability constraint so that the Debye length does not have to be resolved and the stability constraint for the time step is based on the thermal speed instead of the speed of light. Based on our numerical experiments, we found the stability of the PIC code is extremely important for a successful MHD-EPIC/MHD-AEPIC simulation. 
To improve the stability, we designed the Gauss's Law satisfying Energy-Conserving Semi-Implicit Method (GL-ECSIM)~\citep{Chen:2019}, which is based on the Energy-Conserving Semi-Implicit Method (ECSIM) by~\citet{Lapenta:2017}. GL-ECSIM shares the same energy conservation property as ECSIM, i.e., the total energy of the system can be exactly conserved with proper parameters. In practice, we found the code is more stable with parameters that slowly dissipate the total energy numerically. In addition, satisfying Gauss's law (charge conservation) is also crucial for the stability and accuracy of the PIC code. GL-ECSIM applies a novel method to satisfy Gauss's lay by adjusting particle positions at the end of each cycle. The details of GL-ECSIM can be found in~\citet{Chen:2019}.

In a long MHD-AEPIC simulation, the macro-particle number per cell may vary significantly due to the transport of particles. The uneven distribution of particle numbers can cause load imbalance and reduce computational efficiency. To alleviate this problem, we designed particle splitting and merging algorithms for FLEKS. A particle splitting (merging) algorithm is applied to split (merge) particles when the number of particles per cell is below (above) a threshold~\citep{Chen2023}. 

\subsubsection{Kinetic Region Adaptation}
The most important improvement of MHD-AEPIC over MHD-EPIC is the adaptive PIC region. Although the PIC grid is still Cartesian, its cells can be switched on or off so that the active cells can fit any shape of kinetic regions. We note that the PIC cells can be activated or deactivated dynamically during a simulation.
The active PIC region can be defined either based on geometric or physical criteria. For physics-based adaptation, BATS-R-US calculates the physical criteria and sends the corresponding grid information to FLEKS to turn on or turn off cells. 

\subsubsection{Kinetic Scaling}

In some applications, the difference between kinetic and global spatial and temporal scales makes it difficult, if not impossible, to resolve the kinetic scales in an MHD-EPIC, or even MHD-AEPIC simulation. Fortunately, the large separation of scales can be exploited, and the kinetic scales can be increased by changing the mass per charge ratio without affecting the global dynamics~\citep{Toth:2017}. This technique is not needed or even applicable for Ganymede and Mercury simulations, where the kinetic and global scales are not very different. On the other hand, kinetic scaling is applicable and extremely useful for modeling Earth's magnetosphere. We typically increase the kinetic scales by a factor of $4$ to $16$. See~\citet{Toth:2017} for more detail.

\subsection{Applications}

\begin{figure}
\centering
    \includegraphics[width=1.\textwidth, trim=0cm 0cm 0cm 0cm,clip,angle=0]{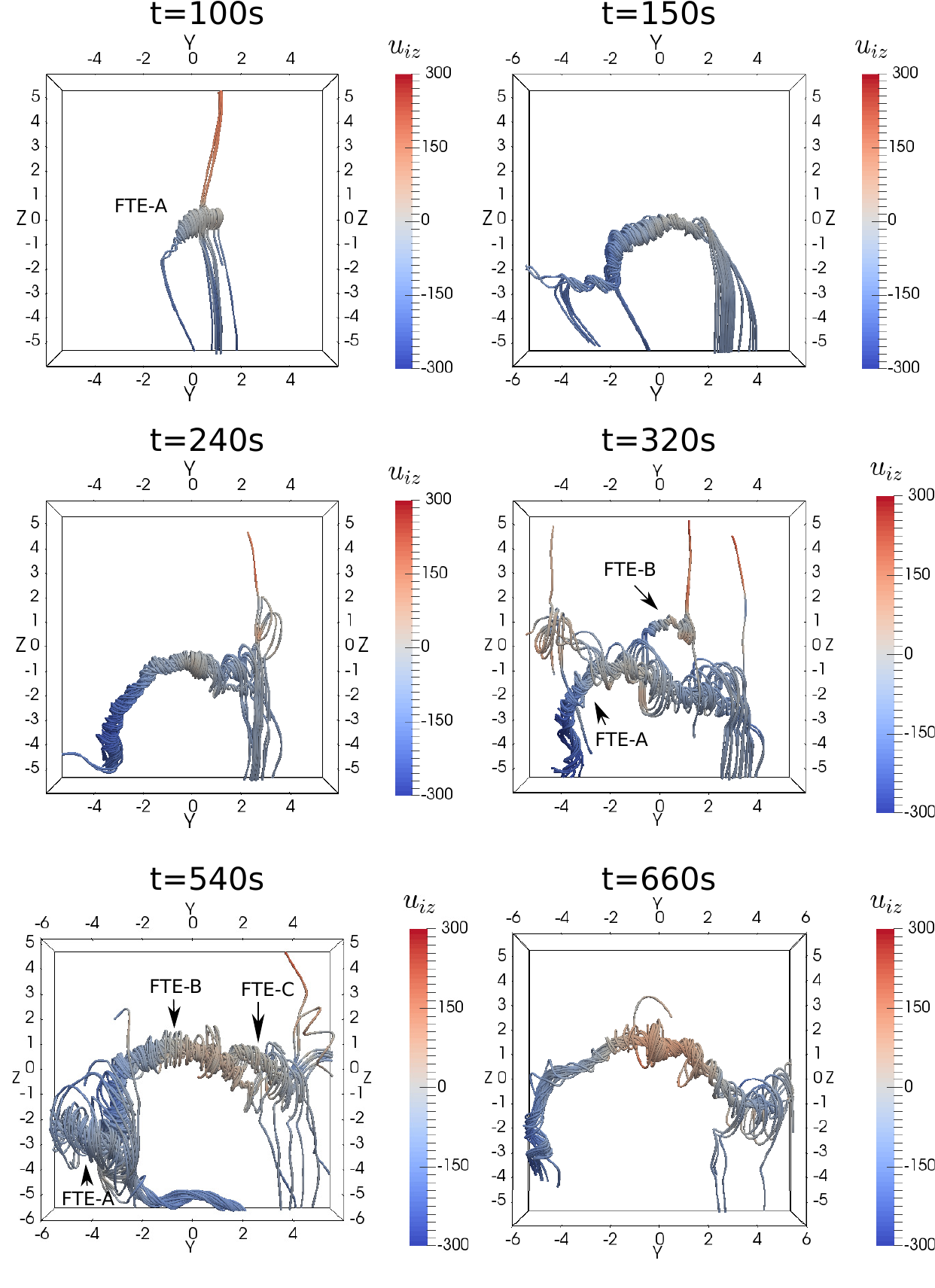}
  \caption{Evolution of FTEs. Viewed from the Sun, a series of snapshots are shown with magnetic field lines colored by ion velocity $u_{iz}[\mathrm{km/s}]$. Adapted from~\citet{Chen:2017}.}
  \label{fig:fr-3d}
\end{figure}

The MHD-EPIC/MHD-AEPIC model has been applied to investigate the physical processes and consequences of both magnetopause and magnetotail reconnection. In these simulations, the PIC code is usually used to cover either the magnetopause or the magnetotail current sheet, where reconnection happens. Since the initial and boundary conditions of the PIC code are obtained from the MHD model, the physical parameters inside the PIC region, such as the plasma quantities and the shape of the current sheet, are more realistic than those in a standalone PIC simulation. On the other hand, the information from the PIC code is also fed back to the MHD model so that we can evaluate the global consequences of the kinetic magnetic reconnection. 

Here we briefly describe a few applications of the MHD-EPIC/MHD-AEPIC model to study Earth's magnetosphere. \citet{Chen:2017} studied both the kinetic features of magnetopause reconnection and the evolution of flux transfer events (FTEs) show in Figure~\ref{fig:fr-3d}. 
Near the reconnection site, the simulation successfully produced key kinetic features of asymmetric magnetic reconnection, such as the crescent electron phase space distribution and the lower hybrid drift instability. Due to the multiple X-line reconnections inside the PIC code, FTEs are generated quasi-periodically at low latitudes, then propagate toward the cusps. We briefly describe the evolution of the FTEs here:
\begin{itemize}
    \item During the growth of an FTE, its cross-section increase, and its length extends along the dawn-dusk direction (from $t=100s$ to $t=150s$ in Figure~\ref{fig:fr-3d}).
    \item Since its ambient plasma flow speed varies, an FTE may become tilted ($t=240s$ in Figure~\ref{fig:fr-3d}).
    \item There may be multiple FTEs on the magnetopause, and a few FTEs can merge into one (from $t=320$ to $t=660s$ in Figure~\ref{fig:fr-3d}).
    \item FTEs can be dissipated at high latitudes due to the reconnection between the FTE magnetic field and the cusp field lines (Figure~5 of~\citet{Chen:2017}).      
\end{itemize}

~\citet{Chen2020} simulated the GEM dayside kinetic processes challenge event, and compared simulation results with both MMS observations and ground-based SuperDARN observations. The MHD-EPIC simulation shows there are usually multiple X-lines at the magnetopause, and the expanding speed of the X-line endpoints is comparable with SuperDARN observations.

The MHD-EPIC model has also been applied to study magnetotail reconnection. ~\citet{Wang2022} found the MHD-EPIC simulation can produce global-scale magnetospheric sawtooth-like oscillations periodically even under steady solar wind conditions, while the ideal- and Hall-MHD simulations do not produce such variations. It suggests that kinetic reconnection physics may play an important role in driving sawtooth oscillations. Recently, the development of the MHD-AEPIC model has enabled us to simulate storm events with a dynamic PIC region that covers the highly dynamic magnetotail reconnection sites.

Our current work focuses on modeling extreme geomagnetic storm events with MHD-AEPIC. Extreme events occur infrequently, which makes it difficult to validate MHD models employing simple numerical diffusion to approximate reconnection physics. Using a higher-fidelity model, such as MHD-AEPIC, can improve the reliability of simulations of extreme events.

\section{Kglobal: Particle acceleration self-consistently embedded in MHD}
\label{sec:kglobal}

Solar flares convert magnetic energy into particle energy via magnetic
reconnection.  Observations of power-law tails in particle
distribution functions imply that a large fraction of the released
energy goes to energetic (i.e., non-thermal) electrons and ions~\citep{warmuth2016constraints}.  However, the particle spectra found in particle-in-cell (PIC) simulations of reconnection in the relevant
regime typically do not form power-laws, except in the limit of
extremely low upstream plasma $\beta$~\citep{dahlin2015electron, dahlin2017_role,zhang2021efficient}.  Why?  With structures extending
$\sim 10^4$ km and a Debye length of $\sim 1$ cm (for $n \sim 10^{10}$
cm$^{-3}$ and $T_e \sim 100$ eV), the corona spans ten orders of
magnitude in physical scale.  Explicit PIC models must resolve kinetic
scales and hence can only simulate a tiny fraction of the macroscopic
domain.  The dependence of the Larmor radius on energy means
nonthermal particles can quickly acquire orbits that approach the size
of the simulation domain, halting further energy gains.

In contrast, MHD simulations study macroscopic domains with a fluid
description that averages over small spatial and temporal scales.
Following test particles in the MHD fields produces information about
how particles gain energy but, without feedback coupling the particles
and the fields, runaway energy gain can occur so that the system as a
whole does not conserve total energy. It is possible to embed PIC
models into MHD descriptions at selected locations, but such models
presume that particle energy gain occurs in the vicinity of magnetic
nulls, which is not consistent with the development and interaction of
macroscale magnetic islands or the development of turbulence in
large-scale current layers.

The {\em kglobal} model incorporates the physics necessary to explore
particle energization from both the PIC and MHD descriptions
~\citep{drake2019computational,arnold2019large}.  The fundamental question is whether
kinetic-scale boundary layers play an essential role in particle
energy gain -- or if they can be ordered out of the equations to
facilitate simulations of macroscale systems.  Kinetic boundary layers
control the regions where $E_{\parallel}$, the component of the
electric field parallel to the magnetic field, is non-zero. However,
Fermi reflection rather than $E_{\parallel}$ is the dominant driver of
energetic particles~\citep{dahlin2016parallel,li2019particle}. Particle energy gain from
Fermi reflection takes place over macro-scale regions and occurs even
where $E_{\parallel} = 0$.  As a consequence, kinetic-scale boundary
layers are not required to describe the non-thermal energization in
macroscale systems.

\begin{figure}
        \vspace{-0.075in}
\begin{center}
 \includegraphics[width=4.5in]{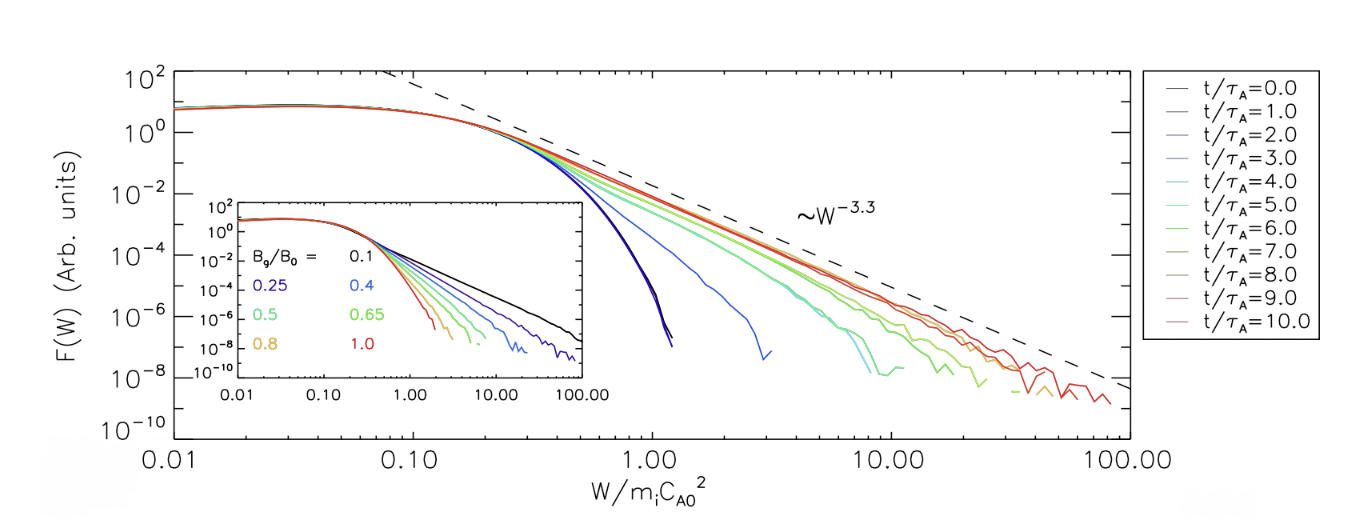}
 
        \vspace{-0.15in}
\caption{{Energetic electron spectra from {\em
          kglobal}.} A log-log plot of the electron differential
      density F(W) versus energy ($W$) at multiple times from a
      reconnection simulation with a guide field $B_g/B_0 = 0.25$. A
      power-law develops after $t/\tau_{A} \sim 3-5$. Inset: The
      late-time F(W) for several guide fields, illustrating the
      dependence on the \textsl{ratio} of the guide-to-ambient
      magnetic field. Adapted from~\citet{arnold2021electron}.}\label{fig:kglobal}
                \vspace{-.3in}
\end{center}
\end{figure}

Hence, {\em kglobal} combines a MHD description of the plasma dynamics
with a macro-particle description. Guiding center particles populate
the MHD grid, are advanced in parallel with the fluid equations, and
feed back on the MHD fluid through their gyrotropic pressure
tensor. They can be small in number density but can contribute a
pressure comparable to the pressure of the reconnecting magnetic
field.  The entire system conserves total energy, which includes that
of the MHD fluid, the magnetic field, and the kinetic energy of the
macro-particles.

The basic version of {\em kglobal} includes three species: fluid ions,
fluid electrons, and particle electrons (the latter of which form the
nonthermal population).  The fields follow the usual MHD equations
\begin{equation}
\frac{\partial \mathbf{B}}{\partial t} =
-c\boldsymbol{\nabla\times}\mathbf{E}_{\perp} \qquad
\mathbf{E}_{\perp} =
-\frac{1}{c}\mathbf{v}_i\boldsymbol{\times}\mathbf{B}
\end{equation}
and the ion fluid satisfies the usual MHD continuity equation
\begin{equation}
\frac{\partial n_i}{\partial t}+
  \boldsymbol{\nabla\cdot}\,n_i\mathbf{v}_i =  0
\end{equation}
and energy equation
\begin{equation}
\frac{d}{dt}\left(\frac{P_i}{n_i^{\gamma}}\right)
  = 0
\end{equation}
The ion momentum equation takes the form
\begin{equation}
\rho_i\frac{d\mathbf{v}_i}{dt} =
\frac{1}{c}\mathbf{J}\boldsymbol{\times}\mathbf{B} -
\boldsymbol{\nabla}P_i -\boldsymbol{\nabla}_{\perp} P_{ef} -
m_en_{ef}v^2_{\parallel ef}\boldsymbol{\kappa} +
en_iE_{\parallel}\mathbf{b}- (\boldsymbol{\nabla
  \cdot}\,\mathrm{T}_{ep})_{\perp}
\end{equation}
in which the left-hand side and first terms on the right-hand side are
the same as in MHD ($P_i$ and $P_{ef}$ are the ion and fluid electron
pressure, respectively).  However the final terms on the right-hand
side include the curvature $\boldsymbol{\kappa} =
\mathbf{b}\boldsymbol{\cdot\nabla}\mathbf{b}$, large-scale parallel
electric field $E_{\parallel}$, and particle electron stress tensor
$\mathrm{T_{ep}}$ and quantify the self-consistent back-reaction of
the particles on the system.

The perpendicular motion of the particle electrons is given by the
conservation of the first adiabatic invariant
\begin{equation}
 \mu_{ep} =
  \frac{p_{ep\perp}^2}{2B}=\text{const.}
\end{equation}
while the parallel motion satisfies
\begin{equation}
\frac{d}{dt}p_{e\parallel} =
  p_{e\parallel}\mathbf{v_E}\boldsymbol{\cdot \kappa} -
  \frac{\mu_e}{\gamma_e}\mathbf{b}\boldsymbol{\cdot\nabla}B -
  eE_{\parallel}
\end{equation}
This equation includes a contribution from the parallel electric field given by
\begin{equation}
E_{\parallel} =
  -\frac{1}{n_ie}\left(\mathbf{B}\boldsymbol{\cdot\nabla}\left(
  \frac{m_en_{ef}v^2_{ef\parallel}} {B}\right) +
  \mathbf{b}\boldsymbol{\cdot\nabla}P_c +
  \mathbf{b}\boldsymbol{\cdot\nabla\cdot}\mathrm{T}_{ep}\right)
\end{equation}
Finally, the fluid electron density enforces quasi-neutrality
\begin{equation}
n_{ef} = n_i - n_{ep}
\end{equation}
the parallel flow eliminates parallel currents
\begin{equation}
 n_{ef}v_{ef\parallel} = n_iv_{i\parallel} - n_{ep}v_{ep\parallel}
\end{equation}
and the pressure equation takes the usual form
\begin{equation}
\frac{d}{dt}\left(\frac{P_{ef}}{n_{ef}^{\gamma}}\right) = 0
\end{equation}
A full derivation of these equations is given in~\citet{drake2019computational} and~\citet{arnold2019large}.

Simulations with these equations pass several tests.  They describe
the linear propagation of stable, circularly polarized Alfv\'en waves
and the linear growth of firehose modes.  The latter plays an
important role in controlling the feedback of energetic particles
during magnetic reconnection since magnetic tension is suppressed on
the approach to firehose marginal stability.  In addition, they
accurately capture the dynamics of electron acoustic waves and
describe the suppression of transport of hot electrons parallel to the
ambient magnetic field.  The inclusion of the large scale
$E_{\parallel}$ is important in describing the development of return
currents that form as hot electrons escape from regions of electron
acceleration in macroscale energy release events such as flares
~\citep{egedal2012large}. 

Reconnection simulations with {\em kglobal} have produced power-law
spectra of energetic electrons that extend nearly three decades in
energy, while simultaneously generating the super-hot thermal
electrons characteristic of flare observations~\citep{arnold2021electron}. Fig.~\ref{fig:kglobal} shows the electron energy
spectrum for a typical simulation.  Electrons in the initial
Maxwellian distribution (black curve) transform into a nonthermal
spectrum in a few Alfv\'en crossing times ($\tau_A$).  Consistent with
observations, the total energy content of the nonthermal electrons can
exceed that of the hot thermal electrons even though the number
density does not. The strength of the ambient out-of-plane guide field
strongly impacts the energy content and power-law index of the
nonthermal electrons (see inset of Figure~\ref{fig:kglobal}): the
guide field increases the radius of curvature of a reconnected field
line, thereby weakening Fermi reflection~\citep{drake2006electron}.  In
contrast, the size of the global system has relatively little
influence.

The governing equations of {\em kglobal} can be extended to include
the contributions of non-thermal (particle) ions.  Unlike for
electrons, whose small mass can be used to simplify the equations, the
ion inertia can not be neglected and must be included.  Recent work
has incorporated these equations into the computational model and
early results reveal the simultaneous development and evolution of
extended electron and proton power law distributions~\citep{yin2024computational}.

\section{Vlasov}
\label{sec:vlasov}
\subsection{Overview}
Eulerian Vlasov-Maxwell numerical simulations are a useful tool for investigating basic kinetic-scale plasma processes, such as turbulence and magnetic reconnection, as well as the interaction between the solar wind and planetary magnetospheres. 

Thanks to the clean description of the plasma dynamics in the entire phase space at the expense of a larger computational cost, Eulerian algorithms complement well Particle-In-Cell (PIC) codes. The almost noise-free description of velocity space is generally guaranteed by the discretization of the plasma distribution function on a six-dimensional phase-space grid characterized by collocation points in both physical and velocity space. On the other hand, PIC methods suffer from the intrinsic stochastic shot noise which becomes especially relevant at small scales and in cases when the number of particles per cell is not large. However, in Eulerian methods, setting a six-dimensional grid in the entire phase space dramatically increases the computational cost. The bottleneck is generally constituted by the memory necessary to store the plasma distribution function, as shown in the following simple example.

\begin{figure*}[htbp!]
\begin{center}
\includegraphics[width=\textwidth]{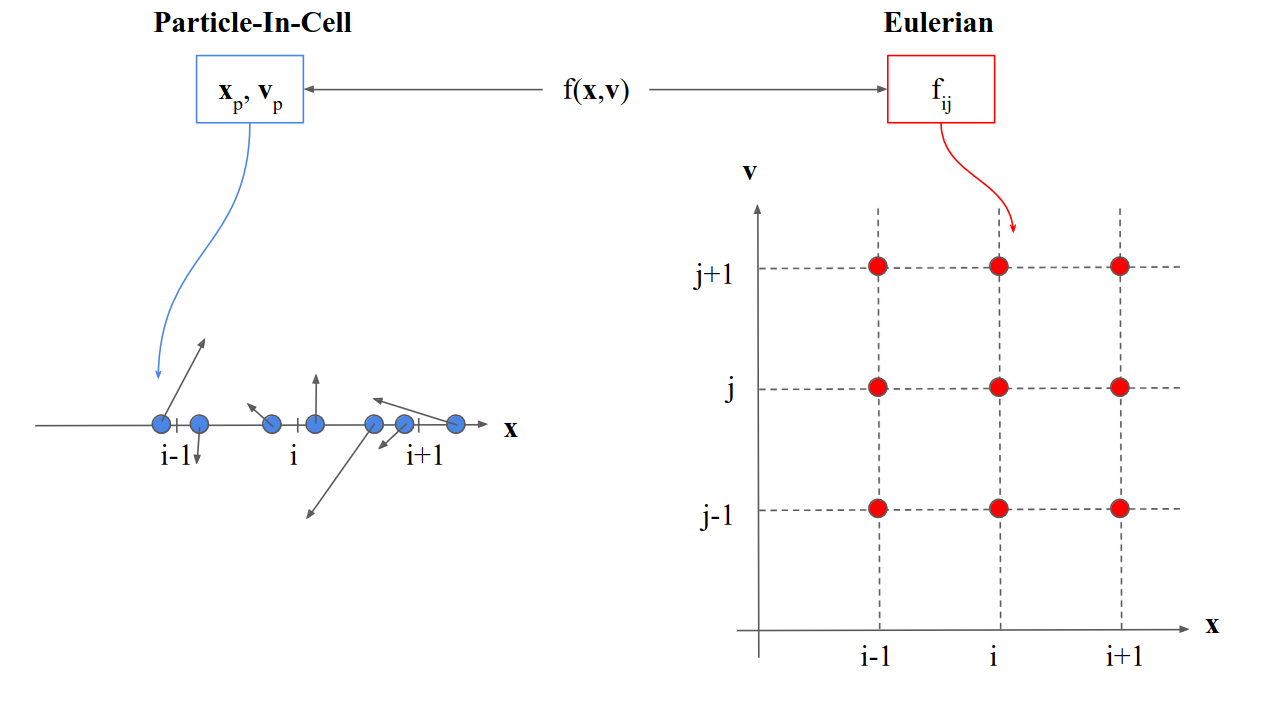}
\caption{Sketch of the typical sampling of the plasma distribution function $f({\bf x},{\bf v})$ adopted in PIC (left) and Eulerian (methods). Adapted from~\cite{finellithesis}.}
\label{fig:PICEulerian}
\end{center}
\end{figure*}

The main difference when sampling the plasma distribution function in PIC and Eulerian approaches is depicted in Fig. \ref{fig:PICEulerian}. In PIC methods, the grid is defined only on the physical space ${\bf x}$, and the distribution function is sampled through macroparticles (blue circles), each one representative of a large number of effective plasma particles. In the Eulerian approach, the grid is defined on the entire phase space $({\bf x}, {\bf v})$ and the distribution function is known on this ensemble of grid points (red circles). In both PIC and Eulerian methods, the electromagnetic fields and the moments of the distribution function (e.g., density, bulk speed, etc.) are defined on the physical-space grid ${\bf x}$.

Let's imagine a generic phenomenon occurring in a plasma composed of protons and electrons that requires a physical-space grid discretized with $512^3$ points. Since the electromagnetic fields have the same memory requirements in both PIC and Eulerian methods, we neglect them here for simplicity. PIC simulations indicate that the plasma dynamics is overall well described with $\sim 1000$ particles per cell. The memory (assuming double-precision variables) required to store particles' positions and velocities is $\sim 6$ TB. Similarly, Eulerian Vlasov simulations generally adopt at least $\sim 51^3$ velocity-space points to well describe the details of the velocity plasma distribution function, thus requiring about $\sim 260$ TB of memory. Hence, PIC simulations can usually be performed in larger physical-space computational boxes compared to Eulerian ones. Moreover, 3D simulations are more easily achievable with PIC methods while they remain often prohibitive for Eulerian codes, or require more advanced methods such as adaptive mesh refinement or sparse velocity space techniques to become tractable.

In the following, we will introduce two models that have been intensively used by the community and, then, specialize the discussion on global and local Vlasov-Maxwell simulations of magnetic reconnection. 

\subsection{Models and Algorithms}
Picturing a virtual journey from large to small scales, the first relevant Vlasov-Maxwell model widely adopted for performing Eulerian simulations is represented by the hybrid Vlasov-Maxwell one. The hybrid model considers protons as a kinetic species, while electrons are a background fluid. It is a low-frequency approximation of the full Vlasov-Maxwell system of equations that assumes quasi-neutrality and neglects the displacement current \citep{mangeney2002}. Faraday's law is used to evolve the magnetic field, while the electric field is provided by the generalized Ohm's law which includes the Hall term, the electron pressure gradient term, and possibly the terms related to electron inertia (see, e.g.,~\citet{valentini2007} for further details about the generalized Ohm's law).

The hybrid Vlasov-Maxwell model has been adopted for global simulations of the interaction between the solar wind and the Earth's magnetosphere through the Vlasiator algorithm \citep{vonalfthan2014, palmroth2018_LR}. The Vlasiator code adopts a splitting algorithm to decompose the six-dimensional Vlasov equation into a set of two three-dimensional advection equations \citep{Strang1968} in physical and velocity space, respectively. The solution of each advection equation is obtained by a semi-Lagrangian method \citep{ZerroukatAllen2012}, which relieves from the strict limitations to the time step length posed by the CFL condition in velocity space acceleration in regions of strong magnetic field. Position space is discretized on a cell-adaptive Cartesian grid \citep{Honkonen2013,ganse2023} and at each position in space the velocity-space grid is stored on a uniform, Cartesian grid. Vlasiator developed a sparse velocity-space method in which only regions of the velocity distribution function above a set phase-space density are stored and propagated \citep{vonalfthan2014}, yielding a gain of two orders of magnitude in terms of memory and computations. This technique made two-dimensional (2D position space periodic in the third dimension, 3D velocity space) magnetospheric simulations possible, as well as quasi-three dimensional simulations with a very limited extent in the third dimension \citep{pfaukempf2020}. The implementation of adaptive mesh refinement in position space, allowing to focus resolution on regions of interest while saving computations in less-resolved regions, is what made full three-dimensional, global magnetospheric simulations achievable with Vlasiator on modern, bleeding-edge supercomputers \citep{grandin2023,ganse2023,palmroth2023}. The electric and magnetic fields are propagated using an upwind constrained transport method \citep{Londrillo2004JCP} with divergence-free magnetic field reconstruction \citep{Balsara2009bJCP} on a uniform Cartesian grid matching the finest refinement level of the Vlasov spatial grid, requiring a dedicated coupling scheme between the grids \citep{Papadakis2022}.

The hybrid Vlasov-Maxwell model has been also adopted for local simulations of plasma turbulence at sub-proton scales by means of the HVM algorithm \citep{valentini2007} (see also \citet{califano2023} for a recent review). The HVM code reduces the six-dimensional Vlasov equation to a set of six one-dimensional advection equations. Each equation is then solved through the van Leer method \citep{vanleer1977towards}. Fields are computed through the Current-Advance Method (CAM). The grid is homogeneous in both physical and velocity space. Periodic boundary conditions are implemented in physical space, while in velocity space the proton distribution function is set to zero after a large number of thermal speeds. 

Moving towards smaller scales, fully-kinetic Vlasov-Maxwell Eulerian algorithms have been recently implemented to describe electron-scale dynamics. Given the larger computational cost of Eulerian simulations with respect to PIC methods, the former are generally more recent than the latter and possibly implement different assumptions to simplify the Maxwell equations. The full Maxwell system has been retained in several codes \citep{umeda2009, umeda2010, delzanno2015, ghizzo2017, juno2018, allmannrahn2022}. However, different approximations of the Maxwell equations have been proposed to alleviate the CFL constraint which sets a very small time step when the wave phase speed approaches the speed of light. In this regard, \citet{wiegelmannbuchner2001} neglect the displacement current while allowing for charge separation, while \citet{tronci2015} ignore both the displacement current and the charge separation. Yet a different approach neglects only the transverse part of the displacement current responsible for ordinary mode propagating at the speed of lights \citep{schmitzgrauer2006a, pezzi2019vida, shiroto2023improved}. Finally, an original approach has been developed based on the Vlasiator model, whereby a small section of interest from an ion-hybrid Vlasiator run is used to initialize an electron-hybrid setup in which the ions are kept static while the electron distribution function evolves \citep{battarbee2021}. This so-called eVlasiator approach has successfully reproduced properties of electron distributions observed in the vicinity of reconnection diffusion regions \citep{alho2022}.

\subsection{Examples of Applications Focused on the Study of Magnetic Reconnection}
\subsubsection{Local simulations}
In this section, we will present key results relevant to magnetic reconnection that have been obtained with the Hybrid Vlasov Maxwell code (HVM). Further treatments, based on fully-kinetic Eulerian Vlasov-Maxwell simulations, will not be covered in detail in this chapter, but the reader is referred to the works of \citet{schmitzgrauer2006b,Inglebert2011,ZenitaniUmeda2014,Sarrat2017,pezzi2019vida}, as well as Table 2 in the review by \citet{palmroth2018_LR} which lists works using Vlasov-based methods in space and astrophysics. 

The HVM code, which retains alpha particles \citep{perrone2013,valentini2016differential} and inter-particle collisions \citep{pezzi2019proton}, has been used for years to investigate plasma processes occurring at ion kinetic scales. It has been massively employed to study the properties of plasma turbulence \citep{valentini2010, servidio2012, servidio2014, servidio2015kinetic, cerri2017kinetic}, showing that turbulent fluctuations generate manifestly non-Maxwellian proton distribution functions \citep{greco2012inhomogeneous}. This emergent velocity-space complexity has been envisioned as a cascade process occurring in velocity space \citep[e.g.,][]{tatsuno2009nonlinear, schekochihin2016phase, servidio2017magnetospheric}: HVM results have allowed to characterize it in a full Vlasov system rather than in the gyrokinetic approximation \citep{cerri2018dual, pezzi2018velocity, pezzi2021current}. Characterizing non-equilibrium plasma distribution functions is significant to understanding energy transfer and dissipation processes occurring at ion kinetic scales in nearly-collisionless plasmas such as the solar wind \citep{matthaeus2020pathways, cassak2023quantifying}, as also reported in different studies based on the HVM code \citep{sorrisovalvo2018, pezzi2019energy, pezzzi2021dissipation, fadanelli2021}. In the perspective of the \citet{holloway1991undamped} work showing that non-Maxwellian plasmas can support the propagation of undamped plasma waves, the HVM code has been adopted to study the onset of a novel type of electrostatic fluctuations triggered by trapped ions \citep{valentini2011new,valentini2011short,valentini2014nonlinear}.  

One of the first studies employing the HVM code for investigating magnetic reconnection reported the onset of a fast reconnection process obtained as a result of magnetic islands developed by the electromagnetic current filamentation \citep{califano2001}. In the following years, despite the large number of studies adopting the HVM code, the vast majority of the research work mostly focused on the investigation of fully developed plasma turbulence. 
More recently, \citet{finelli2021} investigated the magnetic reconnection in a similar manner as PIC-based studies discussed in Section~\ref{sec:PIC}, that is, modeling an isolated Harris-like current sheet \citep{harris1962} in equilibrium or pressure balance. Such a current sheet, usually doubled in the physical-space domain to accommodate for periodic boundary conditions, quickly starts reconnecting thanks to an initial perturbation of proton density and/or current (thus magnetic field). 

\begin{figure*}[htbp!]
\begin{center}
\includegraphics[width=\textwidth]{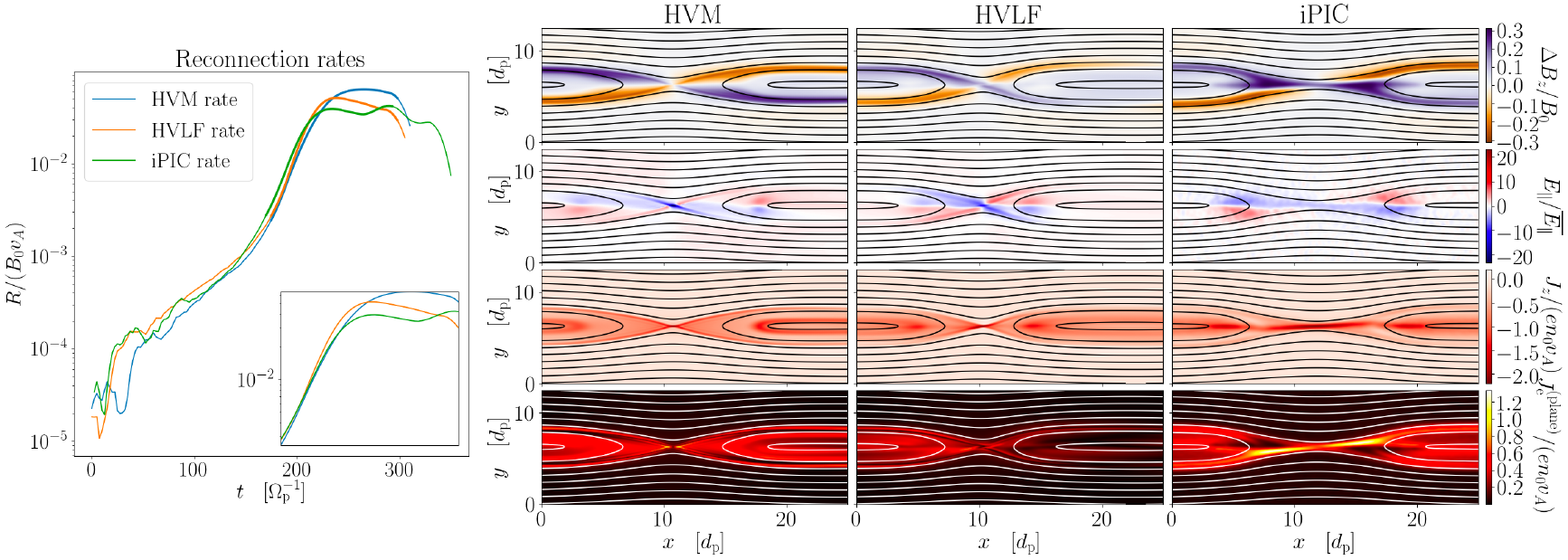}
\caption{Magnetic reconnection modeled using three different codes (HVM, HVLF and PIC). Left: normalized reconnection rate $R/B_0 v_\mathrm{A}$. The inset shows the time interval corresponding to the ticker curves in the main plot. To ease the comparison, the curves in the inset are shifted in time. Right: (first row) out-of-reconnection-plane magnetic field $\Delta B_z/B_0 = (B_z - B_z(t=0))/B_0$ showing the expected Hall quadrupolar pattern; (second row) electric field parallel to the ambient magnetic field $E_{\parallel}/\overline{E_{\parallel}}$, where $\overline{E_{\parallel}}$ is the root mean square of $E_{\parallel}$ in the shown region; (third row) current density in the out-of-plane direction $J_z$; (fourth row) electron current density in the plane $J_\mathrm{e}^{\rm (in-plane)}$. The superposed black or white curves are the magnetic field lines. The three columns show results from the three different models. The left column show results from HVM at the simulation time $t = 237.5 ~\Omega_\mathrm{cp}^{-1}$; the center column show results from the HVLF code at the simulation time $t = 232.5 ~\Omega_\mathrm{cp}^{-1}$; the right column show results from the PIC code at the simulation time $t = 235.0 ~\Omega_\mathrm{cp}^{-1}$. $\Omega_\mathrm{cp}$ is the proton cyclotron frequency. Adapted from \cite{finelli2021}.}
\label{fig:HVM_reconnection_fig1}
\end{center}
\end{figure*}

In particular, \citet{finelli2021} compared results from three different models (i) the HVM model with (isotropic) isothermal electrons including finite electron-inertia, (ii) a modified HVM model, called hybrid-Vlasov-Landau-fluid (HVLF); (iii) a fully-kinetic PIC code (iPIC3D \cite{markidis2010}). The HVLF model is equipped to include anisotropies of the gyrotropic electron pressure with a Landau-fluid (LF) closure for the transport of the gyrotropic electron thermal energy along magnetic field
lines \citep{sulempassot2015}. Using these three models, \citet{finelli2021} performed 2D-3V magnetic reconnection simulations with moderate guide field ($B_g = 0.25 ~B_0$, where $B_0$ is the asymptotic magnetic field) and with reduced mass ratio $m_\mathrm{p}/m_\mathrm{e} = 100$. The initial setup consists of a double Harris current sheet \citep{harris1962} perturbed by long wavelength magnetic field fluctuations with random phase (with $1 < |\mathbf{k}| d_\mathrm{p} < 9$, where $\mathbf{k}$ is the wave vector of the fluctuations and $d_\mathrm{p}$ is the proton inertial length). The size of the simulations domains is $L_x \times L_y = 24 \pi d_\mathrm{p} \times 12 \pi d_\mathrm{p} $ discretized with $N_x \times N_y = 1024 \times 512$ grid points. In the HVM and HVLF simulations, the velocity space domain in each direction ($x$, $y$ and $z$) is [$-$6.4, +6.4] $v_{th,p}$, where $v_{th,p}$ is the proton thermal speed, and it is discretized by $51^3$ grid points. Figure \ref{fig:HVM_reconnection_fig1} shows results comparing the three models. 

While the reconnection linear phase evolution, as well as the overall reconnection signatures and patterns, are quite similar for all three models, \citet{finelli2021} report that the region of intense current at the centre of the current sheet is more elongated in the case of the HVLF and PIC simulations than in the HVM run. Also, 
the normalized reconnection rate $R/B_0 v_A$ computed in the quasi-steady state is higher for the HVM 
($R/B_0 v_A \sim 0.06$, where $B_0$ is the upstream magnetic field and $v_A$ is the Alfv\'en speed) than for the HVLF and PIC simulations ($R/B_0 v_A \sim 0.04$). Despite these differences, the results of all three simulations agree qualitatively. In terms of electron dynamics, which is not captured by the HVM code, the HVLF model reproduces the main features obtained with the fully kinetic treatment of the PIC code.

Magnetic reconnection and turbulence are intricately coupled in plasmas (Stawarz et al., 2024, this issue), where coherent structures such as magnetic holes, magnetic islands, and current sheets naturally develop (see, e.g.,~\citet{matthaeus2015intermittency}). Then, current sheet widths tend to approach the kinetic scale, leading to magnetic reconnection. Hybrid-Vlasov simulations of turbulent plasmas have been successful in modelling both turbulence-induced ``standard'' reconnection with ion-coupling and electron-only magnetic reconnection~\citep{Califano2020,arro2020}. A key result in the context of the interplay between reconnection and turbulence is the fact that turbulence is mediated by magnetic reconnection. More specifically, reconnection plays a key role in driving the onset of sub-ion turbulent cascade \citep{franci2017, cerri2017kinetic, manzini2023,adhikari2024scale}. Hybrid-Vlasov simulations with the HVM code have played a crucial role in providing evidence for the role played by reconnection in this context.

\subsubsection{Global simulations}
The main goal of Vlasiator is to model the solar wind--magnetosphere interaction with a hybrid-Vlasov approach. For this reason, the computational efforts have been mainly directed toward performing global simulations of the entire magnetosphere. As a consequence, a broad variety of magnetospheric plasma phenomena have been investigated using Vlasiator (notably collisionless shock \citep[e.g.,][]{johlander2022} and foreshock physics \citep[e.g.][]{turc2023}, magnetosheath jets \citep[e.g.][]{suni2021}, auroral proton precipitation \citep[e.g.][]{grandin2020, grandin2023} to mention a few). As magnetic reconnection plays a key role in magnetosphere dynamics, it has been investigated in several Vlasiator studies, both at the magnetopause \citep{pfaukempf2016, hoilijoki2017, hoilijoki2019, akhavantafti2020, pfaukempf2020} and in the magnetotail \citep{palmroth2017, juusola2018, runov2021, palmroth2023}. Since Vlasiator does not include an explicit resistive term, magnetic reconnection is enabled by the numerical diffusivity or resistivity.

In this section, we present key results from selected Vlasiator studies in two subsections, one devoted to magnetopause reconnection and the other focusing on magnetotail reconnection. We start with discussing 2D-3V simulations, and then present 3D-3V simulations since recent algorithmic improvements allowed running global, three-dimensional (3D-3V) hybrid-Vlasov simulations of Earth's magnetosphere \citep{ganse2023}.

\emph{Magnetopause Reconnection:}
Global simulations allow us to study the interaction between the solar wind and the magnetosphere, including how and to which extent dayside magnetic reconnection is affected by the solar wind and magnetosheath dynamics. \citet{hoilijoki2017} investigate this topic by using a 2D-3V Vlasiator global simulation in the GSE polar $xz$ plane, focusing in particular on the laminar or bursty nature of magnetic reconnection during steady solar wind conditions. 

The simulation domain covers $x = [-94, \ +48]\,R_\mathrm{E}$ and $z = [-56, \ +56]\,R_\mathrm{E}$ ($R_\mathrm{E} = 6371\,\mathrm{km}$ is the Earth's radius) and it features a 2D line dipole centered at the origin modelling the Earth's magnetosphere dipole and scaled to match the geomagnetic dipole strength. The steady solar wind has a density $n = 1\,\mathrm{cm}^{-3}$, a constant velocity $\mathbf{v}_{SW} = - 750\,\mathrm{km/s} \ \hat{\mathbf{x}}$ and a proton temperature of 0.5\,MK. The interplanetary magnetic field (IMF) is directed purely southward and it has a magnitude of 5\,nT. The resolution is uniform in the simulation domain; the spatial resolution is 300\,km ($\sim 0.047\,R_\mathrm{E} \sim 1.3 \ d_{p,SW}$, where $d_{p,SW}$ is the proton inertial length in the solar wind) and the velocity space resolution is 30\,km/s ($\sim 0.33 \ v_\mathrm{th,p,SW}$, where $v_\mathrm{th,p,SW}$ is the solar wind proton thermal speed). 
The solar wind flows into the simulation domain from the boundary at $x = +48\,R_\mathrm{E}$ with constant parameters.  The boundaries of the simulation box are periodic in the out-of-plane $y$ direction while the $-x$ and $\pm z$ boundaries apply copy boundary conditions. The inner boundary of the magnetosphere is a circle of radius $4.7\,R_\mathrm{E}$ centered at the origin and it enforces a static Maxwellian proton velocity distribution and perfect conductor field boundary conditions.

\citet{hoilijoki2017} reported that, despite the steady solar wind conditions, magnetic reconnection at the subsolar magnetopause does not reach a steady state and it is very dynamic. Indeed, magnetic islands are constantly produced and the presence of multiple X-points is observed. The motion of the X-points appears to be mostly dictated by the outflow produced by the neighboring X-points. \citet{hoilijoki2017} suggest that including the ion kinetic physics in the model promotes the development of a dynamic and bursty reconnection process at the dayside. 

This study investigates also the reconnection rate at the multiple simultaneous X-points and how the rate is affected by the local plasma conditions near the X-point. In particular, the presence of mirror modes in the magnetosheath appears to affect the reconnection rate, in agreement with spacecraft observations \citep{laitinen2010}. The dependence of the magnetopause reconnection rate upon the IMF direction is further investigated in global Vlasiator simulation by \cite{hoilijoki2019}. In particular, the run presented in \citep{hoilijoki2017} is compared to a run with similar parameters but with a positive component of the IMF ($B_{IMF} = [3.54, \ 0, \ -3.54] \ nT$). The Sun-ward tilt of the IMF results in a smaller tangential field at the magnetopause, leading to a reduction of the reconnection rate with respect to the purely southward-directed IMF case. The presence of a non-zero $B_{x,IMF}$ introduces an asymmetry that impacts the reconnection process in terms of flux transfer events (FTE) size, speed and occurrence rate. In particular, FTEs are observed more frequently in the Northern Hemisphere and they are smaller in size with respect to the  Southern Hemisphere.

\begin{figure*}[htbp!]
\begin{center}
\includegraphics[width=0.7\textwidth]{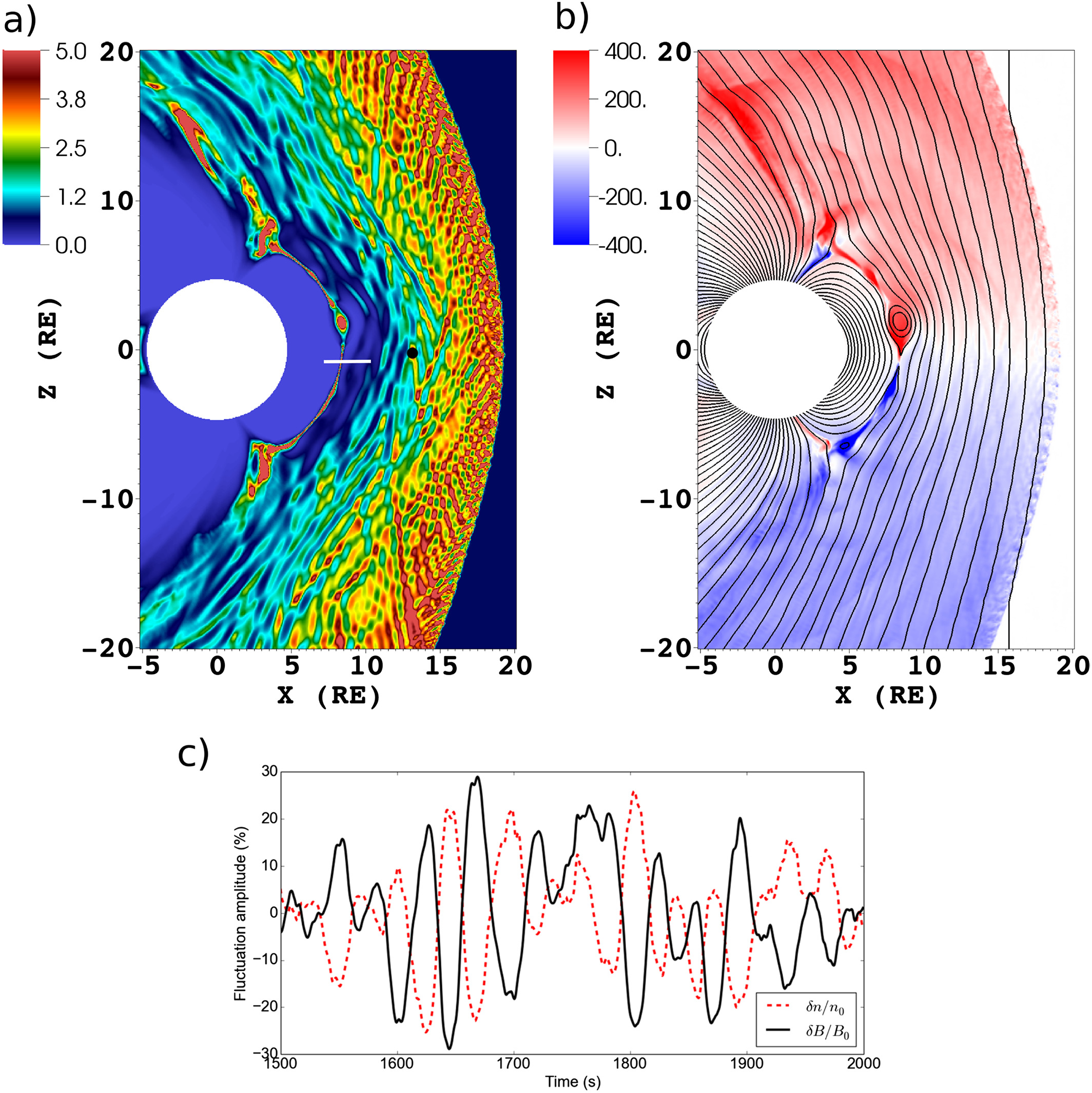}
\caption{(a) Plasma $\beta$; (b) proton $V_z$. The black lines show the magnetic field lines. (c) Magnetic field strength (black solid) and plasma density (red dashed) fluctuations from the virtual spacecraft location indicated with the black dot in panel (a). The anticorrelation between magnetic field and density fluctuations is compatible with mirror mode waves. Adapted from \cite{hoilijoki2017}.}
\label{fig:HVM_reconnection_fig4}
\end{center}
\end{figure*}

The findings of \citet{hoilijoki2017} have been confirmed by \citet{pfaukempf2020}, who analyze the reconnection process in a three-dimensional setup reproducing the magnetopause surface. In particular, \citet{pfaukempf2020} report a Vlasiator simulation of the noon–midnight meridional plane which is extended to cover 7\,$R_\mathrm{E}$ in the dawn–dusk direction. The study by \citet{pfaukempf2020} is the first example of a 3D-3V Vlasiator simulation of a cylindrical geometry mimicking the subsolar dayside magnetosphere. While the dimensionality is increased, the cylindrical geometry and the limited extent in the $y$ direction allow keeping the computational cost affordable and much lower than a global 3D-3V simulation modeling the entire magnetosphere.

The simulation domain covers $x = [-16, \ +31]\,R_\mathrm{E}$, $x = [-3.5, \ +3.5]\,R_\mathrm{E}$ and $z = [-35, \ +35]\,R_\mathrm{E}$ and the solar wind and IMF parameters are the same used for the 2D-3V run reported in \citet{hoilijoki2017} and discussed above. The spatial resolution is $\sim 0.24\,R_\mathrm{E}$, which is larger than the resolution of 2D-3V Vlasiator simulations because of the increased computational cost of 3D-3V runs. 

Identifying the magnetic reconnection site in 3D settings is not as straightforward as in 2D-3V simulations, where the local behaviour of the flux function allows to identify saddle points corresponding to X points that are associated with reconnection sites. Hence, in this study the X-line location is estimated by combining the four-field junction method \citep{Laitinen2006AnGeo} with the identification of the locations exhibiting a flow reversal in the $z$ direction. However, the four-field junction method is insufficient for identifying multiple reconnection X-lines. \cite{pfaukempf2020} find that, despite the uniform initial condition and the cylindrical symmetry in $y$, the X-line is not a straight line and it exhibits variations along the $y$ direction. It is suggested that structures in the magnetosheath break the translation symmetry along $y$. Analogously to \citet{hoilijoki2017}, \citet{pfaukempf2020} point out that reconnection is bursty and patchy, with multiple reconnection sites being present at the various $z$ and $y$ locations across the magnetopause, despite the homogeneous and steady-state solar wind conditions. 

\emph{Magnetotail Reconnection:}
Recently, magnetotail reconnection has been investigated in the context of a 3D-3V Vlasiator simulation investigating the dynamics of plasma eruptions \citep{palmroth2023}. The three-dimensional simulation in both ordinary and velocity space is made possible by technological advances, notably by enabling static adaptive mesh refinement (AMR) for ordinary space \citep{ganse2023}. With AMR, regions of high scientific interest such as the magnetotail plasma sheet are sampled with higher resolution (0.16 $R_E$) with respect to other regions in the simulations, the coarser resolution is 1.26 $R_E$. 

The 3D-3V simulation domain covers $x = [-111, \ +50]\,R_\mathrm{E}$ and $y, z = [-58, \ +58]\,R_\mathrm{E}$. The simulation parameters and initial conditions (IMF, solar wind density and speed) are the same adopted in \citep{hoilijoki2017}. However, since this is a 3D-3V run, the Earth's dipole is 3D and the inner boundary is a sphere of radius 4.7 $R_E$, while the $\pm y$ boundaries apply copy boundary conditions, as the $\pm z$ boundaries. Differently from  \citep{hoilijoki2017, palmroth2018_LR, juusola2018}, where Ohm's law included only the Hall term, this run includes the electron pressure gradient term as well. A polytropic closure is adopted for electrons, $\mathbf{P}_e = p_e \mathbf{I}$ and $p_e = n^\gamma T_e$, where $\mathbf{P}_e$ is the electron pressure tensor, $p_e$ is the scalar pressure, $T_e$ is the electron temperature and $n$ is the density. The polytropic index $\gamma$ is set to $5/3$ (adiabatic).

\begin{figure*}[htbp!]
\begin{center}
\includegraphics[width=\textwidth]{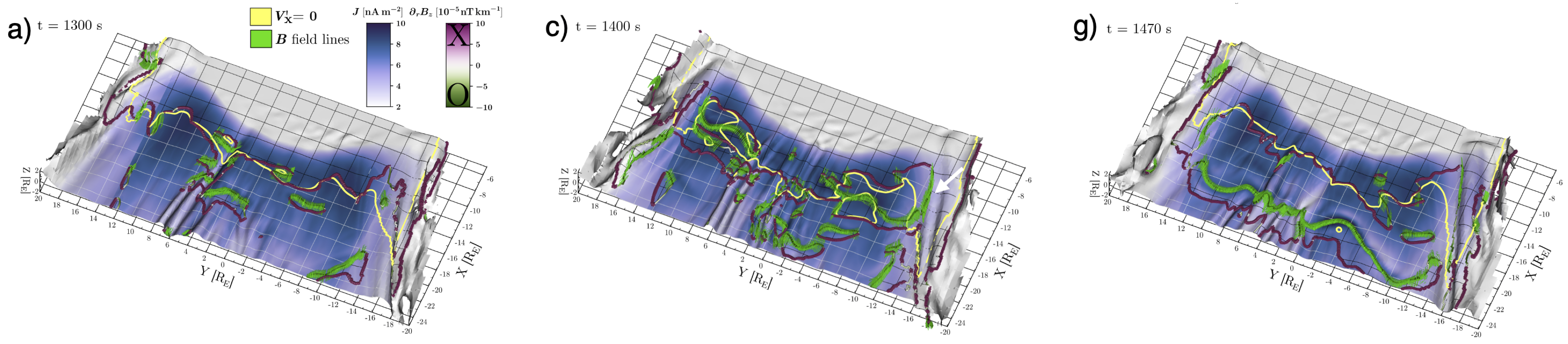}
\caption{Evolution of the magnetotail current sheet in a 3D-3V Vlasiator simulation. The panels show the current sheet surface (defined as $B_r = 0$) at different times, $t =  1300 \ s$ (a), $t = 1400 \ s$ (c), $t = 1470 \ s$ (g). The color of the surface corresponds to the current density $J$. The yellow line indicates the flow reversal between the Earthward and tailward reconnection outflow. The magenta and green lines are locations where $B_r = 0$ and  $B_z = 0$ and correspond to X-lines and O-lines (differentiated using the sign of $\partial B_z/\partial r$, which is positive at the X-lines and negative at the O-lines). The primary reconnection line is where the X-line (magenta) and flow reversal (yellow) contours are approximately co-located. The background grid shows the coordinates but also the magnetic-field topology: the black grid shows areas where the magnetic field is directed northward, and the white grid shows the areas where it is southward-directed. Adapted from \cite{palmroth2023}.}
\label{fig:vlasov-tail}
\end{center}
\end{figure*}

\citet{palmroth2023} focus on the investigation of magnetotail, revealing  complex dynamics where magnetic reconnection and kinking instability co-exist in the magnetotail current sheet. In particular, in Fig.~\ref{fig:vlasov-tail} it is shown that both processes are required to induce a global topological reconfiguration of the magnetotail, with the formation of a tail-wide plasmoid which is released and rapidly moves tailward. As mentioned above, the identification of the reconnection sites is challenging in 3D systems since we cannot rely on the identification based on the flux function. The reconnection site (X-line) in the magnetotail is identified by a combination of magnetic field and velocity proxies. X-lines and O-lines are identified as the locations where both $B_r = B_z = 0$, where  $B_r$ is the radial magnetic field component. The quantity $\partial B_z/\partial r$ allows us to distinguish between X-lines  ($\partial B_z/\partial r>0$ in the magnetotail) and O-lines ($\partial B_z/\partial r<0$ in the magnetotail). The locations where an X-line is co-located with a $\rm v_x$ reversal (diverging plasma flow) are identified as reconnection sites. \cite{palmroth2023} further confirm that reconnection is ongoing at those locations by showing reconnection signatures such as the Hall electric field and ion demagnetization. In the 3D-3V run, a dominant tail-wide reconnection X-line is found at $X \sim -15 \ R_E$ throughout the simulation. The dominant X-line is very dynamic and new X-lines and O-lines with limited extent in the Y-direction are constantly formed. 

\section{The Rice Convection Model} \label{sec:RCM}
\subsection{Introduction:}
The Rice Convection Model (RCM), by definition, never includes the physics of reconnection. However, reconnection is a microscale/mesoscale process. Only a small fraction of the magnetic flux in the magnetosphere is included in reconnection at any given time. Including the RCM allows us to discuss how the reconnection process impacts the rest of the magnetosphere, specifically the inner magnetosphere, where the magnetic field lines are closed. Since the RCM’s outer boundary condition comes from the plasmasheet, any process in the tail, such as reconnection, can impact the inner magnetosphere. Specifically, reconnection can generate low entropy bubbles that move toward the Earth at high speeds and can be a significant source of transport of plasma and magnetic field from the tail to the inner magnetosphere. In addition, the RCM helps quantitatively predict the plasmas from the inner magnetosphere which can reconnect on the dayside.

\subsection{Assumptions and Equations}
The physics behind the RCM can be found in detail in~\citet{toffoletto2003inner} and ~\citet{wolf1983quasi}, and a detailed discussion of the use of the RCM can be found in~\citet{wolf2016forty} and~\citet{toffoletto2003inner}, and ring current models are described by~\citet{toffolleto2020}. In the RCM, the distribution of magnetospheric particles is assumed to be isotropic. A key variable is the isotropic energy invariant
\begin{equation}
    \left| \lambda_k \right| = W\left( \lambda_k, \mathbf{x}, t \right) V^{2/3}
\end{equation}
where $W$ is the particle kinetic energy, including bounce and gyro motion and the sign of $\lambda_k$ is positive for positive ions and negative for electrons. The flux tube volume is
\begin{equation}
    V=\int^{nh}_{sh} \frac{ds}{B(\mathbf{x},t)}
\end{equation}
where the integral extends along the field line from the southern to the northern ionosphere.

The motions of magnetospheric particles in the inner magnetosphere are assumed to be governed by
\[
 \text{Drift velocity} \ll \text{bounce motion} \ll \text{gyro motion}
\]
The RCM calculates the bounce-averaged drift velocity, including gradient, curvature, and $E \times B$ drifts, i.e.,
\begin{equation}
    \mathbf{v}_k= \frac{\left(\mathbf{E} - \frac{1}{q_k}\nabla W\left( \lambda_k, \mathbf{x}, t \right) \right)\times \mathbf{B}(\mathbf{x},t)}{B(\mathbf{x},t)^2}
\end{equation}
where $q_k$ is the charge of a particle of species k. Inertial drift is assumed to be negligible. The quantity $\eta_k(\mathbf{x},t)$ is defined as the number of particles per unit magnetic flux for particles of a specific chemical species and a specific value of energy invariant. It follows a conservation law~\citep{wolf1983quasi}
\begin{equation}\label{eqn:bounce_drift_velocity}
    \bigg[ \frac{\partial}{\partial t} + \mathbf{v}_k\big(\lambda_k, \mathbf{x},t \big)\cdot \nabla \bigg]\eta_k = -L\big( \eta_k \big)
\end{equation}
where $L$ is the loss rate of particles due to precipitation and charge exchange. The classic RCM neglects particles flowing up from the ionosphere to the magnetosphere. In the early 2000’s Stan Sazykin and Darren DeZeeuw implemented a grid-based scheme using the CLAWPAK package~\citep{mandli2016clawpack} that was more robust but a bit more diffusive than the earlier Lagrangian scheme.

The flux tube content $\eta_k$ is related to the thermodynamic pressure $P$
\begin{equation}
    PV^{5/3} = \frac{2}{3} \sum_{k} \eta_k \lvert \lambda_k \rvert
\end{equation}
while the flux tube content is related to the plasma distribution function $f_k(\lambda)$ as
\begin{equation}
    \eta_k = \frac{4\pi 2^{1/2}}{m^{3/2}_k} \int^{\lambda_{max}}_{\lambda_{min}}\lvert \lambda \rvert^{1/2} f_k(\lambda) d\lambda
\end{equation}
where $(\lambda_{max}-\lambda_{min})$ is the width of the invariant energy channel associated with species $k$. The species $k$ is defined for a given chemical species (usually $\text{e}^{-}$, $\text{H}^{+}$, $\text{O}^{+}$), and the specific value of the energy invariant.

The electric field can be expressed as the sum of a potential component and an inductive component
\begin{equation}
    \mathbf{E} = -\nabla \Phi -\mathbf{v}_{induction}\times \mathbf{B}
\end{equation}
In the RCM, the inductive electric field is included implicitly through time-dependent magnetic field mappings. The inductive magnetic field in the ionosphere is assumed to be zero there; however, it is not zero in the magnetosphere.

There are two more complications in the electric field:
\begin{enumerate}
    \item In the classic RCM, we assume the electric field is perpendicular to the magnetic field.
    \item There are two coordinate systems used in the classic RCM. One moves with the Earth as it rotates, and the potential in that system is labeled $\Phi_i$. The other does not rotate with the Earth, is approximately an inertial system, and is labeled $\Phi$.
\end{enumerate}

We can translate from one system to the other in the ionosphere using the formula
\begin{equation}\label{eqn:potential}
    \Phi = \Phi_i - \frac{\omega_E B_0 R^3_E sin^2(\theta_i)}{R_i}
\end{equation}
where $\omega_E$ is the angular rotation rate of the Earth, $B_0$ is the magnetic field at the Earth's equator, $R_E$ is the radius of the Earth, $\theta_i$ is the colatitude, and $R_i$ is the radius of the ionosphere. Equation~\ref{eqn:potential} applies to the equatorial plane, but it applies only to a dipole magnetic field. To compute $\Phi$ in the equatorial plane of the magnetosphere, the RCM calculates $\Phi$ by mapping between the ionosphere to the equatorial plane, assuming $\Phi$ is constant along each field line.

In the thin-shell approximation, the equation for the conservation of current is $(\nabla \cdot J =0)$ can be written
\begin{equation}\label{eqn:currentconservation_thinshell}
    \nabla_i \cdot \Big[ \overleftrightarrow{\Sigma} \cdot \Big(\nabla_i \Phi_i \Big) \Big]= \Big( j_{\parallel nh} - j_{\parallel sh}\Big) sin(I)
\end{equation}
where $\overleftrightarrow{\Sigma}$ is the field-line integrated conductivity tensor due to both hemispheres, $I$ is the dip angle of the magnetic field in the ionosphere, and $j_{\parallel nh} - j_{\parallel sh}$  is the ionospheric field-aligned current density.

The~\citet{vasyliunas1970mathematical} equation, which is based on force balance
\begin{equation}\label{eqn:forcebalance_basic_eqn}
    \mathbf{J} \times \mathbf{B} - \nabla P = 0
\end{equation}
is given by
\begin{equation}\label{eqn:forcebalance}
    \frac{j_{\parallel nh} - j_{\parallel sh}}{B_i}=\frac{\hat{b}}{B}\cdot \nabla V \times \nabla P
\end{equation}
which relates field-aligned currents in the ionosphere to pressure gradients in the magnetosphere, and $B_i$ is the magnetic field at the southern- and northern-ionospheric footprints of the field line (assumed the same). The derivation makes use of the fact the right-hand side of Eqn.~\ref{eqn:forcebalance} can be evaluated anywhere along the field line. The RCM equations are solved on a fixed ionospheric grid that has variable grid spacing in latitude to better resolve the auroral zone. The RCM grid is time-dependent in the equatorial plane, ranging from just inside the magnetopause on the dayside to $10-20 R_E$ on the night side.

\subsection{Inputs, Boundary, and Initial conditions and Outputs}
\textit{The magnetic field model:} For many years, the RCM assumed a constant magnetic field, but, beginning about 2000 the RCM used a time-dependent semi-empirical model such as the Tsgyanenko models (1989, 1995, 2003). The RCM can also use the~\citet{hilmer1995magnetospheric} magnetic field model. In classic RCM runs, the magnetic field is not designed to be consistent with Ampere’s law and equation~\ref{eqn:forcebalance_basic_eqn}. Ways of including eqn.~\ref{eqn:forcebalance_basic_eqn} consistently in the RCM are described in Section~\ref{subection:Generalization_RCM}. In the ionosphere, where the magnetic field is assumed to be dipolar, the magnetic field is
\begin{equation}
    \mathbf{B}_i = -\frac{\mu_0}{4\pi}\frac{\mathbf{M}_E-3\hat{r}\big( \mathbf{M}_E \cdot \hat{r} \big)}{R^3_i}
\end{equation}
where $\hat{r}$ is the radial unit vector in the ionosphere, and $\mathbf{M}_E$ is the magnetic moment, which is in the southern direction, and
\begin{equation}
    B_i = \hat{r} \cdot \mathbf{B}_i
\end{equation}

\textit{Ionospheric conductance:} Ionospheric conductance has two major drivers: Solar heating (e.g., the Sheffield University Ionosphere Plasmasphere (SUPIM) model, ~\citet{bailey1997sheffield}). The second is Auroral heating. The standard treatment uses the electron precipitating energy flux and average energy~\citep{robinson1987calculating}.

\textit{Loss models:} Separate models are needed for electrons and ions. The simplest electron loss model assumes a fixed fraction of strong pitch-angle scattering, often between $33\%$ and $67\%$~\citep{schumaker1989relationship}. That procedure is reasonable for the plasma sheet but overestimates the electron loss rate in the inner magnetosphere. A slightly more sophisticated model~\citep{chen2001simulations} is somewhat more realistic. For the Ion loss model, there are many theoretical models of ion charge exchange. The overall ion loss rate for an energy and $L$ value is calculated using an algorithm developed by James Bishop~\citep{freeman1993magnetospheric, bishop1996multiple}.

The RCM needs boundary conditions both at its outer (large $L$) boundary and at its lowest (low $L$) boundary. The large$-L$ boundary depends on MLT as well as UT. Note that the large$-L$ boundary can’t be aligned with the grid, except in a few cases. At this boundary, the number $\eta_s$ which is the number of particles for a given type $s$ per weber of magnetic flux, is needed as well as the potential distribution, which can be a simple function of solar wind conditions or an empirical model such as the Weimer model (1985).

The low-L boundary (low-latitude) is set at least a few degrees latitude from the equator. Given the aligned-dipole assumption, the latitudinal current density should, in principle, be zero at the equator. The RCM and many other models use a thin-wire approximation to represent the region near the equator, to provide a boundary condition
\begin{equation}
    \frac{\partial J_\theta}{\partial \theta} + S \frac{\partial J_\phi}{\partial \phi} = 0
\end{equation}
which was derived by~\citet{blanc1980ionospheric}. Here $J$ is the current density, $\theta$ is the colatitude coordinate, and $\phi$ is the longitudinal coordinate. $S$ is a function of $\phi$ that was defined by~\citet{blanc1980ionospheric}.

\emph{Initial conditions:} These are needed to provide the value of the initial value of $\eta_k$, which is a function of grid location and time. The RCM can be initialized with an empty value and run for a period of time to fill in the inner magnetosphere. Alternatively, the RCM uses the~\citet{spence1989magnetospheric} model for the initial pressure distribution. Earlier versions of the RCM assumed a Maxwellian distribution, but more recent versions have the option to assume a kappa distribution (e.g.,~\citet{yang2015contribution}).

\emph{RCM outputs:} The main RCM outputs are the electric potential $(\Phi)$, field-aligned currents $(j_{\parallel nh}-j_{\parallel sh})$, the distribution function $(\eta_s)$ and moments (pressure and density) within the RCM modeling region in the ionosphere and the magnetospheric equatorial plane.

\subsection{Generalizations of the RCM}\label{subection:Generalization_RCM}
There have been many modifications to the RCM, particularly since 2000.

\emph{Other planets:} Tom Hill and several of his students modified the RCM to be appropriate for Jupiter and Saturn. Jupiter’s moon Io has volcanoes that loft neutrals and positive ions into the inner magnetosphere, and there is also a similar effect at Saturn’s moon Enceladus. Centrifugal force is stronger than gravity near Io, and the region beyond Io is consequently interchange unstable. In the simulations, plasma develops finger-like structures, moving outward because of centrifugal force and azimuthally because of Coriolis force. The clearest magnetospheric signature of interchange transport occurs in the inner magnetosphere of Saturn, where the hot plasma injection-dispersion structures are evident (\citet{hill2005evidence} and references therein).

\emph{CRCM (Comprehensive Ring Current Model):} This model~\citep{fok2001comprehensive} was similar to the classic RCM, except that it used a much more complete equation for the distribution function. Whereas the classic RCM assumed an isotropic pitch-angle distribution, CRCM assumes conservation of the first and second invariant. Additional terms account for precipitation and charge-exchange losses and pitch-angle scattering. CIMI (Comprehensive Inner Magnetosphere Model) includes radiation belt electrons and the plasmasphere~\citep{fok2014comprehensive}.

\emph{RCM-E:} In the classic RCM, the magnetic field was not required to satisfy the force balance equation $\mathbf{J}\times \mathbf{B} = \nabla P$ but $P=\big( 2/3 \big) V^{-5/3} \sum_k \lvert \lambda_k \rvert \eta_k $ and $\eta_k$ was based on the theoretical equation~\ref{eqn:bounce_drift_velocity}. The RCM-E (equilibrium) is run for a small time step (typically $1-5$ minutes), and a modified MHD code, called the “friction code”, recalculates the magnetic field in order to make it approximately consistent with $\mathbf{J}\times \mathbf{B} = \nabla P$~\citep{lemon2003computing}. For conditions of strong convection, the time development of the RCM-E would cause the magnetic field to be tail-like and more like a substorm growth phase. If $PV^\gamma$ was constant on the nightside large $L-$ boundary, it became difficult to form a realistic strong ring current~\citep{lemon2004magnetic}. The RCM-E usually exhibited the pressure balance inconsistency~\citep{erickson1980steady}. In other words, the more theoretically consistent model became a less realistic representation of observations. Other modelers have developed models that are variants of the RCM-E (e.g.,~\citet{chen2012comparison} or the RAM-SCB model~\citep{zaharia2006self}) that use an alternative ring current and force balance model.
The solution to the pressure balance inconsistency turned out to be bursty bulk flows (BBFs), which are localized regions of the inner and middle plasma sheet~\citep{angelopoulos1992bursty} that flow rapidly earthward. The flow bursts, which are also often called “bubbles”, often move very fast (typically $400 km/s$). These flows correspond to regions of low $PV^\gamma$~\citep{pontius1990transient}. ~\citet{angelopoulos1992bursty} found that BBFs account for a large fraction of the total earthward flow in the plasma sheet. BBFs usually terminate about the inner edge of the plasma sheet, although some of the fast flows penetrate the ring current ~\citep{gkioulidou2014role,yang2016comparison}. ~\citet{lemon2004magnetic} produced a substantial ring current injection by reducing the $PV^\gamma$ at the RCM’s outer boundary over a limited region of local time, simulating a ring current injection during a storm. ~\citet{yang2014rcma} showed a possible relation to the streamers observed in the polar cap and bubbles in the plasma sheet, and~\citet{yang2014rcmb} argued that this effect could account for pressure balance inconsistency and that during storms low entropy flux tubes could account for up to $60\%$ of the ring current~\citep{yang2015contribution}. See also the section~\ref{sec:large_coupled_include_RCM} “Large, coupled models that include RCM”.

\emph{RCM-I (RCM-Inertial):} The main problem with using RCM-E to represent BBFs is they move so fast that the assumption of force balance is not valid. ~\citet{yang2019inertialized} developed a more complex version of the RCM that includes inertial effects in a very approximate way. Equation~\ref{eqn:currentconservation_thinshell} is replaced by a much more complex expression that involves a $\partial \Phi/ \partial t$ term.

\subsection{Large, coupled models that include RCM}\label{sec:large_coupled_include_RCM}
Single fluid, global MHD models have become powerful tools in recent years (e.g.,~\citet{lyon2004lyon};~\citet{raeder2001global},~\citet{zhang2019gamera}). However, these models do not capture all the important physics in the inner magnetosphere where gradient and curvature drifts become important but are neglected in MHD. Coupling these models is a daunting task as the modeling regions overlap in space and information is fed back and forth between them. There are different physics assumptions associated with each model: the RCM model assumes slow flow and force balance and neglects waves, while MHD does not. However, MHD does not include energy-dependent drifts that are important in the inner magnetosphere. There have been several successful efforts to couple the RCM with global MHD, which provides many of the inputs used by the RCM (boundary and initial conditions) such as the magnetic field, plasma density, and pressure as well as the ionospheric potential. In return, the RCM provides the density and pressure that are derived from computing the moments from the RCM distribution function that have been subject to energy-dependent drifts.

\emph{SMWF:} The earliest successful coupling effort was~\citet{de2004coupling}  which merged the BATS-R-US Global MHD~\citep{toth2005space} code with the RCM. The RCM was embedded in the Global MHD code as a subroutine that later became part of the Space Weather Modeling Framework (SWMF)~\citep{toth2005space}. Each coupling exchange requires computing many field-line integrals to obtain the flux tube volume (equation 1), which is used by the RCM, and the mapping of the 2D RCM quantities into the 3D domain of the MHD code, which is then used to update the MHD. This field line tracing requires using a parallelized and efficient field line tracer that exploits the nested adaptive grid used in the MHD code (e.g.,~\citet{de2004coupling}. This version of the model demonstrated that including the RCM increased the pressure in the inner magnetosphere ring current region as compared to standalone MHD. The RCM also was able to model the region-2 currents in the ionosphere. Later versions of the SWMF also include other inner magnetosphere models such as the Comprehensive Ring current model (CRCM)~\citep{glocer2013crcm+} and other models~\citep{toth2012adaptive}, becoming the first coupled magnetosphere model to be used in the Community Coordinated Modeling Center (CCMC).

\emph{LFM-RCM:} ~\citet{pembroke2012initial} coupled the RCM to the Lyon Fedder Mobary (LFM) global MHD code~\citep{lyon2004lyon} that included the MIX ionosphere model~\citep{merkin2010effects}. This approach used a loose coupling scheme where the models (LFM, MIX, and RCM) ran independently as separate processes and used the InterComm software package to exchange information at pre-set intervals~\citep{lee2004efficient}. In this model, RCM returned both pressure and density to the MHD code and included a simple static plasmasphere based on the ~\citet{gallagher2000global} empirical model. Since the RCM was only tracking the distribution function and not computing the potential as in the standalone RCM, the assumption of zero dipole tilt could be relaxed in the coupled model, allowing for more realistic simulations. The resulting coupled model was very dynamic, especially during geomagnetic storm simulations, and significantly impacted the ring current region (e.g.,~\citet{wiltberger2017effects}). To keep the code stable, the RCM boundary was restricted to regions where the field line average plasma beta was less than $1$. With moderately strong solar wind driving, the coupled model produced a strong ring current and region-2 currents.

\emph{OpenGGCM-RCM:} ~\citet{hu2010one} and~\citet{cramer2017plasma} coupled the OpenGGCM global MHD code to RCM~\citep{raeder2001global} that also includes the Coupled Thermosphere-Ionosphere Model (CTIM)~\citep{fuller1996coupled}. The RCM is embedded within the MHD code, where the feedback to the MHD code used a configurable ramp-up region based on the strength of the magnetic field for numerical stability. ~\citet{cramer2017plasma} found that most of the transport of plasma to the inner magnetosphere is via low entropy bubbles, consistent with~\citet{yang2015contribution}. ~\citet{raeder2016sub} also used the coupled model to simulate a geomagnetic storm and showed that it developed subauroral polarization streams (SAPS) from electron precipitation computed from the MHD code. ~\citet{hu2011consequences} examined the entropy profile in an idealized Open GGCM simulation and found that violations of the frozen-in-flux in MHD could lead to an entropy profile that produced a low entropy bubble that was earthward of an entropy enhancement. Such a configuration causes the bubble/blob pair to move earthward/tailward, which thins the current sheet in the region between them and can ultimately result in tearing or other configuration changes.

\emph{MAGE:} The newest version of a global magnetosphere model is the Multiscale Atmosphere Geospace Environment Model (MAGE) that couples the RCM to the Grid Agnostic MHD for Extended Research Application (GAMERA) global MHD code~\citep{zhang2019gamera, sorathia2021role}, the ReMIX ionosphere model~\citep{merkin2010effects} which is a revised version of the MIX solver, and the NCAR Thermosphere-Ionosphere- Electrodynamics General Circulation Model (TIE-GCM)~\cite{roble1988coupled}. The GAMERA MHD model is derived from the LFM model but with improved numerical algorithms and updated software designed for efficient use on modern supercomputers. The coupling to the RCM has also been significantly improved and modernized, for example, it uses a highly configurable and customizable parallel field line tracer. Other improvements include moving the RCM boundary in MAGE further from the Earth compared to the coupled LFM-RCM code, allowing more plasma from the plasmasheet to move into the RCM modeling region, and the option of a Maxwellian distribution to compute RCM distribution functions replaced with a Kappa distribution~\citep{sciola2023}. The new model also includes improved loss rate mechanisms for electrons~\cite{bao2023} where the electron precipitation model is based on RCM-computed electron energy fluxes that are used to modify the ionospheric conductances~\citep{lin2022ionospheric}. The use of this conductance model was found to influence the formation of the SAPS channel~\citep{Lin2021}. The model also includes a dynamic plasmasphere density that is tracked using a zero-energy channel in the RCM that is fed back to the MHD model~\citep{bao2023}. ~\citet{pham2022thermospheric} used the coupled MAGE model to investigate the impact on thermospheric density perturbations produced by traveling ionospheric disturbances. ~\citet{sciola2023} found that in the MAGE model, fast magnetospheric flows associated with low entropy channels can contribute over $50\%$ of the ring current population, consistent with~\citet{yang2015contribution} and~\citet{cramer2017plasma}.

\subsection{Summary}
While the RCM does not model reconnection, it is impacted by it. Reconnection in the tail produces low entropy flux tubes that rapidly interchange their way toward the Earth (e.g.,~\citet{wiltberger2015high, sorathia2021role}). Some of these flux tubes make it into the inner magnetosphere and play an important role in the formation and structure of the ring current region. It can also have ionospheric effects such as the formation of streamers. Over the years, the RCM has helped illuminate the impact of processes in the tail on the inner magnetosphere. This is especially true using the new generation coupled models that have been developed in recent years. However, there are several limitations in the models that present challenges. One is the modeling of the region in the tail where the magnetic field is transitioning from a stretched tail-like configuration to a dipole. When fast flows appear in this region, as they often do, neither MHD, which neglects gradient/curvature drifts, nor the RCM, which assumes slow flow, are applicable. 

Furthermore, including the RCM in MHD models can also affect the location and effectiveness of dayside reconnection as well.  Since the Region 2 FACs modeled by the RCM shields the inner magnetosphere from convection, it also strongly affects the distribution of return flow back to the dayside boundary.  The RCM has been successful in modeling the plasmaspheric plumes that can bring dense plasmas to the dayside reconnection regions~(\cite{goldstein2002efieldovershielding,huba2014rcm,bao2023}). Dayside SWMF runs that include RCM all successfully place MMS within 1 RE of at an X-line (or separatrix) whenever clear EDRs are observed (e.g.~\cite{reiff2017ediffusionregion}). 

The next generation coupled models will need to add the effect of ionospheric plasma sources (e.g.,~\cite{glocer2009modeling,varney2016influence}), which will ultimately require a multi-fluid MHD model coupled to a ring current model and includes a model of ionospheric outflow to track all the species in the magnetosphere.

\section{Conclusions} \label{sec:Conclusion} 

In this paper, we have presented a ``brief'' overview of the large collection of computational methods that are used to study magnetic reconnection. It should be clear to the reader of this text, that simulating magnetic reconnection is nearly a separate field of study in and of itself. It should be and has been~\citep{buchner2003space} the topic of entire books. The single element to take away from this paper is that simulating a multifaceted and multiscale problem like magnetic reconnection is not a simple endeavor. Both plasma models and simulation initial conditions must be tuned carefully to match the goals of the study. 

We have reviewed simulation methods for magnetic reconnection in space plasmas, from macroscopic MHD scales to microscopic kinetic scales. Basically, macroscopic plasma behaviors can be simulated based on fluid modes, and as we resolve smaller scale physics, kinetic models need to be implemented in simulations. In addition, we have reviewed novel approaches to incorporate multi-scale physics.

MHD simulations are useful to study large scale physics, including performing global simulations for planetary magnetospheres. We have discussed the basic algorithm for MHD simulations, and also how to implement test particles that follow MHD fields. Hall MHD simulations contain the Hall physics, which allows kinetic scale waves to propagate, mediating fast reconnection. We have reviewed recent progresses of Hall MHD studies, and also the effect of the electron inertia term and EMHD.

Hybrid PIC simulations and full PIC simulations include particle kinetic physics, where particle motions are directly solved by equations of motion of particles. For hybrid simulations, we have reviewed techniques to overcome the limitation of spatiotemporal resolutions in global models, and also the implementation of electron kinetic physics into hybrid simulations. For full PIC simulations, we have reviewed simulation studies of magnetic reconnection in magnetotail and magnetopause, particle acceleration, and shock driven reconnection. 

Next, we have reviewed two novel approaches to address multi-scale physics in magnetic reconnection: Embedded PIC, and kglobal. In the embedded PIC approach, the macro-scale region is solved using MHD equations, and local kinetic domains are embedded in the MHD domain, where full PIC techniques are employed. In kglobal, on the other hand, equations for the ion and electron fluids are combined with particles’ equations for electrons, and the macro-scale evolution is modified by the kinetic physics. 

Finally, we have reviewed two types of other simulation techniques: Vlasov simulations and the Rice convection model. In Vlasov simulations, kinetic effects are implemented in simulations by solving 2D-3V or 3D-3V Vlasov equations. We have discussed recent progresses of studies by Vlasov models from local reconnection simulations to global simulations. The Rice convection model is a kinetic approach to simulate physics of the inner magnetosphere, where bounce averaged particle drift motion is taken into account.

\par




\bibliographystyle{spbasic} 

\bibliography{abbrev2,refs_PIC,refs_turb,refs_dg,refs_hybrid,refs_outflows,refs_shocks,refs_RCM,refs_particle_tracing,refs_Vlasov,refs_MHDAEPIC,refs_RegionalMHD,refs_kglobal,refs_hallMHD}

\begin{thebibliography}{513}
\providecommand{\natexlab}[1]{#1}
\providecommand{\url}[1]{{#1}}
\providecommand{\urlprefix}{URL }
\expandafter\ifx\csname urlstyle\endcsname\relax
  \providecommand{\doi}[1]{DOI~\discretionary{}{}{}#1}\else
  \providecommand{\doi}{DOI~\discretionary{}{}{}\begingroup
  \urlstyle{rm}\Url}\fi
\providecommand{\eprint}[2][]{\url{#2}}

\bibitem[{Adhikari et~al.(2021)Adhikari, Parashar, Shay, Matthaeus, Pyakurel,
  Fordin, Stawarz, and Eastwood}]{adhikari2021}
Adhikari S, Parashar TN, Shay MA, Matthaeus WH, Pyakurel PS, Fordin S, Stawarz
  JE, Eastwood JP (2021) Energy transfer in reconnection and turbulence.
  Physical Review E 104:065206

\bibitem[{Adhikari et~al.(2024)Adhikari, Yang, Matthaeus, Cassak, Parashar, and
  Shay}]{adhikari2024scale}
Adhikari S, Yang Y, Matthaeus WH, Cassak PA, Parashar TN, Shay MA (2024) Scale
  filtering analysis of kinetic reconnection and its associated turbulence.
  Physics of Plasmas 31(2)

\bibitem[{Akhavan-Tafti et~al.(2020)Akhavan-Tafti, Palmroth, Slavin, Battarbee,
  Ganse, Grandin, Le, Gershman, Eastwood, and Stawarz}]{akhavantafti2020}
Akhavan-Tafti M, Palmroth M, Slavin JA, Battarbee M, Ganse U, Grandin M, Le G,
  Gershman D, Eastwood JP, Stawarz J (2020) Comparative analysis of the
  vlasiator simulations and mms observations of multiple x-line reconnection
  and flux transfer events. Journal of Geophysical Research: Space Physics
  125(7):e2019JA027410, \doi{10.1029/2019JA027410}

\bibitem[{von Alfthan et~al.(2014)von Alfthan, Pokhotelov, Kempf, Hoilijoki,
  Honkonen, Sandroos, and Palmroth}]{Vlasiator2014}
von Alfthan S, Pokhotelov D, Kempf Y, Hoilijoki S, Honkonen I, Sandroos A,
  Palmroth M (2014) Vlasiator: First global hybrid-{V}lasov simulations of
  {E}arth's foreshock and magnetosheath. J Atmospheric and Solar-Terrest Phys
  120:24--35

\bibitem[{Alho et~al.(2019)Alho, Wedlund, Nilsson, Kallio, Jarvinen, and
  Pulkkinen}]{Alho:2019}
Alho M, Wedlund CS, Nilsson H, Kallio E, Jarvinen R, Pulkkinen T (2019) {Hybrid
  modeling of cometary plasma environments}. Astronomy \& Astrophysics 630:A45,
  \doi{10.1051/0004-6361/201834863},
  \urlprefix\url{https://doi.org/10.1051/0004-6361/201834863}

\bibitem[{Alho et~al.(2022)Alho, Battarbee, Pfau-Kempf, Khotyaintsev, Nakamura,
  Cozzani, Ganse, Turc, Johlander, Horaites, Tarvus, Zhou, Grandin, Dubart,
  Papadakis, Suni, George, Bussov, and Palmroth}]{alho2022}
Alho M, Battarbee M, Pfau-Kempf Y, Khotyaintsev YV, Nakamura R, Cozzani G,
  Ganse U, Turc L, Johlander A, Horaites K, Tarvus V, Zhou H, Grandin M, Dubart
  M, Papadakis K, Suni J, George H, Bussov M, Palmroth M (2022) Electron
  signatures of reconnection in a global evlasiator simulation. Geophysical
  Research Letters 49(14):e2022GL098329,
  \doi{https://doi.org/10.1029/2022GL098329}, e2022GL098329 2022GL098329,
  \eprint{https://agupubs.onlinelibrary.wiley.com/doi/pdf/10.1029/2022GL098329}

\bibitem[{Allmann-Rahn et~al.(2022)Allmann-Rahn, Grauer, and
  Kormann}]{allmannrahn2022}
Allmann-Rahn F, Grauer R, Kormann K (2022) A parallel low-rank solver for the
  six-dimensional vlasov–maxwell equations. Journal of Computational Physics
  469:111562, \doi{https://doi.org/10.1016/j.jcp.2022.111562},
  \urlprefix\url{https://www.sciencedirect.com/science/article/pii/S0021999122006246}

\bibitem[{Amano et~al.(2014)Amano, Higashimori, and Shirakawa}]{Amano2014}
Amano T, Higashimori K, Shirakawa K (2014) {A robust method for handling low
  density regions in hybrid simulations for collisionless plasmas}. Journal of
  Computational Physics 275:197--212

\bibitem[{Angelopoulos et~al.(1992)Angelopoulos, Baumjohann, Kennel, Coroniti,
  Kivelson, Pellat, Walker, L{\"u}hr, and Paschmann}]{angelopoulos1992bursty}
Angelopoulos V, Baumjohann W, Kennel C, Coroniti FV, Kivelson M, Pellat R,
  Walker R, L{\"u}hr H, Paschmann G (1992) Bursty bulk flows in the inner
  central plasma sheet. Journal of Geophysical Research: Space Physics
  97(A4):4027--4039

\bibitem[{Arencibia et~al.(2021)Arencibia, Cassak, Shay, and
  Priest}]{Arencibia21}
Arencibia M, Cassak PA, Shay MA, Priest ER (2021) Scaling theory of
  three-dimensional magnetic reconnection spreading. Physics of Plasmas 28(8)

\bibitem[{Arencibia et~al.(2022)Arencibia, Cassak, Shay, Qiu, Petrinec, and
  Liang}]{Arencibia22}
Arencibia M, Cassak PA, Shay MA, Qiu J, Petrinec SM, Liang H (2022)
  Three-dimensional magnetic reconnection spreading in current sheets of
  non-uniform thickness. submitted

\bibitem[{Arnold et~al.(2019)Arnold, Drake, Swisdak, and
  Dahlin}]{arnold2019large}
Arnold H, Drake J, Swisdak M, Dahlin J (2019) Large-scale parallel electric
  fields and return currents in a global simulation model. Physics of Plasmas
  26(10)

\bibitem[{Arnold et~al.(2021)Arnold, Drake, Swisdak, Guo, Dahlin, Chen,
  Fleishman, Glesener, Kontar, Phan et~al.}]{arnold2021electron}
Arnold H, Drake JF, Swisdak M, Guo F, Dahlin JT, Chen B, Fleishman G, Glesener
  L, Kontar E, Phan T, et~al. (2021) Electron acceleration during macroscale
  magnetic reconnection. Physical review letters 126(13):135101

\bibitem[{{Arr\`o, G.} et~al.(2020){Arr\`o, G.}, {Califano, F.}, and {Lapenta,
  G.}}]{arro2020}
{Arr\`o, G}, {Califano, F}, {Lapenta, G} (2020) Statistical properties of
  turbulent fluctuations associated with electron-only magnetic reconnection.
  A\&A 642:A45, \doi{10.1051/0004-6361/202038696},
  \urlprefix\url{https://doi.org/10.1051/0004-6361/202038696}

\bibitem[{Ashour-Abdalla et~al.(2011)Ashour-Abdalla, El-Alaoui, Goldstein,
  Zhou, Schriver, Richard, Walker, Kivelson, and
  Hwang}]{ashour2011observations}
Ashour-Abdalla M, El-Alaoui M, Goldstein ML, Zhou M, Schriver D, Richard R,
  Walker R, Kivelson MG, Hwang KJ (2011) Observations and simulations of
  non-local acceleration of electrons in magnetotail magnetic reconnection
  events. Nature Physics 7(4):360--365

\bibitem[{Attico et~al.(2000)Attico, Califano, and Pegoraro}]{Attico2000}
Attico N, Califano F, Pegoraro F (2000) Fast collisionless reconnection in the
  whistler frequency range. Physics of Plasmas 7(6):2381--2387,
  \doi{10.1063/1.874076}

\bibitem[{Aydemir(1992)}]{Aydemir92}
Aydemir AY (1992) Nonlinear studies of {$m = 1$} modes in high-temperature
  plasmas. Phys~Fluids B 4:3469

\bibitem[{Bailey et~al.(1997)Bailey, Balan, and Su}]{bailey1997sheffield}
Bailey G, Balan N, Su Y (1997) The sheffield university plasmasphere ionosphere
  model—a review. Journal of Atmospheric and Solar-Terrestrial Physics
  59(13):1541--1552

\bibitem[{Balsara(2009)}]{Balsara2009bJCP}
Balsara DS (2009) Divergence-free reconstruction of magnetic fields and {WENO}
  schemes for magnetohydrodynamics. Journal of Computational Physics
  228(14):5040 -- 5056, \doi{10.1016/j.jcp.2009.03.038}

\bibitem[{Bao et~al.(2023)Bao, Toffoletto, Merkin, Wang, Sorathia, Lin, Pham,
  Garretson, Wiltberger, Lyon, and Michael}]{bao2023}
Bao S, Toffoletto F, Merkin S, Wang W, Sorathia K, Lin D, Pham K, Garretson J,
  Wiltberger M, Lyon J, Michael A (2023) Drivers of the geospace plume: Mage
  simulation of the march 31 2001 super storm. In preparation

\bibitem[{Battarbee et~al.(2021)Battarbee, Brito, Alho, Pfau-Kempf, Grandin,
  Ganse, Papadakis, Johlander, Turc, Dubart, and Palmroth}]{battarbee2021}
Battarbee M, Brito T, Alho M, Pfau-Kempf Y, Grandin M, Ganse U, Papadakis K,
  Johlander A, Turc L, Dubart M, Palmroth M (2021) Vlasov simulation of
  electrons in the context of hybrid global models: an evlasiator approach.
  Annales Geophysicae 39(1):85--103, \doi{10.5194/angeo-39-85-2021},
  \urlprefix\url{https://angeo.copernicus.org/articles/39/85/2021/}

\bibitem[{Bessho et~al.(2019)Bessho, Chen, Wang, Hesse, and III}]{bessho2019}
Bessho N, Chen LJ, Wang S, Hesse M, III LBW (2019) Magnetic reconnection in a
  quasi-parallel shock. Geophys Rev Lett 46:9352

\bibitem[{Bessho et~al.(2020)Bessho, Chen, Wang, Hesse, III, and
  Ng}]{bessho2020}
Bessho N, Chen LJ, Wang S, Hesse M, III LBW, Ng J (2020) Magnetic reconnection
  and kinetic waves generated in the earth's quasi-parallel bow shock. Phys
  Plasmas 27:092901

\bibitem[{Bessho et~al.(2022)Bessho, Chen, Stawarz, Wang, Hesse, III, and
  Ng}]{bessho2022}
Bessho N, Chen LJ, Stawarz JE, Wang S, Hesse M, III LBW, Ng J (2022) Strong
  reconnection electric fields in shock-driven turbulence. Phys Plasmas
  29:042304

\bibitem[{Bessho et~al.(2023)Bessho, Chen, Hesse, Ng, III, and
  Stawarz}]{bessho2023}
Bessho N, Chen LJ, Hesse M, Ng J, III LBW, Stawarz JE (2023) Electron
  acceleration and heating during magnetic reconnection in the earth’s
  quasi-parallel bow shock. The Astrophysical Journal 954:25

\bibitem[{Biermann(1950)}]{Biermann50}
Biermann L (1950) {\"U}ber den ursprung der magnetfelder auf sternen und im
  interstellaren raum (miteinem anhang von a. schl{\"u}ter). Zeitschrift
  Naturforschung Teil A 5:65

\bibitem[{Birdsall and Langdon(1991)}]{birdsall1991}
Birdsall CK, Langdon AB (1991) Plasma physics via computer simulation. Adam
  Hilger, Bristol

\bibitem[{Birn and Hesse(1991)}]{birn1991substorm}
Birn J, Hesse M (1991) The substorm current wedge and field-aligned currents in
  mhd simulations of magnetotail reconnection. Journal of Geophysical Research:
  Space Physics 96(A2):1611--1618

\bibitem[{Birn and Hesse(1994)}]{birn1994particle}
Birn J, Hesse M (1994) Particle acceleration in the dynamic magnetotail: Orbits
  in self-consistent three-dimensional mhd fields. Journal of Geophysical
  Research: Space Physics 99(A1):109--119

\bibitem[{Birn and Hesse(2000)}]{birn2000large}
Birn J, Hesse M (2000) Large-scale stability of the magnetotail. In: Fifth
  International Conference on Substorms, vol 443, p~15

\bibitem[{Birn and Hesse(2014)}]{birn2014substorm}
Birn J, Hesse M (2014) The substorm current wedge: Further insights from mhd
  simulations. Journal of Geophysical Research: Space Physics 119(5):3503--3513

\bibitem[{Birn and Hones~Jr(1981)}]{birn1981three}
Birn J, Hones~Jr EW (1981) Three-dimensional computer modeling of dynamic
  reconnection in the geomagnetic tail. Journal of Geophysical Research: Space
  Physics 86(A8):6802--6808

\bibitem[{Birn and Priest(2007)}]{Birn2007}
Birn J, Priest ER (eds)  (2007) {Reconnection of Magnetic Fields}. Cambridge
  University Press, Cambridge, \doi{10.1017/CBO9780511536151},
  \urlprefix\url{https://www.cambridge.org/core/product/identifier/9780511536151/type/book}

\bibitem[{Birn and Schindler(2002)}]{birn2002thin}
Birn J, Schindler K (2002) Thin current sheets in the magnetotail and the loss
  of equilibrium. Journal of Geophysical Research: Space Physics
  107(A7):SMP--18

\bibitem[{Birn et~al.(2001{\natexlab{a}})Birn, Drake, Shay, Rogers, Denton,
  Hesse, Kuznetsova, Ma, Bhattacharjee, Otto, and Pritchett}]{Birn01}
Birn J, Drake JF, Shay MA, Rogers BN, Denton RE, Hesse M, Kuznetsova M, Ma ZW,
  Bhattacharjee A, Otto A, Pritchett PL (2001{\natexlab{a}}) {GEM} magnetic
  reconnection challenge. J Geophys Res 106:3715

\bibitem[{Birn et~al.(2001{\natexlab{b}})Birn, Drake, Shay, Rogers, Denton,
  Hesse, Kuznetsova, Ma, Bhattacharjee, Otto, and Pritchett}]{birn:2001}
Birn J, Drake JF, Shay MA, Rogers BN, Denton RE, Hesse M, Kuznetsova M, Ma ZW,
  Bhattacharjee A, Otto A, Pritchett PL (2001{\natexlab{b}}) Geospace
  {E}nvironmental {M}odeling ({GEM}) magnetic reconnection challenge. J Geophys
  Res 106(A3):3715--3720, \doi{10.1029/1999JA900449}

\bibitem[{Birn et~al.(2004)Birn, Thomsen, and Hesse}]{birn2004electron}
Birn J, Thomsen M, Hesse M (2004) Electron acceleration in the dynamic
  magnetotail: Test particle orbits in three-dimensional magnetohydrodynamic
  simulation fields. Physics of Plasmas 11(5):1825--1833

\bibitem[{Birn et~al.(2018)Birn, Merkin, Sitnov, and Otto}]{birn2018mhd}
Birn J, Merkin V, Sitnov M, Otto A (2018) Mhd stability of magnetotail
  configurations with a bz hump. Journal of Geophysical Research: Space Physics
  123(5):3477--3492

\bibitem[{Birn et~al.(2022)Birn, Hesse, and Runov}]{birn2022electron}
Birn J, Hesse M, Runov A (2022) Electron anisotropies in magnetotail
  dipolarization events. Frontiers in Astronomy and Space Sciences 9:908730

\bibitem[{Bishop(1996)}]{bishop1996multiple}
Bishop J (1996) Multiple charge exchange and ionization collisions within the
  ring current-geocorona-plasmasphere system: Generation of a secondary ring
  current on inner l shells. Journal of Geophysical Research: Space Physics
  101(A8):17325--17336

\bibitem[{Biskamp(1986)}]{Biskamp86}
Biskamp D (1986) Magnetic reconnection via current sheets. Phys Fluids 29:1520

\bibitem[{Biskamp(1996)}]{biskamp1996magnetic}
Biskamp D (1996) Magnetic reconnection in plasmas. Astrophysics and Space
  Science 242:165--207

\bibitem[{Biskamp(2000)}]{Biskamp2000}
Biskamp D (2000) {Magnetic Reconnection in Plasmas}. Cambridge University
  Press, Cambridge, \doi{10.1017/CBO9780511599958},
  \urlprefix\url{http://ebooks.cambridge.org/ref/id/CBO9780511599958}

\bibitem[{Blanc and Richmond(1980)}]{blanc1980ionospheric}
Blanc M, Richmond A (1980) The ionospheric disturbance dynamo. Journal of
  Geophysical Research: Space Physics 85(A4):1669--1686

\bibitem[{Blanco-Cano et~al.(2011)Blanco-Cano, Kajdi{\v{c}}, Omidi, and
  Russell}]{Blanco2011}
Blanco-Cano X, Kajdi{\v{c}} P, Omidi N, Russell C (2011) Foreshock cavitons for
  different interplanetary magnetic field geometries: Simulations and
  observations. Journal of Geophysical Research: Space Physics 116(A9)

\bibitem[{Bohdan et~al.(2017)Bohdan, Niemiec, Kobzar, and Pohl}]{bohdan2017}
Bohdan A, Niemiec J, Kobzar O, Pohl M (2017) Electron pre-acceleration at
  nonrelativistic high-mach-number perpendicular shocks. Astrophys J 847:71

\bibitem[{Bohdan et~al.(2020)Bohdan, Pohl, Niemiec, Vafin, Matsumoto, Amano,
  and Hoshino}]{bohdan2020}
Bohdan A, Pohl M, Niemiec J, Vafin S, Matsumoto Y, Amano T, Hoshino M (2020)
  Kinetic simulations of nonrelativistic perpendicular shocks of young
  supernova remnants. Astrophys J 893:6

\bibitem[{Boris(1970)}]{boris1970}
Boris JP (1970) Relativistic plasma simulation-optimization of a hybrid code.
  Proceeding of the 4th Conference on Numerical Simulation of Plasmas

\bibitem[{Brackbill and Lapenta(2008)}]{Brackbill:2008}
Brackbill JU, Lapenta G (2008) Magnetohydrodynamics with implicit plasma
  simulation. Comm Comput Phys 4:433--456

\bibitem[{Braginskii(1965)}]{braginskii65a}
Braginskii SI (1965) Transport processes in a plasma. In: Leontovich MA (ed)
  Reviews of Plasma Physics, vol~1, Consultants Bureau, New York, pp 205--311

\bibitem[{Brambles et~al.(2013)Brambles, Lotko, Zhang, Ouellette, Lyon, and
  Wiltberger}]{Brambles:2013}
Brambles O, Lotko W, Zhang B, Ouellette J, Lyon J, Wiltberger M (2013) The
  effects of ionospheric outflow on icme and sir driven sawtooth events.
  Journal of Geophysical Research: Space Physics 118(10):6026--6041

\bibitem[{Buchner and Zelenyi(1989)}]{buchner:1989}
Buchner J, Zelenyi L (1989) Regular and chaotic charged-particle motion in
  magnetotail-like field reversals .1. basic theory of trapped motion. Journal
  of Geophysical Research 94(A9):11821--11842

\bibitem[{B{\"u}chner et~al.(2003)B{\"u}chner, Dum, and
  Scholer}]{buchner2003space}
B{\"u}chner J, Dum C, Scholer M (2003) Space plasma simulation, vol 615.
  Springer Science \& Business Media

\bibitem[{Bulanov et~al.(1992)Bulanov, Pegoraro, and Sakharov}]{Bulanov1992}
Bulanov SV, Pegoraro F, Sakharov AS (1992) Magnetic reconnection in electron
  magnetohydrodynamics. Physics of Fluids B: Plasma Physics 4(8):2499--2508,
  \doi{10.1063/1.860467}

\bibitem[{Burch et~al.(2016)Burch, Torbert, Phan, Chen, Moore, Ergun, Eastwood,
  Gershman, Cassak, Argal, Wang, Hesse, Pollock, Giles, Nakamura, Mauk,
  Fuselier, Russell, Strangeway, Drake, Shay, Khotyaintsev, Lindqvist,
  Marklund, Wilder, Young, Torkar, Goldstein, Dorelli, Avanov, Oka, Baker,
  Jaynes, Goodrich, Cohen, Turner, Fennell, Blake, Clem-mons, Goldman, Newman,
  Petrinec, Trattner, Lavraud, Reiff, Baumjohann, Magnes, Steller, Lewis,
  Saito, Coffey, , and Chandler}]{burch16a}
Burch JL, Torbert RB, Phan TD, Chen LJ, Moore TE, Ergun RE, Eastwood JP,
  Gershman DJ, Cassak PA, Argal MR, Wang S, Hesse M, Pollock CJ, Giles BL,
  Nakamura R, Mauk BH, Fuselier SA, Russell CT, Strangeway RJ, Drake JF, Shay
  MA, Khotyaintsev YV, Lindqvist PA, Marklund G, Wilder FD, Young DT, Torkar K,
  Goldstein J, Dorelli JC, Avanov LA, Oka M, Baker DN, Jaynes AN, Goodrich KA,
  Cohen IJ, Turner DL, Fennell JF, Blake JB, Clem-mons J, Goldman M, Newman D,
  Petrinec SM, Trattner K, Lavraud B, Reiff PH, Baumjohann W, Magnes W, Steller
  M, Lewis W, Saito Y, Coffey V, , Chandler M (2016) Electron-scale
  measurements of magnetic reconnection in space. Science 352(6290):aaf2939

\bibitem[{Burch et~al.(2020)Burch, Webster, Hesse, Genestreti, Denton, Phan,
  Hasegawa, Cassak, Torbert, Giles, Gershman, Ergun, Russell, Strangeway,
  Le~Contel, Pritchard, Marshall, Hwang, Dokgo, Fuselier, Chen, Wang, Swisdak,
  Drake, Argall, Trattner, Yamada, and Paschmann}]{burch2020}
Burch JL, Webster JM, Hesse M, Genestreti KJ, Denton RE, Phan TD, Hasegawa H,
  Cassak PA, Torbert RB, Giles BL, Gershman DJ, Ergun RE, Russell CT,
  Strangeway RJ, Le~Contel O, Pritchard KR, Marshall AT, Hwang KJ, Dokgo K,
  Fuselier SA, Chen LJ, Wang S, Swisdak M, Drake JF, Argall MR, Trattner KJ,
  Yamada M, Paschmann G (2020) Electron {{Inflow Velocities}} and
  {{Reconnection Rates}} at {{Earth}}'s {{Magnetopause}} and {{Magnetosheath}}.
  Geophysical Research Letters 47(17):e2020GL089082, \doi{10.1029/2020GL089082}

\bibitem[{Buzulukova et~al.(2010)Buzulukova, Fok, Pulkkinen, Kuznetsova, Moore,
  Glocer, Brandt, T\'oth, and Rastatter}]{Buzulukova:2010}
Buzulukova N, Fok MC, Pulkkinen A, Kuznetsova M, Moore TE, Glocer A, Brandt PC,
  T\'oth G, Rastatter L (2010) Dynamics of ring current and electric fields in
  the inner magnetosphere during disturbed periods: {CRCM}--{BATS-R-US} coupled
  model. J Geophys Res 115:A05210, \doi{doi:10.1029/2009JA014621}

\bibitem[{Califano and Cerri(2022)}]{califano2023}
Califano F, Cerri SS (2022) Eulerian approach to solve the vlasov equation and
  hybrid-vlasov simulations. In: Space and Astrophysical Plasma Simulation:
  Methods, Algorithms, and Applications, Springer, pp 123--161

\bibitem[{Califano et~al.(2001)Califano, Attico, Pegoraro, Bertin, and
  Bulanov}]{califano2001}
Califano F, Attico N, Pegoraro F, Bertin G, Bulanov SV (2001) Fast formation of
  magnetic islands in a plasma in the presence of counterstreaming electrons.
  Phys Rev Lett 86:5293--5296, \doi{10.1103/PhysRevLett.86.5293},
  \urlprefix\url{https://link.aps.org/doi/10.1103/PhysRevLett.86.5293}

\bibitem[{Califano et~al.(2020)Califano, Cerri, Faganello, Laveder, Sisti, and
  Kunz}]{Califano2020}
Califano F, Cerri SS, Faganello M, Laveder D, Sisti M, Kunz MW (2020)
  Electron-only reconnection in plasma turbulence. Frontiers in Physics 8:317

\bibitem[{Caprioli(2014)}]{Caprioli:2014}
Caprioli D (2014) {Hybrid Simulations of Particle Acceleration at Shocks}.
  Nuclear Physics B (Proc Suppl) 256--257:48--55,
  \doi{10.1016/j.nuclphysbps.2014.10.005},
  \urlprefix\url{https://doi.org/10.1016/j.nuclphysbps.2014.10.005}

\bibitem[{Cassak and Shay(2007)}]{cassak07a}
Cassak PA, Shay MA (2007) Scaling of asymmetric magnetic reconnection: General
  theory and collisional simulations. Phys Plasmas 14:102114,
  \doi{10.1063/1.2795630}

\bibitem[{Cassak and Shay(2012)}]{Cassak12}
Cassak PA, Shay MA (2012) Magnetic reconnection for coronal conditions:
  Reconnection rates, secondary islands and onset. Space Sci~Rev 172:283

\bibitem[{Cassak et~al.(2005)Cassak, Shay, and Drake}]{cassak05a}
Cassak PA, Shay MA, Drake JF (2005) Catastrophe model for fast magnetic
  reconnection onset. Phys Rev Lett 95:235002,
  \doi{10.1103/PhysRevLett.95.235002}

\bibitem[{Cassak et~al.(2015)Cassak, Baylor, Fermo, Beidler, Shay, Swisdak,
  Drake, and Karimabadi}]{Cassak15}
Cassak PA, Baylor RN, Fermo RL, Beidler MT, Shay MA, Swisdak M, Drake JF,
  Karimabadi H (2015) Fast magnetic reconnection due to anisotropic electron
  pressure. Phys~Plasmas 22:020705

\bibitem[{{Cassak} et~al.(2023){Cassak}, {Barbhuiya}, {Liang}, and
  {Argall}}]{cassak2023quantifying}
{Cassak} PA, {Barbhuiya} MH, {Liang} H, {Argall} MR (2023) {Quantifying Energy
  Conversion in Higher-Order Phase Space Density Moments in Plasmas}. Physical
  Review Letters 130(8):085201, \doi{10.1103/PhysRevLett.130.085201},
  \eprint{2306.01106}

\bibitem[{Cerri et~al.(2017)Cerri, Servidio, and Califano}]{cerri2017kinetic}
Cerri SS, Servidio S, Califano F (2017) Kinetic cascade in solar-wind
  turbulence: 3d3v hybrid-kinetic simulations with electron inertia. The
  Astrophysical Journal Letters 846(2):L18, \doi{10.3847/2041-8213/aa87b0},
  \urlprefix\url{https://dx.doi.org/10.3847/2041-8213/aa87b0}

\bibitem[{{Cerri} et~al.(2018){Cerri}, {Kunz}, and {Califano}}]{cerri2018dual}
{Cerri} SS, {Kunz} MW, {Califano} F (2018) {Dual Phase-space Cascades in 3D
  Hybrid-Vlasov-Maxwell Turbulence}. The Astrophysical Journal Letters
  856(1):L13, \doi{10.3847/2041-8213/aab557}, \eprint{1802.06133}

\bibitem[{Chac{\'o}n et~al.(2007)Chac{\'o}n, Simakov, and Zocco}]{Chacon2007}
Chac{\'o}n L, Simakov AN, Zocco A (2007) Steady-{{State Properties}} of
  {{Driven Magnetic Reconnection}} in {{2D Electron Magnetohydrodynamics}}.
  Physical Review Letters 99(23):235001, \doi{10.1103/PhysRevLett.99.235001}

\bibitem[{Chen et~al.(2021)Chen, Ng, Omelchenko, and Wang}]{Chen2021}
Chen LJ, Ng J, Omelchenko Y, Wang S (2021) Magnetopause reconnection and
  indents induced by foreshock turbulence. Geophysical Research Letters
  48(11):e2021GL093029

\bibitem[{Chen and Schulz(2001)}]{chen2001simulations}
Chen MW, Schulz M (2001) Simulations of diffuse aurora with plasma sheet
  electrons in pitch angle diffusion less than everywhere strong. Journal of
  Geophysical Research: Space Physics 106(A12):28949--28966

\bibitem[{Chen et~al.(2012)Chen, Lemon, Guild, Schulz, Roeder, and
  Le}]{chen2012comparison}
Chen MW, Lemon CL, Guild TB, Schulz M, Roeder JL, Le G (2012) Comparison of
  self-consistent simulations with observed magnetic field and ion plasma
  parameters in the ring current during the 10 august 2000 magnetic storm.
  Journal of Geophysical Research: Space Physics 117(A9)

\bibitem[{Chen and T\'oth(2019)}]{Chen:2019}
Chen Y, T\'oth G (2019) Gauss's law satisfying energy-conserving semi-implicit
  particle-in-cell method. J Comput Phys 386:632,
  \doi{10.1016/j.jcp.2019.02.032}

\bibitem[{Chen et~al.(2017)Chen, T\'oth, Cassak, Jia, Gombosi, Slavin,
  Markidis, and Peng}]{Chen:2017}
Chen Y, T\'oth G, Cassak P, Jia X, Gombosi TI, Slavin J, Markidis S, Peng B
  (2017) Global three-dimensional simulation of earth's dayside reconnection
  using a two-way coupled magnetohydrodynamics with embedded particle-in-cell
  model: initial results. J Geophys Res 122:10318, \doi{10.1002/2017JA024186}

\bibitem[{Chen et~al.(2019)Chen, T\'oth, Jia, Slavin, Sun, Markidis, Gombosi,
  and Raines}]{Chen_mercury:2019}
Chen Y, T\'oth G, Jia X, Slavin J, Sun W, Markidis S, Gombosi T, Raines J
  (2019) Studying dawn-dusk asymmetries of mercury's magnetotail using mhd-epic
  simulations. J Geophys Res 124:8954, \doi{10.1029/2019JA026840}

\bibitem[{Chen et~al.(2020)Chen, T{\'{o}}th, Hietala, Vines, Zou, Nishimura,
  Silveira, Guo, Lin, and Markidis}]{Chen2020}
Chen Y, T{\'{o}}th G, Hietala H, Vines SK, Zou Y, Nishimura Y, Silveira MV, Guo
  Z, Lin Y, Markidis S (2020) {Magnetohydrodynamic with embedded
  particle-in-cell simulation of the Geospace Environment Modeling dayside
  kinetic processes challenge event}. Earth and Space Science
  \doi{10.1029/2020ea001331}

\bibitem[{Chen et~al.(2023)Chen, Tóth, Zhou, and Wang}]{Chen2023}
Chen Y, Tóth G, Zhou H, Wang X (2023) {FLEKS}: {A} flexible particle-in-cell
  code for multi-scale plasma simulations. Computer Physics Communications
  287:108714, \doi{10.1016/j.cpc.2023.108714}

\bibitem[{Cheng et~al.(2020)Cheng, Lin, Perez, Johnson, and Wang}]{Cheng2020}
Cheng L, Lin Y, Perez J, Johnson JR, Wang X (2020) Kinetic alfv{\'e}n waves
  from magnetotail to the ionosphere in global hybrid simulation associated
  with fast flows. Journal of Geophysical Research: Space Physics
  125(2):e2019JA027062

\bibitem[{Chew et~al.(1956)Chew, Goldberger, and Low}]{CGL}
Chew GF, Goldberger ML, Low FE (1956) Boltzmann equation and the one-fluid
  hydrodynamic equations in the absence of particle collisions. Proc Roy Soc
  A236:112

\bibitem[{Choudhuri(1998)}]{Choudhuri98}
Choudhuri AR (1998) The Physics of Fluids and Plasmas. Cambridge University
  Press

\bibitem[{Cramer et~al.(2017)Cramer, Raeder, Toffoletto, Gilson, and
  Hu}]{cramer2017plasma}
Cramer WD, Raeder J, Toffoletto F, Gilson M, Hu B (2017) Plasma sheet
  injections into the inner magnetosphere: Two-way coupled openggcm-rcm model
  results. Journal of Geophysical Research: Space Physics 122(5):5077--5091

\bibitem[{Curran and Goertz(1989)}]{curran1989particle}
Curran DB, Goertz C (1989) Particle distributions in a two-dimensional
  reconnection field geometry. Journal of Geophysical Research: Space Physics
  94(A1):272--286

\bibitem[{Dahlin et~al.(2015)Dahlin, Drake, and Swisdak}]{dahlin2015electron}
Dahlin J, Drake J, Swisdak M (2015) Electron acceleration in three-dimensional
  magnetic reconnection with a guide field. Physics of Plasmas 22(10)

\bibitem[{Dahlin et~al.(2016)Dahlin, Drake, and Swisdak}]{dahlin2016parallel}
Dahlin J, Drake J, Swisdak M (2016) Parallel electric fields are inefficient
  drivers of energetic electrons in magnetic reconnection. Physics of Plasmas
  23(12)

\bibitem[{Dahlin et~al.(2017)Dahlin, Drake, and Swisdak}]{dahlin2017_role}
Dahlin J, Drake J, Swisdak M (2017) The role of three-dimensional transport in
  driving enhanced electron acceleration during magnetic reconnection. Physics
  of Plasmas 24(9)

\bibitem[{{Dahlin} et~al.(2014){Dahlin}, {Drake}, and {Swisdak}}]{Dahlin2014}
{Dahlin} JT, {Drake} JF, {Swisdak} M (2014) {The mechanisms of electron heating
  and acceleration during magnetic reconnection}. Physics of Plasmas
  21(9):092304, \doi{10.1063/1.4894484}, \eprint{1406.0831}

\bibitem[{{Dahlin} et~al.(2015){Dahlin}, {Drake}, and {Swisdak}}]{Dahlin2015}
{Dahlin} JT, {Drake} JF, {Swisdak} M (2015) {Electron acceleration in
  three-dimensional magnetic reconnection with a guide field}. Physics of
  Plasmas 22(10):100704, \doi{10.1063/1.4933212}, \eprint{1503.02218}

\bibitem[{Daldorff et~al.(2014{\natexlab{a}})Daldorff, T\'oth, Gombosi,
  Lapenta, Amaya, Markidis, and Brackbill}]{daldorff}
Daldorff LKS, T\'oth G, Gombosi TI, Lapenta G, Amaya J, Markidis S, Brackbill
  JU (2014{\natexlab{a}}) Two-way coupling of a global {H}all
  magnetohydrodynamics model with a local implicit particle-in-cell model.
  Journal of Computational Physics 268:236--254

\bibitem[{Daldorff et~al.(2014{\natexlab{b}})Daldorff, T\'oth, Gombosi,
  Lapenta, Amaya, Markidis, and Brackbill}]{Daldorff:2014}
Daldorff LKS, T\'oth G, Gombosi TI, Lapenta G, Amaya J, Markidis S, Brackbill
  JU (2014{\natexlab{b}}) Two-way coupling of a global {Hall}
  magnetohydrodynamics model with a local implicit {Particle-in-Cell} model. J
  Comput Phys 268:236, \doi{10.1016/j.jcp.2014.03.009}

\bibitem[{Daughton and J.Scudder(2006)}]{daughton2006}
Daughton W, JScudder (2006) Fully kinetic simulations of undriven magnetic
  reconnection with open boundary conditions. Physics of Plasmas 13:072101

\bibitem[{Daughton et~al.(2006)Daughton, Scudder, and
  Karimabadi}]{daughton:2006}
Daughton W, Scudder J, Karimabadi H (2006) Fully kinetic simulations of
  undriven magnetic reconnection with open boundary conditions. Physic of
  Plasmas 13(7):072101, \doi{10.1063/1.2218817}

\bibitem[{Daughton et~al.(2009)Daughton, Roytershteyn, Albright, Karimabadi,
  Yin, and Bowers}]{Daughton09}
Daughton W, Roytershteyn V, Albright BJ, Karimabadi H, Yin L, Bowers KJ (2009)
  Transition from collisional to kinetic regimes in large-scale reconnection
  layers. Phys~Rev~Lett 103:065004

\bibitem[{{Daughton} et~al.(2011){Daughton}, {Roytershteyn}, {Karimabadi},
  {Yin}, {Albright}, {Bergen}, and {Bowers}}]{daughton:2011}
{Daughton} W, {Roytershteyn} V, {Karimabadi} H, {Yin} L, {Albright} BJ,
  {Bergen} B, {Bowers} KJ (2011) {Role of electron physics in the development
  of turbulent magnetic reconnection in collisionless plasmas}. Nature Physics
  7:539--542, \doi{10.1038/nphys1965}

\bibitem[{{De~Zeeuw} et~al.(2004){De~Zeeuw}, Sazykin, Wolf, Gombosi, Ridley,
  and T{\'o}th}]{DeZeeuw:2004}
{De~Zeeuw} D, Sazykin S, Wolf R, Gombosi T, Ridley A, T{\'o}th G (2004)
  Coupling of a global {MHD} code and an inner magnetosphere model: {I}nitial
  results. J Geophys Res 109(A12):219, \doi{10.1029/2003JA010366}

\bibitem[{De~Zeeuw et~al.(2004)De~Zeeuw, Sazykin, Wolf, Gombosi, Ridley, and
  T{\'o}th}]{de2004coupling}
De~Zeeuw DL, Sazykin S, Wolf RA, Gombosi TI, Ridley AJ, T{\'o}th G (2004)
  Coupling of a global mhd code and an inner magnetospheric model: Initial
  results. Journal of Geophysical Research: Space Physics 109(A12)

\bibitem[{Delamere(2009)}]{Delamere:2009}
Delamere PA (2009) {Hybrid code simulations of the solar wind interaction with
  Pluto}. Journal of Geophysical Research 114:A03220,
  \doi{10.1029/2008JA013756},
  \urlprefix\url{https://doi.org/10.1029/2008JA013756}

\bibitem[{Delcourt and Sauvaud(1994)}]{delcourt1994plasma}
Delcourt D, Sauvaud J (1994) Plasma sheet ion energization during
  dipolarization events. Journal of Geophysical Research: Space Physics
  99(A1):97--108

\bibitem[{Delzanno(2015)}]{delzanno2015}
Delzanno G (2015) Multi-dimensional, fully-implicit, spectral method for the
  vlasov–maxwell equations with exact conservation laws in discrete form.
  Journal of Computational Physics 301:338 -- 356,
  \doi{https://doi.org/10.1016/j.jcp.2015.07.028},
  \urlprefix\url{http://www.sciencedirect.com/science/article/pii/S0021999115004738}

\bibitem[{Dong et~al.(2019)Dong, Wang, Hakim, Bhattacharjee, Slavin, DiBraccio,
  and Germaschewski}]{Dong19}
Dong C, Wang L, Hakim A, Bhattacharjee A, Slavin JA, DiBraccio GA,
  Germaschewski K (2019) Global ten-moment multifluid simulations of the solar
  wind interaction with mercury: From the planetary conducting core to the
  dynamic magnetosphere. Geophysical Research Letters 46(21):11584--11596

\bibitem[{Dong et~al.(2021)Dong, Le, Wang, Stanier, Wetherton, Daughton,
  Bhattacharjee, Slavin, and DiBraccio}]{Dong_atal:2021}
Dong C, Le A, Wang L, Stanier A, Wetherton B, Daughton W, Bhattacharjee A,
  Slavin J, DiBraccio G (2021) {Global Hybrid-VPIC Simulations of the Solar
  Wind Interaction with Murcury's Dynamic Magnetosphere: Reconnection and
  Foreshock}. In: EGU General Assembly, 19–30 Apr 2021, vol EGU-12954,
  \doi{10.5194/egusphere-egu21-12954},
  \urlprefix\url{https://doi.org/10.5194/egusphere-egu21-12954}

\bibitem[{Dorelli and Birn(2003)}]{Dorelli:2003}
Dorelli J, Birn J (2003) {Whistler-mediated magnetic reconnection in large
  systems: Magnetic flux pileup and the formation of thin current sheets}. J
  Geophys Res Space Physics 108(A3):1133, \doi{10.1029/2001JA009180},
  \urlprefix\url{https://doi.org/10.1029/2001JA009180}

\bibitem[{Dorelli et~al.(2015)Dorelli, Glocer, Collinson, and
  T{\'o}th}]{Dorelli15}
Dorelli JC, Glocer A, Collinson G, T{\'o}th G (2015) The role of the hall
  effect in the global structure and dynamics of planetary magnetospheres:
  Ganymede as a case study. J~Geophys~Res 120:5377

\bibitem[{Drake et~al.(2006)Drake, Swisdak, Che, and Shay}]{drake2006electron}
Drake J, Swisdak M, Che H, Shay M (2006) Electron acceleration from contracting
  magnetic islands during reconnection. Nature 443(7111):553--556

\bibitem[{Drake et~al.(2019)Drake, Arnold, Swisdak, and
  Dahlin}]{drake2019computational}
Drake J, Arnold H, Swisdak M, Dahlin J (2019) A computational model for
  exploring particle acceleration during reconnection in macroscale systems.
  Physics of Plasmas 26(1)

\bibitem[{Drake and Shay(2007)}]{Drake07}
Drake JF, Shay MA (2007) The fundamentals of collisionless reconnection. In:
  Birn J, Priest E (eds) Reconnection of Magnetic Fields: Magnetohydrodynamics
  and Collisionless Theory and Observations, Cambridge University Press

\bibitem[{Drake et~al.(1994)Drake, Kleva, and Mandt}]{Drake1994}
Drake JF, Kleva RG, Mandt ME (1994) Structure of {{Thin Current Layers}}:
  {{Implications}} for {{Magnetic Reconnection}}. Physical Review Letters
  73(9):1251--1254, \doi{10.1103/PhysRevLett.73.1251}

\bibitem[{Drake et~al.(1997)Drake, Biskamp, and Zeiler}]{Drake1997}
Drake JF, Biskamp D, Zeiler A (1997) Breakup of the electron current layer
  during 3-{{D}} collisionless magnetic reconnection. Geophysical Research
  Letters 24(22):2921--2924, \doi{10.1029/97GL52961}

\bibitem[{{Drake} et~al.(2006){Drake}, {Swisdak}, {Che}, and
  {Shay}}]{Drake2006}
{Drake} JF, {Swisdak} M, {Che} H, {Shay} MA (2006) {Electron acceleration from
  contracting magnetic islands during reconnection}. Nature 443(7111):553--556,
  \doi{10.1038/nature05116}

\bibitem[{{Drake} et~al.(2008){Drake}, {Shay}, and {Swisdak}}]{Drake:2008}
{Drake} JF, {Shay} MA, {Swisdak} M (2008) {The Hall fields and fast magnetic
  reconnection}. Physics of Plasmas 15(4):042306, \doi{10.1063/1.2901194}

\bibitem[{Dreher et~al.(1996)Dreher, Arendt, and Schindler}]{dreher1996}
Dreher J, Arendt U, Schindler K (1996) Particle simulations of collisionless
  reconnection in magnetotail configuration including electron dynamics.
  Journal of Geophysical Research: Space Physics 101(A12):27375--27381

\bibitem[{{Du} et~al.(2018){Du}, {Guo}, {Zank}, {Li}, and {Stanier}}]{Du2018}
{Du} S, {Guo} F, {Zank} GP, {Li} X, {Stanier} A (2018) {Plasma Energization in
  Colliding Magnetic Flux Ropes}. Astrophysical Journal 867(1):16,
  \doi{10.3847/1538-4357/aae30e}, \eprint{1809.08357}

\bibitem[{Dungey(1961)}]{Dungey:1961}
Dungey J (1961) Interplanetary magnetic field and the auroral zones. Phys Rev
  Lett 93:47, \doi{10.1103/PhysRevLett.6.47}

\bibitem[{Dungey(1954)}]{Dungey54}
Dungey JW (1954) The attenuation of alfvén waves. Journal of Geophysical
  Research (1896-1977) 59(3):323--328,
  \doi{https://doi.org/10.1029/JZ059i003p00323}

\bibitem[{Dyadechkin et~al.(2013)Dyadechkin, Kallio, and
  Jarvinen}]{Dyadechkin2013}
Dyadechkin S, Kallio E, Jarvinen R (2013) {A new 3-D spherical hybrid model for
  solar wind interaction studies}. Journal of Geophysical Research: Space
  Physics 118:5157--5168, \doi{10.1002/jgra.50497},
  \urlprefix\url{https://doi.org/10.1002/jgra.50497}

\bibitem[{Edwards et~al.(1986)Edwards, Campbell, Engelhardt, Farhbach, Gill,
  Granetz, Tsuji, Tubbing, Weller, Wesson, and Zasche}]{Edwards86}
Edwards AW, Campbell DJ, Engelhardt WW, Farhbach HU, Gill RD, Granetz RS, Tsuji
  S, Tubbing BJD, Weller A, Wesson J, Zasche D (1986) Rapid collapse of a
  plasma sawtooth oscillation in the jet tokamak. Phys~Rev~Lett 57:210--213

\bibitem[{Egedal et~al.(2012)Egedal, Daughton, and Le}]{egedal2012large}
Egedal J, Daughton W, Le A (2012) Large-scale electron acceleration by parallel
  electric fields during magnetic reconnection. Nature Physics 8(4):321--324

\bibitem[{Egedal et~al.(2013)Egedal, Le, and Daughton}]{Egedal13}
Egedal J, Le A, Daughton W (2013) A review of pressure anisotropy caused by
  electron trapping in collisionless plasma, and its implications for magnetic
  reconnection. Phys Plasmas 20:061201

\bibitem[{Erickson and Wolf(1980)}]{erickson1980steady}
Erickson GM, Wolf R (1980) Is steady convection possible in the earth's
  magnetotail? Geophysical Research Letters 7(11):897--900

\bibitem[{Eriksson et~al.(2016)Eriksson, Wilder, Ergun, Schwartz, Cassak,
  Burch, Chen, Torbert, Phan, Lavraud, Goodrich, Holmes, Stawarz, Sturner,
  Malaspina, Usanova, Trattner, Strangeway, Russell, Pollock, Giles, Hesse,
  Lindqvist, Drake, Shay, Nakamura, and Marklund}]{Eriksson2016}
Eriksson S, Wilder FD, Ergun RE, Schwartz SJ, Cassak PA, Burch JL, Chen LJ,
  Torbert RB, Phan TD, Lavraud B, Goodrich KA, Holmes JC, Stawarz JE, Sturner
  AP, Malaspina DM, Usanova ME, Trattner KJ, Strangeway RJ, Russell CT, Pollock
  CJ, Giles BL, Hesse M, Lindqvist PA, Drake JF, Shay MA, Nakamura R, Marklund
  GT (2016) Magnetospheric {{Multiscale Observations}} of the {{Electron
  Diffusion Region}} of {{Large Guide Field Magnetic Reconnection}}. Physical
  Review Letters 117(1):015001, \doi{10.1103/PhysRevLett.117.015001}

\bibitem[{Fadanelli et~al.(2021)Fadanelli, Lavraud, Califano, Cozzani, Finelli,
  and Sisti}]{fadanelli2021}
Fadanelli S, Lavraud B, Califano F, Cozzani G, Finelli F, Sisti M (2021) Energy
  conversions associated with magnetic reconnection. Journal of Geophysical
  Research: Space Physics 126(1):e2020JA028333,
  \doi{https://doi.org/10.1029/2020JA028333}, e2020JA028333 2020JA028333,
  \eprint{https://agupubs.onlinelibrary.wiley.com/doi/pdf/10.1029/2020JA028333}

\bibitem[{Farrugia et~al.(2021)Farrugia, Rogers, Torbert, Genestreti, Nakamura,
  Lavraud, Montag, Egedal, Payne, Keesee, Ahmadi, Ergun, Reiff, Argall, Matsui,
  Wilson~III, Lugaz, Burch, Russell, Fuselier, and Dors}]{Farrugia2021}
Farrugia CJ, Rogers AJ, Torbert RB, Genestreti KJ, Nakamura TKM, Lavraud B,
  Montag P, Egedal J, Payne D, Keesee A, Ahmadi N, Ergun R, Reiff P, Argall M,
  Matsui H, Wilson~III LB, Lugaz N, Burch JL, Russell CT, Fuselier SA, Dors I
  (2021) An {{Encounter With}} the {{Ion}} and {{Electron Diffusion Regions}}
  at a {{Flapping}} and {{Twisted Tail Current Sheet}}. Journal of Geophysical
  Research: Space Physics 126(3):e2020JA028903, \doi{10.1029/2020JA028903}

\bibitem[{Fatemi et~al.(2017)Fatemi, Poppe, Delory, and Farrell}]{Fatemi:2017}
Fatemi S, Poppe AR, Delory GT, Farrell WM (2017) {AMITIS: A 3D GPU-Based
  Hybrid-PIC Model for Space and Plasma Physics}. J Phys: Conf Ser 837:012017,
  \doi{10.1088/1742-6596/837/1/012017},
  \urlprefix\url{https://iopscience.iop.org/article/10.1088/1742-6596/837/1/012017}

\bibitem[{Finelli(2022)}]{finellithesis}
Finelli F (2022) Magnetic reconnection in space plasmas: advanced numerical
  models and detection in turbulence. Phd thesis, University of Pisa, Pisa,
  Italy, \urlprefix\url{https://etd.adm.unipi.it/t/etd-06062022-180752/}

\bibitem[{Finelli et~al.(2021)Finelli, Cerri, Califano, Pucci, Laveder,
  Lapenta, and Passot}]{finelli2021}
Finelli F, Cerri SS, Califano F, Pucci F, Laveder D, Lapenta G, Passot T (2021)
  Bridging hybrid- and full-kinetic models with landau-fluid electrons.
  Astronomy \& Astrophysics 653:A156, \doi{10.1051/0004-6361/202140279}

\bibitem[{Fok et~al.(2001)Fok, Wolf, Spiro, and Moore}]{fok2001comprehensive}
Fok MC, Wolf R, Spiro R, Moore T (2001) Comprehensive computational model of
  earth's ring current. Journal of Geophysical Research: Space Physics
  106(A5):8417--8424

\bibitem[{Fok et~al.(2014)Fok, Buzulukova, Chen, Glocer, Nagai, Valek, and
  Perez}]{fok2014comprehensive}
Fok MC, Buzulukova N, Chen SH, Glocer A, Nagai T, Valek P, Perez J (2014) The
  comprehensive inner magnetosphere-ionosphere model. Journal of Geophysical
  Research: Space Physics 119(9):7522--7540

\bibitem[{Fox et~al.(2012)Fox, Bhattacharjee, and Germaschewski}]{Fox12}
Fox W, Bhattacharjee A, Germaschewski K (2012) Magnetic reconnection in
  high-energy-density laser-produced plasmas. Physics of Plasmas 19(5)

\bibitem[{Franci et~al.(2017)Franci, Cerri, Califano, Landi, Papini, Verdini,
  Matteini, Jenko, and Hellinger}]{franci2017}
Franci L, Cerri SS, Califano F, Landi S, Papini E, Verdini A, Matteini L, Jenko
  F, Hellinger P (2017) Magnetic reconnection as a driver for a sub-ion-scale
  cascade in plasma turbulence. The Astrophysical Journal Letters 850(1):L16,
  \doi{10.3847/2041-8213/aa93fb},
  \urlprefix\url{https://dx.doi.org/10.3847/2041-8213/aa93fb}

\bibitem[{Franci et~al.(2018)Franci, Hellinger, Guarrasi, Chen, Papini,
  Verdini, Matteini, and Landi}]{franci2018}
Franci L, Hellinger P, Guarrasi M, Chen CH, Papini E, Verdini A, Matteini L,
  Landi S (2018) {Three-dimensional simulations of solar wind turbulence with
  the hybrid code CAMELIA}. Journal of Physics: Conference Series 1031(1),
  \doi{10.1088/1742-6596/1031/1/012002}

\bibitem[{Freeman et~al.(1993)Freeman, Wolf, Spiro, Hausman, Bales, Hilmer,
  Nagai, and Lambour}]{freeman1993magnetospheric}
Freeman J, Wolf R, Spiro R, Hausman B, Bales B, Hilmer R, Nagai A, Lambour R
  (1993) Magnetospheric specification model development code documentation,
  scientific description, and software documentation. contract
  F19628-90-K-0012, Rice Univ for Air Force Geophys Lab, Hanscom Air Force
  Base, Mass, July

\bibitem[{{French} et~al.(2022){French}, {Guo}, {Zhang}, and
  {Uzdensky}}]{French2022}
{French} O, {Guo} F, {Zhang} Q, {Uzdensky} D (2022) {Particle Injection and
  Nonthermal Particle Acceleration in Relativistic Magnetic Reconnection}.
  arXiv e-prints arXiv:2210.08358, \eprint{2210.08358}

\bibitem[{{Fu} et~al.(2006){Fu}, {Lu}, and {Wang}}]{Fu2006}
{Fu} XR, {Lu} QM, {Wang} S (2006) {The process of electron acceleration during
  collisionless magnetic reconnection}. Physics of Plasmas 13(1):012309,
  \doi{10.1063/1.2164808}

\bibitem[{Fuller-Rowell et~al.(1996)Fuller-Rowell, Rees, Quegan, Moffett,
  Codrescu, and Millward}]{fuller1996coupled}
Fuller-Rowell T, Rees D, Quegan S, Moffett R, Codrescu M, Millward G (1996) A
  coupled thermosphere-ionosphere model (ctim). STEP report 239(4)

\bibitem[{Gabrielse et~al.(2012)Gabrielse, Angelopoulos, Runov, and
  Turner}]{gabrielse2012effects}
Gabrielse C, Angelopoulos V, Runov A, Turner D (2012) The effects of transient,
  localized electric fields on equatorial electron acceleration and transport
  toward the inner magnetosphere. Journal of Geophysical Research: Space
  Physics 117(A10)

\bibitem[{Gallagher et~al.(2000)Gallagher, Craven, and
  Comfort}]{gallagher2000global}
Gallagher DL, Craven PD, Comfort RH (2000) Global core plasma model. Journal of
  Geophysical Research: Space Physics 105(A8):18819--18833

\bibitem[{Ganse et~al.(2023)Ganse, Koskela, Battarbee, Pfau-Kempf, Papadakis,
  Alho, Bussov, Cozzani, Dubart, George, Gordeev, Grandin, Horaites, Suni,
  Tarvus, Kebede, Turc, Zhou, and Palmroth}]{ganse2023}
Ganse U, Koskela T, Battarbee M, Pfau-Kempf Y, Papadakis K, Alho M, Bussov M,
  Cozzani G, Dubart M, George H, Gordeev E, Grandin M, Horaites K, Suni J,
  Tarvus V, Kebede FT, Turc L, Zhou H, Palmroth M (2023) Enabling technology
  for global {3D} + {3V} hybrid-vlasov simulations of near-earth space. Physics
  of Plasmas 30(4):042902

\bibitem[{{Ghizzo} et~al.(2017){Ghizzo}, {Sarrat}, and {Del
  Sarto}}]{ghizzo2017}
{Ghizzo} A, {Sarrat} M, {Del Sarto} D (2017) {Vlasov models for kinetic
  Weibel-type instabilities}. Journal of Plasma Physics 83(1):705830101,
  \doi{10.1017/S0022377816001215}

\bibitem[{Gingell et~al.(2019)Gingell, Schwartz, Eastwood, Burch, Ergun,
  Fuselier, Gershman, Giles, Khotyaintsev, Lavraud, Lindqvist, Paterson, Phan,
  Russell, Stawarz, Strangeway, Torbert, and Wilder}]{Gingell2019}
Gingell I, Schwartz SJ, Eastwood JP, Burch JL, Ergun RE, Fuselier S, Gershman
  DJ, Giles BL, Khotyaintsev YV, Lavraud B, Lindqvist PA, Paterson WR, Phan TD,
  Russell CT, Stawarz JE, Strangeway RJ, Torbert RB, Wilder F (2019)
  Observations of {{Magnetic Reconnection}} in the {{Transition Region}} of
  {{Quasi-Parallel Shocks}}. Geophysical Research Letters 46(3):1177--1184,
  \doi{10.1029/2018GL081804}

\bibitem[{Gingell et~al.(2020)Gingell, Schwartz, Eastwood, E.Stawarz, Burch,
  Ergun, Fuselier, Gershman, Giles, Khotyaintsev, Lavraud, Lindqvist, Paterson,
  Phan, Russell, Strangeway, Torbert, and Wilder}]{Gingell2020}
Gingell I, Schwartz SJ, Eastwood JP, EStawarz J, Burch JL, Ergun RE, Fuselier
  SA, Gershman DJ, Giles BL, Khotyaintsev YV, Lavraud B, Lindqvist PA, Paterson
  WR, Phan TD, Russell CT, Strangeway RJ, Torbert RB, Wilder F (2020)
  Statistics of reconnecting current sheets in the transition region of earth's
  bow shock. Journal of Geophysical Research 125:e2019JA027119

\bibitem[{Gingell et~al.(2021)Gingell, Schwartz, H.Kucharek, Farrugia, and
  J.Trattner}]{gingell2021}
Gingell I, Schwartz SJ, HKucharek, Farrugia CJ, JTrattner K (2021) Observing
  the prevalence of thin current sheets downstream of earth's bow shock.
  Physics of Plasmas 28:102902

\bibitem[{Gkioulidou et~al.(2014)Gkioulidou, Ukhorskiy, Mitchell, Sotirelis,
  Mauk, and Lanzerotti}]{gkioulidou2014role}
Gkioulidou M, Ukhorskiy A, Mitchell D, Sotirelis T, Mauk B, Lanzerotti L (2014)
  The role of small-scale ion injections in the buildup of earth's ring current
  pressure: Van allen probes observations of the 17 march 2013 storm. Journal
  of Geophysical Research: Space Physics 119(9):7327--7342

\bibitem[{Glasser et~al.(1999)Glasser, Sovinec, Nebel, Gianakon, Plimpton, Chu,
  Schnack, and the NIMROD~Team}]{Glasser99}
Glasser AH, Sovinec CR, Nebel RA, Gianakon TA, Plimpton SJ, Chu MS, Schnack DD,
  the NIMROD~Team (1999) The nimrod code: a new approach to numerical plasma
  physics. Plasma Physics and Controlled Fusion 41(3A):A747,
  \doi{10.1088/0741-3335/41/3A/067},
  \urlprefix\url{https://dx.doi.org/10.1088/0741-3335/41/3A/067}

\bibitem[{Glocer et~al.(2009{\natexlab{a}})Glocer, T{\'o}th, Gombosi, and
  Welling}]{glocer2009modeling}
Glocer A, T{\'o}th G, Gombosi T, Welling D (2009{\natexlab{a}}) Modeling
  ionospheric outflows and their impact on the magnetosphere, initial results.
  Journal of Geophysical Research: Space Physics 114(A5)

\bibitem[{Glocer et~al.(2009{\natexlab{b}})Glocer, T\'oth, Ma, Gombosi, Zhang,
  and Kistler}]{Glocer:2009multifluid}
Glocer A, T\'oth G, Ma YJ, Gombosi T, Zhang JC, Kistler LM (2009{\natexlab{b}})
  Multifluid {B}lock-{A}daptive-{T}ree {S}olar wind {R}oe-type {U}pwind
  {S}cheme: Magnetospheric composition and dynamics during geomagnetic storms
  -- initial results. J Geophys Res 114:A12203, \doi{10.1029/2009JA014418}

\bibitem[{Glocer et~al.(2013)Glocer, Fok, Meng, Toth, Buzulukova, Chen, and
  Lin}]{glocer2013crcm+}
Glocer A, Fok M, Meng X, Toth G, Buzulukova N, Chen S, Lin K (2013) Crcm+
  bats--r--us two--way coupling. Journal of Geophysical Research: Space Physics
  118(4):1635--1650

\bibitem[{Godfrey(1980)}]{Godfrey1980}
Godfrey BB (1980) Time-biased field solver for electromagnetic pic code.
  Proceedings of the Ninth Conference on Numerical Simulation of Plasmas p
  OD–4

\bibitem[{{Goldstein} et~al.(2002){Goldstein}, {Spiro}, {Reiff}, {Wolf},
  {Sandel}, {Freeman}, and {Lambour}}]{goldstein2002efieldovershielding}
{Goldstein} J, {Spiro} RW, {Reiff} PH, {Wolf} RA, {Sandel} BR, {Freeman} JW,
  {Lambour} RL (2002) {IMF-driven overshielding electric field and the origin
  of the plasmaspheric shoulder of May 24, 2000}. Geophysical Research Letters
  29(16):1819, \doi{10.1029/2001GL014534}

\bibitem[{Gomez(2006)}]{Gomez06}
Gomez D (2006) Parallel simulations of hall-mhd plasmas. Space Sci~Rev
  122:231--238

\bibitem[{Gonzalez and Parker(2016)}]{Gonzalez2016}
Gonzalez W, Parker E (eds)  (2016) {Magnetic Reconnection}, Astrophysics and
  Space Science Library, vol 427. Springer International Publishing, Cham,
  \doi{10.1007/978-3-319-26432-5},
  \urlprefix\url{http://link.springer.com/10.1007/978-3-319-26432-5}

\bibitem[{Gordeev et~al.(1994)Gordeev, Kingsep, and Rudakov}]{Gordeev1994}
Gordeev AV, Kingsep AS, Rudakov LI (1994) Electron magnetohydrodynamics.
  Physics Reports 243(5):215--315, \doi{10.1016/0370-1573(94)90097-3}

\bibitem[{Grandin et~al.(2020)Grandin, Turc, Battarbee, Ganse, Johlander,
  Pfau-Kempf, Dubart, and Palmroth}]{grandin2020}
Grandin M, Turc L, Battarbee M, Ganse U, Johlander A, Pfau-Kempf Y, Dubart M,
  Palmroth M (2020) Hybrid-vlasov simulation of auroral proton precipitation in
  the cusps: Comparison of northward and southward interplanetary magnetic
  field driving. J Space Weather Space Clim 10:51, \doi{10.1051/swsc/2020053},
  \urlprefix\url{https://doi.org/10.1051/swsc/2020053}

\bibitem[{{Grandin} et~al.(2023){Grandin}, {Luttikhuis}, {Battarbee},
  {Cozzani}, {Zhou}, {Turc}, {Pfau-Kempf}, {George}, {Horaites}, {Gordeev},
  {Ganse}, {Papadakis}, {Alho}, {Tesema}, {Suni}, {Dubart}, {Tarvus}, and
  {Palmroth}}]{grandin2023}
{Grandin} M, {Luttikhuis} T, {Battarbee} M, {Cozzani} G, {Zhou} H, {Turc} L,
  {Pfau-Kempf} Y, {George} H, {Horaites} K, {Gordeev} E, {Ganse} U, {Papadakis}
  K, {Alho} M, {Tesema} F, {Suni} J, {Dubart} M, {Tarvus} V, {Palmroth} M
  (2023) First 3d hybrid-vlasov global simulation of auroral proton
  precipitation and comparison with satellite observations. Journal of Space
  Weather and Space Climate p TBD, \doi{10.1051/swsc/2023017}

\bibitem[{{Greco} et~al.(2012){Greco}, {Valentini}, {Servidio}, and
  {Matthaeus}}]{greco2012inhomogeneous}
{Greco} A, {Valentini} F, {Servidio} S, {Matthaeus} WH (2012) {Inhomogeneous
  kinetic effects related to intermittent magnetic discontinuities}. Physical
  Review E 86(6):066405, \doi{10.1103/PhysRevE.86.066405}

\bibitem[{Greess et~al.(2021)Greess, Egedal, Stanier, Daughton, Olson, L{\^e},
  Myers, {Millet-Ayala}, Clark, Wallace, Endrizzi, and Forest}]{Greess2021}
Greess S, Egedal J, Stanier A, Daughton W, Olson J, L{\^e} A, Myers R,
  {Millet-Ayala} A, Clark M, Wallace J, Endrizzi D, Forest C (2021) Laboratory
  {{Verification}} of {{Electron-Scale Reconnection Regions Modulated}} by a
  {{Three-Dimensional Instability}}. Journal of Geophysical Research: Space
  Physics 126(7):e2021JA029316, \doi{10.1029/2021JA029316}

\bibitem[{Guo and Giacalone(2013)}]{Guo_Gia:2013}
Guo F, Giacalone J (2013) {THE ACCELERATION OF THERMAL PROTONS AT PARALLEL
  COLLISIONLESS SHOCKS: THREE-DIMENSIONAL HYBRID SIMULATIONS}. The
  Astrophysical Journal 773:158, \doi{10.1088/0004-637X/773/2/158},
  \urlprefix\url{https://doi.org/10.1088/0004-637X/773/2/158}

\bibitem[{{Guo} et~al.(2014){Guo}, {Li}, {Daughton}, and {Liu}}]{Guo2014}
{Guo} F, {Li} H, {Daughton} W, {Liu} YH (2014) {Formation of Hard Power Laws in
  the Energetic Particle Spectra Resulting from Relativistic Magnetic
  Reconnection}. Physical Review Letters 113(15):155005,
  \doi{10.1103/PhysRevLett.113.155005}, \eprint{1405.4040}

\bibitem[{{Guo} et~al.(2015){Guo}, {Liu}, {Daughton}, and {Li}}]{Guo2015}
{Guo} F, {Liu} YH, {Daughton} W, {Li} H (2015) {Particle Acceleration and
  Plasma Dynamics during Magnetic Reconnection in the Magnetically Dominated
  Regime}. Astrophysical Journal 806(2):167, \doi{10.1088/0004-637X/806/2/167},
  \eprint{1504.02193}

\bibitem[{{Guo} et~al.(2019){Guo}, {Li}, {Daughton}, {Kilian}, {Li}, {Liu},
  {Yan}, and {Ma}}]{Guo2019}
{Guo} F, {Li} X, {Daughton} W, {Kilian} P, {Li} H, {Liu} YH, {Yan} W, {Ma} D
  (2019) {Determining the Dominant Acceleration Mechanism during Relativistic
  Magnetic Reconnection in Large-scale Systems}. Astrophysical Journal Letters
  879(2):L23, \doi{10.3847/2041-8213/ab2a15}, \eprint{1901.08308}

\bibitem[{{Guo} et~al.(2021){Guo}, {Li}, {Daughton}, {Li}, {Kilian}, {Liu},
  {Zhang}, and {Zhang}}]{Guo2021}
{Guo} F, {Li} X, {Daughton} W, {Li} H, {Kilian} P, {Liu} YH, {Zhang} Q, {Zhang}
  H (2021) {Magnetic Energy Release, Plasma Dynamics, and Particle Acceleration
  in Relativistic Turbulent Magnetic Reconnection}. Astrophysical Journal
  919(2):111, \doi{10.3847/1538-4357/ac0918}, \eprint{2008.02743}

\bibitem[{{Guo} et~al.(2023){Guo}, {Liu}, {Zenitani}, and {Hoshino}}]{Guo2023}
{Guo} F, {Liu} YH, {Zenitani} S, {Hoshino} M (2023) {Magnetic Reconnection and
  Associated Particle Acceleration in High-energy Astrophysics}. arXiv e-prints
  arXiv:2309.13382, \doi{10.48550/arXiv.2309.13382}, \eprint{2309.13382}

\bibitem[{Guo et~al.(2021{\natexlab{a}})Guo, Lu, Lu, Lin, Wang, Huang, Wang,
  and Wang}]{GuoJ2021}
Guo J, Lu S, Lu Q, Lin Y, Wang X, Huang K, Wang R, Wang S (2021{\natexlab{a}})
  Re-reconnection processes of magnetopause flux ropes: Three-dimensional
  global hybrid simulations. Journal of Geophysical Research: Space Physics
  126(6):e2021JA029388

\bibitem[{Guo et~al.(2021{\natexlab{b}})Guo, Lu, Lu, Lin, Wang, Zhang, Xing,
  Huang, Wang, and Wang}]{Guo2021three}
Guo J, Lu S, Lu Q, Lin Y, Wang X, Zhang Q, Xing Z, Huang K, Wang R, Wang S
  (2021{\natexlab{b}}) Three-dimensional global hybrid simulations of high
  latitude magnetopause reconnection and flux ropes during the northward imf.
  Geophysical Research Letters 48(21):e2021GL095003

\bibitem[{Guo et~al.(2020)Guo, Lin, Wang, Vines, Lee, and Chen}]{Guo2020}
Guo Z, Lin Y, Wang X, Vines SK, Lee S, Chen Y (2020) Magnetopause reconnection
  as influenced by the dipole tilt under southward imf conditions: Hybrid
  simulation and mms observation. Journal of Geophysical Research: Space
  Physics 125(9):e2020JA027795

\bibitem[{Guo et~al.(2021{\natexlab{c}})Guo, Lin, and Wang}]{Guo_Lin:2021}
Guo Z, Lin Y, Wang X (2021{\natexlab{c}}) {Investigation of the Interaction
  between Magnetosheath Reconnection and Magnetopause Reconnection Driven by
  Oblique Interplanetary Tangential Discontinuity Using Three-Dimensional
  Global Hybrid Simulation}. J Geophys Res Space Physics 126:e2020JA028558,
  \doi{10.1029/2020JA028558},
  \urlprefix\url{https://doi.org/10.1029/2020JA028558}

\bibitem[{Guo et~al.(2021{\natexlab{d}})Guo, Lin, Wang, and Du}]{Guo2021_JGR}
Guo Z, Lin Y, Wang X, Du A (2021{\natexlab{d}}) Magnetic reconnection inside
  solar wind rotational discontinuity during its interaction with the
  quasi-perpendicular bow shock and magnetosheath. Journal of Geophysical
  Research: Space Physics 126(12):e2021JA029979

\bibitem[{Haggerty et~al.(2017)Haggerty, Parashar, Matthaeus, Shay, Yang, Wan,
  Wu, and Servidio}]{Hhggerty2017}
Haggerty CC, Parashar TN, Matthaeus WH, Shay MA, Yang Y, Wan M, Wu P, Servidio
  S (2017) Exploring the statistics of magnetic reconnection x-points in
  kinetic particle-in-cell turbulence. Phys Plasmas 24:102308

\bibitem[{Haiducek et~al.(2020)Haiducek, Welling, Morley, Ga\~nushkina, and
  Chu}]{Haiducek:2020}
Haiducek JD, Welling DT, Morley SK, Ga\~nushkina NY, Chu X (2020) Using
  multiple signatures to improve accuracy of substorm identificati\ on. J
  Geophys Res 125(4):e2019JA027559, \doi{https://doi.org/10.1029/2019JA027559},
  \urlprefix\url{https://agupubs.onlinelibrary.wiley.com/doi/abs/10.1029/2019JA027559}

\bibitem[{Hall et~al.(1879)}]{hall1879new}
Hall EH, et~al. (1879) On a new action of the magnet on electric currents.
  American Journal of Mathematics 2(3):287--292

\bibitem[{Harris(1962{\natexlab{a}})}]{harris1962equilibrium}
Harris EG (1962{\natexlab{a}}) The equilibrium of oppositely directed magnetic
  fields. Nuovo Cimento 23:115--121

\bibitem[{Harris(1962{\natexlab{b}})}]{harris1962}
Harris EG (1962{\natexlab{b}}) On a plasma sheath separating regions of
  oppositely directed magnetic field. Il Nuovo Cimento (1955-1965)
  23(1):115--121

\bibitem[{Her{\v c}\'ik et~al.(2013)Her{\v c}\'ik, Tr\'avn\'i{\v c}ek, Johnson,
  Kim, and Hellinger}]{Hercik2013}
Her{\v c}\'ik D, Tr\'avn\'i{\v c}ek PM, Johnson R, Kim EH, Hellinger P (2013)
  {Mirror mode structures in the asymmetric Hermean magnetosheath: Hybrid
  simulations}. Journal of Geophysical Research: Space Physics 118:405--417,
  \doi{10.1029/2012JA018083},
  \urlprefix\url{https://doi.org/10.1029/2012JA018083}

\bibitem[{Hesse and Winske(1998)}]{Hesse1998}
Hesse M, Winske D (1998) Electron dissipation in collisionless magnetic
  reconnection. Journal of Geophysical Research: Space Physics
  103(A11):26479--26486, \doi{10.1029/98JA01570}

\bibitem[{Hesse et~al.(1996)Hesse, Birn, Baker, and Slavin}]{hesse1996}
Hesse M, Birn J, Baker DN, Slavin JA (1996) Mhd simulations of the transition
  of magnetic reconnection from closed to open field lines. Journal of
  Geophysical Research: Space Physics 101(A5):10805--10816,
  \doi{https://doi.org/10.1029/95JA02857},
  \urlprefix\url{https://agupubs.onlinelibrary.wiley.com/doi/abs/10.1029/95JA02857},
  \eprint{https://agupubs.onlinelibrary.wiley.com/doi/pdf/10.1029/95JA02857}

\bibitem[{Hesse et~al.(2016)Hesse, Liu, Chen, Bessho, Kuznetsova, Birn, and
  Burch}]{Hesse2016}
Hesse M, Liu YH, Chen LJ, Bessho N, Kuznetsova M, Birn J, Burch JL (2016) On
  the electron diffusion region in asymmetric reconnection with a guide
  magnetic field. Geophysical Research Letters 43(6):2359--2364,
  \doi{10.1002/2016GL068373}

\bibitem[{Hesse et~al.(2018a)Hesse, Norgren, Tenfjord, Burch, Liu, Chen,
  Bessho, Wang, Nakamura, Eastwood, Hoshino, Torbert, and Ergun}]{hesse2018a}
Hesse M, Norgren C, Tenfjord P, Burch JL, Liu YH, Chen LJ, Bessho N, Wang S,
  Nakamura R, Eastwood JP, Hoshino M, Torbert RB, Ergun RE (2018a) On the role
  of separatrix instabilities in heating the reconnection outflow region.
  Physics of Plasmas 25:122902

\bibitem[{Hesse et~al.(2018b)Hesse, Liu, Chen, Bessho, Wang, Burch, Moretto,
  Norgren, Genestreti, Phan et~al.}]{hesse2018b}
Hesse M, Liu YH, Chen LJ, Bessho N, Wang S, Burch J, Moretto T, Norgren C,
  Genestreti K, Phan T, et~al. (2018b) The physical foundation of the
  reconnection electric field. Physics of Plasmas 25(3)

\bibitem[{Hesse et~al.(2019)Hesse, Norgren, Tenfjord, Burch, Liu, Chen, Bessho,
  Wang, Nakamura, Eastwood, Hoshino, Torbert, and Ergun}]{hesse2019}
Hesse M, Norgren C, Tenfjord P, Burch JL, Liu YH, Chen LJ, Bessho N, Wang S,
  Nakamura R, Eastwood JP, Hoshino M, Torbert RB, Ergun RE (2019) Erratum:``on
  the role of separatrix instabilities in heating the reconnection outflow
  region". Physics of Plasmas 26:049901

\bibitem[{Hill et~al.(2005)Hill, Rymer, Burch, Crary, Young, Thomsen, Delapp,
  Andr{\'e}, Coates, and Lewis}]{hill2005evidence}
Hill T, Rymer A, Burch J, Crary F, Young D, Thomsen M, Delapp D, Andr{\'e} N,
  Coates A, Lewis G (2005) Evidence for rotationally driven plasma transport in
  saturn's magnetosphere. Geophysical Research Letters 32(14)

\bibitem[{Hilmer and Voigt(1995)}]{hilmer1995magnetospheric}
Hilmer RV, Voigt GH (1995) A magnetospheric magnetic field model with flexible
  current systems driven by independent physical parameters. Journal of
  Geophysical Research: Space Physics 100(A4):5613--5626

\bibitem[{Hoilijoki et~al.(2017)Hoilijoki, Ganse, Pfau-Kempf, Cassak, Walsh,
  Hietala, von Alfthan, and Palmroth}]{hoilijoki2017}
Hoilijoki S, Ganse U, Pfau-Kempf Y, Cassak PA, Walsh BM, Hietala H, von Alfthan
  S, Palmroth M (2017) Reconnection rates and x line motion at the
  magnetopause: Global 2d-3v hybrid-vlasov simulation results. Journal of
  Geophysical Research: Space Physics 122(3):2877--2888,
  \doi{https://doi.org/10.1002/2016JA023709},
  \urlprefix\url{https://agupubs.onlinelibrary.wiley.com/doi/abs/10.1002/2016JA023709},
  \eprint{https://agupubs.onlinelibrary.wiley.com/doi/pdf/10.1002/2016JA023709}

\bibitem[{Hoilijoki et~al.(2019)Hoilijoki, Ganse, Sibeck, Cassak, Turc,
  Battarbee, Fear, Blanco-Cano, Dimmock, Kilpua, Jarvinen, Juusola, Pfau-Kempf,
  and Palmroth}]{hoilijoki2019}
Hoilijoki S, Ganse U, Sibeck DG, Cassak PA, Turc L, Battarbee M, Fear RC,
  Blanco-Cano X, Dimmock AP, Kilpua EKJ, Jarvinen R, Juusola L, Pfau-Kempf Y,
  Palmroth M (2019) Properties of magnetic reconnection and ftes on the dayside
  magnetopause with and without positive imf bx component during southward imf.
  Journal of Geophysical Research: Space Physics 124(6):4037--4048,
  \doi{https://doi.org/10.1029/2019JA026821},
  \urlprefix\url{https://agupubs.onlinelibrary.wiley.com/doi/abs/10.1029/2019JA026821},
  \eprint{https://agupubs.onlinelibrary.wiley.com/doi/pdf/10.1029/2019JA026821}

\bibitem[{{Holloway} and {Dorning}(1991)}]{holloway1991undamped}
{Holloway} JP, {Dorning} JJ (1991) {Undamped plasma waves}. Physical Review A
  44(6):3856--3868, \doi{10.1103/PhysRevA.44.3856}

\bibitem[{Honkonen et~al.(2013)Honkonen, {von Alfthan}, Sandroos, Janhunen, and
  Palmroth}]{Honkonen2013}
Honkonen I, {von Alfthan} S, Sandroos A, Janhunen P, Palmroth M (2013) Parallel
  grid library for rapid and flexible simulation development. Computer Physics
  Communications 184(4):1297--1309, \doi{10.1016/j.cpc.2012.12.017}

\bibitem[{Horiuchi and Sato(1994)}]{horiuchi:1994}
Horiuchi R, Sato T (1994) Particle simulation study of driven magnetic
  reconnection in a collisionless plasma. Physics of Plasmas 1(11):3587--3597

\bibitem[{Hoshino(1987)}]{hoshino1987}
Hoshino M (1987) The electrostatic effect for the collisionless tearing mode.
  Journal of Geophysical Research: Space Physics 1:7368--7380

\bibitem[{{Hoshino} et~al.(2001){Hoshino}, {Mukai}, {Terasawa}, and
  {Shinohara}}]{Hoshino2001}
{Hoshino} M, {Mukai} T, {Terasawa} T, {Shinohara} I (2001) {Suprathermal
  electron acceleration in magnetic reconnection}. Journal of Geophysical
  Research 106(A11):25979--25998, \doi{10.1029/2001JA900052}

\bibitem[{Hsieh and Otto(2014)}]{hsieh2014influence}
Hsieh MS, Otto A (2014) The influence of magnetic flux depletion on the
  magnetotail and auroral morphology during the substorm growth phase. Journal
  of Geophysical Research: Space Physics 119(5):3430--3443

\bibitem[{Hsieh and Otto(2015)}]{hsieh2015thin}
Hsieh MS, Otto A (2015) Thin current sheet formation in response to the loading
  and the depletion of magnetic flux during the substorm growth phase. Journal
  of Geophysical Research: Space Physics 120(6):4264--4278

\bibitem[{Hu et~al.(2010)Hu, Toffoletto, Wolf, Sazykin, Raeder, Larson, and
  Vapirev}]{hu2010one}
Hu B, Toffoletto F, Wolf R, Sazykin S, Raeder J, Larson D, Vapirev A (2010)
  One-way coupled openggcm/rcm simulation of the 23 march 2007 substorm event.
  Journal of Geophysical Research: Space Physics 115(A12)

\bibitem[{Hu et~al.(2011)Hu, Wolf, Toffoletto, Yang, and
  Raeder}]{hu2011consequences}
Hu B, Wolf R, Toffoletto F, Yang J, Raeder J (2011) Consequences of violation
  of frozen-in-flux: Evidence from openggcm simulations. Journal of Geophysical
  Research: Space Physics 116(A6)

\bibitem[{Huang et~al.(2020)Huang, Liu, Lu, and Hesse}]{huang2020}
Huang K, Liu YH, Lu Q, Hesse M (2020) Scaling of magnetic reconnection with a
  limited x-line extent. Geophysical Research Letters 47:e2020GL088147

\bibitem[{Huang et~al.(2011)Huang, Bhattacharjee, and Sullivan}]{Huang11}
Huang YM, Bhattacharjee A, Sullivan BP (2011) Onset of fast reconnection in
  {H}all magnetohydrodynamics mediated by the plasmoid instability.
  Phys~Plasmas 18:072109

\bibitem[{Huang et~al.(2019)Huang, T\'oth, van~der Holst, Chen, and
  Gombosi}]{Huang:2019}
Huang Z, T\'oth G, van~der Holst B, Chen Y, Gombosi T (2019) A six-moment
  multi-fluid plasma model. J Comput Phys 387:134,
  \doi{10.1016/j.jcp.2019.02.023}

\bibitem[{Huba(1995)}]{Huba95}
Huba JD (1995) Hall magnetohydrodynamics in space and laboratory plasmas.
  Phys~Plasmas 2(6):2504--2513

\bibitem[{Huba(1998)}]{huba1998nrl}
Huba JD (1998) NRL plasma formulary. 98-358, Naval Research Laboratory

\bibitem[{Huba(2003)}]{Huba03c}
Huba JD (2003) A tutorial on {H}all magnetohydrodynamics. In: B\"{u}chner J,
  Dum CT, Scholer M (eds) Space Plasma Simulation, Springer, New York, p 170

\bibitem[{Huba and Rudakov(2002)}]{Huba02}
Huba JD, Rudakov LI (2002) Three-dimensional {H}all magnetic reconnection. Phys
  Plasmas 9:4435

\bibitem[{Huba and Rudakov(2003)}]{Huba03}
Huba JD, Rudakov LI (2003) Hall magnetohydrodynamics of neutral layers.
  Phys~Plasmas 10:3139

\bibitem[{{Huba} and {Sazykin}(2014)}]{huba2014rcm}
{Huba} JD, {Sazykin} S (2014) {Storm time ionosphere and plasmasphere
  structuring: SAMI3-RCM simulation of the 31 March 2001 geomagnetic storm}.
  Geophysical Research Letters 41(23):8208--8214, \doi{10.1002/2014GL062110}

\bibitem[{Hubbert et~al.(2022)Hubbert, Russell, Qi, Lu, Burch, Giles, and
  Moore}]{Hubbert2022}
Hubbert M, Russell CT, Qi Y, Lu S, Burch JL, Giles BL, Moore TE (2022)
  Electron-{{Only Reconnection}} as a {{Transition Phase From Quiet Magnetotail
  Current Sheets}} to {{Traditional Magnetotail Reconnection}}. Journal of
  Geophysical Research: Space Physics 127(3):e2021JA029584,
  \doi{10.1029/2021JA029584}

\bibitem[{Hwang et~al.(2019)Hwang, Choi, Dokgo, Burch, Sibeck, Giles,
  Goldstein, Paterson, Pollock, Shi, Fu, Hasegawa, Gershman, Khotyaintsev,
  Torbert, Ergun, Dorelli, Avanov, Russell, and Strangeway}]{Hwang2019}
Hwang KJ, Choi E, Dokgo K, Burch JL, Sibeck DG, Giles BL, Goldstein ML,
  Paterson WR, Pollock CJ, Shi QQ, Fu H, Hasegawa H, Gershman DJ, Khotyaintsev
  Y, Torbert RB, Ergun RE, Dorelli JC, Avanov L, Russell CT, Strangeway RJ
  (2019) Electron {{Vorticity Indicative}} of the {{Electron Diffusion Region}}
  of {{Magnetic Reconnection}}. Geophysical Research Letters 46(12):6287--6296,
  \doi{10.1029/2019GL082710}

\bibitem[{Hyman(1983)}]{hyman1983accurate}
Hyman JM (1983) Accurate monotonicity preserving cubic interpolation. SIAM
  Journal on Scientific and Statistical Computing 4(4):645--654

\bibitem[{Inglebert et~al.(2011)Inglebert, Ghizzo, Reveille, Sarto, Bertrand,
  and Califano}]{Inglebert2011}
Inglebert A, Ghizzo A, Reveille T, Sarto DD, Bertrand P, Califano F (2011) A
  multi-stream vlasov modeling unifying relativistic weibel-type instabilities.
  Europhysics Letters 95(4):45002, \doi{10.1209/0295-5075/95/45002}

\bibitem[{Jain and B{\"u}chner(2014{\natexlab{a}})}]{Jain2014}
Jain N, B{\"u}chner J (2014{\natexlab{a}}) Nonlinear evolution of
  three-dimensional instabilities of thin and thick electron scale current
  sheets: {{Plasmoid}} formation and current filamentation. Physics of Plasmas
  21(7):072306, \doi{10.1063/1.4887279}

\bibitem[{Jain and B{\"u}chner(2014{\natexlab{b}})}]{Jain2014a}
Jain N, B{\"u}chner J (2014{\natexlab{b}}) Three dimensional instabilities of
  an electron scale current sheet in collisionless magnetic reconnection.
  Physics of Plasmas 21(6):062116, \doi{10.1063/1.4885636}

\bibitem[{Jain and B{\"u}chner(2015)}]{Jain2015}
Jain N, B{\"u}chner J (2015) Effect of guide field on three-dimensional
  electron shear flow instabilities in electron current sheets. Journal of
  Plasma Physics 81(6):905810606, \doi{10.1017/S0022377815001257}

\bibitem[{Jain and Sharma(2009)}]{Jain2009}
Jain N, Sharma AS (2009) Electron scale structures in collisionless magnetic
  reconnection. Physics of Plasmas 16(5):050704, \doi{10.1063/1.3134045}

\bibitem[{Jain and Sharma(2015{\natexlab{a}})}]{Jain2015a}
Jain N, Sharma AS (2015{\natexlab{a}}) Electron-scale nested quadrupole
  {{Hall}} field in {{Cluster}} observations of magnetic reconnection. Annales
  Geophysicae 33(6):719--724, \doi{10.5194/angeo-33-719-2015}

\bibitem[{Jain and Sharma(2015{\natexlab{b}})}]{Jain2015b}
Jain N, Sharma AS (2015{\natexlab{b}}) Evolution of electron current sheets in
  collisionless magnetic reconnection. Physics of Plasmas 22(10):102110,
  \doi{10.1063/1.4933120}

\bibitem[{Jain et~al.(2012)Jain, Sharma, Zelenyi, and Malova}]{Jain2012}
Jain N, Sharma AS, Zelenyi LM, Malova HV (2012) Electron scale structures of
  thin current sheets in magnetic reconnection. Annales Geophysicae
  30(4):661--666, \doi{10.5194/angeo-30-661-2012}

\bibitem[{Jain et~al.(2013)Jain, B{\"u}chner, Dorfman, Ji, and
  Surjalal~Sharma}]{Jain13}
Jain N, B{\"u}chner J, Dorfman S, Ji H, Surjalal~Sharma A (2013) Current
  disruption and its spreading in collisionless magnetic reconnection. Physics
  of Plasmas 20(11)

\bibitem[{Jain et~al.(2017{\natexlab{a}})Jain, B{\"u}chner, and
  Mu{\~n}oz}]{jain2017a}
Jain N, B{\"u}chner J, Mu{\~n}oz PA (2017{\natexlab{a}}) Nonlinear evolution of
  electron shear flow instabilities in the presence of an external guide
  magnetic field. Physics of Plasmas 24(3):032303, \doi{10.1063/1.4977528}

\bibitem[{Jain et~al.(2017{\natexlab{b}})Jain, {von Stechow}, Mu{\~n}oz,
  B{\"u}chner, Grulke, and Klinger}]{Jain2017}
Jain N, {von Stechow} A, Mu{\~n}oz PA, B{\"u}chner J, Grulke O, Klinger T
  (2017{\natexlab{b}}) Electron-magnetohydrodynamic simulations of electron
  scale current sheet dynamics in the {{Vineta}}.{{II}} guide field
  reconnection experiment. Physics of Plasmas 24(9):092312,
  \doi{10.1063/1.5004564}

\bibitem[{Jain et~al.(2022)Jain, Mu{\~n}oz, Farzalipour~Tabriz, Rampp, and
  B{\"u}chner}]{Jain2022}
Jain N, Mu{\~n}oz PA, Farzalipour~Tabriz M, Rampp M, B{\"u}chner J (2022)
  Importance of accurate consideration of the electron inertia in
  hybrid-kinetic simulations of collisionless plasma turbulence: {{The 2D}}
  limit. Physics of Plasmas 29(5):053902, \doi{10.1063/5.0087103}

\bibitem[{Jain et~al.(2023)Jain, Mu{\~{n}}oz, and B{\"u}chner}]{Jain2023}
Jain N, Mu{\~{n}}oz PA, B{\"u}chner J (2023) Hybrid-kinetic approach: Inertial
  electrons. In: B{\"u}chner J (ed) Space and Astrophysical Plasma Simulation:
  Methods, Algorithms, and Applications, Springer International Publishing, pp
  283--311

\bibitem[{Jardin et~al.(2008)Jardin, Ferraro, Luo, Chen, Breslau, Jansen, and
  Shephard}]{Jardin08}
Jardin S, Ferraro N, Luo X, Chen J, Breslau J, Jansen K, Shephard M (2008) The
  m3d-c1 approach to simulating 3d 2-fluid magnetohydrodynamics in magnetic
  fusion experiments. In: Journal of Physics: Conference Series, IOP
  Publishing, vol 125, p 012044

\bibitem[{Jarvinen et~al.(2020)Jarvinen, Alho, Kallio, and
  Pulkkinen}]{Jarvinen:2020}
Jarvinen R, Alho M, Kallio E, Pulkkinen TI (2020) {Ultra-low-frequency waves in
  the foreschock of Mercury: a global hybrid modelling study}. MNRAS
  491:4147--4161, \doi{10.1093/mnras/stz3257},
  \urlprefix\url{https://doi.org/10.1093/mnras/stz3257}

\bibitem[{Johlander et~al.(2022)Johlander, Battarbee, Turc, Ganse, Pfau-Kempf,
  Grandin, Suni, Tarvus, Bussov, Zhou, Alho, Dubart, George, Papadakis, and
  Palmroth}]{johlander2022}
Johlander A, Battarbee M, Turc L, Ganse U, Pfau-Kempf Y, Grandin M, Suni J,
  Tarvus V, Bussov M, Zhou H, Alho M, Dubart M, George H, Papadakis K, Palmroth
  M (2022) Quasi-parallel shock reformation seen by magnetospheric multiscale
  and ion-kinetic simulations. Geophysical Research Letters
  49(2):e2021GL096335, \doi{https://doi.org/10.1029/2021GL096335},
  e2021GL096335 2021GL096335,
  \eprint{https://agupubs.onlinelibrary.wiley.com/doi/pdf/10.1029/2021GL096335}

\bibitem[{Jordanova et~al.(1994)Jordanova, Kozyra, Khazanov, Nagy, Rasmussen,
  and Fok}]{Jordanova:1994}
Jordanova VK, Kozyra JU, Khazanov GV, Nagy AF, Rasmussen CE, Fok MC (1994) A
  bounce-averaged kinetic model of the ring current ion population. Geophys Res
  Lett 21:2785

\bibitem[{Juno et~al.(2018)Juno, Hakim, TenBarge, Shi, and Dorland}]{juno2018}
Juno J, Hakim A, TenBarge J, Shi E, Dorland W (2018) Discontinuous galerkin
  algorithms for fully kinetic plasmas. Journal of Computational Physics
  353:110 -- 147, \doi{https://doi.org/10.1016/j.jcp.2017.10.009},
  \urlprefix\url{http://www.sciencedirect.com/science/article/pii/S0021999117307477}

\bibitem[{Juusola et~al.(2018)Juusola, Hoilijoki, Pfau-Kempf, Ganse, Jarvinen,
  Battarbee, Kilpua, Turc, and Palmroth}]{juusola2018}
Juusola L, Hoilijoki S, Pfau-Kempf Y, Ganse U, Jarvinen R, Battarbee M, Kilpua
  E, Turc L, Palmroth M (2018) Fast plasma sheet flows and x line motion in the
  earth's magnetotail: results from a global hybrid-vlasov simulation. Annales
  Geophysicae 36(5):1183--1199, \doi{10.5194/angeo-36-1183-2018},
  \urlprefix\url{https://angeo.copernicus.org/articles/36/1183/2018/}

\bibitem[{Kallio et~al.(2019)Kallio, Dyadechkin, Wurz, and
  Khodachenko}]{Kallio:2019}
Kallio E, Dyadechkin S, Wurz P, Khodachenko M (2019) {Space weathering on the
  Moon: Farside-nearside solar wind precipitation asymmetry}. Planet Space
  Science 166:9--22, \doi{10.1016/j.pss.2018.07.013},
  \urlprefix\url{https://doi.org/10.1016/j.pss.2018.07.013}

\bibitem[{Karimabadi et~al.(2004{\natexlab{a}})Karimabadi, Krauss-Varban, Huba,
  and Vu}]{Karimabadi2004}
Karimabadi H, Krauss-Varban D, Huba D, Vu H (2004{\natexlab{a}}) {On magnetic
  reconnection regimes and associated three-dimensional asymmetries: Hybrid,
  Hall-less hybrid, and Hall-MHD simulations}. Journal of Geophysical Research
  109(A9):A09205, \doi{10.1029/2004JA010478},
  \urlprefix\url{http://doi.wiley.com/10.1029/2004JA010478}

\bibitem[{Karimabadi et~al.(2004{\natexlab{b}})Karimabadi, Krauss-Varban, Huba,
  and Vu}]{Karimabadi04}
Karimabadi H, Krauss-Varban D, Huba JD, Vu HX (2004{\natexlab{b}}) On magnetic
  reconnection regimes and associated three-dimensional asymmetries: {H}ybrid,
  {H}all-less hybrid, and {H}all-{MHD} simulations. J~Geophys~Res 109:A09205

\bibitem[{Karimabadi et~al.(2007)Karimabadi, Daughton, and
  Scudder}]{Karimabadi07}
Karimabadi H, Daughton W, Scudder J (2007) Multi-scale structure of the
  electron diffusion region. Geophys~Res~Lett 34:L13104

\bibitem[{Karimabadi et~al.(2014)Karimabadi, Roytershteyn, Vu, Omelchenko,
  Scudder, Daughton, Dimmock, Nykyri, Wan, Sibeck, Tatineni, Majumdar, Loring,
  and Geveci}]{homa14}
Karimabadi H, Roytershteyn V, Vu HX, Omelchenko YA, Scudder J, Daughton W,
  Dimmock A, Nykyri K, Wan M, Sibeck D, Tatineni M, Majumdar A, Loring B,
  Geveci B (2014) The link between shocks, turbulence, and magnetic
  reconnection in collisionless plasmas. Physics of Plasmas 21(6):062308,
  \doi{10.1063/1.4882875}

\bibitem[{Kepko et~al.(2015)Kepko, McPherron, Amm, Apatenkov, Baumjohann, Birn,
  Lester, Nakamura, Pulkkinen, and Sergeev}]{kepko2015substorm}
Kepko L, McPherron R, Amm O, Apatenkov S, Baumjohann W, Birn J, Lester M,
  Nakamura R, Pulkkinen TI, Sergeev V (2015) Substorm current wedge revisited.
  Space Science Reviews 190:1--46

\bibitem[{{Kilian} et~al.(2020){Kilian}, {Li}, {Guo}, and {Li}}]{Kilian2020}
{Kilian} P, {Li} X, {Guo} F, {Li} H (2020) {Exploring the Acceleration
  Mechanisms for Particle Injection and Power-law Formation during
  Transrelativistic Magnetic Reconnection}. Astrophysical Journal 899(2):151,
  \doi{10.3847/1538-4357/aba1e9}, \eprint{2001.02732}

\bibitem[{Kingsep et~al.(1990)Kingsep, Chukbar, and Yan'kov}]{Kingsep90}
Kingsep AS, Chukbar KV, Yan'kov YY (1990) Electron magnetohydrodynamics. In:
  Reviews of Plasma Physics, Consultants Bureau, New York, vol~16

\bibitem[{Kleva et~al.(1995)Kleva, Drake, and Waelbroeck}]{Kleva95}
Kleva R, Drake J, Waelbroeck F (1995) Fast reconnection in high temperature
  plasma. Phys Plasma 2:23

\bibitem[{Kuznetsova(2000)}]{Kuznetsova2000}
Kuznetsova M (2000) Toward a transport model of collisionless magnetic
  reconnection. Journal of Geophysical Research: Space Physics
  105(A4):7601--7616, \doi{10.1029/1999JA900396}

\bibitem[{Kuznetsova et~al.(1998)Kuznetsova, Hesse, and
  Winske}]{Kuznetsova1998}
Kuznetsova M, Hesse M, Winske D (1998) {Kinetic quasi-viscous and bulk flow
  inertia effects in collisionless magnetotail reconnection}. Journal of
  Geophysical Research 103:199--213

\bibitem[{Kuznetsova et~al.(2001)Kuznetsova, Hesse, and
  Winske}]{Kuznetsova2001}
Kuznetsova MM, Hesse M, Winske D (2001) Collisionless reconnection supported by
  nongyrotropic pressure effects in hybrid and particle simulations. Journal of
  Geophysical Research: Space Physics 106(A3):3799--3810,
  \doi{10.1029/1999JA001003}

\bibitem[{Laitinen et~al.(2006)Laitinen, Janhunen, Pulkkinen, Palmroth, and
  Koskinen}]{Laitinen2006AnGeo}
Laitinen TV, Janhunen P, Pulkkinen TI, Palmroth M, Koskinen HEJ (2006) On the
  characterization of magnetic reconnection in global {MHD} simulations.
  Annales Geophysicae 24(11):3059--3069, \doi{10.5194/angeo-24-3059-2006}

\bibitem[{Laitinen et~al.(2010)Laitinen, Khotyaintsev, Andr{\'e}, Vaivads, and
  Reme}]{laitinen2010}
Laitinen TV, Khotyaintsev YV, Andr{\'e} M, Vaivads A, Reme H (2010) Local
  influence of the magnetosheath plasma beta fluctuations on magnetopause
  reconnection. Ann Geophys 28:1053

\bibitem[{Lapenta(2017)}]{Lapenta:2017}
Lapenta G (2017) Exactly energy conserving semi-implicit particle in cell
  formulation. J Comput Phys 334:349, \doi{10.1016/j.jcp.2017.01.002}

\bibitem[{Lapenta et~al.(2006)Lapenta, Krauss-Varban, Karimabadi, Huba,
  Rudakov, and Ricci}]{Lapenta06}
Lapenta G, Krauss-Varban D, Karimabadi H, Huba JD, Rudakov LI, Ricci P (2006)
  Kinetic simulations of x-line expansion in 3{D} reconnection.
  Geophys~Res~Lett 33:L10102

\bibitem[{Leclercq et~al.(2016)Leclercq, Modolo, Leblanc, Hess, and
  Mancini}]{Leclercq:2016}
Leclercq L, Modolo R, Leblanc F, Hess S, Mancini M (2016) {3D magnetospheric
  parallel hybrid multi-grid method applied to planet–plasma interactions}. J
  Comp Phys 309:295--313, \doi{10.1016/j.jcp.2016.01.005},
  \urlprefix\url{https://doi.org/10.1016/j.jcp.2016.01.005}

\bibitem[{Lee and Sussman(2004)}]{lee2004efficient}
Lee JY, Sussman A (2004) Efficient communication between parallel programs with
  intercomm. Tech. rep., Citeseer

\bibitem[{Lemon et~al.(2003)Lemon, Toffoletto, Hesse, and
  Birn}]{lemon2003computing}
Lemon C, Toffoletto F, Hesse M, Birn J (2003) Computing magnetospheric force
  equilibria. Journal of Geophysical Research: Space Physics 108(A6)

\bibitem[{Lemon et~al.(2004)Lemon, Wolf, Hill, Sazykin, Spiro, Toffoletto,
  Birn, and Hesse}]{lemon2004magnetic}
Lemon C, Wolf R, Hill T, Sazykin S, Spiro R, Toffoletto F, Birn J, Hesse M
  (2004) Magnetic storm ring current injection modeled with the rice convection
  model and a self-consistent magnetic field. Geophysical research letters
  31(21)

\bibitem[{Li et~al.(2023)Li, Jia, Chen, Toth, Zhou, Slavin, Sun, and
  Poh}]{li2023global}
Li C, Jia X, Chen Y, Toth G, Zhou H, Slavin JA, Sun W, Poh G (2023) Global hall
  mhd simulations of mercury's magnetopause dynamics and ftes under different
  solar wind and imf conditions. Journal of Geophysical Research: Space Physics
  128(5):e2022JA031206

\bibitem[{Li et~al.(1998)Li, Baker, Temerin, Reeves, and
  Belian}]{li1998simulation}
Li X, Baker D, Temerin M, Reeves G, Belian R (1998) Simulation of
  dispersionless injections and drift echoes of energetic electrons associated
  with substorms. Geophysical Research Letters 25(20):3763--3766

\bibitem[{{Li} et~al.(2015){Li}, {Guo}, {Li}, and {Li}}]{Li2015}
{Li} X, {Guo} F, {Li} H, {Li} G (2015) {Nonthermally Dominated Electron
  Acceleration during Magnetic Reconnection in a Low-{\ensuremath{\beta}}
  Plasma}. Astrophysical Journal Letters 811(2):L24,
  \doi{10.1088/2041-8205/811/2/L24}, \eprint{1505.02166}

\bibitem[{{Li} et~al.(2017){Li}, {Guo}, {Li}, and {Li}}]{Li2017}
{Li} X, {Guo} F, {Li} H, {Li} G (2017) {Particle Acceleration during Magnetic
  Reconnection in a Low-beta Plasma}. Astrophysical Journal 843(1):21,
  \doi{10.3847/1538-4357/aa745e}

\bibitem[{{Li} et~al.(2018){Li}, {Guo}, {Li}, and {Birn}}]{Li2018}
{Li} X, {Guo} F, {Li} H, {Birn} J (2018) {The Roles of Fluid Compression and
  Shear in Electron Energization during Magnetic Reconnection}. Astrophysical
  Journal 855(2):80, \doi{10.3847/1538-4357/aaacd5}, \eprint{1801.02255}

\bibitem[{Li et~al.(2019)Li, Guo, and Li}]{li2019particle}
Li X, Guo F, Li H (2019) Particle acceleration in kinetic simulations of
  nonrelativistic magnetic reconnection with different ion--electron mass
  ratios. The Astrophysical Journal 879(1):5

\bibitem[{{Li} et~al.(2019){Li}, {Guo}, {Li}, {Stanier}, and {Kilian}}]{Li2019}
{Li} X, {Guo} F, {Li} H, {Stanier} A, {Kilian} P (2019) {Formation of Power-law
  Electron Energy Spectra in Three-dimensional Low-{\ensuremath{\beta}}
  Magnetic Reconnection}. Astrophysical Journal 884(2):118,
  \doi{10.3847/1538-4357/ab4268}, \eprint{1909.01911}

\bibitem[{{Li} et~al.(2023){Li}, {Guo}, {Liu}, and {Li}}]{Li2023}
{Li} X, {Guo} F, {Liu} YH, {Li} H (2023) {A Model for Nonthermal Particle
  Acceleration in Relativistic Magnetic Reconnection}. Astrophysical Journal
  Letters 954(2):L37, \doi{10.3847/2041-8213/acf135}, \eprint{2302.12737}

\bibitem[{Liemohn et~al.(2018)Liemohn, Ganushkina, Zeeuw, Rastaetter,
  Kuznetsova, Welling, T\'oth, Ilie, Gombosi, and van~der Holst}]{Liemohn:2018}
Liemohn M, Ganushkina N, Zeeuw DD, Rastaetter L, Kuznetsova M, Welling D,
  T\'oth G, Ilie R, Gombosi T, van~der Holst B (2018) Real-time swmf at ccmc:
  assessing the dst output from continuous operational simulations. Space
  Weather 16:1583, \doi{10.1029/2018SW001953}

\bibitem[{Liemohn et~al.(2001)Liemohn, Kozyra, Clauer, and
  Ridley}]{Liemohn:2001b}
Liemohn MW, Kozyra JU, Clauer CR, Ridley AJ (2001) Computational analysis of
  the near-{Earth} magnetospheric current system during two-phase decay storms.
  J Geophys Res 106:29,531

\bibitem[{Lin et~al.(2022{\natexlab{a}})Lin, Wang, Merkin, Huang, Oppenheim,
  Sorathia, Pham, Michael, Bao, Wu et~al.}]{lin2022ionospheric}
Lin D, Wang W, Merkin VG, Huang C, Oppenheim MM, Sorathia K, Pham KH, Michael
  A, Bao S, Wu Q, et~al. (2022{\natexlab{a}}) Ionospheric dawnside subauroral
  polarization streams: A unique feature of major geomagnetic storms. Authorea
  Preprints

\bibitem[{Lin(2002)}]{Lin2002}
Lin Y (2002) Global hybrid simulation of hot flow anomalies near the bow shock
  and in the magnetosheath. Planetary and space science 50(5-6):577--591

\bibitem[{Lin and Wang(2006)}]{Lin2006}
Lin Y, Wang X (2006) Formation of dayside low-latitude boundary layer under
  northward interplanetary magnetic field. Geophysical research letters 33(21)

\bibitem[{Lin and Wang(2005)}]{Lin2005}
Lin Y, Wang XY (2005) {Three-dimensional global hybrid simulation of dayside
  dynamics associated with the quasi-parallel bow shock}. Journal of
  Geophysical Research: Space Physics 110(A12):1--13,
  \doi{10.1029/2005JA011243},
  \urlprefix\url{http://dx.doi.org/10.1029/2005JA011243}

\bibitem[{Lin et~al.(1996)Lin, Lee, and Yan}]{Lin1996}
Lin Y, Lee L, Yan M (1996) Generation of dynamic pressure pulses downstream of
  the bow shock by variations in the interplanetary magnetic field orientation.
  Journal of Geophysical Research: Space Physics 101(A1):479--493

\bibitem[{Lin et~al.(2007)Lin, Wang, and Chang}]{Lin2007}
Lin Y, Wang XY, Chang SW (2007) {Connection between bow shock and cusp
  energetic ions}. Geophys Res Lett 34:L11107, \doi{10.1029/2007GL030038},
  \urlprefix\url{https://doi.org/10.1029/2007GL030038}

\bibitem[{Lin et~al.(2008)Lin, Wang, Brown, Schaffer, and Cothran}]{Lin2008}
Lin Y, Wang XY, Brown MR, Schaffer MJ, Cothran CD (2008) {Modeling Swarthmore
  spheromak reconnection experiment using hybrid code}. Plasma Phys Control
  Fusion 50:074012, \doi{10.1088/0741-3335/50/7/074012},
  \urlprefix\url{https://doi.org/10.1088/0741-3335/50/7/074012}

\bibitem[{Lin et~al.(2012)Lin, Johnson, and Wang}]{Lin2012}
Lin Y, Johnson JR, Wang X (2012) Three-dimensional mode conversion associated
  with kinetic alfv{\'e}n waves. Physical review letters 109(12):125003

\bibitem[{Lin et~al.(2014)Lin, Wang, Lu, Perez, and Lu}]{Lin2014}
Lin Y, Wang XY, Lu S, Perez JD, Lu Q (2014) Investigation of storm time
  magnetotail and ion injection using three-dimensional global hybrid
  simulation. Journal of Geophysical Research (Space Physics)
  119(9):7413--7432, \doi{10.1002/2014JA020005},
  \urlprefix\url{https://ui.adsabs.harvard.edu/abs/2014JGRA..119.7413L}

\bibitem[{Lin et~al.(2017)Lin, Wing, Johnson, Wang, Perez, and Cheng}]{Lin2017}
Lin Y, Wing S, Johnson JR, Wang X, Perez JD, Cheng L (2017) Formation and
  transport of entropy structures in the magnetotail simulated with a 3-d
  global hybrid code. Geophysical Research Letters 44(12):5892--5899

\bibitem[{Lin et~al.(2021{\natexlab{a}})Lin, Wang, Fok, Buzulukova, Perez,
  Cheng, and Chen}]{Lin2021}
Lin Y, Wang X, Fok MC, Buzulukova N, Perez J, Cheng L, Chen LJ
  (2021{\natexlab{a}}) Magnetotail-inner magnetosphere transport associated
  with fast flows based on combined global-hybrid and cimi simulation. Journal
  of Geophysical Research: Space Physics 126(3):e2020JA028405

\bibitem[{Lin et~al.(2021{\natexlab{b}})Lin, Wang, Fok, Buzulukova, Perez,
  Cheng, and Chen}]{Lin_Wang:2021}
Lin Y, Wang X, Fok MC, Buzulukova N, Perez JD, Cheng L, Chen LJ
  (2021{\natexlab{b}}) {Magnetotail-Inner Magnetosphere Transport Associated
  With Fast Flows Based on Combined Global-Hybrid and CIMI Simulation}. J
  Geophys Res Space Physics 126:e2020JA028405, \doi{10.1029/2020JA028405},
  \urlprefix\url{https://doi.org/10.1029/2020JA028405}

\bibitem[{Lin et~al.(2022{\natexlab{b}})Lin, Wang, Sibeck, Wang, and
  Lee}]{Lin2022}
Lin Y, Wang X, Sibeck DG, Wang CP, Lee SH (2022{\natexlab{b}}) Global
  asymmetries of hot flow anomalies. Geophysical Research Letters
  49(4):e2021GL096970

\bibitem[{Lin et~al.(2022{\natexlab{c}})Lin, Wang, Sibeck, Wang, and
  Lee}]{Lin_Wang:2022}
Lin Y, Wang XY, Sibeck DG, Wang CP, Lee SH (2022{\natexlab{c}}) {Global
  Asymmetries of Hot Flow Anomalies}. Geophys Res Lett 49:e2021GL096970,
  \doi{10.1029/2021GL096970},
  \urlprefix\url{https://doi.org/10.1029/2021GL096970}

\bibitem[{Lipatov(2002{\natexlab{a}})}]{Lipatov2002}
Lipatov AS (2002{\natexlab{a}}) The {{Hybrid Multiscale Simulation
  Technology}}. Scientific {{Computation}}, {Springer}, {Berlin, Heidelberg},
  \doi{10.1007/978-3-662-05012-5}

\bibitem[{Lipatov(2002{\natexlab{b}})}]{Lipatov:2002}
Lipatov AS (2002{\natexlab{b}}) {The Hybrid Multiscale Simulation Technology}.
  Scientific Computation, Springer Berlin Heidelberg, Berlin, Heidelberg,
  \doi{10.1007/978-3-662-05012-5}

\bibitem[{Liu et~al.(2015)Liu, V.~Angelopoulos, Yao, and Runov}]{liu2015}
Liu J, V~Angelopoulos XZZ, Yao ZH, Runov A (2015) Cross-tail expansion of
  dipolarizing flux bundles. Journal of Geophysical Research 120:2516

\bibitem[{Liu et~al.(2013)Liu, Daughton, Karimabadi, Li, and
  Roytershteyn}]{Liu2013}
Liu YH, Daughton W, Karimabadi H, Li H, Roytershteyn V (2013) {Bifurcated
  Structure of the Electron Diffusion Region in Three-Dimensional Magnetic
  Reconnection}. Physical Review Letters 110(26):265004,
  \doi{10.1103/PhysRevLett.110.265004},
  \urlprefix\url{http://link.aps.org/doi/10.1103/PhysRevLett.110.265004}

\bibitem[{Liu et~al.(2014)Liu, Birn, Daughton, Hesse, and
  Schindler}]{liu2014onset}
Liu YH, Birn J, Daughton W, Hesse M, Schindler K (2014) Onset of reconnection
  in the near magnetotail: Pic simulations. Journal of Geophysical Research:
  Space Physics 119(12):9773--9789

\bibitem[{Liu et~al.(2019)Liu, Li, Hesse, and Huang}]{liu2019}
Liu YH, Li TC, Hesse M, Huang K (2019) Three-dimensional magnetic reconnection
  with a spatially confined x-line extent: Implications for dipolarizing flux
  bundles and the dawn-dusk asymmetry. Journal of Geophysical Research: Space
  Physics 128:2819--2830

\bibitem[{Liu et~al.(2022)Liu, Cassak, Li, Hesse, Lin, and Genestreti}]{Liu22}
Liu YH, Cassak P, Li X, Hesse M, Lin SC, Genestreti K (2022) First-principles
  theory of the rate of magnetic reconnection in magnetospheric and solar
  plasmas. Comms Phys 5:97, \doi{10.1038/s42005-022-00854-x}

\bibitem[{Liu et~al.(2024)Liu, Hesse, Genestreti, Nakamura, Burch, Cassak,
  Bessho, Eastwood, Phan, Swisdak et~al.}]{liu2024ohm}
Liu YH, Hesse M, Genestreti K, Nakamura R, Burch J, Cassak P, Bessho N,
  Eastwood J, Phan T, Swisdak M, et~al. (2024) Ohm's law, the reconnection
  rate, and energy conversion in collisionless magnetic reconnection. arXiv
  preprint arXiv:240600875

\bibitem[{Londrillo and Del~Zanna(2004)}]{Londrillo2004JCP}
Londrillo P, Del~Zanna L (2004) On the divergence-free condition in
  {Godunov}-type schemes for ideal magnetohydrodynamics: the upwind constrained
  transport method. Journal of Computational Physics 195(1):17--48,
  \doi{10.1016/j.jcp.2003.09.016}

\bibitem[{Lottermoser and Scholar(1997)}]{Lottermoser97}
Lottermoser RF, Scholar M (1997) Undriven magnetic reconnection in
  magnetohydrodynamics and hall magnetohydrodynamics. Journal of Geophysical
  Research: Space Physics 102(A3):4875--4892,
  \doi{https://doi.org/10.1029/96JA03634}

\bibitem[{Loureiro et~al.(2007)Loureiro, Schekochihin, and Cowley}]{Loureiro07}
Loureiro NF, Schekochihin AA, Cowley SC (2007) Instability of current sheets
  and formation of plasmoid chains. Phys~Plasmas 14:100703

\bibitem[{Lu et~al.(2016)Lu, Lin, Angelopoulos, Artemyev, Pritchett, Lu, and
  Wang}]{Lu2016}
Lu S, Lin Y, Angelopoulos V, Artemyev A, Pritchett P, Lu Q, Wang X (2016) Hall
  effect control of magnetotail dawn-dusk asymmetry: A three-dimensional global
  hybrid simulation. Journal of Geophysical Research: Space Physics
  121(12):11--882

\bibitem[{Lu et~al.(2020)Lu, Wang, Lu, Angelopoulos, Nakamura, Artemyev,
  Pritchett, Liu, Zhang, Baumjohann, Gonzalez, Rager, Torbert, Giles, Gershman,
  Russell, Strangeway, Qi, Ergun, Lindqvist, Burch, and Wang}]{Lu2020}
Lu S, Wang R, Lu Q, Angelopoulos V, Nakamura R, Artemyev AV, Pritchett PL, Liu
  TZ, Zhang XJ, Baumjohann W, Gonzalez W, Rager AC, Torbert RB, Giles BL,
  Gershman DJ, Russell CT, Strangeway RJ, Qi Y, Ergun RE, Lindqvist PA, Burch
  JL, Wang S (2020) Magnetotail reconnection onset caused by electron kinetics
  with a strong external driver. Nature Communications 11(1):5049,
  \doi{10.1038/s41467-020-18787-w}

\bibitem[{Lu et~al.(2022)Lu, Lu, Wang, Pritchett, Hubbert, Qi, Huang, Li, and
  Russell}]{Lu2022}
Lu S, Lu Q, Wang R, Pritchett PL, Hubbert M, Qi Y, Huang K, Li X, Russell CT
  (2022) Electron-{{Only Reconnection}} as a {{Transition From Quiet Current
  Sheet}} to {{Standard Reconnection}} in {{Earth}}'s {{Magnetotail}}:
  {{Particle-In-Cell Simulation}} and {{Application}} to {{MMS Data}}.
  Geophysical Research Letters 49(11):e2022GL098547, \doi{10.1029/2022GL098547}

\bibitem[{Lyon et~al.(2004)Lyon, Fedder, and Mobarry}]{lyon2004lyon}
Lyon J, Fedder J, Mobarry C (2004) The lyon--fedder--mobarry (lfm) global mhd
  magnetospheric simulation code. Journal of Atmospheric and Solar-Terrestrial
  Physics 66(15-16):1333--1350

\bibitem[{Ma et~al.(2018)Ma, Russell, T\'oth, Chen, Nagy, Harada, McFadden,
  Halekas, Lillis, Connerney, Espley, DiBraccio, Markidis, Peng, Fang, and
  Jakosky}]{Ma:2018}
Ma Y, Russell C, T\'oth G, Chen Y, Nagy A, Harada Y, McFadden J, Halekas J,
  Lillis R, Connerney J, Espley J, DiBraccio G, Markidis S, Peng IB, Fang X,
  Jakosky B (2018) Reconnection in the martian magnetotail: Hall-mhd with
  embedded particle-in-cell simulations. J Geophys Res 123:3742,
  \doi{10.1029/2017JA024729}

\bibitem[{Ma and Bhattacharjee(1996)}]{Ma96}
Ma ZW, Bhattacharjee A (1996) Fast impulsive reconnection and current sheet
  intensfication due to electron pressure gradients in semi-collisional
  plasmas. Geophys Res Lett 23:1673

\bibitem[{Ma and Bhattacharjee(2001)}]{Ma01}
Ma ZW, Bhattacharjee A (2001) Hall magnetohydrodynamic reconnection: The
  geospace environment modeling challenge. Journal of Geophysical Research:
  Space Physics 106(A3):3773--3782, \doi{https://doi.org/10.1029/1999JA001004}

\bibitem[{Makwana et~al.(2017)Makwana, Keppens, and Lapenta}]{Makwana:2017}
Makwana K, Keppens R, Lapenta G (2017) Two-way coupling of magnetohydrodynamic
  simulations with embedded particle-in-cell simulations. Computer Phys Comm
  221:81, \doi{10.1016/j.cpc.2017.08.003}

\bibitem[{Man et~al.(2020)Man, Zhou, Yi, Zhong, Tian, Deng, Khotyaintsev,
  Russell, and Giles}]{Man2020}
Man HY, Zhou M, Yi YY, Zhong ZH, Tian AM, Deng XH, Khotyaintsev Y, Russell CT,
  Giles BL (2020) Observations of {{Electron-Only Magnetic Reconnection
  Associated With Macroscopic Magnetic Flux Ropes}}. Geophysical Research
  Letters 47(19):e2020GL089659, \doi{10.1029/2020GL089659}

\bibitem[{Mandli et~al.(2016)Mandli, Ahmadia, Berger, Calhoun, George,
  Hadjimichael, Ketcheson, Lemoine, and LeVeque}]{mandli2016clawpack}
Mandli KT, Ahmadia AJ, Berger M, Calhoun D, George DL, Hadjimichael Y,
  Ketcheson DI, Lemoine GI, LeVeque RJ (2016) Clawpack: building an open source
  ecosystem for solving hyperbolic pdes. PeerJ Computer Science 2:e68

\bibitem[{Mandt et~al.(1994{\natexlab{a}})Mandt, Denton, and Drake}]{Mandt94}
Mandt ME, Denton RE, Drake JF (1994{\natexlab{a}}) Transition to whistler
  mediated magnetic reconnection. Geophys Res Lett 21:73

\bibitem[{Mandt et~al.(1994{\natexlab{b}})Mandt, Denton, and Drake}]{Mandt1994}
Mandt ME, Denton RE, Drake JF (1994{\natexlab{b}}) Transition to whistler
  mediated magnetic reconnection. Geophysical Research Letters 21(1):73--76,
  \doi{10.1029/93GL03382}

\bibitem[{Mangeney et~al.(2002)Mangeney, Califano, Cavazzoni, and
  Travnicek}]{mangeney2002}
Mangeney A, Califano F, Cavazzoni C, Travnicek P (2002) A numerical scheme for
  the integration of the vlasov–maxwell system of equations. Journal of
  Computational Physics 179(2):495--538,
  \doi{https://doi.org/10.1006/jcph.2002.7071},
  \urlprefix\url{https://www.sciencedirect.com/science/article/pii/S0021999102970713}

\bibitem[{Manzini et~al.(2023)Manzini, Sahraoui, and Califano}]{manzini2023}
Manzini D, Sahraoui F, Califano F (2023) Subion-scale turbulence driven by
  magnetic reconnection. Phys Rev Lett 130:205201,
  \doi{10.1103/PhysRevLett.130.205201},
  \urlprefix\url{https://link.aps.org/doi/10.1103/PhysRevLett.130.205201}

\bibitem[{Markidis et~al.(2010{\natexlab{a}})Markidis, Lapenta, and
  Rizwan-Uddin}]{Markidis:2010}
Markidis S, Lapenta G, Rizwan-Uddin (2010{\natexlab{a}}) Multi-scale
  simulations of plasma with ipic3d. Mathematics and Computers in Simulation
  80:1509--1519, \doi{10.1016/j.matcom.2009.08.038}

\bibitem[{Markidis et~al.(2010{\natexlab{b}})Markidis, Lapenta, and
  Rizwan-uddin}]{markidis2010}
Markidis S, Lapenta G, Rizwan-uddin (2010{\natexlab{b}}) Multi-scale
  simulations of plasma with ipic3d. Mathematics and Computers in Simulation
  80(7):1509--1519, \doi{https://doi.org/10.1016/j.matcom.2009.08.038},
  \urlprefix\url{https://www.sciencedirect.com/science/article/pii/S0378475409002444},
  multiscale modeling of moving interfaces in materials

\bibitem[{Matsumoto et~al.(2015)Matsumoto, Amano, Kato, and
  Hoshino}]{matsumoto2015}
Matsumoto Y, Amano T, Kato TN, Hoshino M (2015) Stochastic electron
  acceleration during spontaneous turbulent reconnection in a strong shock
  wave. Science 347:974

\bibitem[{{Matthaeus} et~al.(2015){Matthaeus}, {Wan}, {Servidio}, {Greco},
  {Osman}, {Oughton}, and {Dmitruk}}]{matthaeus2015intermittency}
{Matthaeus} WH, {Wan} M, {Servidio} S, {Greco} A, {Osman} KT, {Oughton} S,
  {Dmitruk} P (2015) {Intermittency, nonlinear dynamics and dissipation in the
  solar wind and astrophysical plasmas}. Philosophical Transactions of the
  Royal Society of London Series A 373(2041):20140154--20140154,
  \doi{10.1098/rsta.2014.0154}

\bibitem[{Matthaeus et~al.(2016)Matthaeus, Parashar, Wan, and
  Wu}]{Matthaeus2016}
Matthaeus WH, Parashar TN, Wan M, Wu P (2016) Turbulence and proton–electron
  heating in kinetic plasma. Astrophysical Journal Letters 827:L7

\bibitem[{{Matthaeus} et~al.(2020){Matthaeus}, {Yang}, {Wan}, {Parashar},
  {Bandyopadhyay}, {Chasapis}, {Pezzi}, and
  {Valentini}}]{matthaeus2020pathways}
{Matthaeus} WH, {Yang} Y, {Wan} M, {Parashar} TN, {Bandyopadhyay} R, {Chasapis}
  A, {Pezzi} O, {Valentini} F (2020) {Pathways to Dissipation in Weakly
  Collisional Plasmas}. The Astrophysical Journal 891(1):101,
  \doi{10.3847/1538-4357/ab6d6a}

\bibitem[{McPherron et~al.(1973)McPherron, Russell, and
  Aubry}]{mcpherron1973satellite}
McPherron RL, Russell CT, Aubry MP (1973) Satellite studies of magnetospheric
  substorms on august 15, 1968: 9. phenomenological model for substorms.
  Journal of Geophysical Research 78(16):3131--3149

\bibitem[{Meng et~al.(2013)Meng, T\'oth, Glocer, Fok, and Gombosi}]{Meng:2013}
Meng X, T\'oth G, Glocer A, Fok MC, Gombosi TI (2013) Pressure anisotropy in
  global magnetospheric simulations: Coupling with ring current models. J
  Geophys Res 118:5639, \doi{10.1002/jgra.50539}

\bibitem[{Merkin and Lyon(2010)}]{merkin2010effects}
Merkin V, Lyon J (2010) Effects of the low-latitude ionospheric boundary
  condition on the global magnetosphere. Journal of Geophysical Research: Space
  Physics 115(A10)

\bibitem[{Merkin and Sitnov(2016)}]{merkin2016stability}
Merkin V, Sitnov M (2016) Stability of magnetotail equilibria with a tailward
  bz gradient. Journal of Geophysical Research: Space Physics
  121(10):9411--9426

\bibitem[{Mouikis et~al.(2021)Mouikis, Omelchenko, and
  Roytershteyn}]{Mouikis:2021}
Mouikis C, Omelchenko Y, Roytershteyn V (2021) {Comparative study of the
  magnetotail loading of ionospheric O+ using 3D Global Hybrid simulations}.
  In: AGU Fall Meeting 2021, held in New Orleans, LA, 13-17 December 2021, vol
  SH35G-02,
  \urlprefix\url{https://ui.adsabs.harvard.edu/abs/2021AGUFMSH35G..02M/abstract}

\bibitem[{M{\"u}ller et~al.(2011)M{\"u}ller, Simon, Motschmann, Sc{\"u}ller,
  and Glassmeier}]{Muller:2011}
M{\"u}ller J, Simon S, Motschmann U, Sc{\"u}ller J, Glassmeier KH (2011)
  {A.I.K.E.F.: Adaptive hybrid model for space plasma simulations}. Comp Phys
  Comm 182:946--966

\bibitem[{Mu{\~n}oz et~al.(2018)Mu{\~n}oz, Jain, Kilian, and
  B{\"u}chner}]{Munoz2018}
Mu{\~n}oz PA, Jain N, Kilian P, B{\"u}chner J (2018) A new hybrid code
  ({{CHIEF}}) implementing the inertial electron fluid equation without
  approximation. Computer Physics Communications 224:245--264,
  \doi{10.1016/j.cpc.2017.10.012}

\bibitem[{Mu{\~n}oz et~al.(2023)Mu{\~n}oz, Jain, Farzalipour~Tabriz, Rampp, and
  B{\"u}chner}]{Munoz2023}
Mu{\~n}oz PA, Jain N, Farzalipour~Tabriz M, Rampp M, B{\"u}chner J (2023)
  {Electron inertia effects in 3D hybrid-kinetic collisionless plasma
  turbulence}. Physics of Plasmas 30(9):092302, \doi{10.1063/5.0148818},
  \urlprefix\url{https://doi.org/10.1063/5.0148818}

\bibitem[{Nakamura and Daughton(2014)}]{nakamura-t2014}
Nakamura TKM, Daughton W (2014) Turbulent plasma transport across the earth’s
  low-latitude boundary layer. Geophys Res Lett 41:8704

\bibitem[{Nakamura et~al.(2012)Nakamura, Nakamura, Alexandrova, Kubota, and
  Nagai}]{Nakamura12}
Nakamura TKM, Nakamura R, Alexandrova A, Kubota Y, Nagai T (2012) Hall
  magnetohydrodynamic effects for three-dimensional magnetic reconnection with
  finite width along the direction of the current. J Geophys Res 117:03220

\bibitem[{Nakamura et~al.(2017)Nakamura, Hasegawa, Daughton, Eriksson, Li, and
  Nakamura}]{nakamura-t2017}
Nakamura TKM, Hasegawa H, Daughton W, Eriksson S, Li W, Nakamura R (2017)
  Turbulent mass transfer caused by vortex induced reconnection in
  collisionless magnetospheric plasmas. Nature Communications 8:1582

\bibitem[{Nakamura et~al.(2022)Nakamura, Blasl, Hasegawa, Umeda, Liu, Peery,
  Plaschke, Nakamura, Holmes, Stawarz, and Nystrom}]{nakamura-t2022}
Nakamura TKM, Blasl KA, Hasegawa H, Umeda T, Liu YH, Peery SA, Plaschke F,
  Nakamura R, Holmes JC, Stawarz JE, Nystrom WD (2022) Multi-scale evolution of
  kelvin–helmholtz waves at the earth’s magnetopause during southward imf
  periods. Phys Plasmas 29:012901

\bibitem[{Ng et~al.(2021)Ng, Chen, and Omelchenko}]{Ng:2021}
Ng J, Chen LJ, Omelchenko YA (2021) {Bursty magnetic reconnection at the
  Earth's magnetopause triggered by high-speed jets}. Phys Plasmas 28:092902,
  \doi{10.1063/5.0054394}, \urlprefix\url{https://doi.org/10.1063/5.0054394}

\bibitem[{Ng et~al.(2022)Ng, Chen, Bessho, Shuster, Burkholder, and
  Yoo}]{ng2022}
Ng J, Chen LJ, Bessho N, Shuster J, Burkholder B, Yoo J (2022) Electron-scale
  reconnection in three-dimensional shock turbulence. Geophysical Research
  Letters 49:e2022GL099544

\bibitem[{Ohia et~al.(2012)Ohia, Egedal, Lukin, Daughton, and Le}]{Ohia12}
Ohia O, Egedal J, Lukin VS, Daughton W, Le A (2012) Demonstration of
  anisotropic fluid closure capturing the kinetic structure of magnetic
  reconnection. Phys Rev Lett 109:115004, \doi{10.1103/PhysRevLett.109.115004},
  \urlprefix\url{https://link.aps.org/doi/10.1103/PhysRevLett.109.115004}

\bibitem[{Ohtani and Horiuchi(2009)}]{ohtani2009}
Ohtani H, Horiuchi R (2009) Open boundary condition for particle simulation in
  magnetic reconnection research. Plasma and Fusion Research 4:24

\bibitem[{{Oka} et~al.(2010){Oka}, {Phan}, {Krucker}, {Fujimoto}, and
  {Shinohara}}]{Oka2010}
{Oka} M, {Phan} TD, {Krucker} S, {Fujimoto} M, {Shinohara} I (2010) {Electron
  Acceleration by Multi-Island Coalescence}. Astrophysical Journal
  714(1):915--926, \doi{10.1088/0004-637X/714/1/915}, \eprint{1004.1154}

\bibitem[{{Oka} et~al.(2023){Oka}, {Birn}, {Egedal}, {Guo}, {Ergun}, {Turner},
  {Khotyaintsev}, {Hwang}, {Cohen}, and {Drake}}]{Oka2023}
{Oka} M, {Birn} J, {Egedal} J, {Guo} F, {Ergun} RE, {Turner} DL, {Khotyaintsev}
  Y, {Hwang} KJ, {Cohen} IJ, {Drake} JF (2023) {Particle Acceleration by
  Magnetic Reconnection in Geospace}. Space Science Reviews 219(8):75,
  \doi{10.1007/s11214-023-01011-8}, \eprint{2307.01376}

\bibitem[{Omelchenko and Karimabadi(2006{\natexlab{a}})}]{Omelchenko:2006a}
Omelchenko Y, Karimabadi H (2006{\natexlab{a}}) Event-driven, hybrid
  particle-in-cell simulation: A new paradigm for multi-scale plasma modeling.
  Journal of Computational Physics 216(1):153 -- 178,
  \doi{http://dx.doi.org/10.1016/j.jcp.2005.11.029},
  \urlprefix\url{http://www.sciencedirect.com/science/article/pii/S0021999105005474}

\bibitem[{Omelchenko and Karimabadi(2006{\natexlab{b}})}]{Omelchenko:2006}
Omelchenko Y, Karimabadi H (2006{\natexlab{b}}) Self-adaptive time integration
  of flux-conservative equations with sources. Journal of Computational Physics
  216(1):179 -- 194, \doi{http://dx.doi.org/10.1016/j.jcp.2005.12.008},
  \urlprefix\url{http://www.sciencedirect.com/science/article/pii/S0021999105005619}

\bibitem[{Omelchenko and Karimabadi(2007)}]{Omelchenko:2007}
Omelchenko Y, Karimabadi H (2007) A time-accurate explicit multi-scale
  technique for gas dynamics. Journal of Computational Physics 226(1):282 --
  300, \doi{http://dx.doi.org/10.1016/j.jcp.2007.04.010},
  \urlprefix\url{http://www.sciencedirect.com/science/article/pii/S002199910700157X}

\bibitem[{Omelchenko and Karimabadi(2012)}]{Omelchenko2012a}
Omelchenko Y, Karimabadi H (2012) {HYPERS: A unidimensional asynchronous
  framework for multiscale hybrid simulations}. Journal of Computational
  Physics 231(4):1766--1780, \doi{10.1016/j.jcp.2011.11.004},
  \urlprefix\url{https://linkinghub.elsevier.com/retrieve/pii/S0021999111006462}

\bibitem[{Omelchenko et~al.(2021{\natexlab{a}})Omelchenko, Chen, and
  Ng}]{Omelchenko2021}
Omelchenko Y, Chen LJ, Ng J (2021{\natexlab{a}}) 3d space-time adaptive hybrid
  simulations of magnetosheath high-speed jets. Journal of Geophysical
  Research: Space Physics 126(7):e2020JA029035

\bibitem[{Omelchenko(2015)}]{Omelchenko:2015}
Omelchenko YA (2015) Formation, spin-up, and stability of field-reversed
  configurations. Phys Rev E 92:023105, \doi{10.1103/PhysRevE.92.023105},
  \urlprefix\url{http://link.aps.org/doi/10.1103/PhysRevE.92.023105}

\bibitem[{Omelchenko and Karimabadi(2022)}]{Omelchenko:2022spr}
Omelchenko YA, Karimabadi H (2022) {EMAPS - An Intelligent Agent-Based
  Technology for Simulation of Multiscale Systems}. In: B\"{u}chner J (ed)
  {Space and Astrophysical Plasma Simulation}, Springer,
  \urlprefix\url{https://doi.org/10.1007/978-3-031-11870-8\_13}

\bibitem[{Omelchenko and Sudan(1997)}]{Omelchenko:1997}
Omelchenko YA, Sudan R (1997) {A 3-D Darwin-EM Hybrid, PIC code for ion ring
  studies}. J Comp Phys 133:146--159, \doi{10.1006/jcph.1997.5670},
  \urlprefix\url{https://doi.org/10.1006/jcph.1997.5670}

\bibitem[{Omelchenko et~al.(2021{\natexlab{b}})Omelchenko, Chen, and
  Ng}]{Omelchenko:2021a}
Omelchenko YA, Chen LJ, Ng J (2021{\natexlab{b}}) {3D Space-Time Adaptive
  Hybrid Simulations of Magnetosheath High-Speed Jets}. J Geophys Res Space
  Physics 126:e2020JA029035, \doi{10.1029/2020JA029035},
  \urlprefix\url{https://doi.org/10.1029/2020JA029035}

\bibitem[{Omelchenko et~al.(2021{\natexlab{c}})Omelchenko, Roytershtyen, Chen,
  Ng, and Hietala}]{Omelchenko:2021b}
Omelchenko YA, Roytershtyen V, Chen LJ, Ng J, Hietala H (2021{\natexlab{c}})
  {HYPERS simulations of solar wind interactions with the Earth's magnetosphere
  and the Moon}. J Atmos Sol-Terr Phys 215:105581,
  \doi{10.1016/j.jastp.2021.105581},
  \urlprefix\url{https://doi.org/10.1016/j.jastp.2021.105581}

\bibitem[{Omelchenko et~al.(2021{\natexlab{d}})Omelchenko, Rudakov, Ng,
  Crabtree, and Ganguli}]{Omelchenko:2021c}
Omelchenko YA, Rudakov LI, Ng J, Crabtree C, Ganguli G (2021{\natexlab{d}}) {On
  the rate of energy deposition by an ion ring velocity ring}. Phys Plasmas
  28:052102, \doi{10.1063/5.0046309},
  \urlprefix\url{https://doi.org/10.1063/5.0046309}

\bibitem[{Omelchenko et~al.(2022)Omelchenko, Mouikis, Ng, Roytershteyn, and
  Chen}]{Omelchenko:2023}
Omelchenko YA, Mouikis C, Ng J, Roytershteyn V, Chen LJ (2022) {Multiscale
  Hybrid Modeling of the Impact Response of the Earth’s Magnetotail to
  Ionospheric O+ Outflow}. Frontiers in Astronomy and Space Science Submitted

\bibitem[{Omidi and Sibeck(2007)}]{Omidi2007}
Omidi N, Sibeck D (2007) Flux transfer events in the cusp. Geophysical Research
  Letters 34(4)

\bibitem[{Omidi et~al.(2004)Omidi, Blanco-Cano, Russell, and
  Karimabadi}]{Omidi2004}
Omidi N, Blanco-Cano X, Russell C, Karimabadi H (2004) {Dipolar magnetospheres
  and their characterization as a function of magnetic moment}. Advances in
  Space Research 33:1996--2003, \doi{10.1016/j.asr.2003.08.041},
  \urlprefix\url{https://doi.org/10.1016/j.asr.2003.08.041}

\bibitem[{Omidi et~al.(2009)Omidi, Phan, and Sibeck}]{Omidi2009}
Omidi N, Phan T, Sibeck D (2009) Hybrid simulations of magnetic reconnection
  initiated in the magnetosheath. Journal of Geophysical Research: Space
  Physics 114(A2)

\bibitem[{Omidi et~al.(2010)Omidi, Eastwood, and Sibeck}]{Omidi2010}
Omidi N, Eastwood J, Sibeck D (2010) Foreshock bubbles and their global
  magnetospheric impacts. Journal of Geophysical Research: Space Physics
  115(A6)

\bibitem[{Palmroth et~al.(2017)Palmroth, Hoilijoki, Juusola, Pulkkinen,
  Hietala, Pfau-Kempf, Ganse, von Alfthan, Vainio, and Hesse}]{palmroth2017}
Palmroth M, Hoilijoki S, Juusola L, Pulkkinen TI, Hietala H, Pfau-Kempf Y,
  Ganse U, von Alfthan S, Vainio R, Hesse M (2017) Tail reconnection in the
  global magnetospheric context: Vlasiator first results. Annales Geophysicae
  35(6):1269--1274, \doi{10.5194/angeo-35-1269-2017},
  \urlprefix\url{https://angeo.copernicus.org/articles/35/1269/2017/}

\bibitem[{Palmroth et~al.(2018{\natexlab{a}})Palmroth, Ganse, Pfau-Kempf,
  Battarbee, Turc, Brito, Grandin, Hoilijoki, Sandroos, and von
  Alfthan}]{palmroth2018_LR}
Palmroth M, Ganse U, Pfau-Kempf Y, Battarbee M, Turc L, Brito T, Grandin M,
  Hoilijoki S, Sandroos A, von Alfthan S (2018{\natexlab{a}}) Vlasov methods in
  space physics and astrophysics. Living Reviews in Computational Astrophysics
  4(1):1

\bibitem[{Palmroth et~al.(2018{\natexlab{b}})Palmroth, Hietala, Plaschke,
  Archer, Karlsson, Blanco-Cano, Sibeck, Kajdič, Ganse, Pfau-Kempf, Battarbee,
  and Turc}]{palmroth2018}
Palmroth M, Hietala H, Plaschke F, Archer M, Karlsson T, Blanco-Cano X, Sibeck
  D, Kajdič P, Ganse U, Pfau-Kempf Y, Battarbee M, Turc L (2018{\natexlab{b}})
  {Magnetosheath jet properties and evolution as determined by a global
  hybrid-Vlasov simulation}. Annales Geophysicae 36:1171--1182,
  \doi{10.5194/angeo-36-1171-2018},
  \urlprefix\url{https://doi.org/10.5194/angeo-36-1171-2018}

\bibitem[{Palmroth et~al.(2023)Palmroth, Pulkkinen, Ganse, Pfau-Kempf, Koskela,
  Zaitsev, Alho, Cozzani, Turc, Battarbee, Dubart, George, Gordeev, Grandin,
  Horaites, Osmane, Papadakis, Suni, Tarvus, Zhou, and Nakamura}]{palmroth2023}
Palmroth M, Pulkkinen TI, Ganse U, Pfau-Kempf Y, Koskela T, Zaitsev I, Alho M,
  Cozzani G, Turc L, Battarbee M, Dubart M, George H, Gordeev E, Grandin M,
  Horaites K, Osmane A, Papadakis K, Suni J, Tarvus V, Zhou H, Nakamura R
  (2023) Magnetotail plasma eruptions driven by magnetic reconnection and
  kinetic instabilities. Nature Geoscience 16(7):570--576

\bibitem[{Pang et~al.(2010)Pang, Lin, Deng, Wang, and Tan}]{Pang2010}
Pang Y, Lin Y, Deng X, Wang X, Tan B (2010) Three-dimensional hybrid simulation
  of magnetosheath reconnection under northward and southward interplanetary
  magnetic field. Journal of Geophysical Research: Space Physics 115(A3)

\bibitem[{Papadakis et~al.(2022)Papadakis, Pfau-Kempf, Ganse, Battarbee, Alho,
  Grandin, Dubart, Turc, Zhou, Horaites, Zaitsev, Cozzani, Bussov, Gordeev,
  Tesema, George, Suni, Tarvus, and Palmroth}]{Papadakis2022}
Papadakis K, Pfau-Kempf Y, Ganse U, Battarbee M, Alho M, Grandin M, Dubart M,
  Turc L, Zhou H, Horaites K, Zaitsev I, Cozzani G, Bussov M, Gordeev E, Tesema
  F, George H, Suni J, Tarvus V, Palmroth M (2022) {Spatial filtering in a 6D
  hybrid-Vlasov scheme to alleviate adaptive mesh refinement artifacts: a case
  study with Vlasiator (versions 5.0, 5.1, and 5.2.1)}. Geoscientific Model
  Development 15(20):7903--7912, \doi{10.5194/gmd-15-7903-2022}

\bibitem[{Parker(1957{\natexlab{a}})}]{PhysRev.107.924}
Parker EN (1957{\natexlab{a}}) Newtonian development of the dynamical
  properties of ionized gases of low density. Phys Rev 107:924--933,
  \doi{10.1103/PhysRev.107.924},
  \urlprefix\url{https://link.aps.org/doi/10.1103/PhysRev.107.924}

\bibitem[{Parker(1957{\natexlab{b}})}]{Parker57}
Parker EN (1957{\natexlab{b}}) Sweet's mechanism for merging magnetic fields in
  conducting fluids. J Geophys Res 62:509

\bibitem[{Parker(1963)}]{Parker63}
Parker EN (1963) The solar-flare phenomenon and the theory of reconnection and
  annihilation of magnetic fields. Ap~J 8:177

\bibitem[{Parker(1973)}]{Parker73}
Parker EN (1973) The reconnection rate of magnetic fields. Ap~J 180:247

\bibitem[{Paty and Winglee(2004)}]{Paty04}
Paty C, Winglee R (2004) Multi-fluid simulations of ganymede's magnetosphere.
  Geophys~Res~Lett 31:{L}24806

\bibitem[{Pembroke et~al.(2012)Pembroke, Toffoletto, Sazykin, Wiltberger, Lyon,
  Merkin, and Schmitt}]{pembroke2012initial}
Pembroke A, Toffoletto F, Sazykin S, Wiltberger M, Lyon J, Merkin V, Schmitt P
  (2012) Initial results from a dynamic coupled magnetosphere-ionosphere-ring
  current model. Journal of Geophysical Research: Space Physics 117(A2)

\bibitem[{Peroomian and El-Alaoui(2008)}]{peroomian2008storm}
Peroomian V, El-Alaoui M (2008) The storm-time access of solar wind ions to the
  nightside ring current and plasma sheet. Journal of Geophysical Research:
  Space Physics 113(A6)

\bibitem[{Perrone et~al.(2012)Perrone, Valentini, Servidio, Dalena, and
  Veltri}]{perrone2013}
Perrone D, Valentini F, Servidio S, Dalena S, Veltri P (2012) Vlasov
  simulations of multi-ion plasma turbulence in the solar wind. The
  Astrophysical Journal 762(2):99, \doi{10.1088/0004-637X/762/2/99},
  \urlprefix\url{https://dx.doi.org/10.1088/0004-637X/762/2/99}

\bibitem[{{Pezzi} et~al.(2018){Pezzi}, {Servidio}, {Perrone}, {Valentini},
  {Sorriso-Valvo}, {Greco}, {Matthaeus}, and {Veltri}}]{pezzi2018velocity}
{Pezzi} O, {Servidio} S, {Perrone} D, {Valentini} F, {Sorriso-Valvo} L, {Greco}
  A, {Matthaeus} WH, {Veltri} P (2018) {Velocity-space cascade in magnetized
  plasmas: Numerical simulations}. Physics of Plasmas 25(6):060704,
  \doi{10.1063/1.5027685}, \eprint{1803.01633}

\bibitem[{Pezzi et~al.(2019)Pezzi, Cozzani, Califano, Valentini, Guarrasi,
  Camporeale, Brunetti, Retinò, and Veltri}]{pezzi2019vida}
Pezzi O, Cozzani G, Califano F, Valentini F, Guarrasi M, Camporeale E, Brunetti
  G, Retinò A, Veltri P (2019) Vida: a vlasov–darwin solver for plasma
  physics at electron scales. Journal of Plasma Physics 85(5):905850506,
  \doi{10.1017/S0022377819000631}

\bibitem[{{Pezzi} et~al.(2019){Pezzi}, {Perrone}, {Servidio}, {Valentini},
  {Sorriso-Valvo}, and {Veltri}}]{pezzi2019proton}
{Pezzi} O, {Perrone} D, {Servidio} S, {Valentini} F, {Sorriso-Valvo} L,
  {Veltri} P (2019) {Proton-Proton Collisions in the Turbulent Solar Wind:
  Hybrid Boltzmann-Maxwell Simulations}. The Astrophysical Journal 887(2):208,
  \doi{10.3847/1538-4357/ab5285}, \eprint{1903.03398}

\bibitem[{Pezzi et~al.(2019)Pezzi, Yang, Valentini, Servidio, Chasapis,
  Matthaeus, and Veltri}]{pezzi2019energy}
Pezzi O, Yang Y, Valentini F, Servidio S, Chasapis A, Matthaeus WH, Veltri P
  (2019) {Energy conversion in turbulent weakly collisional plasmas: Eulerian
  hybrid Vlasov-Maxwell simulations}. Physics of Plasmas 26(7),
  \doi{10.1063/1.5100125}, \urlprefix\url{https://doi.org/10.1063/1.5100125},
  072301,
  \eprint{https://pubs.aip.org/aip/pop/article-pdf/doi/10.1063/1.5100125/13820638/072301\_1\_online.pdf}

\bibitem[{{Pezzi} et~al.(2021{\natexlab{a}}){Pezzi}, {Liang}, {Juno}, {Cassak},
  {V{\'a}sconez}, {Sorriso-Valvo}, {Perrone}, {Servidio}, {Roytershteyn},
  {TenBarge}, and {Matthaeus}}]{pezzzi2021dissipation}
{Pezzi} O, {Liang} H, {Juno} JL, {Cassak} PA, {V{\'a}sconez} CL,
  {Sorriso-Valvo} L, {Perrone} D, {Servidio} S, {Roytershteyn} V, {TenBarge}
  JM, {Matthaeus} WH (2021{\natexlab{a}}) {Dissipation measures in weakly
  collisional plasmas}. Monthly Notices of the Royal Astronomical Society
  505(4):4857--4873, \doi{10.1093/mnras/stab1516}, \eprint{2101.00722}

\bibitem[{{Pezzi} et~al.(2021{\natexlab{b}}){Pezzi}, {Pecora}, {Le Roux},
  {Engelbrecht}, {Greco}, {Servidio}, {Malova}, {Khabarova}, {Malandraki},
  {Bruno}, {Matthaeus}, {Li}, {Zelenyi}, {Kislov}, {Obridko}, and
  {Kuznetsov}}]{pezzi2021current}
{Pezzi} O, {Pecora} F, {Le Roux} J, {Engelbrecht} NE, {Greco} A, {Servidio} S,
  {Malova} HV, {Khabarova} OV, {Malandraki} O, {Bruno} R, {Matthaeus} WH, {Li}
  G, {Zelenyi} LM, {Kislov} RA, {Obridko} VN, {Kuznetsov} VD
  (2021{\natexlab{b}}) {Current Sheets, Plasmoids and Flux Ropes in the
  Heliosphere. Part II: Theoretical Aspects}. Space Science Reviews 217(3):39,
  \doi{10.1007/s11214-021-00799-7}, \eprint{2101.05007}

\bibitem[{Pfau-Kempf et~al.(2016)Pfau-Kempf, Hietala, Milan, Juusola,
  Hoilijoki, Ganse, von Alfthan, and Palmroth}]{pfaukempf2016}
Pfau-Kempf Y, Hietala H, Milan SE, Juusola L, Hoilijoki S, Ganse U, von Alfthan
  S, Palmroth M (2016) Evidence for transient, local ion foreshocks caused by
  dayside magnetopause reconnection. Annales Geophysicae 34(11):943--959,
  \doi{10.5194/angeo-34-943-2016},
  \urlprefix\url{https://angeo.copernicus.org/articles/34/943/2016/}

\bibitem[{Pfau-Kempf et~al.(2020{\natexlab{a}})Pfau-Kempf, Palmroth, Johlander,
  Turc, Alho, Battarbee, Dubart, Grandin, and Ganse}]{Pfau:2020}
Pfau-Kempf Y, Palmroth M, Johlander A, Turc L, Alho M, Battarbee M, Dubart M,
  Grandin M, Ganse U (2020{\natexlab{a}}) {Hybrid-Vlasov modeling of
  three-dimensional dayside magnetopause reconnection}. Phys Plasmas 27:092903,
  \doi{https://doi.org/10.1063/5.0020685},
  \urlprefix\url{https://aip.scitation.org/doi/10.1063/5.0020685}

\bibitem[{Pfau-Kempf et~al.(2020{\natexlab{b}})Pfau-Kempf, Palmroth, Johlander,
  Turc, Alho, Battarbee, Dubart, Grandin, and Ganse}]{pfaukempf2020}
Pfau-Kempf Y, Palmroth M, Johlander A, Turc L, Alho M, Battarbee M, Dubart M,
  Grandin M, Ganse U (2020{\natexlab{b}}) {Hybrid-Vlasov} modeling of
  three-dimensional dayside magnetopause reconnection. Physics of Plasmas
  27(9):092903

\bibitem[{Pham et~al.(2022)Pham, Zhang, Sorathia, Dang, Wang, Merkin, Liu, Lin,
  Wiltberger, Lei et~al.}]{pham2022thermospheric}
Pham KH, Zhang B, Sorathia K, Dang T, Wang W, Merkin V, Liu H, Lin D,
  Wiltberger M, Lei J, et~al. (2022) Thermospheric density perturbations
  produced by traveling atmospheric disturbances during august 2005 storm.
  Journal of Geophysical Research: Space Physics 127(2):e2021JA030071

\bibitem[{Phan et~al.(2007)Phan, Drake, Shay, Mozer, and Eastwood}]{Phan07}
Phan TD, Drake JF, Shay MA, Mozer FS, Eastwood JP (2007) Evidence for an
  elongated ($>$60 ion skin depths) electron diffusion region during fast
  magnetic reconnection. Phys~Rev~Lett 99:255002

\bibitem[{Phan et~al.(2018)Phan, Eastwood, Shay, Drake, Sonnerup, Fujimoto,
  Cassak, {\O}ieroset, Burch, Torbert, Rager, Dorelli, Gershman, Pollock,
  Pyakurel, Haggerty, Khotyaintsev, Lavraud, Saito, Oka, Ergun, Retino,
  Le~Contel, Argall, Giles, Moore, Wilder, Strangeway, Russell, Lindqvist, and
  Magnes}]{Phan2018}
Phan TD, Eastwood JP, Shay MA, Drake JF, Sonnerup BU{\"O}, Fujimoto M, Cassak
  PA, {\O}ieroset M, Burch JL, Torbert RB, Rager AC, Dorelli JC, Gershman DJ,
  Pollock C, Pyakurel PS, Haggerty CC, Khotyaintsev Y, Lavraud B, Saito Y, Oka
  M, Ergun RE, Retino A, Le~Contel O, Argall MR, Giles BL, Moore TE, Wilder FD,
  Strangeway RJ, Russell CT, Lindqvist PA, Magnes W (2018) Electron magnetic
  reconnection without ion coupling in {{Earth}}'s turbulent magnetosheath.
  Nature 557(7704):202--206, \doi{10.1038/s41586-018-0091-5}

\bibitem[{Pontius~Jr and Wolf(1990)}]{pontius1990transient}
Pontius~Jr D, Wolf R (1990) Transient flux tubes in the terrestrial
  magnetosphere. Geophysical research letters 17(1):49--52

\bibitem[{Poppe(2019)}]{Poppe2019}
Poppe AR (2019) Comment on ''the dominant role of energetic ions in solar wind
  interaction with the moon'' by omidi et al. J Geophys Res: Space Physics
  124:6927--6932, \doi{https://doi.org/10.1029/2019JA026692}

\bibitem[{Powell et~al.(1999)Powell, Roe, Linde, Gombosi, and {De
  Zeeuw}}]{Powell:1999}
Powell K, Roe P, Linde T, Gombosi T, {De Zeeuw} DL (1999) A solution-adaptive
  upwind scheme for ideal magnetohydrodynamics. J Comput Phys 154:284--309,
  \doi{10.1006/jcph.1999.6299}

\bibitem[{Powell(1994)}]{Powell:1994}
Powell KG (1994) An approximate {R}iemann solver for magnetohydrodynamics (that
  works in more than one dimension). Tech. Rep. 94-24, Inst. for Comput. Appl.
  in Sci. and Eng., NASA Langley Space Flight Center, Hampton, Va.

\bibitem[{Press et~al.(1992)Press, Teukolsky, Vetterling, and
  Flannery}]{press92a}
Press WH, Teukolsky SA, Vetterling WT, Flannery BP (1992) Numerical Recipes in
  {F}ortran 77, Cambridge University Press, chap 7.3, pp 281--286

\bibitem[{Priest and Forbes(2000)}]{Priest00}
Priest E, Forbes T (2000) Magnetic Reconnection. Cambridge University Press

\bibitem[{Pritchett et~al.(1991)Pritchett, Coroniti, and Pella}]{pritchett1991}
Pritchett P, Coroniti FV, Pella R (1991) Collisionless reconnection in
  two-dimension magnetotail equilibria. Journal of Geophysical Research: Space
  Physics 96(A7):11523--11538

\bibitem[{Pritchett(2005)}]{Pritchett2005}
Pritchett PL (2005) Onset and saturation of guide-field magnetic reconnection.
  Physics of Plasmas 12(6):062301, \doi{10.1063/1.1914309}

\bibitem[{Raeder(2006)}]{Raeder:2006}
Raeder J (2006) {Flux Transfer Events: 1. generation mechanism for strong
  southward IMF}. Ann Geophys 24(1):381--392, \doi{10.5194/angeo-24-381-2006}

\bibitem[{Raeder et~al.(1997)Raeder, Berchem, Ashour-Abdalla, Frank, Paterson,
  Ackerson, Kokubun, Yamamoto, and Slavin}]{Raeder:1997}
Raeder J, Berchem J, Ashour-Abdalla M, Frank L, Paterson W, Ackerson K, Kokubun
  S, Yamamoto T, Slavin J (1997) Boundary layer formation in the magnetotail:
  Geotail observations and comparisons with a global {MHD} simulation. Geophys
  Res Lett 24:951, \doi{10.1029/97GL00218}

\bibitem[{Raeder et~al.(2001)Raeder, McPherron, Frank, Kokubun, Lu, Mukai,
  Paterson, Sigwarth, Singer, and Slavin}]{raeder2001global}
Raeder J, McPherron R, Frank L, Kokubun S, Lu G, Mukai T, Paterson W, Sigwarth
  J, Singer H, Slavin J (2001) Global simulation of the geospace environment
  modeling substorm challenge event. Journal of Geophysical Research: Space
  Physics 106(A1):381--395

\bibitem[{Raeder et~al.(2016)Raeder, Cramer, Jensen, Fuller-Rowell, Maruyama,
  Toffoletto, and Vo}]{raeder2016sub}
Raeder J, Cramer WD, Jensen J, Fuller-Rowell T, Maruyama N, Toffoletto F, Vo H
  (2016) Sub-auroral polarization streams: A complex interaction between the
  magnetosphere, ionosphere, and thermosphere. In: Journal of Physics:
  Conference Series, IOP Publishing, vol 767, p 012021

\bibitem[{Reiff et~al.(2017)Reiff, Webster, Daou, Marshall, Sazykin,
  Rastaetter, Welling, DeZeeuw, Kuznetsova, Glocer, and
  Russell}]{reiff2017ediffusionregion}
Reiff PH, Webster JM, Daou AG, Marshall A, Sazykin SY, Rastaetter L, Welling
  DT, DeZeeuw D, Kuznetsova MM, Glocer A, Russell CT (2017) {CCMC} modeling of
  magnetic reconnection in electron diffusion region events. In: Foullon C,
  Malandraki OE (eds) Proceedings of the International Astronomical Union,
  vol~13, pp 142--146, doi:10.1017/S1743921317010845

\bibitem[{{Ridley} et~al.(2004){Ridley}, {Gombosi}, and
  {Dezeeuw}}]{Ridley:2004}
{Ridley} A, {Gombosi} T, {Dezeeuw} D (2004) {Ionospheric control of the
  magnetosphere: conductance}. Annales Geophysicae 22:567--584,
  \doi{10.5194/angeo-22-567-2004}

\bibitem[{Rieke et~al.(2015)Rieke, Trost, and Grauer}]{Rieke2015}
Rieke M, Trost T, Grauer R (2015) {Coupled Vlasov and two-fluid codes on GPUs}.
  Journal of Computational Physics 283:436--452,
  \doi{10.1016/j.jcp.2014.12.016},
  \urlprefix\url{http://dx.doi.org/10.1016/j.jcp.2014.12.016}

\bibitem[{Robinson et~al.(1987)Robinson, Vondrak, Miller, Dabbs, and
  Hardy}]{robinson1987calculating}
Robinson R, Vondrak R, Miller K, Dabbs T, Hardy D (1987) On calculating
  ionospheric conductances from the flux and energy of precipitating electrons.
  Journal of Geophysical Research: Space Physics 92(A3):2565--2569

\bibitem[{Roble et~al.(1988)Roble, Ridley, Richmond, and
  Dickinson}]{roble1988coupled}
Roble R, Ridley EC, Richmond A, Dickinson R (1988) A coupled
  thermosphere/ionosphere general circulation model. Geophysical Research
  Letters 15(12):1325--1328

\bibitem[{Rogers et~al.(2001)Rogers, Denton, Drake, and Shay}]{Rogers01}
Rogers BN, Denton RE, Drake JF, Shay MA (2001) Role of dispersive waves in
  collisionless magnetic reconnection. Phys Rev Lett 87(19):195004

\bibitem[{Rogers et~al.(2003)Rogers, Denton, and Drake}]{rogers03a}
Rogers BN, Denton RE, Drake JF (2003) Signatures of collisionless magnetic
  reconnection. J Geophys Res 108(A3):10.1029/2002JA009699

\bibitem[{Roytershteyn et~al.(2015)Roytershteyn, Karimabadi, Omelchenko, and
  Germaschewski}]{Royter:2015}
Roytershteyn V, Karimabadi H, Omelchenko Y, Germaschewski K (2015) Kinetic
  simulations of collisionless turbulence across scales. In: Solar Heliospheric
  and Interplanetary Environment (SHINE 2016), The SHINE conference held 5-10
  July, 2015 at The Stoweflake Resort in Stowe, VT,
  \urlprefix\url{https://ui.adsabs.harvard.edu/abs/2015shin.confE.117R/abstract},
  2015shin.confE.117R

\bibitem[{Rueda et~al.(2021)Rueda, Verscharen, T.Wicks, Owen, Nicolaou,
  P.Walsh, Zouganelis, Germaschewski, and Domínguez}]{rueda2021}
Rueda JAA, Verscharen D, TWicks R, Owen CJ, Nicolaou G, PWalsh A, Zouganelis I,
  Germaschewski K, Domínguez SV (2021) Three-dimensional magnetic reconnection
  in particle-in-cell simulations of anisotropic plasma turbulence. J Plasma
  Phys 87:905870228

\bibitem[{Runov et~al.(2021)Runov, Grandin, Palmroth, Battarbee, Ganse,
  Hietala, Hoilijoki, Kilpua, Pfau-Kempf, Toledo-Redondo, Turc, and
  Turner}]{runov2021}
Runov A, Grandin M, Palmroth M, Battarbee M, Ganse U, Hietala H, Hoilijoki S,
  Kilpua E, Pfau-Kempf Y, Toledo-Redondo S, Turc L, Turner D (2021) Ion
  distribution functions in magnetotail reconnection: global hybrid-vlasov
  simulation results. Annales Geophysicae 39(4):599--612,
  \doi{10.5194/angeo-39-599-2021},
  \urlprefix\url{https://angeo.copernicus.org/articles/39/599/2021/}

\bibitem[{Sachsenweger et~al.(1989)Sachsenweger, Scholer, and
  M{\"o}bius}]{sachsenweger1989test}
Sachsenweger D, Scholer M, M{\"o}bius E (1989) Test particle acceleration in a
  magnetotail reconnection configuration. Geophysical Research Letters
  16(9):1027--1030

\bibitem[{Sarrat et~al.(2017)Sarrat, Ghizzo, Del~Sarto, and
  Serrat}]{Sarrat2017}
Sarrat M, Ghizzo A, Del~Sarto D, Serrat L (2017) Parallel implementation of a
  relativistic semi-{L}agrangian {V}lasov–{M}axwell solver. The European
  Physical Journal D 71(11):271, \doi{10.1140/epjd/e2017-80188-4}

\bibitem[{Sato and Hayashi(1979)}]{Sato79}
Sato T, Hayashi T (1979) Externally driven magnetic reconnection and a powerful
  magnetic energy converter. Phys Fluids 22:1189

\bibitem[{{Schekochihin} et~al.(2016){Schekochihin}, {Parker}, {Highcock},
  {Dellar}, {Dorland}, and {Hammett}}]{schekochihin2016phase}
{Schekochihin} AA, {Parker} JT, {Highcock} EG, {Dellar} PJ, {Dorland} W,
  {Hammett} GW (2016) {Phase mixing versus nonlinear advection in drift-kinetic
  plasma turbulence}. Journal of Plasma Physics 82(2):905820212,
  \doi{10.1017/S0022377816000374}, \eprint{1508.05988}

\bibitem[{Schindler(1972)}]{schindler1972self}
Schindler K (1972) A self-consistent theory of the tail of the magnetosphere.
  In: Earth’s Magnetospheric Processes: Proceedings of a Symposium Organized
  by the Summer Advanced Study Institute and Ninth ESRO Summer School, Held in
  Cortina, Italy, August 30-September 10, 1971, Springer, pp 200--209

\bibitem[{Schmitz and Grauer(2006{\natexlab{a}})}]{schmitzgrauer2006a}
Schmitz H, Grauer R (2006{\natexlab{a}}) Darwin–vlasov simulations of
  magnetised plasmas. Journal of Computational Physics 214(2):738--756,
  \doi{https://doi.org/10.1016/j.jcp.2005.10.013},
  \urlprefix\url{https://www.sciencedirect.com/science/article/pii/S0021999105004717}

\bibitem[{Schmitz and Grauer(2006{\natexlab{b}})}]{schmitzgrauer2006b}
Schmitz H, Grauer R (2006{\natexlab{b}}) Kinetic vlasov simulations of
  collisionless magnetic reconnection. Physics of Plasmas 13(9):092309,
  \doi{https://doi.org/10.1063/1.2347101}

\bibitem[{Scholer and Jamitzky(1987)}]{scholer1987particle}
Scholer M, Jamitzky F (1987) Particle orbits during the development of
  plasmoids. Journal of Geophysical Research: Space Physics
  92(A11):12181--12186

\bibitem[{Scholer and Otto(1991)}]{scholer1991magnetotail}
Scholer M, Otto A (1991) Magnetotail reconnection: Current diversion and
  field-aligned currents. Geophysical Research Letters 18(4):733--736

\bibitem[{Schriver et~al.(2005)Schriver, Ashour-Abdalla, Zelenyi, Gombosi,
  Ridley, De~Zeeuw, Toth, and Monostori}]{schriver2005modeling}
Schriver D, Ashour-Abdalla M, Zelenyi L, Gombosi T, Ridley A, De~Zeeuw D, Toth
  G, Monostori G (2005) Modeling the kinetic transport of electrons through the
  earth’s global magnetosphere. In: Proceedings of the 7th International
  School/Symposium for Space Simulations. Research Institute for Sustainable
  Humanosphere, Kyoto University, Kyoto, Japan, pp 345--346

\bibitem[{Schumaker et~al.(1989)Schumaker, Gussenhoven, Hardy, and
  Carovillano}]{schumaker1989relationship}
Schumaker TL, Gussenhoven MS, Hardy DA, Carovillano RL (1989) The relationship
  between diffuse auroral and plasma sheet electron distributions near local
  midnight. Journal of Geophysical Research: Space Physics 94(A8):10061--10078

\bibitem[{Sciola et~al.(2023)Sciola, Merkin, Sorathia, Gkioulidou, Bao,
  F.Toffoletto, Pham, Lin, Michael, Wiltberger, and Ukhorskiy}]{sciola2023}
Sciola A, Merkin V, Sorathia K, Gkioulidou M, Bao S, FToffoletto, Pham K, Lin
  D, Michael A, Wiltberger M, Ukhorskiy A (2023) The buildup and evolution of
  the stormtime ring current via mesoscale plasma sheet flows. Submitted to
  Journal of Geophysical Research

\bibitem[{Sergeev et~al.(2014)Sergeev, Nikolaev, Tsyganenko, Angelopoulos,
  Runov, Singer, and Yang}]{sergeev2014testing}
Sergeev V, Nikolaev A, Tsyganenko N, Angelopoulos V, Runov A, Singer H, Yang J
  (2014) Testing a two-loop pattern of the substorm current wedge (scw2l).
  Journal of Geophysical Research: Space Physics 119(2):947--963

\bibitem[{Servidio et~al.(2012)Servidio, Valentini, Califano, and
  Veltri}]{servidio2012}
Servidio S, Valentini F, Califano F, Veltri P (2012) Local kinetic effects in
  two-dimensional plasma turbulence. Phys Rev Lett 108:045001,
  \doi{10.1103/PhysRevLett.108.045001},
  \urlprefix\url{https://link.aps.org/doi/10.1103/PhysRevLett.108.045001}

\bibitem[{Servidio et~al.(2014)Servidio, Osman, Valentini, Perrone, Califano,
  Chapman, Matthaeus, and Veltri}]{servidio2014}
Servidio S, Osman KT, Valentini F, Perrone D, Califano F, Chapman S, Matthaeus
  WH, Veltri P (2014) Proton kinetic effects in vlasov and solar wind
  turbulence. The Astrophysical Journal Letters 781(2):L27,
  \doi{10.1088/2041-8205/781/2/L27},
  \urlprefix\url{https://dx.doi.org/10.1088/2041-8205/781/2/L27}

\bibitem[{{Servidio} et~al.(2015){Servidio}, {Valentini}, {Perrone}, {Greco},
  {Califano}, {Matthaeus}, and {Veltri}}]{servidio2015kinetic}
{Servidio} S, {Valentini} F, {Perrone} D, {Greco} A, {Califano} F, {Matthaeus}
  WH, {Veltri} P (2015) {A kinetic model of plasma turbulence}. Journal of
  Plasma Physics 81(1):325810107, \doi{10.1017/S0022377814000841}

\bibitem[{{Servidio} et~al.(2017){Servidio}, {Chasapis}, {Matthaeus},
  {Perrone}, {Valentini}, {Parashar}, {Veltri}, {Gershman}, {Russell}, {Giles},
  {Fuselier}, {Phan}, and {Burch}}]{servidio2017magnetospheric}
{Servidio} S, {Chasapis} A, {Matthaeus} WH, {Perrone} D, {Valentini} F,
  {Parashar} TN, {Veltri} P, {Gershman} D, {Russell} CT, {Giles} B, {Fuselier}
  SA, {Phan} TD, {Burch} J (2017) {Magnetospheric Multiscale Observation of
  Plasma Velocity-Space Cascade: Hermite Representation and Theory}. Physical
  Review Letters 119(20):205101, \doi{10.1103/PhysRevLett.119.205101},
  \eprint{1707.08180}

\bibitem[{Sharma~Pyakurel et~al.(2019)Sharma~Pyakurel, Shay, Phan, Matthaeus,
  Drake, TenBarge, Haggerty, Klein, Cassak, Parashar, Swisdak, and
  Chasapis}]{Pyakurel2019}
Sharma~Pyakurel P, Shay MA, Phan TD, Matthaeus WH, Drake JF, TenBarge JM,
  Haggerty CC, Klein KG, Cassak PA, Parashar TN, Swisdak M, Chasapis A (2019)
  Transition from ion-coupled to electron-only reconnection: {{Basic}} physics
  and implications for plasma turbulence. Physics of Plasmas 26(8):082307,
  \doi{10.1063/1.5090403}

\bibitem[{Shay et~al.(1998)Shay, Drake, Denton, and Biskamp}]{Shay1998}
Shay MA, Drake JF, Denton RE, Biskamp D (1998) Structure of the dissipation
  region during collisionless magnetic reconnection. Journal of Geophysical
  Research: Space Physics 103(A5):9165--9176, \doi{10.1029/97JA03528}

\bibitem[{Shay et~al.(1999)Shay, Drake, Rogers, and Denton}]{Shay1999}
Shay MA, Drake JF, Rogers BN, Denton RE (1999) The scaling of collisionless,
  magnetic reconnection for large systems. Geophysical Research Letters
  26(14):2163--2166, \doi{10.1029/1999GL900481}

\bibitem[{Shay et~al.(2003)Shay, Drake, Swisdak, Dorland, and Rogers}]{Shay03}
Shay MA, Drake JF, Swisdak M, Dorland W, Rogers BN (2003) Inherently
  three-dimensional magnetic reconnection: {A} mechanism for bursty bulk flows?
  Geophys Res Lett 30:1345

\bibitem[{Shay et~al.(2004)Shay, Drake, Swisdak, and Rogers}]{shay04a}
Shay MA, Drake JF, Swisdak M, Rogers BN (2004) The scaling of embedded
  collisionless reconnection. Phys Plasmas 11(5):2199--2213

\bibitem[{Shay et~al.(2007)Shay, Drake, and Swisdak}]{shay07a}
Shay MA, Drake JF, Swisdak M (2007) Two-scale structure of the electron
  dissipation region during collisionless magnetic reconnection. Phys Rev Lett
  99:155002, \doi{10.1103/PhysRevLett.99.155002}

\bibitem[{Shay et~al.(2018)Shay, Haggerty, Matthaeus, Parashar, Wan, and
  Wu}]{Shay2018}
Shay MA, Haggerty CC, Matthaeus WH, Parashar TN, Wan M, Wu P (2018) Turbulent
  heating due to magnetic reconnection. Phys Plasmas 25:012304

\bibitem[{Shepherd and Cassak(2010)}]{Shepherd10}
Shepherd LS, Cassak PA (2010) Comparison of secondary islands in collisional
  reconnection to {H}all reconnection. Phys~Rev~Lett 105:015004

\bibitem[{Shepherd and Cassak(2012)}]{Shepherd12}
Shepherd LS, Cassak PA (2012) Guide field dependence of 3{D} {X}-line spreading
  during collisionless magnetic reconnection. J~Geophys~Res 117:A10101

\bibitem[{Shi et~al.(2013)Shi, Lin, and Wang}]{Shi2013}
Shi F, Lin Y, Wang X (2013) Global hybrid simulation of mode conversion at the
  dayside magnetopause. Journal of Geophysical Research: Space Physics
  118(10):6176--6187

\bibitem[{Shi et~al.(2021)Shi, Lin, Wang, Wang, and Nishimura}]{Shi2021}
Shi F, Lin Y, Wang X, Wang B, Nishimura Y (2021) 3-d global hybrid simulations
  of magnetospheric response to foreshock processes. Earth, Planets and Space
  73(1):1--17

\bibitem[{{Shiroto}(2023)}]{shiroto2023improved}
{Shiroto} T (2023) {An improved Darwin approximation in the classical
  electromagnetism}. Physics of Plasmas 30(4):044501, \doi{10.1063/5.0138048}

\bibitem[{Shou et~al.(2021)Shou, Tenishev, Chen, T\'oth, and
  Ganushkina}]{Shou2021}
Shou Y, Tenishev V, Chen Y, T\'oth G, Ganushkina N (2021) {Magnetohydrodynamic
  with Adaptively Embedded Particle-in-Cell model: MHD-AEPIC}. Journal of
  Computational Physics 446:110656, \doi{10.1016/j.jcp.2021.110656}

\bibitem[{Sitnov et~al.(2019)Sitnov, Birn, Ferdousi, Gordeev, Khotyaintsev,
  Merkin, Motoba, Otto, Panov, Pritchett et~al.}]{sitnov2019explosive}
Sitnov M, Birn J, Ferdousi B, Gordeev E, Khotyaintsev Y, Merkin V, Motoba T,
  Otto A, Panov E, Pritchett P, et~al. (2019) Explosive magnetotail activity.
  Space science reviews 215:1--95

\bibitem[{Sonnerup(1979)}]{sonnerup79a}
Sonnerup BU{\"O} (1979) Magnetic field reconnection. In: Lanzerotti LJ, Kennel
  CF, Parker EN (eds) Solar System Plasma Physics, vol~3, North Holland
  Publishing, Amsterdam, p~46

\bibitem[{Sonnerup et~al.(1986)Sonnerup, Paschmann, Papamastorakis, Sckopke,
  Haerendel, Bame, Asbridge, Gosling, and Russell}]{sonnerup81a}
Sonnerup BU{\"O}, Paschmann G, Papamastorakis I, Sckopke N, Haerendel G, Bame
  SJ, Asbridge JR, Gosling JT, Russell CT (1986) Evidence for magnetic field
  reconnection at the earth's magnetopause. Journal of Geophysical Research:
  Space Physics 86(A12):10049--10067

\bibitem[{Sorathia et~al.(2017)Sorathia, Merkin, Ukhorskiy, Mauk, and
  Sibeck}]{sorathia2017energetic}
Sorathia K, Merkin V, Ukhorskiy A, Mauk B, Sibeck D (2017) Energetic particle
  loss through the magnetopause: A combined global mhd and test-particle study.
  Journal of Geophysical Research: Space Physics 122(9):9329--9343

\bibitem[{Sorathia et~al.(2021)Sorathia, Michael, Merkin, Ukhorskiy, Turner,
  Lyon, Garretson, Gkioulidou, and Toffoletto}]{sorathia2021role}
Sorathia K, Michael A, Merkin VG, Ukhorskiy AY, Turner DL, Lyon J, Garretson J,
  Gkioulidou M, Toffoletto F (2021) The role of mesoscale plasma sheet dynamics
  in ring current formation. Frontiers in Astronomy and Space Sciences 8:761875

\bibitem[{Sorriso-Valvo et~al.(2018)Sorriso-Valvo, Perrone, Pezzi, Valentini,
  Servidio, Zouganelis, and Veltri}]{sorrisovalvo2018}
Sorriso-Valvo L, Perrone D, Pezzi O, Valentini F, Servidio S, Zouganelis I,
  Veltri P (2018) Local energy transfer rate and kinetic processes: the fate of
  turbulent energy in two-dimensional hybrid vlasov–maxwell numerical
  simulations. Journal of Plasma Physics 84(2):725840201,
  \doi{10.1017/S0022377818000302}

\bibitem[{Spence et~al.(1989)Spence, Kivelson, Walker, and
  McComas}]{spence1989magnetospheric}
Spence HE, Kivelson MG, Walker RJ, McComas DJ (1989) Magnetospheric plasma
  pressures in the midnight meridian: Observations from 2.5 to 35 re. Journal
  of Geophysical Research: Space Physics 94(A5):5264--5272

\bibitem[{Stanier et~al.(2015)Stanier, Daughton, Chac{\'{o}}n, Karimabadi, Ng,
  Huang, Hakim, and Bhattacharjee}]{Stanier2015}
Stanier A, Daughton W, Chac{\'{o}}n L, Karimabadi H, Ng J, Huang YM, Hakim A,
  Bhattacharjee A (2015) {Role of Ion Kinetic Physics in the Interaction of
  Magnetic Flux Ropes}. Physical Review Letters 115(17):175004,
  \doi{10.1103/PhysRevLett.115.175004},
  \urlprefix\url{http://link.aps.org/doi/10.1103/PhysRevLett.115.175004}

\bibitem[{Stawarz et~al.(2022)Stawarz, Eastwood, Phan, Gingell, Pyakurel, Shay,
  Robertson, Russell, and Le~Contel}]{Stawarz2022}
Stawarz JE, Eastwood JP, Phan TD, Gingell IL, Pyakurel PS, Shay MA, Robertson
  SL, Russell CT, Le~Contel O (2022) Turbulence-driven magnetic reconnection
  and the magnetic correlation length: {{Observations}} from {{Magnetospheric
  Multiscale}} in {{Earth}}'s magnetosheath. Physics of Plasmas 29(1):012302,
  \doi{10.1063/5.0071106}

\bibitem[{Strang(1968)}]{Strang1968}
Strang G (1968) {On the Construction and Comparison of Difference Schemes}.
  SIAM Journal on Numerical Analysis 5(3):506--517, \doi{10.1137/0705041}

\bibitem[{Sulem and Passot(2015)}]{sulempassot2015}
Sulem PL, Passot T (2015) Landau fluid closures with nonlinear large-scale
  finite larmor radius corrections for collisionless plasmas. Journal of Plasma
  Physics 81(1):325810103, \doi{10.1017/S0022377814000671}

\bibitem[{Sun et~al.(2022)Sun, Dewey, Aizawa, Huang, Slavin, Fu, Wei, and
  Bowers}]{Sun2022}
Sun W, Dewey RM, Aizawa S, Huang J, Slavin JA, Fu S, Wei Y, Bowers CF (2022)
  Review of mercury’s dynamic magnetosphere: Post-messenger era and
  comparative magnetospheres. Science China Earth Sciences 65:25

\bibitem[{Sun et~al.(2016)Sun, Fu, Slavin, Raines, Zong, Poh, and
  Zurbuchen}]{Sun2016}
Sun WJ, Fu SY, Slavin JA, Raines JM, Zong QG, Poh GK, Zurbuchen TH (2016)
  Spatial distribution of mercury's flux ropes and reconnection fronts:
  Messenger observations. Journal of Geophysical Research 121:7590

\bibitem[{Suni et~al.(2021)Suni, Palmroth, Turc, Battarbee, Johlander, Tarvus,
  Alho, Bussov, Dubart, Ganse, Grandin, Horaites, Manglayev, Papadakis,
  Pfau-Kempf, and Zhou}]{suni2021}
Suni J, Palmroth M, Turc L, Battarbee M, Johlander A, Tarvus V, Alho M, Bussov
  M, Dubart M, Ganse U, Grandin M, Horaites K, Manglayev T, Papadakis K,
  Pfau-Kempf Y, Zhou H (2021) Connection between foreshock structures and the
  generation of magnetosheath jets: Vlasiator results. Geophysical Research
  Letters 48(20):e2021GL095655, \doi{https://doi.org/10.1029/2021GL095655},
  e2021GL095655 2021GL095655,
  \eprint{https://agupubs.onlinelibrary.wiley.com/doi/pdf/10.1029/2021GL095655}

\bibitem[{Sweet(1958)}]{Sweet58}
Sweet PA (1958) The neutral point theory of solar flares. In: Lehnert B (ed)
  Electromagnetic Phenomena in Cosmical Physics, Cambridge University Press,
  New York, p 123

\bibitem[{Swift(1996)}]{Swift:1996}
Swift DW (1996) {Use of a Hybrid Code for Global-Scale Plasma Simulation}.
  Journal of Computational Physics 126:109--121

\bibitem[{Swift and Lin(2001)}]{swift2001substorm}
Swift DW, Lin Y (2001) Substorm onset viewed by a two-dimensional, global-scale
  hybrid code. Journal of Atmospheric and Solar-Terrestrial Physics
  63(7):683--704

\bibitem[{Swisdak et~al.(2018)Swisdak, Drake, Price, Burch, Cassak, and
  Phan}]{swisdak18a}
Swisdak M, Drake JF, Price L, Burch JL, Cassak PA, Phan TD (2018) Localized and
  intense energy conversion in the diffusion region of asymmetric magnetic
  reconnection. Geophysical Research Letters 45(11):5260--5267

\bibitem[{Tan et~al.(2011)Tan, Lin, Perez, and Wang}]{Tan2011}
Tan B, Lin Y, Perez J, Wang X (2011) Global-scale hybrid simulation of dayside
  magnetic reconnection under southward imf: Structure and evolution of
  reconnection. Journal of Geophysical Research: Space Physics 116(A2)

\bibitem[{Tan et~al.(2012)Tan, Lin, Perez, and Wang}]{Tan2012}
Tan B, Lin Y, Perez J, Wang X (2012) Global-scale hybrid simulation of cusp
  precipitating ions associated with magnetopause reconnection under southward
  imf. Journal of Geophysical Research: Space Physics 117(A3)

\bibitem[{{Tatsuno} et~al.(2009){Tatsuno}, {Dorland}, {Schekochihin}, {Plunk},
  {Barnes}, {Cowley}, and {Howes}}]{tatsuno2009nonlinear}
{Tatsuno} T, {Dorland} W, {Schekochihin} AA, {Plunk} GG, {Barnes} M, {Cowley}
  SC, {Howes} GG (2009) {Nonlinear Phase Mixing and Phase-Space Cascade of
  Entropy in Gyrokinetic Plasma Turbulence}. Physical Review Letters
  103(1):015003, \doi{10.1103/PhysRevLett.103.015003}, \eprint{0811.2538}

\bibitem[{Thoma et~al.(2013)Thoma, Welch, and Hsu}]{Thoma:2013}
Thoma C, Welch DR, Hsu SC (2013) {Particle-in-cell simulations of collisionless
  shock formation via head-on merging of two laboratory supersonic plasma
  jets}. Phys Plasmas 20:082128, \doi{10.1063/1.4819063},
  \urlprefix\url{http://dx.doi.org/10.1063/1.4819063}

\bibitem[{Toffoletto(2020)}]{toffolleto2020}
Toffoletto F (2020) "Modelling Techniques" in Ring current investigations: The
  quest for space weather prediction. Elsevier, edited by Jordanova, Vania K
  and Ilie, Raluca and Chen, Margaret W

\bibitem[{Toffoletto et~al.(2003{\natexlab{a}})Toffoletto, Sazykin, Spiro, and
  Wolf}]{Toffoletto:2003}
Toffoletto F, Sazykin S, Spiro R, Wolf R (2003{\natexlab{a}}) Inner
  magnetospheric modeling with the {R}ice {C}onvection {M}odel. Space Sci Rev
  107:175--196, \doi{10.1023/A:1025532008047}

\bibitem[{Toffoletto et~al.(2003{\natexlab{b}})Toffoletto, Sazykin, Spiro, and
  Wolf}]{toffoletto2003inner}
Toffoletto F, Sazykin S, Spiro R, Wolf R (2003{\natexlab{b}}) Inner
  magnetospheric modeling with the rice convection model. Space science reviews
  107:175--196

\bibitem[{Toledo-Redondo et~al.(2021)Toledo-Redondo, Andr{\'e}, Aunai,
  Chappell, Dargent, Fuselier, Glocer, Graham, Haaland, Hesse, Kistler,
  Lavraud, Li, Moore, Tenfjord, and Vines}]{Toledo:2021}
Toledo-Redondo S, Andr{\'e} M, Aunai N, Chappell CR, Dargent J, Fuselier SA,
  Glocer A, Graham DB, Haaland S, Hesse M, Kistler LM, Lavraud B, Li W, Moore
  TE, Tenfjord P, Vines SK (2021) {Impacts of Ionospheric Ions on Magnetic
  Reconnection and Earth's Magnetosphere Dynamics}. Reviews of Geophysics
  59:e2020RG000707, \doi{10.1029/2020RG000707},
  \urlprefix\url{https://doi.org/10.1029/2020RG000707}

\bibitem[{T\'oth et~al.(2005)T\'oth, Sokolov, Gombosi, Chesney, Clauer, Zeeuw,
  Hansen, Kane, Manchester, Powell, Ridley, Roussev, Stout, Volberg, Wolf,
  Sazykin, Chan, Yu, and K\'ota}]{Toth:2005swmf}
T\'oth G, Sokolov IV, Gombosi TI, Chesney DR, Clauer C, Zeeuw DLD, Hansen KC,
  Kane KJ, Manchester WB, Powell KG, Ridley AJ, Roussev II, Stout QF, Volberg
  O, Wolf RA, Sazykin S, Chan A, Yu B, K\'ota J (2005) {Space Weather Modeling
  Framework}: A new tool for the space science community. J Geophys Res
  110:A12226, \doi{10.1029/2005JA011126}

\bibitem[{T{\'o}th et~al.(2005)T{\'o}th, Sokolov, Gombosi, Chesney, Clauer,
  De~Zeeuw, Hansen, Kane, Manchester, Oehmke et~al.}]{toth2005space}
T{\'o}th G, Sokolov IV, Gombosi TI, Chesney DR, Clauer CR, De~Zeeuw DL, Hansen
  KC, Kane KJ, Manchester WB, Oehmke RC, et~al. (2005) Space weather modeling
  framework: A new tool for the space science community. Journal of Geophysical
  Research: Space Physics 110(A12)

\bibitem[{T{\'o}th et~al.(2008)T{\'o}th, Ma, and Gombosi}]{Toth08}
T{\'o}th G, Ma YJ, Gombosi TI (2008) Hall magnetohydrodynamics on block
  adaptive grids. J~Comput~Phys 227:6967

\bibitem[{T{\'o}th et~al.(2012)T{\'o}th, Van~der Holst, Sokolov, De~Zeeuw,
  Gombosi, Fang, Manchester, Meng, Najib, Powell et~al.}]{toth2012adaptive}
T{\'o}th G, Van~der Holst B, Sokolov IV, De~Zeeuw DL, Gombosi TI, Fang F,
  Manchester WB, Meng X, Najib D, Powell KG, et~al. (2012) Adaptive numerical
  algorithms in space weather modeling. Journal of Computational Physics
  231(3):870--903

\bibitem[{T\'oth et~al.(2012)T\'oth, van~der Holst, Sokolov, Zeeuw, Gombosi,
  Fang, Manchester, Meng, Najib, Powell, Stout, Glocer, Ma, and
  Opher}]{Toth:2012swmf}
T\'oth G, van~der Holst B, Sokolov IV, Zeeuw DLD, Gombosi TI, Fang F,
  Manchester WB, Meng X, Najib D, Powell KG, Stout QF, Glocer A, Ma YJ, Opher M
  (2012) Adaptive numerical algorithms in space weather modeling. J Comput Phys
  231:870--903, \doi{10.1016/j.jcp.2011.02.006}

\bibitem[{T\'oth et~al.(2016)T\'oth, Jia, Markidis, Peng, Chen, Daldorff,
  Tenishev, Borovikov, Haiducek, Gombosi, Glocer, and Dorelli}]{Toth:2016}
T\'oth G, Jia X, Markidis S, Peng B, Chen Y, Daldorff L, Tenishev V, Borovikov
  D, Haiducek J, Gombosi T, Glocer A, Dorelli J (2016) Extended
  magnetohydrodynamics with embedded particle-in-cell simulation of ganymede's
  magnetosphere. J Geophys Res 121, \doi{10.1002/2015JA021997}

\bibitem[{T\'oth et~al.(2017)T\'oth, Chen, Gombosi, Cassak, , Markidis, and
  Peng}]{Toth:2017}
T\'oth G, Chen Y, Gombosi TI, Cassak P, , Markidis S, Peng B (2017) Scaling the
  ion inertial length and its implications for modeling reconnection in global
  simulations. J Geophys Res 122:10336, \doi{10.1002/2017JA024189}

\bibitem[{Tronci and Camporeale(2015)}]{tronci2015}
Tronci C, Camporeale E (2015) Neutral vlasov kinetic theory of magnetized
  plasmas. Physics of Plasmas 22(2):020704, \doi{10.1063/1.4907665},
  \urlprefix\url{https://doi.org/10.1063/1.4907665},
  \eprint{https://doi.org/10.1063/1.4907665}

\bibitem[{Trottenberg et~al.(2000)Trottenberg, Oosterlee, and
  Schuller}]{Trottenberg00}
Trottenberg U, Oosterlee CW, Schuller A (2000) Multigrid. Academic Press, San
  Diego

\bibitem[{Turc et~al.(2015)Turc, Fontaine, Savoini, and Modolo}]{Turc:2015}
Turc L, Fontaine D, Savoini P, Modolo R (2015) 3d hybrid simulations of the
  interaction of a magnetic cloud with a bow shock. J Geophys Res: Space
  Physics 120:6133--6151, \doi{10.1002/2015JA021318},
  \urlprefix\url{https://doi.org/10.1002/2015JA021318}

\bibitem[{Turc et~al.(2023)Turc, Roberts, Verscharen, Dimmock, Kajdi{\v c},
  Palmroth, Pfau-Kempf, Johlander, Dubart, Kilpua, Soucek, Takahashi,
  Takahashi, Battarbee, and Ganse}]{turc2023}
Turc L, Roberts OW, Verscharen D, Dimmock AP, Kajdi{\v c} P, Palmroth M,
  Pfau-Kempf Y, Johlander A, Dubart M, Kilpua EKJ, Soucek J, Takahashi K,
  Takahashi N, Battarbee M, Ganse U (2023) Transmission of foreshock waves
  through earth's bow shock. Nature Physics 19(1):78--86

\bibitem[{Ugai and Tsuda(1977)}]{Ugai77}
Ugai M, Tsuda T (1977) Magnetic field line reconnexion by localized enhancement
  of resistivity, 1, evolution in a compressible mhd fluid. J~Plasma Phys
  17:337

\bibitem[{Ukhorskiy et~al.(2017)Ukhorskiy, Sitnov, Merkin, Gkioulidou, and
  Mitchell}]{ukhorskiy2017ion}
Ukhorskiy A, Sitnov M, Merkin V, Gkioulidou M, Mitchell D (2017) Ion
  acceleration at dipolarization fronts in the inner magnetosphere. Journal of
  Geophysical Research: Space Physics 122(3):3040--3054

\bibitem[{Ukhorskiy et~al.(2018)Ukhorskiy, Sorathia, Merkin, Sitnov, Mitchell,
  and Gkioulidou}]{ukhorskiy2018ion}
Ukhorskiy AY, Sorathia KA, Merkin VG, Sitnov MI, Mitchell DG, Gkioulidou M
  (2018) Ion trapping and acceleration at dipolarization fronts:
  High-resolution mhd and test-particle simulations. Journal of Geophysical
  Research: Space Physics 123(7):5580--5589

\bibitem[{{Umeda} et~al.(2009){Umeda}, {Togano}, and {Ogino}}]{umeda2009}
{Umeda} T, {Togano} K, {Ogino} T (2009) {Two-dimensional full-electromagnetic
  Vlasov code with conservative scheme and its application to magnetic
  reconnection}. Computer Physics Communications 180(3):365--374,
  \doi{10.1016/j.cpc.2008.11.001}

\bibitem[{{Umeda} et~al.(2010){Umeda}, {Miwa}, {Matsumoto}, {Nakamura},
  {Togano}, {Fukazawa}, and {Shinohara}}]{umeda2010}
{Umeda} T, {Miwa} Ji, {Matsumoto} Y, {Nakamura} TKM, {Togano} K, {Fukazawa} K,
  {Shinohara} I (2010) {Full electromagnetic Vlasov code simulation of the
  Kelvin-Helmholtz instability}. Physics of Plasmas 17(5):052311,
  \doi{10.1063/1.3422547}

\bibitem[{Valentini et~al.(2007)Valentini, Trávníček, Califano, Hellinger,
  and Mangeney}]{valentini2007}
Valentini F, Trávníček P, Califano F, Hellinger P, Mangeney A (2007) A
  hybrid-vlasov model based on the current advance method for the simulation of
  collisionless magnetized plasma. Journal of Computational Physics
  225(1):753--770, \doi{https://doi.org/10.1016/j.jcp.2007.01.001},
  \urlprefix\url{https://www.sciencedirect.com/science/article/pii/S0021999107000022}

\bibitem[{Valentini et~al.(2010)Valentini, Califano, and
  Veltri}]{valentini2010}
Valentini F, Califano F, Veltri P (2010) Two-dimensional kinetic turbulence in
  the solar wind. Phys Rev Lett 104:205002,
  \doi{10.1103/PhysRevLett.104.205002},
  \urlprefix\url{https://link.aps.org/doi/10.1103/PhysRevLett.104.205002}

\bibitem[{{Valentini} et~al.(2011{\natexlab{a}}){Valentini}, {Califano},
  {Perrone}, {Pegoraro}, and {Veltri}}]{valentini2011new}
{Valentini} F, {Califano} F, {Perrone} D, {Pegoraro} F, {Veltri} P
  (2011{\natexlab{a}}) {New Ion-Wave Path in the Energy Cascade}. Physical
  Review Letters 106(16):165002, \doi{10.1103/PhysRevLett.106.165002}

\bibitem[{{Valentini} et~al.(2011{\natexlab{b}}){Valentini}, {Perrone}, and
  {Veltri}}]{valentini2011short}
{Valentini} F, {Perrone} D, {Veltri} P (2011{\natexlab{b}}) {Short-wavelength
  Electrostatic Fluctuations in the Solar Wind}. The Astrophysical Journal
  739(1):54, \doi{10.1088/0004-637X/739/1/54}

\bibitem[{{Valentini} et~al.(2014){Valentini}, {Vecchio}, {Donato}, {Carbone},
  {Briand}, {Bougeret}, and {Veltri}}]{valentini2014nonlinear}
{Valentini} F, {Vecchio} A, {Donato} S, {Carbone} V, {Briand} C, {Bougeret} J,
  {Veltri} P (2014) {The Nonlinear and Nonlocal Link between Macroscopic
  Alfv{\'e}nic and Microscopic Electrostatic Scales in the Solar Wind}. The
  Astrophysical Journal Letters 788(1):L16, \doi{10.1088/2041-8205/788/1/L16}

\bibitem[{{Valentini} et~al.(2016){Valentini}, {Perrone}, {Stabile}, {Pezzi},
  {Servidio}, {De Marco}, {Marcucci}, {Bruno}, {Lavraud}, {De Keyser},
  {Consolini}, {Brienza}, {Sorriso-Valvo}, {Retin{\`o}}, {Vaivads}, {Salatti},
  and {Veltri}}]{valentini2016differential}
{Valentini} F, {Perrone} D, {Stabile} S, {Pezzi} O, {Servidio} S, {De Marco} R,
  {Marcucci} F, {Bruno} R, {Lavraud} B, {De Keyser} J, {Consolini} G, {Brienza}
  D, {Sorriso-Valvo} L, {Retin{\`o}} A, {Vaivads} A, {Salatti} M, {Veltri} P
  (2016) {Differential kinetic dynamics and heating of ions in the turbulent
  solar wind}. New Journal of Physics 18(12):125001,
  \doi{10.1088/1367-2630/18/12/125001}, \eprint{1611.04802}

\bibitem[{{van Leer}(1977)}]{vanleer1977towards}
{van Leer} B (1977) {Towards the Ultimate Conservative Difference Scheme. IV. A
  New Approach to Numerical Convection}. Journal of Computational Physics
  23:276, \doi{10.1016/0021-9991(77)90095-X}

\bibitem[{Varney et~al.(2016)Varney, Wiltberger, Zhang, Lotko, and
  Lyon}]{varney2016influence}
Varney R, Wiltberger M, Zhang B, Lotko W, Lyon J (2016) Influence of ion
  outflow in coupled geospace simulations: 1. physics-based ion outflow model
  development and sensitivity study. Journal of Geophysical Research: Space
  Physics 121(10):9671--9687

\bibitem[{Vasyliunas(1970)}]{vasyliunas1970mathematical}
Vasyliunas VM (1970) Mathematical models of magnetospheric convection and its
  coupling to the ionosphere. In: Particles and Fields in the Magnetosphere:
  Proceedings of a Symposium Organized by the Summer Advanced Study Institute,
  Held at the University of California, Santa Barbara, Calif., August 4--15,
  1969, Springer, pp 60--71

\bibitem[{Vasyliunas(1975)}]{Vasyliunas75}
Vasyliunas VM (1975) Theoretical models of magnetic field line merging, 1. Rev
  Geophys 13(1):303

\bibitem[{Vega et~al.(2020)Vega, Roytershteyn, Delzanno, and
  Boldyrev}]{vega2020}
Vega C, Roytershteyn V, Delzanno GL, Boldyrev S (2020) Electron-only
  {{Reconnection}} in {{Kinetic-Alfv\'en Turbulence}}. The Astrophysical
  Journal Letters 893(1):L10, \doi{10.3847/2041-8213/ab7eba}

\bibitem[{Villasenor and Bunemann(1992)}]{villasenor1992}
Villasenor J, Bunemann O (1992) Rigorous charge conservation for local
  electromagnetic field solvers. Computer Physics Communication 63:306

\bibitem[{{von Alfthan} et~al.(2014){von Alfthan}, Pokhotelov, Kempf,
  Hoilijoki, Honkonen, Sandroos, and Palmroth}]{vonalfthan2014}
{von Alfthan} S, Pokhotelov D, Kempf Y, Hoilijoki S, Honkonen I, Sandroos A,
  Palmroth M (2014) Vlasiator: First global hybrid-vlasov simulations of
  earth's foreshock and magnetosheath. Journal of Atmospheric and
  Solar-Terrestrial Physics 120:24--35,
  \doi{https://doi.org/10.1016/j.jastp.2014.08.012},
  \urlprefix\url{https://www.sciencedirect.com/science/article/pii/S1364682614001916}

\bibitem[{Wang et~al.(2020{\natexlab{a}})Wang, Wang, Liu, and
  Lin}]{Wang2020_foreshock}
Wang CP, Wang X, Liu TZ, Lin Y (2020{\natexlab{a}}) Evolution of a foreshock
  bubble in the midtail foreshock and impact on the magnetopause: 3-d global
  hybrid simulation. Geophysical Research Letters 47(22):e2020GL089844

\bibitem[{Wang et~al.(2019)Wang, Lin, Wang, and Guo}]{Wang2019}
Wang H, Lin Y, Wang X, Guo Z (2019) Generation of kinetic alfv{\'e}n waves in
  dayside magnetopause reconnection: A 3-d global-scale hybrid simulation.
  Physics of Plasmas 26(7):072102

\bibitem[{Wang et~al.(2018)Wang, Germaschewski, Hakim, Dong, Raeder, and
  Bhattacharjee}]{Wang18}
Wang L, Germaschewski K, Hakim A, Dong C, Raeder J, Bhattacharjee A (2018)
  Electron physics in 3-d two-fluid 10-moment modeling of ganymede's
  magnetosphere. Journal of Geophysical Research: Space Physics
  123(4):2815--2830, \doi{https://doi.org/10.1002/2017JA024761}

\bibitem[{Wang et~al.(2020{\natexlab{b}})Wang, Lu, Lu, Russell, Burch,
  Gershman, Gonzalez, and Wang}]{Wang2020}
Wang R, Lu Q, Lu S, Russell CT, Burch JL, Gershman DJ, Gonzalez W, Wang S
  (2020{\natexlab{b}}) Physical {{Implication}} of {{Two Types}} of
  {{Reconnection Electron Diffusion Regions With}} and {{Without Ion-Coupling}}
  in the {{Magnetotail Current Sheet}}. Geophysical Research Letters
  47(21):e2020GL088761, \doi{10.1029/2020GL088761}

\bibitem[{Wang and Bhattacharjee(1993)}]{Wang93}
Wang X, Bhattacharjee A (1993) Nonlinear dynamics of the {$m = 1$} instability
  and fast sawtooth collapse in high-temperature plasmas. Phys~Rev~Lett
  70(11):1627--1630

\bibitem[{Wang et~al.(2009)Wang, Lin, and Chang}]{Wang2009}
Wang X, Lin Y, Chang SW (2009) Hybrid simulation of foreshock waves and ion
  spectra and their linkage to cusp energetic ions. Journal of Geophysical
  Research: Space Physics 114(A6)

\bibitem[{Wang et~al.(2022{\natexlab{a}})Wang, Chen, and Toth}]{Wang:2022a}
Wang X, Chen Y, Toth G (2022{\natexlab{a}}) Global magnetohydrodynamic
  magnetosphere simulation with an adaptively\ embedded particle-in-cell model.
  J Geophys Res 127:e2021JA030091, \doi{10.1029/2021JA030091}

\bibitem[{Wang et~al.(2022{\natexlab{b}})Wang, Chen, and T{\'o}th}]{Wang2021a}
Wang X, Chen Y, T{\'o}th G (2022{\natexlab{b}}) Global magnetohydrodynamic
  magnetosphere simulation with an adaptively embedded particle-in-cell model.
  Journal of Geophysical Research: Space Physics 127(8):e2021JA030091

\bibitem[{Wang et~al.(2022{\natexlab{c}})Wang, Chen, and Toth}]{Wang:2022b}
Wang X, Chen Y, Toth G (2022{\natexlab{c}}) Simulation of magnetospheric
  sawtooth oscillations: the role of kineti\ c reconnection in the magnetotail.
  Geophys Res Lett 49:e2022GL099638, \doi{10.1029/2022GL099638}

\bibitem[{Wang et~al.(2022{\natexlab{d}})Wang, Chen, and Tóth}]{Wang2022}
Wang X, Chen Y, Tóth G (2022{\natexlab{d}}) Simulation of {Magnetospheric}
  {Sawtooth} {Oscillations}: {The} {Role} of {Kinetic} {Reconnection} in the
  {Magnetotail}. Geophysical Research Letters 49(15):e2022GL099638,
  \doi{10.1029/2022GL099638}

\bibitem[{Warmuth and Mann(2016)}]{warmuth2016constraints}
Warmuth A, Mann G (2016) Constraints on energy release in solar flares from
  rhessi and goes x-ray observations-ii. energetics and energy partition.
  Astronomy \& Astrophysics 588:A116

\bibitem[{Weimer(1996)}]{Weimer:1996}
Weimer D (1996) A flexible, {IMF} dependent model of high-latitude electric
  potential having "space weather" applications. Geophys Res Lett 23:2549

\bibitem[{Weimer(2001)}]{Weimer:2001}
Weimer D (2001) An improved model of ionospheric electric potentials including
  substorm perturbations and application to the {Geosphace Environment Modeling
  November} 24, 1996, event. J Geophys Res 106:407

\bibitem[{Wiegelmann and B\"uchner(2001)}]{wiegelmannbuchner2001}
Wiegelmann T, B\"uchner J (2001) Evolution of magnetic helicity in the course
  of kinetic magnetic reconnection. Nonlinear Processes in Geophysics
  8(3):127--140, \doi{10.5194/npg-8-127-2001},
  \urlprefix\url{https://npg.copernicus.org/articles/8/127/2001/}

\bibitem[{Wilder et~al.(2018)Wilder, Ergun, Burch, Ahmadi, Eriksson, Phan,
  Goodrich, Shuster, Rager, Torbert, Giles, Strangeway, Plaschke, Magnes,
  Lindqvist, and Khotyaintsev}]{Wilder2018}
Wilder FD, Ergun RE, Burch JL, Ahmadi N, Eriksson S, Phan TD, Goodrich KA,
  Shuster J, Rager AC, Torbert RB, Giles BL, Strangeway RJ, Plaschke F, Magnes
  W, Lindqvist PA, Khotyaintsev YV (2018) The {{Role}} of the {{Parallel
  Electric Field}} in {{Electron-Scale Dissipation}} at {{Reconnecting
  Currents}} in the {{Magnetosheath}}. Journal of Geophysical Research: Space
  Physics 123(8):6533--6547, \doi{10.1029/2018JA025529}

\bibitem[{Wiltberger et~al.(2015)Wiltberger, Merkin, Lyon, and
  Ohtani}]{wiltberger2015high}
Wiltberger M, Merkin V, Lyon J, Ohtani S (2015) High-resolution global
  magnetohydrodynamic simulation of bursty bulk flows. Journal of Geophysical
  Research: Space Physics 120(6):4555--4566

\bibitem[{Wiltberger et~al.(2017)Wiltberger, Merkin, Zhang, Toffoletto,
  Oppenheim, Wang, Lyon, Liu, Dimant, Sitnov et~al.}]{wiltberger2017effects}
Wiltberger M, Merkin V, Zhang B, Toffoletto F, Oppenheim M, Wang W, Lyon J, Liu
  J, Dimant Y, Sitnov M, et~al. (2017) Effects of electrojet turbulence on a
  magnetosphere-ionosphere simulation of a geomagnetic storm. Journal of
  Geophysical Research: Space Physics 122(5):5008--5027

\bibitem[{Winglee(submitted, 2004)}]{Winglee04}
Winglee RM (submitted, 2004) Ion cyclotron and heavy ion effects on
  reconnection in a global magnetotail. J Geophys Res

\bibitem[{Winske et~al.(2003)Winske, Yin, Omidi, Karimabadi, and
  Quest}]{Winske2003}
Winske D, Yin L, Omidi N, Karimabadi H, Quest K (2003) Hybrid simulation codes:
  Past, present and future - a tutorial. In: B{\"{u}}chner J, Scholer M, Dum CT
  (eds) Space Plasma Simulation, Lecture Notes in Physics, vol 615, Springer
  Berlin Heidelberg, Berlin, Heidelberg, pp 136--165,
  \doi{10.1007/3-540-36530-3\_8},
  \urlprefix\url{https://doi.org/10.1007/3-540-36530-3\_8}

\bibitem[{Wolf(1983)}]{wolf1983quasi}
Wolf R (1983) The quasi-static (slow-flow) region of the magnetosphere. In:
  Solar-Terrestrial Physics: Principles and Theoretical Foundations Based Upon
  the Proceedings of the Theory Institute Held at Boston College, August 9--26,
  1982, Springer, pp 303--368

\bibitem[{Wolf et~al.(2016)Wolf, Spiro, Sazykin, Toffoletto, and
  Yang}]{wolf2016forty}
Wolf R, Spiro R, Sazykin S, Toffoletto F, Yang J (2016) Forty-seven years of
  the rice convection model. Magnetosphere-ionosphere coupling in the solar
  system pp 215--225

\bibitem[{Wolf et~al.(1982)Wolf, Harel, Spiro, Voigt, Reiff, and
  Chen}]{Wolf:1982}
Wolf RA, Harel M, Spiro RW, Voigt G, Reiff PH, Chen CK (1982) Computer
  simulation of inner magnetospheric dynamics for the magnetic storm of {J}uly
  29, 1977. J Geophys Res 87:5949--5962, \doi{10.1029/JA087iA08p05949}

\bibitem[{Wu et~al.(2013)Wu, Wan, Matthaeus, Shay, and Swisdak}]{wu2013}
Wu P, Wan M, Matthaeus WH, Shay MA, Swisdak M (2013) von kármán energy decay
  and heating of protons and electrons in a kinetic turbulent plasma. Physical
  Review Letters 111:121105

\bibitem[{Yamada et~al.(1994)Yamada, Levinton, Pomphrey, Budny, Manickam, and
  Nagayama}]{Yamada94}
Yamada M, Levinton FM, Pomphrey N, Budny R, Manickam J, Nagayama Y (1994)
  Investigation of magnetic reconnection during a sawtooth crash in a
  high-temperature tokamak plasma. Phys Plasmas 1:3269--3276

\bibitem[{Yang et~al.(2014{\natexlab{a}})Yang, Toffoletto, and
  Wolf}]{yang2014rcma}
Yang J, Toffoletto FR, Wolf RA (2014{\natexlab{a}}) Rcm-e simulation of a thin
  arc preceded by a north-south-aligned auroral streamer. Geophysical Research
  Letters 41(8):2695--2701

\bibitem[{Yang et~al.(2014{\natexlab{b}})Yang, Wolf, Toffoletto, Sazykin, and
  Wang}]{yang2014rcmb}
Yang J, Wolf RA, Toffoletto FR, Sazykin S, Wang CP (2014{\natexlab{b}}) Rcm-e
  simulation of bimodal transport in the plasma sheet. Geophysical Research
  Letters 41(6):1817--1822

\bibitem[{Yang et~al.(2015)Yang, Toffoletto, Wolf, and
  Sazykin}]{yang2015contribution}
Yang J, Toffoletto FR, Wolf RA, Sazykin S (2015) On the contribution of plasma
  sheet bubbles to the storm time ring current. Journal of Geophysical
  Research: Space Physics 120(9):7416--7432

\bibitem[{Yang et~al.(2016)Yang, Toffoletto, and Wolf}]{yang2016comparison}
Yang J, Toffoletto FR, Wolf RA (2016) Comparison study of ring current
  simulations with and without bubble injections. Journal of Geophysical
  Research: Space Physics 121(1):374--379

\bibitem[{Yang et~al.(2019)Yang, Wolf, Toffoletto, Sazykin, Wang, and
  Cui}]{yang2019inertialized}
Yang J, Wolf R, Toffoletto F, Sazykin S, Wang W, Cui J (2019) The inertialized
  rice convection model. Journal of Geophysical Research: Space Physics
  124(12):10294--10317

\bibitem[{Yee(1966)}]{yee1966}
Yee K (1966) Numerical solution of initial boundary value problems involving
  maxwell’s equations in isotropic media. IEEE Transactions on Antennas and
  Propagation 14:302

\bibitem[{Yi et~al.(2022)Yi, Zhou, Song, Pang, and Deng}]{Yi2022}
Yi Y, Zhou M, Song L, Pang Y, Deng X (2022) Electron-{{Only Magnetic
  Reconnection}}: {{Lessons Learned From Magnetic Island Coalescence}}.
  Geophysical Research Letters 49(6):e2022GL098124, \doi{10.1029/2022GL098124}

\bibitem[{Yin et~al.(2024)Yin, Drake, and Swisdak}]{yin2024computational}
Yin Z, Drake J, Swisdak M (2024) A computational model for ion and electron
  energization during macroscale magnetic reconnection. arXiv preprint
  arXiv:240114500

\bibitem[{Zaharia et~al.(2000)Zaharia, Cheng, and
  Johnson}]{zaharia2000particle}
Zaharia S, Cheng C, Johnson JR (2000) Particle transport and energization
  associated with substorms. Journal of Geophysical Research: Space Physics
  105(A8):18741--18752

\bibitem[{Zaharia et~al.(2006{\natexlab{a}})Zaharia, Jordanova, Thomsen, and
  Reeves}]{zaharia2006self}
Zaharia S, Jordanova V, Thomsen M, Reeves G (2006{\natexlab{a}})
  Self-consistent modeling of magnetic fields and plasmas in the inner
  magnetosphere: Application to a geomagnetic storm. Journal of Geophysical
  Research: Space Physics 111(A11)

\bibitem[{Zaharia et~al.(2006{\natexlab{b}})Zaharia, Jordanova, Thomsen, and
  Reeves}]{Zaharia:2006}
Zaharia S, Jordanova VK, Thomsen MF, Reeves GD (2006{\natexlab{b}})
  Self-consistent modeling of magnetic fields and plasmas in the inner
  magnetosphere: Application to a geomagnetic storm. J Geophys Res 111:A11S14,
  \doi{10.1029/2006JA011619}

\bibitem[{Zalesak(1979)}]{Zalesak79}
Zalesak ST (1979) Fully multidimensional flux-corrected transport algorithms
  for fluids. J~Comput~Phys 31:335

\bibitem[{Zalesak(1981)}]{Zalesak81}
Zalesak ST (1981) High order “zip” differencing of convective terms.
  Journal of Computational Physics 40(2):497--508,
  \doi{https://doi.org/10.1016/0021-9991(81)90225-4},
  \urlprefix\url{https://www.sciencedirect.com/science/article/pii/0021999181902254}

\bibitem[{Zeiler et~al.(2002)Zeiler, Biskamp, Drake, Rogers, Shay, and
  Scholer}]{zeiler02a}
Zeiler A, Biskamp D, Drake JF, Rogers BN, Shay MA, Scholer M (2002)
  Three-dimensional particle simulations of collisionless magnetic
  reconnection. J Geophys Res 107(A9):1230, \doi{10.1029/2001JA000287}

\bibitem[{Zenitani and Umeda(2014)}]{ZenitaniUmeda2014}
Zenitani S, Umeda T (2014) {Some remarks on the diffusion regions in magnetic
  reconnection}. Physics of Plasmas 21(3):034503, \doi{10.1063/1.4869717}

\bibitem[{Zerroukat and Allen(2012)}]{ZerroukatAllen2012}
Zerroukat M, Allen T (2012) {A three-dimensional monotone and conservative
  semi-Lagrangian scheme (SLICE-3D) for transport problems}. Quarterly Journal
  of the Royal Meteorological Society 138(667):1640--1651,
  \doi{10.1002/qj.1902}

\bibitem[{Zhang et~al.(2019)Zhang, Sorathia, Lyon, Merkin, Garretson, and
  Wiltberger}]{zhang2019gamera}
Zhang B, Sorathia KA, Lyon JG, Merkin VG, Garretson JS, Wiltberger M (2019)
  Gamera: A three-dimensional finite-volume mhd solver for non-orthogonal
  curvilinear geometries. The Astrophysical Journal Supplement Series 244(1):20

\bibitem[{Zhang et~al.(2020)Zhang, Brambles, Lotko, and Lyon}]{Zhang:2020}
Zhang B, Brambles OJ, Lotko W, Lyon JG (2020) Is nightside outflow required to
  induce magnetospheric sawtooth oscillations. Geophysical Research Letters
  47(6):e2019GL086419

\bibitem[{Zhang et~al.(2021)Zhang, Guo, Daughton, Li, and
  Li}]{zhang2021efficient}
Zhang Q, Guo F, Daughton W, Li H, Li X (2021) Efficient nonthermal ion and
  electron acceleration enabled by the flux-rope kink instability in 3d
  nonrelativistic magnetic reconnection. Physical Review Letters 127(18):185101

\bibitem[{Zhong et~al.(2018)Zhong, Tang, Zhou, Deng, Pang, Paterson, Giles,
  Burch, Tobert, Ergun, Khotyaintsev, and Lindquist}]{Zhong2018}
Zhong ZH, Tang RX, Zhou M, Deng XH, Pang Y, Paterson WR, Giles BL, Burch JL,
  Tobert RB, Ergun RE, Khotyaintsev YV, Lindquist PA (2018) Evidence for
  {{Secondary Flux Rope Generated}} by the {{Electron Kelvin-Helmholtz
  Instability}} in a {{Magnetic Reconnection Diffusion Region}}. Physical
  Review Letters 120(7):075101, \doi{10.1103/PhysRevLett.120.075101}

\bibitem[{Zhong et~al.(2022)Zhong, Zhou, Liu, Deng, Tang, Graham, Song, Man,
  Pang, and Khotyaintsev}]{Zhong2022}
Zhong ZH, Zhou M, Liu YH, Deng XH, Tang RX, Graham DB, Song LJ, Man HY, Pang Y,
  Khotyaintsev YV (2022) Stacked {{Electron Diffusion Regions}} and {{Electron
  Kelvin}}\textendash{{Helmholtz Vortices}} within the {{Ion Diffusion Region}}
  of {{Collisionless Magnetic Reconnection}}. The Astrophysical Journal Letters
  926(2):L27, \doi{10.3847/2041-8213/ac4dee}

\bibitem[{Zhou et~al.(2019)Zhou, T{\'{o}}th, Jia, Chen, and
  Markidis}]{Zhou2019}
Zhou H, T{\'{o}}th G, Jia X, Chen Y, Markidis S (2019) {Embedded Kinetic
  Simulation of Ganymede's Magnetosphere: Improvements and Inferences}. Journal
  of Geophysical Research: Space Physics 124(7):5441--5460,
  \doi{10.1029/2019JA026643}

\bibitem[{Zhou et~al.(2020)Zhou, Tóth, Jia, and Chen}]{Zhou2020}
Zhou H, Tóth G, Jia X, Chen Y (2020) Reconnection-{Driven} {Dynamics} at
  {Ganymede}'s {Upstream} {Magnetosphere}: 3-{D} {Global} {Hall} {MHD} and
  {MHD}-{EPIC} {Simulations}. Journal of Geophysical Research: Space Physics
  125(8):e2020JA028162, \doi{10.1029/2020JA028162}

\bibitem[{Zhou et~al.(2021)Zhou, Man, Deng, Pang, Khotyaintsev, Lapenta, Yi,
  Zhong, and Ma}]{Zhou2021}
Zhou M, Man HY, Deng XH, Pang Y, Khotyaintsev Y, Lapenta G, Yi YY, Zhong ZH, Ma
  WQ (2021) Observations of {{Secondary Magnetic Reconnection}} in the
  {{Turbulent Reconnection Outflow}}. Geophysical Research Letters
  48(4):e2020GL091215, \doi{10.1029/2020GL091215}

\bibitem[{Zhu and Winglee(1996)}]{Zhu1996}
Zhu Z, Winglee RM (1996) Tearing instability, flux ropes, and the kinetic
  current sheet kink instability in the earth's magnetotail: A
  three-dimensional perspective from particle simulations. Journal of
  Geophysical Research: Space Physics 101(A12):4885--4897

\bibitem[{Zocco et~al.(2009)Zocco, Chac{\'o}n, and Simakov}]{zocco2009}
Zocco A, Chac{\'o}n L, Simakov AN (2009) Current sheet bifurcation and collapse
  in electron magnetohydrodynamics. Physics of Plasmas 16(11):110703,
  \doi{10.1063/1.3264102}

\end{thebibliography}


\end{document}